\documentclass[a4paper,fleqn, 12pt]{cas-sc}

\usepackage{pdflscape}
\usepackage{xcolor}
\usepackage[authoryear,round]{natbib}
\usepackage{threeparttable}
\usepackage[section]{placeins}
\usepackage{amsmath}
\usepackage{color} 
\usepackage{array} 
\usepackage{graphicx} 
\usepackage{booktabs} 
\usepackage{amsmath}
\usepackage{float}
\usepackage{caption} 
\usepackage{subcaption} 
\usepackage{setspace}
\usepackage{xcolor}
\definecolor{ownGreen}{RGB}{0,100,0}
\definecolor{darkgreen}{RGB}{0,100,0}

\def\tsc#1{\csdef{#1}{\textsc{\lowercase{#1}}\xspace}}
\tsc{WGM}
\tsc{QE}
\tsc{EP}
\tsc{PMS}
\tsc{BEC}
\tsc{DE}
\usepackage{makecell}
\usepackage{amsthm}
\newdefinition{rmk}{Remark}

\usepackage{xcolor}
\usepackage{hyperref}
\usepackage{tabularx}
\usepackage{booktabs}
\usepackage{threeparttable}
\usepackage{multirow}
\usepackage{algorithm}
\usepackage{algorithmic}
\def\Appref#1{\textbf{Appendix}~\ref{#1}}
\def\figref#1{\textbf{Fig.}~\ref{#1}}
\def\Figref#1{\textbf{Fig.}~\ref{#1}}
\def\twofigref#1#2{\textbf{Figs.} \ref{#1} and \ref{#2}}
\def\manyfigref#1#2{\textbf{Figs.} \ref{#1} - \ref{#2}}

\def\tabref#1{\textbf{table}~\ref{#1}}
\def\Tabref#1{\textbf{Table}~\ref{#1}}

\def\secref#1{\textbf{Section}~\ref{#1}}
\def\Secref#1{\textbf{Section}~\ref{#1}}

\def\eqref#1{\textbf{Eq.}~(\ref{#1})}

\def\twoeqref#1#2{\textbf{Eqs.} (\ref{#1})$\sim$(\ref{#2})}

\hypersetup{
  colorlinks,
  citecolor=blue,
  linkcolor=blue,
  urlcolor=blue}
\newproof{pf}{Proof}
\usepackage{cleveref}
\Crefname{table}{\textbf{table}}{tables}
\Crefname{table}{\textbf{Table}}{\textbf{Tables}}
\Crefname{figure}{\textbf{figure}}{figures}
\Crefname{figure}{\textbf{Figure}}{\textbf{Figures}}
\Crefname{equation}{\textbf{equation}}{\textbf{equations}}
\Crefname{equation}{\textbf{Equation}}{\textbf{Equations}}
\Crefname{lemma}{\textbf{Lemma}}{\textbf{Lemma}}
\Crefname{theorem}{\textbf{Proposition}}{\textbf{Propositions}}
\usepackage{pdfcomment}
\usepackage[pagewise]{lineno}

\begin{document}
\def\floatpagepagefraction{1}
\def\textpagefraction{.001}
\let\printorcid\relax 

\shorttitle{}

\shortauthors{Jin et~al.}

\title [mode = title]{
Substitution or Complement? Uncovering the Interplay between Ride-hailing Services and Public Transit
}                      

\author[1]{Zhicheng Jin}
\credit{Data curation, Methodology, Formal analysis, Software, Writing-original draft}
\address[1]{Department of Electrical and Electronic Engineering, The Hong Kong Polytechnic University, Hong Kong SAR, China}
    
\author[2]{Xiaotong Sun}
\cormark[1]
\credit{Investigation, Methodology, Formal analysis, Writing - review \& editing}
\address[2]{Intelligent Transportation Thrust, Systems Hub, The Hong Kong University of Science and Technology (Guangzhou), Guangzhou, China}

\author[1]{Li Zhen}
\credit{Investigation,  Visualization}

\author[1]{Weihua Gu}
\cormark[1]
\ead{weihua.gu@polyu.edu.hk}
\credit{Conceptualization, Writing - review \& editing, Supervision, Funding acquisition}
\cortext[cor1]{Corresponding author}

\author[3]{Huizhao Tu}
\credit{Data curation, Resources}
\address[3]{Key Laboratory of Road and Traffic Engineering of the Ministry of Education, College of Transportation Engineering, Tongji University, Shanghai, China}

\begin{abstract}
\doublespacing
\small
The literature on transportation network companies (TNCs), also known as ride-hailing services, has often characterized these service providers as predominantly substitutive to public transit (PT). However, as TNC markets expand and mature, the complementary and substitutive relationships with PT may shift. To explore whether such a transformation is occurring, this study collected travel data from 96,716 ride-hailing vehicles during September 2022 in Shanghai, a city characterized by an increasingly saturated TNC market. An enhanced data-driven framework is proposed to classify TNC-PT relationships into four types: first-mile complementary, last-mile complementary, substitutive, and independent. 
Our findings reveal a substantial increase in the complementary ratio (9.22\%) and a relative decline in the substitutive ratio (9.06\%) compared to previous studies.
Furthermore, to examine the nonlinear impact of various influential factors on these ratios, a machine learning method integrating categorical boosting (CatBoost) and Shapley additive explanations (SHAP) is proposed.
The results show significant nonlinear effects in some variables, including the distance to the nearest metro station and the density of bus stops. Moreover, metro hubs and regular single-line stations exhibit distinct effects on first- or last-mile complementary ratios. These ratios' relation to the distance to single-line stations shows an inverted U-shaped pattern, with effects rising sharply within 1.5 km,  remaining at the peak between 1.5 and 3 km, and then declining as the distance increases to about 15 km. On the other hand, the distance to multi-line hubs initially maintains a positive influence on the complementary ratios within 9 km, which then stabilizes in the 9–18 km range. These findings offer valuable insights for policymakers to promote urban multimodal mobility systems integrating ride-hailing and public transit.
\end{abstract}

\begin{highlights}
    \item The complementary role of TNCs to public transit increases, while their substitutive role decreases, marking a clear shift from patterns observed in early-stage TNC markets.
    \item Nonlinear effects of various determinants on both complementary and substitutive ratios are unveiled using advanced machine learning techniques.
    \item Metro hubs and single-line metro stations exhibit distinct impacts on first- and last-mile complementary ratios.
    \item The distance to bus hubs and bus hub density emerge as the most significant variables affecting the substitutive ratio.
\end{highlights}


\maketitle

\section{Introduction}\label{introduction}
\begin{figure}
    \centering
    \includegraphics[width=0.75\linewidth]{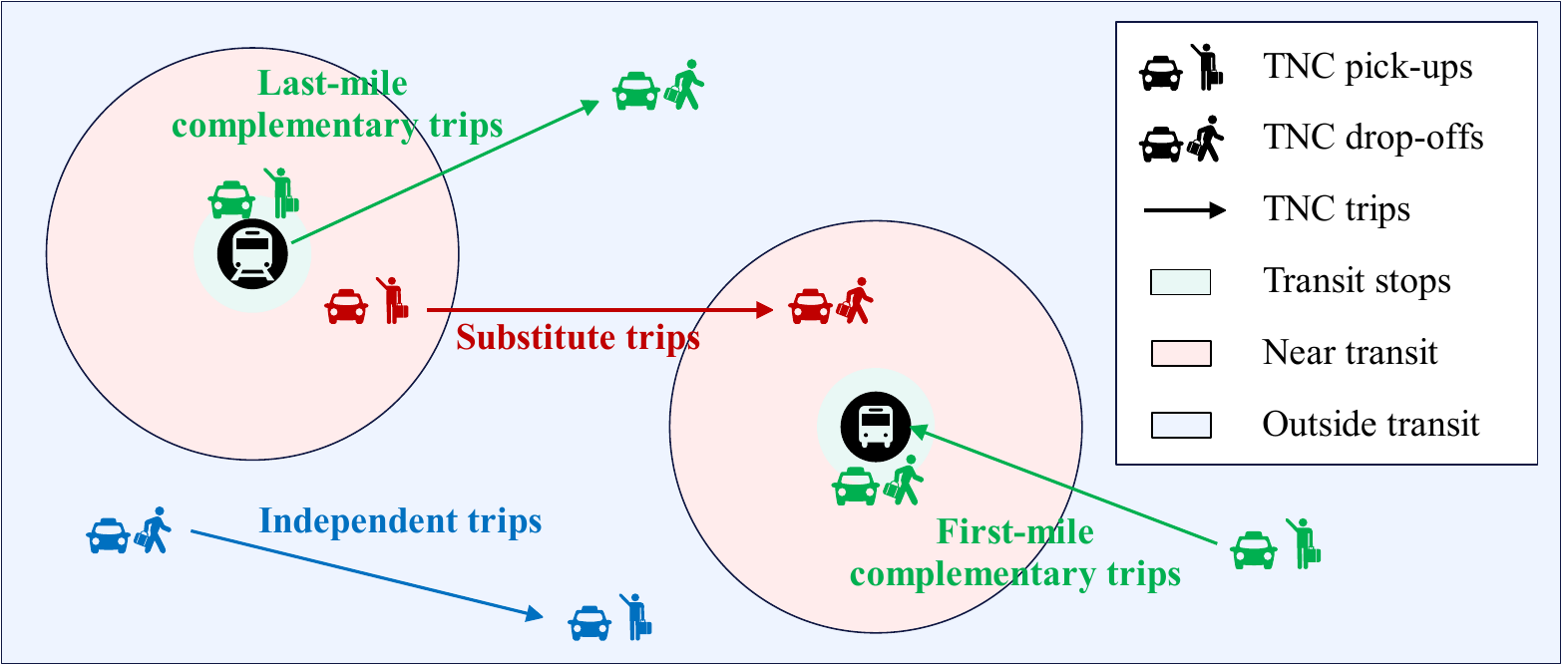}
    \caption{\centering{TNC trip classifications.}}
    \label{fig:trips classification}
\end{figure}

\doublespacing
The relationship between transportation network companies (TNCs, e.g., DiDi Chuxing, Uber) and public transit (PT) remains a subject of ongoing debate, with no clear consensus yet established. On the one hand, TNCs can complement PT by facilitating access to transit stations and enhancing first- and last-mile connectivity \citep{youngMeasuringWhenUber2020, tuffourContestedMobilityInteractions2022, jin2024multi}. On the other hand, ride-sharing services are often more demand-responsive than PT and less costly than taxis, positioning them as potential substitutes for PT \citep{pereiraRidehailingTransitAccessibility2024}. A schematic of the relationships between TNCs and PT is illustrated in \Figref{fig:trips classification}. In general, based on the spatial characteristics of trip origins and destinations, as well as the availability and efficiency of PT alternatives, these relationships can be generally classified into three categories:
\begin{enumerate}
    \item \textbf{Substitutive}: TNC trips whose origins and destinations are both within a comfortable walking distance of transit stations, which can be effectively substituted by PT in terms of both travel costs and travel time.
    \item \textbf{Complementary}: TNC trips that provide connections to PT, effectively bridging passengers into or out of the PT network. 
    \item \textbf{Independent}: TNC trips that operate between OD pairs without PT alternatives. Some studies also considered them as complementary trips due to ride-hailing filling ``transit deserts'' \citep{jiaoSharedMobilityTransitdependent2021}. We follow the notion in \cite{meredith-karamRelationshipRidehailingPublic2021} and define these trips as ``independent trips'' to distinguish them from complementary TNC trips. 
\end{enumerate}

Some empirical studies have attempted to use large-scale operational data to investigate whether competition or complement plays a dominant role in the relationship between ride-hailing services and public transit \citep{kongHowDoesRidesourcing2020, meredith-karamRelationshipRidehailingPublic2021}. These studies, typically based on ride-hailing service data from 2016 to 2018 when TNC markets were still developing, indicate that ride-hailing services primarily substitute for PT, with relatively low complementary ratios  \citep{henao2017impacts, tirachiniRidehailingSantiagoChile2019, meredith-karamRelationshipRidehailingPublic2021}. 
However, these conclusions may be outdated for today's China, given the recent developments in the world's largest TNC market. Since 2022, the ride-hailing market has been approaching saturation in cities like Shanghai and Shenzhen \citep{ridehailing_total_number_2024, ridehailing_suspends_license}, and daily average earnings for ride-hailing drivers have been declining steadily \citep{pakChinasRidehailingTaxi2025}.
At the same time, the PT systems have been continuously expanding. For instance, in Shanghai, metro and bus stops achieved a 64\% service coverage rate within a 500-meter radius in 2022, and PT served nearly four times as many daily trips as TNCs \citep{Shanghai_Transportation_Commission}. Against these backdrops, this study seeks to determine whether the complementary and substitutive effects noted in previous studies remain valid under current TNC and PT market conditions. 

Existing empirical studies typically define the complementary ratio as the proportion of ride-hailing trips serving as connections to PT, and the substitutive ratio as the proportion of trips that PT could potentially replace. Most of them have employed spatial regression models to explore the impact of various explanatory variables (e.g., socio-demographics, built environment, and transportation-related factors) on complementary and substitutive ratios. 
However, such models can only capture linear relationships, whereas interactions among these variables are often highly complex and nonlinear \citep{gaoImpactSharedMobility2024, zheng2025customizedbus_ssrn}. To date, only \cite{liuInvestigatingRelationshipsRidesourcing2025} and \cite{jinWhatPromotesIntegration2025} have applied machine learning methods such as XGBoost to investigate these nonlinear relationships. Yet, these analyses primarily focused on either trip type classification or trip volume. Nonlinear relationship analysis on the spatial distributions of complement and substitutive trips remains an underexplored topic.

This gap limits the ability to analyze the relative patterns of cooperation versus competition between PT and TNC in different regions. For example, a high number of complementary or substitutive trips in a central urban area might simply reflect high overall demand, rather than genuinely indicating a stronger trend towards cooperation or competition. 
Additionally, existing studies have not examined how different types of metro and bus stations, such as single-line stations versus multi-line hubs, or standalone bus stops versus major bus terminals, affect the complementary and substitutive ratios of TNCs to PT in distinct ways. This oversight matters because studies have reported that station types significantly shape travel demand patterns \citep{gan2020examining}. For instance, multi-line metro or bus hubs may encourage more complementary trips due to convenient transfer options. Thus, neglecting these station-type differences in analysis restricts our understanding of how transportation infrastructure influences the competitive or cooperative dynamics between PT and TNCs.

To resolve the aforementioned issues, this study designs an enhanced data-mining framework to identify TNC-PT relationships using actual TNC trip data combined with corresponding PT alternatives  \citep{meredith-karamRelationshipRidehailingPublic2021, liuInvestigatingRelationshipsRidesourcing2025}. 
A total of 16,862,195  trip records associated with 96,716 ride-hailing vehicles were collected in Shanghai in September 2022. The data includes the timestamps, labels, and coordinates of pick-up and drop-off locations.
Utilizing the Amap Route Planning API \citep{amap_amap_2024}, alternative PT routes are simulated for each ride-hailing trip. Subsequently, a framework is developed to classify  TNC trips into four categories: first-mile complementary to PT, last-mile complementary to PT, substitutive for PT, and independent from PT. 
A unique feature of our paper is how we match the pick-up and drop-off location labels input by passengers in TNC apps, rather than the location coordinates, with specific transit stations. The latter is a common approach employed in previous studies \citep{kongHowDoesRidesourcing2020,qiaoRidehailingCompetingComplementing2023, liuInvestigatingRelationshipsRidesourcing2025}. Our novel approach can decrease the rate of mismatches related to trips that are linked to nearby facilities rather than actual transit stations for PT connection \citep{liuInvestigatingRelationshipsRidesourcing2025}. 
CatBoost models are then applied to evaluate the effects of explanatory variables, including population, transit accessibility, road network, TNC travel characteristics, and points of interest (POIs), on complementary and substitutive ratios. Finally, the SHAP method is adopted to rank the importance of different explanatory variables and further explore their nonlinear effects. Partial dependence plots (PDPs) complement the SHAP analysis by illustrating the average marginal effect of individual variables on the complementary and substitutive ratios, thereby offering more insights into the nature of nonlinear relationships.
The key findings of our study are summarized as follows:
\begin{itemize}
    \item The complementary and substitutive ratios derived from the 2022 ride-hailing data in Shanghai are 9.22\% and 5.09\%–22.12\%, respectively, sharply contrasting with previous findings \citep{liuInvestigatingRelationshipsRidesourcing2025}, which reported substitutive ratios of 15–38\% and complementary ratios of 5–10\% using the 2016 ride-hailing data in Shanghai.
    \item In terms of temporal patterns, the first-mile complementary ratio peaks during morning rush hours, while the last-mile complementary ratio exhibits a distinct pattern with peaks during both morning and evening rush periods. In terms of spatial patterns, both first- and last-mile complementary ratios are higher in suburban regions and lower in central urban areas. 
    \item Unlike previous studies \citep{kongHowDoesRidesourcing2020, meredith-karamRelationshipRidehailingPublic2021}, we compute substitutive ratios based on both trip origins (departure substitutive ratio) and destinations (arrival substitutive ratio). Both ratios exhibit similar temporal and spatial patterns. Temporally, they remain high throughout the day without showing pronounced peaks. Spatially, both are high in the urban center and decrease notably toward suburban areas.
    \item According to the CatBoost model results, the first- and last-mile complementary ratios are mostly affected by the distance to the nearest metro station. On the other hand, substitutive ratios are primarily influenced by the distance to the nearest bus hub and the bus hub density.
    \item The SHAP value analysis reveals that the effects of some key variables, such as the distance to the nearest metro stations and bus stop density, are nonlinear. In addition, the distance to different types of metro stations (i.e., single-line metro stations and multi-line metro hubs) has distinct impacts on the first- or last-mile complementary ratios. 
    \item PDPs reveal that both the first- and last-mile complementary ratios increase as the distance to the nearest single-line metro station grows from 0 to 1.5 km, peak between 1.5 to 3 km, and then gradually decrease. Regarding the distance to the nearest multi-line metro hub, the ratios increase steadily from 0 km, then stabilize in the 9–18 km range. 
    \item Also from the PDPs, a three-phase pattern is identified in the impact of the distance to the nearest bus hub on the departure and arrival substitutive ratios: the values remain stable from 0 to 0.5 km, decline sharply from 0.5 to 1 km, and then level off beyond 1 km. 
\end{itemize}

The rest of the paper is arranged as follows. \Secref{sec: literature review} reviews relevant literature on TNC-PT relationships. \Secref{sec: Data processing} introduces the dataset and develops a framework to identify complementary and substitutive TNC-PT relationships. The temporal and spatial patterns of these relationships derived by applying this framework are presented in \Secref{sec: dynamics}. \Secref{sec: methodology} discusses the methods for examining the determinants of these relationships and \Secref{sec: results} presents the results and insights. \Secref{sec: policy} discusses the policy implications of the results. \Secref{sec: conclusion} concludes the paper and points out future research directions.

\section{Literature review}\label{sec: literature review}
\subsection{Relationships between TNCs and PT}
The preference survey is a commonly adopted method for the descriptive statistical analysis of TNC-PT relationships. For instance, \cite{rayle2016just} surveyed 380 respondents in San Francisco and found that 33\% of TNC trips replaced PT. Similarly, \cite{henao2017impacts} reported that 22.2\% of TNC trips in Denver replaced PT, while only 1\% were used for complement. \cite{tirachiniRidehailingSantiagoChile2019} showed that among 1,529 respondents in Santiago, Chile, 37.6\% chose to substitute ride-hailing for PT, whereas only 3.3\% used ride-hailing in connection with PT. Although these survey-based findings offer valuable insights into individual decision-making from the perspective of passengers or drivers, they are limited by their small sample sizes.

With the increasing data accessibility in recent years, big data analytics and statistical models have been widely used to explore TNC-PT relationships. \cite{sadowskyImpactRideHailingServices2017} employed a regression discontinuity design to analyze the impact of TNC market entry on PT use across U.S. urban areas, concluding that TNCs initially complemented but later substituted for PT as their services expanded. Using a sample of 1,578 TNC trip records in Toronto from 2016, \cite{youngMeasuringWhenUber2020} indicated that 31\% of TNC trips have PT alternatives that are no more than 15 minutes longer in travel time. In addition, they built ordered logistic regression models and uncovered that the travel time difference is mainly caused by transfer times in transit, including walking and waiting times. While these studies effectively capture the aggregate TNC-PT relationship within their study areas, they fail to account for granular spatial or temporal trends.

To better understand the temporal and spatial dynamics of TNC-PT relationships, \cite{kongHowDoesRidesourcing2020} utilized 2016 DiDi Chuxing data from Chengdu, China, to assess the substitutive impact of TNC trips on PT. Their findings suggested that 33.1\% of DiDi trips are potentially replacements of PT, with a higher substitutive ratio during daytime hours (8:00–18:00). Using the same dataset, \cite{qiaoRidehailingCompetingComplementing2023} further demonstrated that up to 80\% of TNC trips for commuting could be converted to PT. However, these studies relied on a single day's trip data, which limits the robustness and generalizability of their findings. In contrast, \cite{meredith-karamRelationshipRidehailingPublic2021} studied the TNC-PT relationship in Chicago using trip data from 2019–2020, covering periods before and after the COVID-19 outbreak, allowing for the capture of longer-term dynamics. Their results indicated that complementary trips accounted for only a minor proportion (approximately 2\%), whereas substitutive trips represented a significantly larger share (45–50\%). 

The above studies are summarized in \Tabref{Table:Literature review} by data type, data characteristics, methods, and relationship outcomes. 
\begin{landscape}
    \begin{table}[width=\linewidth,cols=8,pos=h]\scriptsize\rmfamily
    \doublespacing
    \caption{Data-based studies on TNC-PT relationship.}\label{Table:Literature review}
    \begin{tabular*}{\tblwidth}{@{} LLLLLLLLL@{} } 
        \toprule
        \multirow{2}*{Article} & \multirow{2}*{Type of data} & \multicolumn{4}{c}{Sample characteristics} & \multirow{2}*{Methods} & \multirow{2}*{Relationship} & \multirow{2}*{Ratio}\\
        \cmidrule{3-3}\cmidrule{4-4}\cmidrule{5-5} \cmidrule{6-6}  &   & Area  & Year & Period & Data size &  \\
        \midrule
        \cite{rayle2016just} & Surveys & Denver, US & 2014 & 2-month & 380 & Statistical analysis & Sub. $^a$ & 33\% \\
        \cite{henao2017impacts} & Surveys  & Colorado, US & 2016 & 3-month & 311 & Statistical analysis & Com. \& Sub. & 1\% \& 22.2\% \\
        \cite{sadowskyImpactRideHailingServices2017} & PT trips  & US & 2014-2016 & 3-year & 1465 & OLS model & Com. \& Sub. & / \\
        \cite{tirachiniRidehailingSantiagoChile2019} & Surveys  & Santiago, Chile & 2017 & 1-month & 1,529 & Logit model & Com. \& Sub. & 3.3\% \& 37.6\%\\
        \cite{youngMeasuringWhenUber2020} & TNC trips & Toronto, Canada & 2016 & 1-year & 1578 & logistic regression model  & Sub. & 31\% \\
        \cite{kongHowDoesRidesourcing2020} & TNC trips & Chengdu, China & 2016 & 1-day & 181,068 & Spatial regression model  & Sub. & 33.1\% \\
        \cite{meredith-karamRelationshipRidehailingPublic2021} & TNC trips & Chicago, US & 2019-2020 & 8-day & 2,144,888 & Spatial regression model  & Com.\& Sub. \& Inde. & 2.0\%\& 47.2\% \& 50.8\% \\
        \cite{erhardt2022transportation} & TNC \& PT trips  & San Francisco, US & 2010-2015 & 6-week & 7,358 & Panel data model  & Sub. & 10.8\% \\
        \cite{catsDichotomyHowRidehailing2022} & TNC trips & US and Europe & 2018 & 1-month & 3,506,592 & Accessibility analysis & Sub. & 9-15\% \\
        \cite{panHowRideSharingShaping2022} & TNC trips & US & 2010-2015 & 5-year & 35,781 & Panel data model & Sub. & 8.8\% \\
        \cite{liExploringCorrelationRidehailing2022} & TNC \& PT trips & Toronto, Canada & 2016-2018 & 2-year &  / $^b$ & Panel data model  & Com.\& Sub. & /  \\
        \cite{qiaoRidehailingCompetingComplementing2023} & Surveys \& TNC trips & Chengdu, China & 2016 & 1-day & 201,028 & Discrete choice model  & Sub. & 38\%  \\
        \cite{gaoImpactSharedMobility2024} & TNC trips & Shenzhen, China & 2019 & 1-month &  / $^b$ & Random forest model  & Com. & /  \\
        \cite{liuInvestigatingRelationshipsRidesourcing2025} & TNC trips & Shanghai, China & 2016 & 1-month &  339,348 & XGBoost  & Com.\& Sub. & 5\% \& 38\% $^c$ \\
        This paper & TNC trips & Shanghai, China & 2022 & 2-week & 16,862,195 & CatBoost and SHAP & Com.\& Sub. & 9.22\% \& 9.06\%$^d$ \\
        \bottomrule
    \end{tabular*}
        \begin{tablenotes}    
            \scriptsize
            \item[1] $^a$ Sub. = Substitutive, Com. = Complementary, Inde. = Independent.
            \item[2] $^b$ Total data size is not listed in \cite{liExploringCorrelationRidehailing2022} and \cite{gaoImpactSharedMobility2024}.
            \item[3] $^c$ In \cite{liuInvestigatingRelationshipsRidesourcing2025}, the substitutive and complementary ratios in urban areas are 38\% and 5\%, respectively, while in suburban areas, they are 15\% and 10\%, respectively. 
            \item[4] $^d$ The substitutive ratio 9.06\% is computed under baseline thresholds (400 m walking distance, 15-min time difference, 2 transfers, and 0.5 cost ratio), see \Appref{appendix: Parameter elasticity analysis} for elasticity analysis.
        \end{tablenotes}  
    \end{table}
\end{landscape}

While literature provides valuable insights into the relationships between TNCs and PT, several notable research gaps remain unaddressed. First, most studies collected TNC trip data from 2016 to 2018, a period when TNC markets were under development \citep{kongHowDoesRidesourcing2020,qiaoRidehailingCompetingComplementing2023, liuInvestigatingRelationshipsRidesourcing2025}, which may not accurately represent the current, more saturated TNC market, especially in China.  

Moreover, earlier research typically relied on datasets collected from specific companies (e.g., Didi, Uber) or was limited by small sample sizes \citep{catsDichotomyHowRidehailing2022, liuInvestigatingRelationshipsRidesourcing2025, meredith-karamRelationshipRidehailingPublic2021}. Such datasets are often biased due to the specific policies, pricing schemes, and the limited operational regions of those companies, making it difficult to capture a comprehensive view of TNC trips across the entire city, particularly when the city is equipped with an integrated urban transit system. In contrast, our study collected trip data from nearly all major TNC providers in Shanghai, yielding a dataset exceeding ten million trips and a sample coverage rate of over 85\% (see \secref{sec: Data description} for details).

In addition, previous studies typically identified TNC trips as complementary if their origin or destination's geographic coordinates are near transit stations. However, this approach may overestimate the complementary ratio by mistakenly including trips linked to nearby facilities rather than transit stations themselves \citep{liuInvestigatingRelationshipsRidesourcing2025, jinWhatPromotesIntegration2025}. To minimize this overestimation, we use the OD location labels provided by passengers (e.g., ``Gate 1A, Shanghai Hongqiao Railway Station'') to identify complementary trips, thus better reflecting passengers' true travel intentions. To our best knowledge, we are the first to adopt this identification approach. 

\subsection{Determinants of TNC-PT relationships}
When examining the spatial and temporal factors influencing TNC-PT relationships, traditional linear regression-based approaches are primarily adopted, including ordinary least squares (OLS) models and spatial econometric models \citep{kongHowDoesRidesourcing2020, meredith-karamRelationshipRidehailingPublic2021}. \cite{kongHowDoesRidesourcing2020} used the DiDi trip data of November 1, 2016, in Chengdu, China, to build spatial autoregressive models that examine the factors influencing the extent to which TNC trips substitute for PT. 
Their results showed that higher POI density, greater land-use diversity, and a higher number of bus stops positively affect substitution, whereas a greater distance to metro stations has a negative impact. 
From ride-hailing data between May and October 2017 in Haikou, China, \cite{biWhyTheyDont2021} found that ride-hailing activities near bus stops are significantly associated with various land use characteristics, such as catering services, shopping services, and residential buildings. Analyzing data from Chicago before and during the early stages of the COVID-19 pandemic, \cite{meredith-karamRelationshipRidehailingPublic2021} built spatial regression models and found that an increased substitutive ratio was associated with a lower proportion of elderly residents, a higher share of peak-period TNC usage, greater transit network availability, a higher percentage of white residents, and elevated crime rates. 
However, these linear regression models can only capture the linear effects of independent variables on TNC-PT relationships. In reality, the effects of these variables are highly complex and potentially non-linear \citep{liuInvestigatingRelationshipsRidesourcing2025}. Neglecting such non-linearities may lead to misinterpretations and flawed policy recommendations \citep{gaoImpactSharedMobility2024}. 

Advanced machine learning methods, such as extreme gradient boosting (XGBoost) and Shapley additive explanations (SHAP), provide an effective remedy for capturing the nonlinear relationships.
Using these methods, \cite{liuInvestigatingRelationshipsRidesourcing2025} applied an XGBoost classification model to categorize each TNC trip as either competing with or complementing public transit. They further conducted a SHAP analysis to evaluate the relative importance of various contributing factors, indicating that travel cost, distance to the central business district (CBD), and travel time were the three most influential variables.
\cite{jinWhatPromotesIntegration2025} employed XGBoost models to understand the effects of POI variables (i.e., residence density and employment density), metro accessibility, travel distance, and bus route density on predicting the volume of the TNC trips connected to metro stations. 

These two recent studies both investigated different types of TNC trips. The first study focused on classifying the types of individual TNC trips, while the second aimed to predict the number of TNC trips connected to metro stations. However, neither study explored how the ratios of substitutive and complementary TNC trips vary spatially within a city. Their policy implications are limited since they could not faithfully reflect the spatial variation of competitive and cooperative relationships between TNC and PT. For example, the high number of complementary or substitutive trips in inner cities might simply reflect a greater TNC demand rather than TNC's relative strength to PT. 

In addition, previous studies have not differentiated among various transit facility types when selecting independent variables, such as multi-line metro hubs versus single-line metro stations, or bus hubs versus regular bus stops. 
Although it is well established that different station types attract distinct travel demands \citep{gan2020examining}, these differences have rarely been incorporated into models examining TNC-PT relationships. 
In this paper, we consider a finer-scale classification of transit stations by categorizing metro stations and bus stops into transfer and non-transfer types. 

\section{Data processing and variable configurations}\label{sec: Data processing}
This section details the data processing and variable configuration through three subsections. \secref{sec: Data description} describes the ride-hailing and public transit data utilized. \secref{sec: TNC-PT relationship recognition} presents the framework for identifying and classifying the relationship between TNC and PT trips. Finally, \secref{sec: Dependent and independent variables} outlines the selection and definition of the variables used in the analysis.

\subsection{Data description}\label{sec: Data description}
With the help of the Shanghai Electric Vehicle Public Data Collecting, Monitoring, and Research Center (SHEVDC), this study collected 2-week trip data from 96,716 vehicles across 17 TNCs operating in Shanghai, China in September 2022 \footnote{As suggested by \cite{ridehailing_total_number_2024}, the total number of for-hire vehicles in Shanghai, including taxis and ride-hailing vehicles, is around 110,000  by June 2022. Therefore, our sampling rate is quite high, around 85\%, ensuring a broad representation of the behavior of for-hire vehicles in Shanghai. In addition, Shanghai had already lifted the city-wide COVID-19 lockdown by June 2022. Therefore, the trip data collected in September 2022 is not influenced by lockdown measures.}. A sample of the data is shown in \Tabref{Table: Example data samples}, including vehicle plate ID, origin and destination coordinates, travel distance, pick-up location label, pick-up time, drop-off location label, drop-off time, and total costs. 

Travel distance is the actual mileage of a trip, derived from onboard device records. 
Both “pick-up location label” and “drop-off location label” are predefined location labels selected by passengers. 
When a passenger requests a ride by typing the origin or destination in a TNC app, the app will suggest a list of nearby locations (e.g., frequently used or official points of interest (POIs)), which are predefined labels generated according to the electronic map or common inputs by past customers. 
The passenger will choose a pick-up location label from the list and proceed there to wait, and the driver will conclude the trip upon arriving at the passenger’s selected drop-off location. 

In the data processing step, trips with extreme travel times or distances were excluded using the interquartile range (IQR) method \citep{tukey1977exploratory}. Outliers were defined as trips with travel times or distances beyond $Q_3 + 3 \times IQR$ or below $Q_1 - 3 \times IQR$, where $Q_1$ and $Q_3$ are the 25th and 75th percentiles, respectively, and IQR = $Q_3 - Q_1$. After filtering, 16,862,195 valid trips remained. 

\begin{figure}
    \centering
    \begin{subfigure}[b]{0.45\textwidth}
        \centering
        \includegraphics[height = 0.875\textwidth]{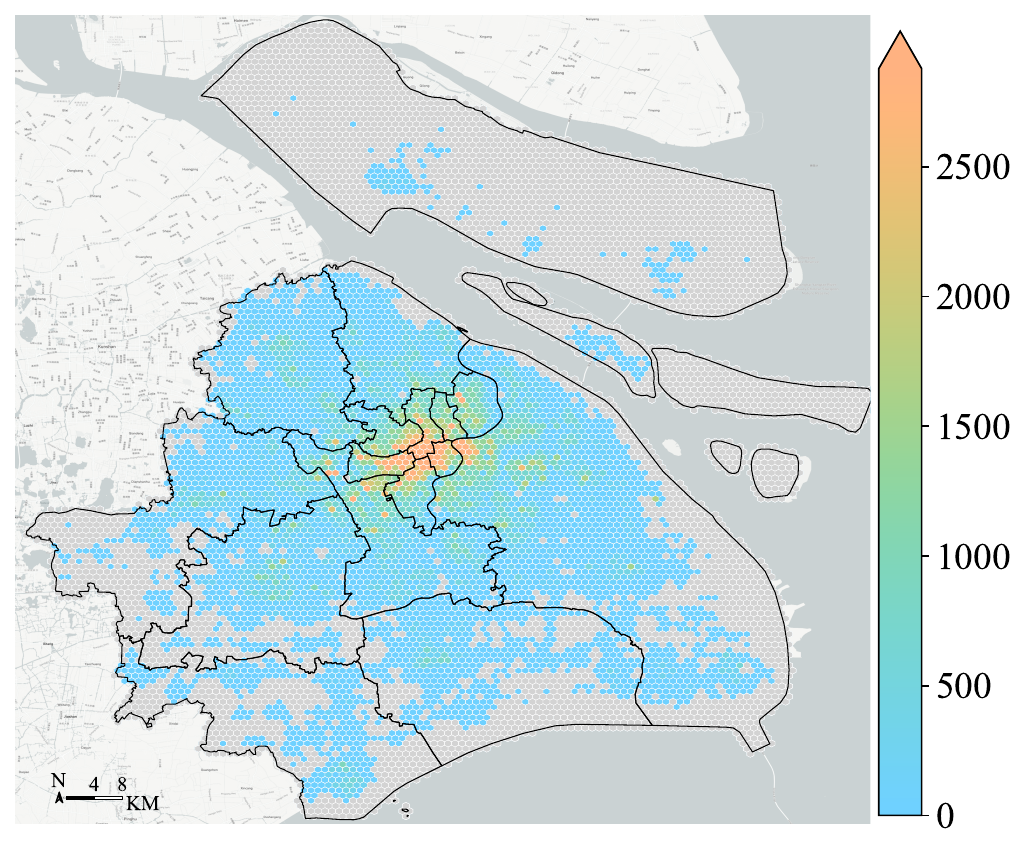}
        \caption{Number of trips (pick-ups)}
        \label{fig:Spatial distribution}
    \end{subfigure}
    \begin{subfigure}[b]{0.45\textwidth}
        \centering
        \includegraphics[height = 0.875\textwidth]{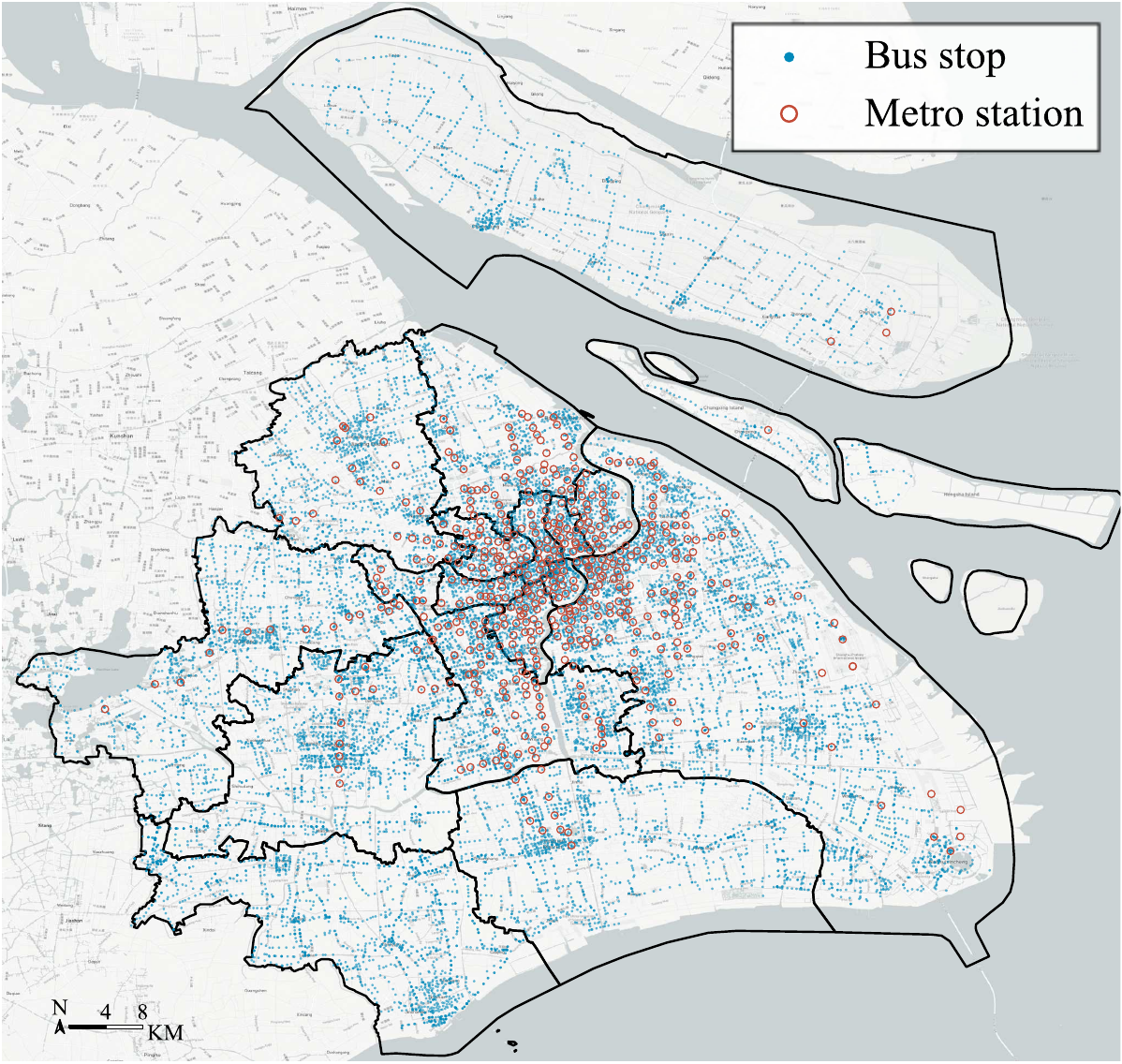}
        \caption{Public transit coverage}
        \label{fig:OD distribution}
    \end{subfigure}
    \caption{\centering{Spatial distribution of ride-hailing trips and public transit stations in Shanghai.}}
    \label{fig:Distribution of ride-hailing}
\end{figure}

To facilitate spatial analysis, the study area was further divided into 9,746 hexagonal grids with 0.5 km sides. Shanghai’s public transport network data was collected from \cite{amap_amap_2024}, featuring a metro system with 20 lines connecting hundreds of metro stations, along with 1,596 bus routes spanning a total length of 25,185 kilometers. The metro system carries millions of passengers daily, while the bus network achieves a 64\% service coverage within a 500-meter radius of bus stops \citep{Shanghai_Transportation_Commission_Bus}. \Figref{fig:Distribution of ride-hailing} illustrates the number of ride-hailing trips in each hexagonal grid \footnote{In this study, unless otherwise specified, the number of ride-hailing or TNC trips in each grid refers to the average daily trips originating from that grid, calculated as the mean number of pick-ups per day over the two-week period.} and public transit coverage. 

\begin{table}[width=.98\linewidth,cols=9,pos=h]\scriptsize\rmfamily
    \centering
    \onehalfspacing
    \caption{Example data samples on September 03, 2022.}
    \begin{tabular*}{\tblwidth}{@{} Lp{1.2cm}p{2.5cm}p{1.2cm}p{1.2cm}p{2.5cm}p{1.2cm}p{0.8cm}p{0.8cm}@{} } 
    \toprule
        Plate ID & Origin & Pick-up location label & Pick-up time & Destination & Drop-off location label & Drop-off time & Distance (km) & Cost (RMB) \\ 
    \midrule    
        ADXXXXX & 121.521568, 31.194264 & Yanggao South Road & 2022/9/3 08:30 & 121.157783, 31.290597 & Anting Station Exit A & 2022/9/3 09:35 & 55.3 & 165.2 \\ 
        ADXXXXX & 121.313393, 31.195234 & Hongqiao Station B1 & 2022/9/3 08:53 & 121.591149, 31.101143 & Zhoupu Hospital & 2022/9/3 09:50 & 40.5 & 146.6 \\ 
        ADXXXXX & 121.471047, 31.234589 & People’s Square Bus Stop & 2022/9/3 09:33 & 121.519348, 31.085276 & No. 328 Wankang Road & 2022/9/3 10:18 & 23.0 & 73.9 \\ 
        ADXXXXX & 121.444195, 31.224326 & No. 1515 Nanjing West Road & 2022/9/3 10:05 & 121.509225, 31.338216 & No. 999 Zhenghe Road & 2022/9/3 10:58 & 23.1 & 62.5\\ 
        ADXXXXX & 121.493394, 31.238409 & Riverside Park Exit 7  & 2022/9/3 10:28 & 121.623444,	31.242152 & Jinkui New City West & 2022/9/3 10:58 & 16.7 & 55.6 \\ 
    \bottomrule
    \end{tabular*}
    \label{Table: Example data samples}
\end{table}

\subsection{TNC-PT relationship recognition}\label{sec: TNC-PT relationship recognition}
To facilitate a better identification of TNC-PT relationships, the framework proposed by \cite{meredith-karamRelationshipRidehailingPublic2021} is adopted with three key extensions.  
First,  the first- and last-mile complementary TNC trips are respectively identified by whether the destination or origin of the TNC ride is connected to a PT station. 
Second, instead of using the geographic coordinates, the origins and destinations are determined by the labels of pick-up and drop-off locations selected by passengers (see \Tabref{Table: Example data samples}), allowing for a more precise identification of transit-connected TNC trips. 
Third, the Amap Route Planning API is used to derive alternative PT routes and corresponding fare and travel times between TNC trip ODs \citep{amap_amap_2024}, enabling a reliable comparison with actual TNC travel times and costs for identifying substitutive trips. 

\begin{figure}
    \centering
\includegraphics[width=\textwidth]{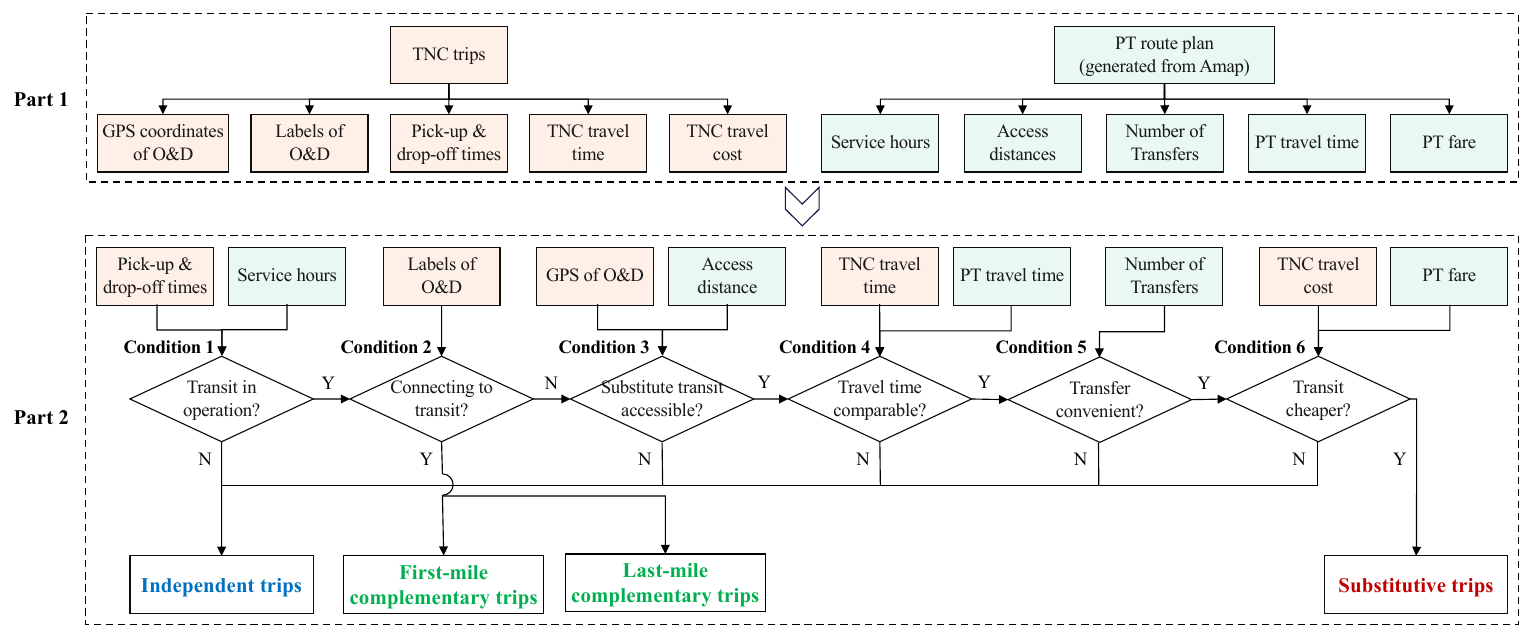}
    \caption{\centering{Framework of recognizing TNC-PT relationships.}}
    \label{fig: Framework recognition}
\end{figure}

The framework for identifying TNC-PT relationships is illustrated in \Figref{fig: Framework recognition} and comprises two parts. Part 1 involves data preparation. The red rectangles represent actual TNC trip data, while the blue rectangles represent simulated PT alternatives. For each ride-hailing trip, the coordinates and labels of the pick-up and drop-off locations are extracted and fed into the Amap Route Planning API \citep{amap_amap_2024}. 
The API then simulates alternative PT routes, returning the corresponding transit service hours, access distances to transit, the number of transfers, travel time, and fare. These values are derived from the integrated transportation network information in the Amap geographic information system, offering realistic approximations \footnote{
Although the Amap simulation was done in 2024, the public transit network layout and operational characteristics in Shanghai remained largely unchanged from 2022 to 2024. Government reports indicate that there were no significant modifications to PT routes, schedules, or service levels during this period \citep{Shanghai_Transportation_Commission_Bus}, ensuring that the Amap API provides reliable PT alternatives for TNC trips in 2022.}.

Part 2 is a condition-based recognition mechanism involving six sequentially checked conditions. First, for each TNC trip, \textbf{Condition 1} checks whether its PT alternative (bus or metro) is in service during the TNC trip period. 
Trips occurring when PT is unavailable are classified as independent. Trips passing Condition 1 proceed to \textbf{Condition 2}, which examines whether passenger-selected pick-up and drop-off location labels are aligned with transit stations or access points. 
Trips with pick-up locations meeting this criterion are categorized as last-mile complementary trips, while those with qualifying drop-off locations are classified as first-mile complementary trips. 
Specifically, the identification process leverages the standardized nature of the Shanghai TNC market, where apps suggest predefined POI labels based on electronic maps. We collected names of all metro stations and bus stops across Shanghai to serve as a ground-truth database. Then, we matched these passenger-provided labels with the database and excluded labels associated with nearby commercial, entertainment, or residential facilities.
Unlike coordinate-only methods, this approach better reflects  passengers’ intended connection with public transit and reduces the overestimation of complementary trips arising from station-adjacent but non-transit destinations \citep{meredith-karamRelationshipRidehailingPublic2021, liuInvestigatingRelationshipsRidesourcing2025, jinWhatPromotesIntegration2025}. Further details and examples are provided in \Appref{appendix: Identification of complementary trips}.

Four additional conditions collectively determine whether a TNC trip can be substituted by PT. 
\textbf{Condition 3} evaluates PT's spatial accessibility by measuring the walking distance from a TNC trip's origin and destination to stations of alternative transit routes. 
In this study, a 400-meter walking distance threshold is adopted, consistent with \cite{murrayPublicTransportationAccess1998} and \cite{kongHowDoesRidesourcing2020}. If no accessible transit alternative within this threshold is found, the trip is categorized as independent. Otherwise, the process advances to \textbf{Condition 4}, which compares TNC and PT travel times. 
For TNC trips lasting 15 minutes or less ($t_{\text{TNC}} \le 15 \text{min}$), PT is considered a substitution only if the travel time difference satisfies $t_{\text{PT}}-t_{\text{TNC}} \le 15 \text{min}$. 
For TNC trips exceeding 15 minutes ($t_{\text{TNC}} > 15 \text{min}$), substitution requires $t_{\text{PT}} \le 2 t_{\text{TNC}}$ \citep{meredith-karamRelationshipRidehailingPublic2021}. 
If the travel time conditions are violated, we consider the PT alternative to be too long to be a substitute, and then categorize the trip as independent. 
\textbf{Condition 5} then assesses transfer convenience, requiring no more than two transfers in the PT alternative to justify it as a substitution.
Finally, given the significant role that trip fare plays in mode choice behaviors \citep{pereiraRidehailingTransitAccessibility2024}, \textbf{Condition 6} compares monetary costs between PT and TNCs. A TNC trip is classified as substitutive only if the fare of its PT alternative is at least 50\% lower than the TNC cost. 
Otherwise, it is classified as independent as the small fare difference renders PT less attractive. 
The 50\% threshold is inspired by the findings from \cite{catsDichotomyHowRidehailing2022} and \cite{pereiraRidehailingTransitAccessibility2024}.  No prior studies have incorporated this criterion into TNC–PT relationship recognition due to the lack of TNC cost data. This study is the first to address this gap by collecting actual TNC costs and retrieving real PT fares via the Amap API. 

It should be noted that the numerical thresholds adopted in Conditions 3-6 may affect the substitutive ratio. 
Although these values are largely based on prior studies, we include an elasticity analysis in \Appref{appendix: Parameter elasticity analysis} of this paper to quantify how responsive the substitutive ratio is to changes in each of these parameters. This allows us to pinpoint the thresholds that can significantly influence our results and to verify the robustness of our findings. 
It is important to note that the substitutive ratio exhibits high elasticity to the 400-meter walking distance and 15-min time difference thresholds.
In contrast, the substitutive ratio is relatively insensitive to the transfer limit (2) and the cost ratio (0.5).
The variations in these thresholds can alter the substitutive ratios (5.09\%–22.12\%). However, even when we adopt thresholds that are highly favorable to substitution (i.e., an 500-meter walking distance and a 30-min time difference), which are larger than those commonly used in prior studies \citep{kongHowDoesRidesourcing2020, meredith-karamRelationshipRidehailingPublic2021, liuInvestigatingRelationshipsRidesourcing2025}, the observed trend—a significant rise in complementarity and a relative decline in substitution compared to early-stage markets—remains unchanged.

\subsection{Variable selection}\label{sec: Dependent and independent variables}
After identifying different types of TNC trips, this study further defines four variables to capture various TNC-PT relationships: the first-mile complementary ratio, the last-mile complementary ratio, the departure substitutive ratio, and the arrival substitutive ratio. These metrics are introduced to systematically quantify and compare how TNC services complement and substitute for PT within each spatial grid.
The formulas for these variables are given below.  The first-mile complementary ratio in grid $i$ is defined as
\begin{equation}\label{eq:first-mile complementary ratio}
    \text{FCR}_i = \frac{\text{FC}_i}{O_i}
\end{equation}
where $O_i$ is the total number of TNC trips originating from grid $i$, and $\text{FC}_i$ is the number of trips in $O_i$ that are identified as first-mile complementary trips, indicating their destinations are connected to PT.  

The last-mile complementary ratio $\text{LCR}_i$ in grid $i$ is defined as 
\begin{equation}\label{eq:last-mile complement destiantion rate}
    \text{LCR}_i = \frac{\text{LC}_i}{A_i}
\end{equation}
where $A_i$ is the total number of TNC trips arriving at grid $i$, and $\text{LC}_i$ is the number of trips in $A_i$ that are identified as last-mile complementary trips with their origins connected to PT and destinations within grid $i$. 
For each grid, the first-mile complementary ratio captures the percentage of trips from the grid that connect to PT, while the last-mile complementary ratio measures the percentage of trips that connect from PT to the grid. 

Unlike previous geospatial studies that did not differentiate the aggregation of substitutive trips by origin from that by destination \citep{meredith-karamRelationshipRidehailingPublic2021, kongHowDoesRidesourcing2020}, this study defines two substitutive ratios: departure and arrival substitutive ratios. 
The departure substitutive ratio measures the percentage of substitutive TNC trips originating in a grid, while the arrival substitutive ratio captures the substitution percentage of TNC trips arriving at a grid.   
The departure substitutive ratio is defined as 
\begin{equation}\label{eq:departure substitutive ratio}
\text{DSR}_i = \frac{\text{DS}_i}{O_i},
\end{equation}
with $\text{DS}_i$ being the number of substitutive TNC trips originating from grid $i$. The arrival substitutive ratio is
\begin{equation}\label{eq:arrival substitutive ratio}
\text{ASR}_i = \frac{\text{AS}_i}{A_i},
\end{equation}
where  $\text{AS}_i$ is the number of substitutive TNC trips arriving at grid $i$. 

To explore factors influencing these ratios, 19 independent variables, classified into three types, are selected as candidates. As shown in Table \ref{Table: variables and definitions}, they are:
\begin{enumerate}
    \item \textbf{Population density}. Collected from the WorldPop dataset \citep{worldpop_spatial_2020}, the original population data is provided in 100 $\times$ 100 meter square grids. This study aggregates this data to obtain population density in each hexagonal grid. 
    \item \textbf{Transportation-related variables}. 
    \begin{enumerate}
        \item Accessibility to transit. 
        For each grid, we differentiate transit stations based on their network connectivity. We classify all metro stations into two distinct types: single-line stations, which primarily serve commuters from the neighborhood, and multi-line hubs, which function as major regional interchange nodes with high transfer passenger volumes. The same classification method is applied to bus stops. Based on this classification, we define four independent variables to measure the proximity to transit: the distances from the center of each grid to the nearest single-line metro station, to the nearest multi-line metro hub, to the nearest single-line bus stop, and to the nearest multi-line bus hub. These distances are estimated using the OpenStreetMap toolbox \citep{openstreetmap_openstreetmap_2024}. Given the extensive bus network coverage in Shanghai, we also include bus stop densities in each grid as independent variables, estimated using data from the Amap API \citep{amap_amap_2024}. These densities are likewise divided by type to align with the station distinctions, i.e., the density of single-line bus stops and the density of multi-line bus hubs.
        \item Road network. Mobility activities are significantly influenced by road length, density, and network topology \citep{cui_how_2023, yuTrafficVolumeRoad2023}. Consequently, a set of road network characteristics, including the density of roadways, network clustering, network centrality, and highway density, is extracted from OpenStreetMap \citep{openstreetmap_openstreetmap_2024}.
        \item TNC travel. For each grid, the average number of daily TNC trips (pick-ups and drop-offs), the average waiting time per TNC trip, the average travel time, and the average fare per km of TNC trips are selected \citep{kongHowDoesRidesourcing2020, meredith-karamRelationshipRidehailingPublic2021}.
    \end{enumerate}
    \item \textbf{Point of interests (POIs)}. Ride-hailing ridership has been shown to correlate with socio-economic opportunities and resources, such as restaurants and enterprises \citep{gao_spatial_2021, karimpour_data-driven_2023}. With the help of Amap API \citep{amap_amap_2024}, the densities of different POIs, encompassing residential neighborhoods, retail outlets, catering establishments, and enterprises, are collected as independent variables. 
\end{enumerate}

\begin{table}[width=.98\linewidth,cols=3,pos=h]\scriptsize\rmfamily
\onehalfspacing
\caption{Definition of independent variables.}\label{Table: variables and definitions}
\begin{tabular*}{\tblwidth}{@{} LLL@{} } 
\toprule
Variables & Definitions & Unit \\
\midrule
\textbf{Population}$^a$ & The density of residents in each grid & persons/$\text{km}^2$ \\
\textbf{Accessibility to transit} &  &  \\
Distance to metro station& The distance from the center of each grid to the nearest single-line metro station & $\text{km}$ \\
Distance to metro hub & The distance from the center of each grid to the nearest multi-line metro hub & $\text{km}$ \\
Distance to bus stop & The distance from the center of each grid to the nearest single-line bus stop & $\text{km}$ \\
Distance to bus hub & The distance from the center of each grid to the nearest multi-line bus hub & $\text{km}$ \\
Bus stop density & The density of single-line bus stops in each grid & number/$\text{km}^2$ \\
Bus hub density & The density of multi-line bus hubs in each grid & number/$\text{km}^2$ \\
\textbf{Road network} &  &  \\
Average clustering & Average value of clustering coefficient$^b$ of all nodes in each grid & - \\
Average centrality & Average degree centrality$^c$ of all nodes in each grid & - \\
Road network density & Total length of roads in each grid divided by total area of each grid & km/$\text{km}^2$ \\
Highway density & Total length of highways in each grid divided by total area of each grid & km/$\text{km}^2$ \\
\textbf{TNC travel-related} &  &  \\
No. Trips & The number of daily TNC trips originating from or arriving at each grid & number \\
Average waiting time & The average waiting time per TNC trip in each grid & $\text{min}$ \\
Average travel time & The average travel time per TNC trip in each grid & $\text{min}$ \\
Average fare per km & The average TNC cost per unit of distance in each grid & $\text{RMB}^d/\text{km}$ \\
\textbf{POIs} &  &  \\
Residence density & The density of residential neighborhoods in each grid & number/$\text{km}^2$ \\
Retail density & The density of shops in each grid & number/$\text{km}^2$ \\
Catering density & The density of catering establishments in each grid & number/$\text{km}^2$ \\
Enterprise density & The density of enterprises in each grid & number/$\text{km}^2$ \\
\bottomrule
\end{tabular*}
\begin{tablenotes}     
        \scriptsize
        \item[1] $^a$ The logarithmic value of population density is used in the following regression analysis for model smoothing.
        \item[2] $^b$ The local clustering coefficient for a node is defined as the ratio of the actual number of edges between its neighbors to the 
        maximum possible number of such edges. This is calculated as $C(v) = \frac{2E_v}{k_v(k_v - 1)}$, where $E_v$ is the number of edges between the neighbors of node $v$, and $k_v$ is the degree of node $v$, representing the number of neighbors (connections) node $v$ has.     
        \item[3] $^c$ Degree centrality is the number of roads directly connected to a given node.
        \item[4] $^d$ 1 RMB $\approx$ 0.14 USD.
  \end{tablenotes}  
\end{table}

\section{Dynamics of TNC-PT relationships}\label{sec: dynamics}
This section presents descriptive statistics of the four key ratios defined in \secref{sec: Dependent and independent variables} for characterizing the TNC-PT relationships. 
For all TNC trips in Shanghai, the first- and last-mile complementary ratios, and the departure substitutive ratio \footnote{The overall arrival substitutive ratio is equal to the departure substitutive ratio for all trips combined.} are 4.43\%, 4.79\%, and 9.06\%, respectively. 
Instead of the 30\%–50\% substitutive ratio and sub-5\% complementary ratio reported in prior studies \citep{meredith-karamRelationshipRidehailingPublic2021, qiaoRidehailingCompetingComplementing2023, liuInvestigatingRelationshipsRidesourcing2025}, our results show that the gap between the two ratios has narrowed significantly.
Note in those previous works that the small complementary ratios were obtained using coordinates matching for identifying complementary trips, which likely overestimated the complementary ratio. This means the actual gap between our results and those of previous studies is even greater.
We believe this large disparity arises from the different time periods and TNC market development stages when the data was collected. 
As most prior studies analyzed data from  2016 to 2018 \citep{kongHowDoesRidesourcing2020, meredith-karamRelationshipRidehailingPublic2021, liuInvestigatingRelationshipsRidesourcing2025}, a period when TNC services were in their early market-entry stage, whereas our study uses data collected in September 2022, by which time the ride-hailing market had become increasingly saturated \citep{ridehailing_total_number_2024, pakChinasRidehailingTaxi2025}, the differing findings likely reflect a substantial shift in TNC-PT relationships over time and across different market contexts. 

In the following subsections, \secref{sec: Characteristics of TNC trips} characterizes the basic attributes of complementary and substitutive TNC trips, such as waiting time, travel time, and fare. Subsequently, \secref{sec: temporal TNC-PT} investigates the temporal variations of these ratios throughout an average day. Finally, \secref{sec: spatial TNC-PT} explores the geographical patterns of these ratios across the urban landscape.

\subsection{Characteristics of TNC trips}\label{sec: Characteristics of TNC trips}
\begin{figure}
    \centering
    \begin{subfigure}[b]{0.99\textwidth}
        \centering
        \includegraphics[width=\textwidth, keepaspectratio]{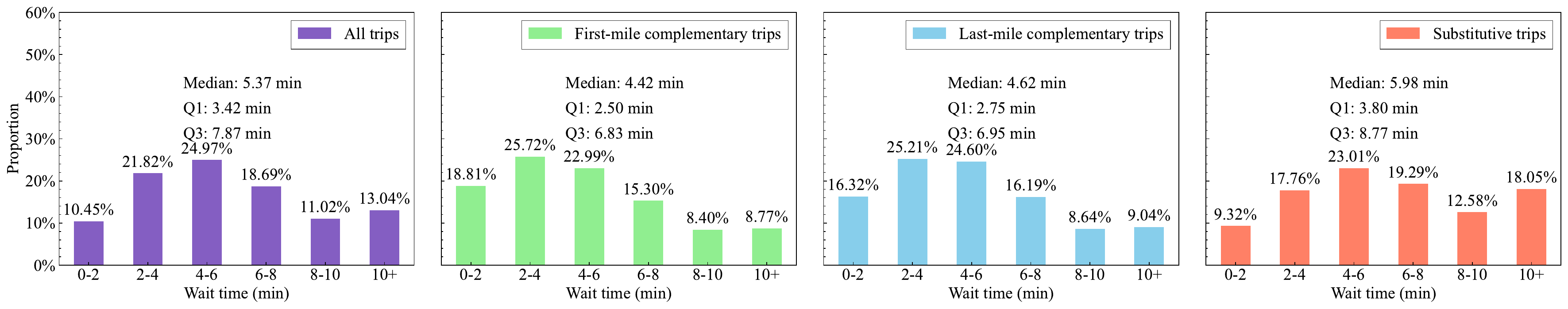}
        \caption{Wait time}
        \label{fig:Wait time}
    \end{subfigure}
    \begin{subfigure}[b]{0.99\textwidth}
        \centering
        \includegraphics[width=\textwidth, keepaspectratio]{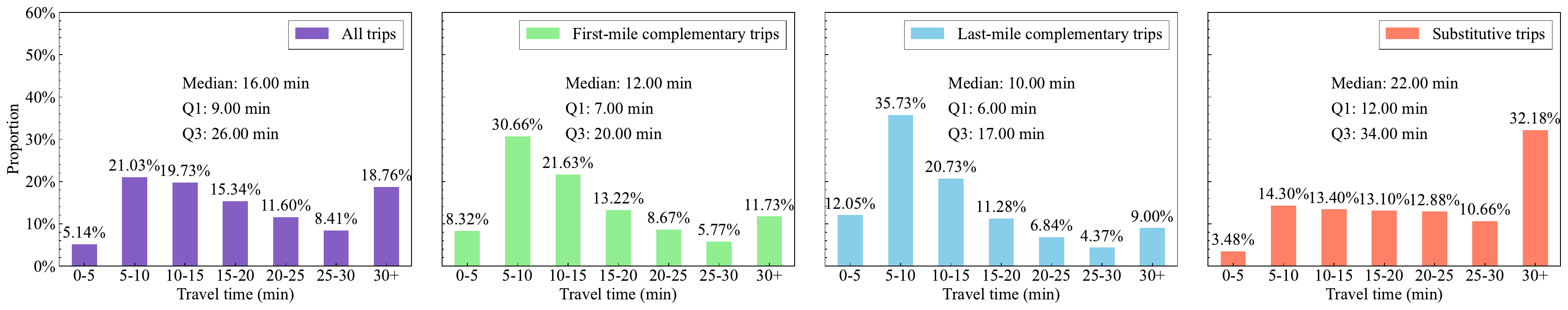}
        \caption{Travel time}
        \label{fig:Travel time}
    \end{subfigure}
    \begin{subfigure}[b]{0.99\textwidth}
        \centering
        \includegraphics[width=\textwidth, keepaspectratio]{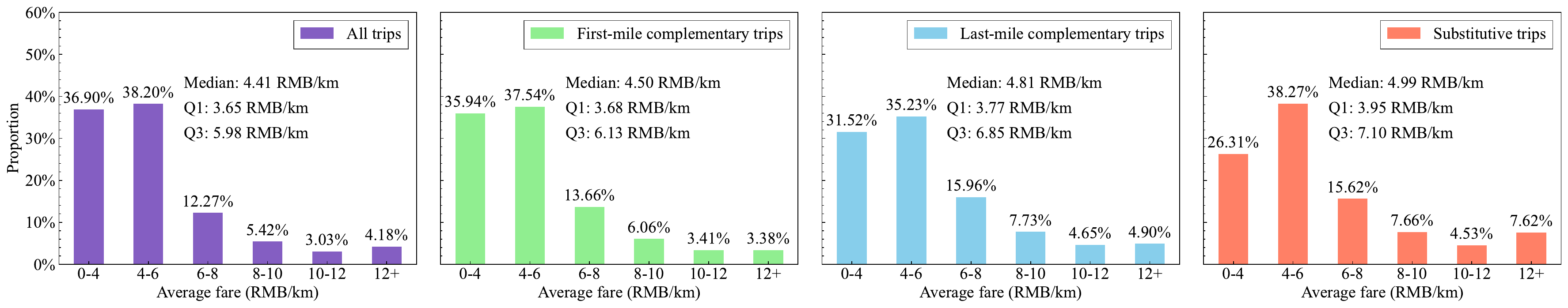}
        \caption{Unit fare}
        \label{fig:Unit fare}
    \end{subfigure}
    \caption{\centering{Statistics of complementary and substitutive trips.}}
    \label{fig:Statistics_trips}
\end{figure}

\Figref{fig:Statistics_trips} characterizes the three types of PT-related TNC trips by three user metrics: waiting time (from order request to pick-up), travel time (from pick-up to drop-off), and average fare per km (total fee divided by travel distance). 
First- and last-mile complementary trips clearly show a distribution of wait times more concentrated around shorter values, with nearly 70\% of trips having wait times under 6 minutes. 
In contrast, substitutive trips demonstrate more evenly distributed wait times across different intervals with significantly higher proportions of longer wait times than complementary trips.
Travel times of both first- and last-mile complementary trips predominantly fall in lower ranges (<10 minutes) and peak in the 5-10 minute range (30.66\% and 35.73\%, respectively), implying that these trips often serve as short-distance shuttles. 
In comparison, travel times of substitutive trips have the highest proportion in the category above 30 minutes (32.18\%), revealing the popularity of TNC services among long-distance travelers. 
At just 4.41 RMB/km, Shanghai’s average ride-hailing fare is less than half of common fares (roughly 10RMB/km) in developed countries \citep{rangel2022exploring}, indirectly reflecting the intense competition within Shanghai's ride-hailing market. 
Compared against complementary trips, substitutive trips exhibit a higher median price (4.99 RMB/km) and a greater proportion (7.62\%) in higher fare ranges (above 12 RMB/km). 

\subsection{Temporal variation of TNC-PT relationships}\label{sec: temporal TNC-PT}
\begin{figure}
    \centering
    \begin{subfigure}[b]{0.45\textwidth}
        \centering
        \includegraphics[width=\textwidth]{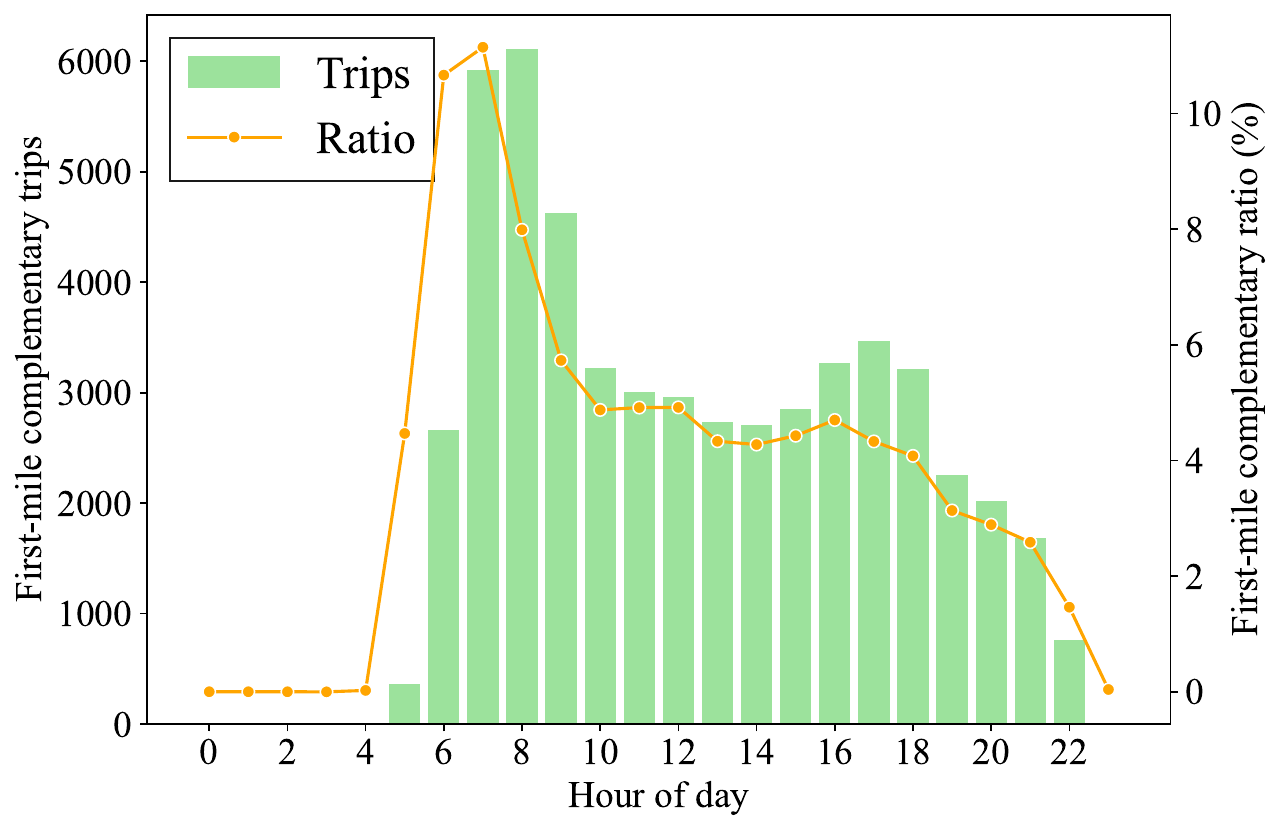}
        \caption{First-mile complementary trips and ratio}
        \label{fig:Temporal First-mile_Complement}
    \end{subfigure}
    \begin{subfigure}[b]{0.45\textwidth}
        \centering
        \includegraphics[width=\textwidth]{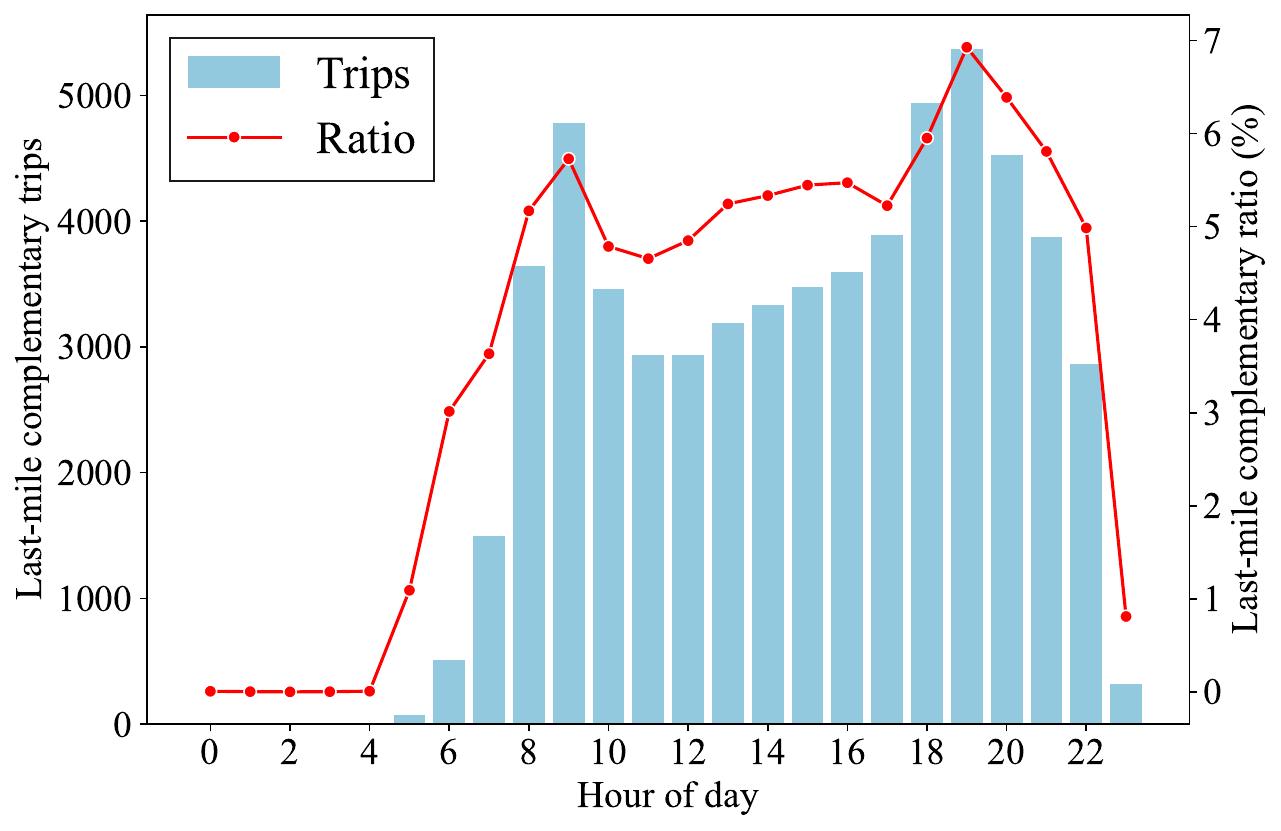 }
        \caption{Last-mile complementary trips and ratio}
        \label{fig:Temporal Last-mile_Complement}
    \end{subfigure}
    \begin{subfigure}[b]{0.45\textwidth}
        \centering
        \includegraphics[width=\textwidth]{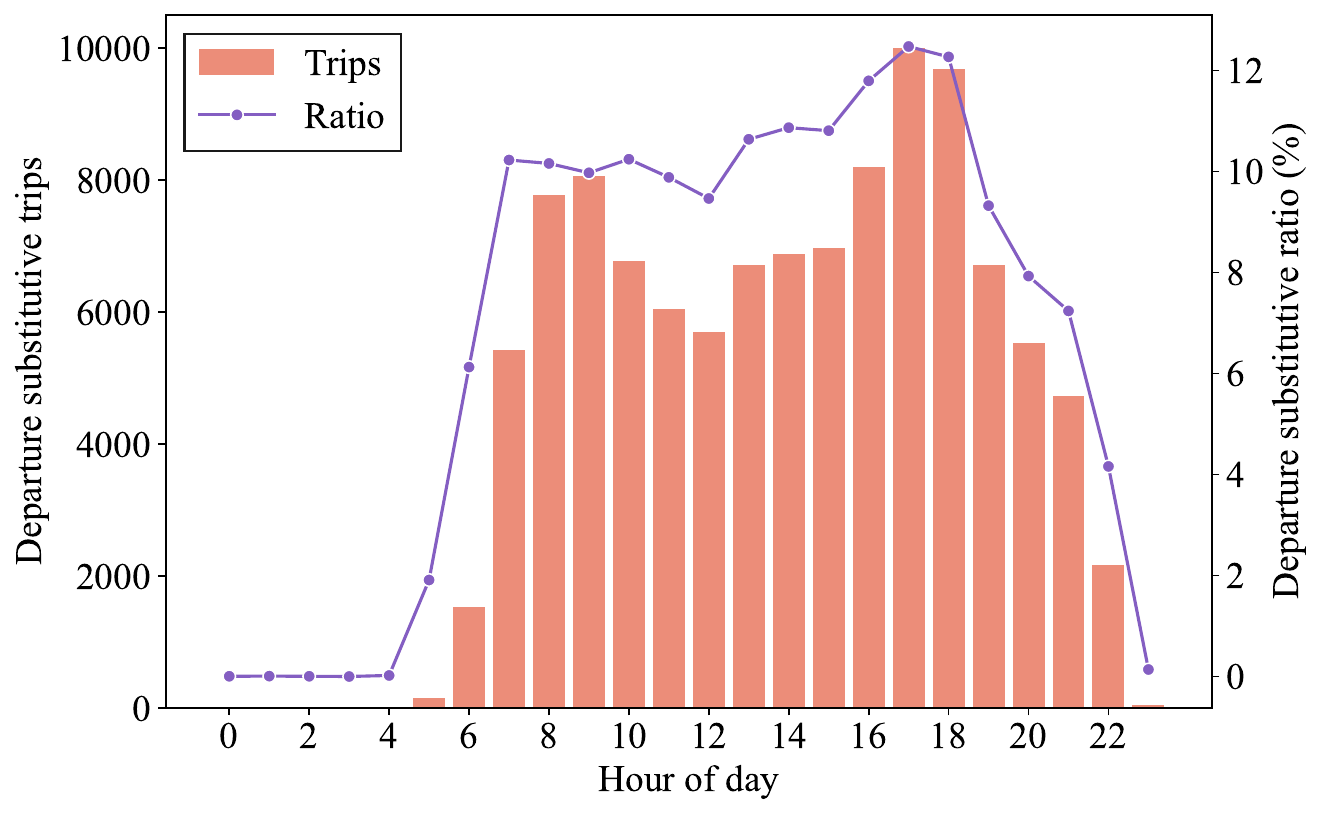}
        \caption{Departure substitutive trips and ratio}
        \label{fig:Temporal dep substitution}
    \end{subfigure}
    \begin{subfigure}[b]{0.45\textwidth}
        \centering
        \includegraphics[width=\textwidth]{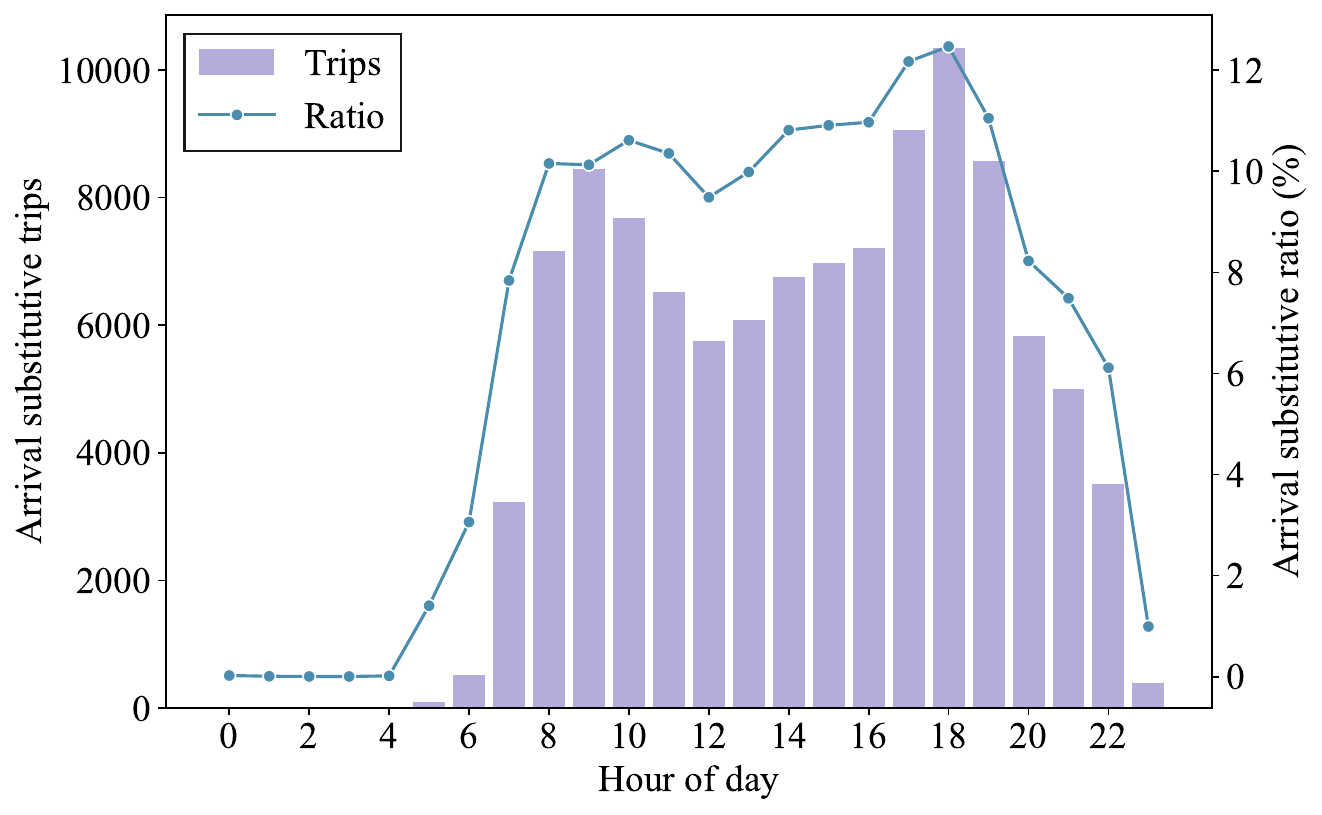}
        \caption{Arrival substitutive trips and ratio}
        \label{fig:Temporal arr substitution}
    \end{subfigure}
    \caption{\centering{Temporal distribution of complementary and substitutive trips and ratios.}}
    \label{fig:Temporal}
\end{figure}

\Figref{fig:Temporal} presents the temporal distributions of PT-related TNC trips and their associated ratios throughout the day. The distributions of first- and last-mile complementary trips and their respective ratios are depicted in \twofigref{fig:Temporal First-mile_Complement}{fig:Temporal Last-mile_Complement}. 
During the early morning hours (0–5 AM), both the number of trips and ratios remain nearly zero due to the absence of PT services.
Sharp increases of these numbers and ratios occur in the morning peak (7–9 AM) for both first- and last-mile complementary trips.
During the mid-day hours (10 AM–4 PM), the trends begin to diverge: first-mile trips decline rapidly and stabilize at moderate levels, while last-mile trips experience a slight drop but remain relatively high. 
A similar divergence is observed in the corresponding ratios. 
The contrast becomes more pronounced during the evening peak, as the first-mile complementary trip number and ratio continue to decline, while those of last-mile trips rise again and even surpass the morning-peak values.
After 10 PM, both decline rapidly as PT services gradually cease.

This divergence between the first- and last-mile complementary trips and their ratios during the morning and evening peaks reflects intriguing commuting patterns and priorities.
During morning peaks, both first- and last-mile complementary ratios are high, indicating that people not only prefer using ride-hailing to connect their homes to PT but also rely on it to reach their workplaces after exiting PT.
The former may be due to commuters living beyond a comfortable walking distance from transit stations, while the latter reflects the urgency to save time and avoid being late for work. 
For example, they may exit PT at a transfer station and hail a ride to their destination, bypassing time-consuming connecting PT trips, particularly less-reliable bus services.
However, the situation changes in the evening peaks. 
In evening peaks, the first-mile complementary trips and ratio remain low, possibly because: (i) most workplaces located in central urban areas are close to transit stations; and (ii) commuters tend to be less time-sensitive in evening peaks compared to in morning peaks, allowing them to choose walking access to PT stations and PT trips with more transfers. In contrast, the continued growth of the last-mile complementary trips and ratio suggests that many commuters still prefer ride-hailing as a more convenient and time-saving solution for the final leg of their returns. Again, this is likely due to the longer distance between PT stations and their homes in the suburbs, which is quite common in mega-cities like Shanghai.

These asymmetric morning-evening patterns are also consistent with prior transportation theory. Existing studies have shown that morning and evening commutes involve different schedule constraints and generalized costs \citep{gonzalesEveningCommuteCars2013}, and that commuters’ mode choices depend on heterogeneous values of time and trip lengths \citep{gonzalesMorningCommuteCompeting2012}.
From this perspective, the morning concentration of both first- and last-mile complementary trips likely reflects stronger schedule pressure and a higher penalty for late arrival during this period. 

The temporal distributions of departure and arrival substitutive trips and ratios (see \secref{sec: Dependent and independent variables} and \twoeqref{eq:departure substitutive ratio}{eq:arrival substitutive ratio} for the definitions of these trips and ratios) are presented in \twofigref{fig:Temporal dep substitution}{fig:Temporal arr substitution}. The temporal patterns of both substitutive ratios exhibit remarkable consistency, indicating that computation of these ratios based on trip origin or destination may yield similar findings. 
During the morning peak period (6-9 AM), substitutive trips demonstrate a rapid escalation, constituting approximately 10\% of total trip volume. 
Following this initial surge, both ratios maintain a relatively stable level during mid-day hours but rise again to a second peak in the evening. 
The peak in the evening may be driven by both return commutes and recreational (e.g., dine-out) trips, for which the convenience and directness of TNCs are most valued. 

\subsection{Spatial variation of TNC-PT relationships}\label{sec: spatial TNC-PT}
\begin{figure}
    \centering
    \begin{subfigure}[b]{0.4\textwidth}
        \centering
        \includegraphics[height=0.875\textwidth, keepaspectratio]{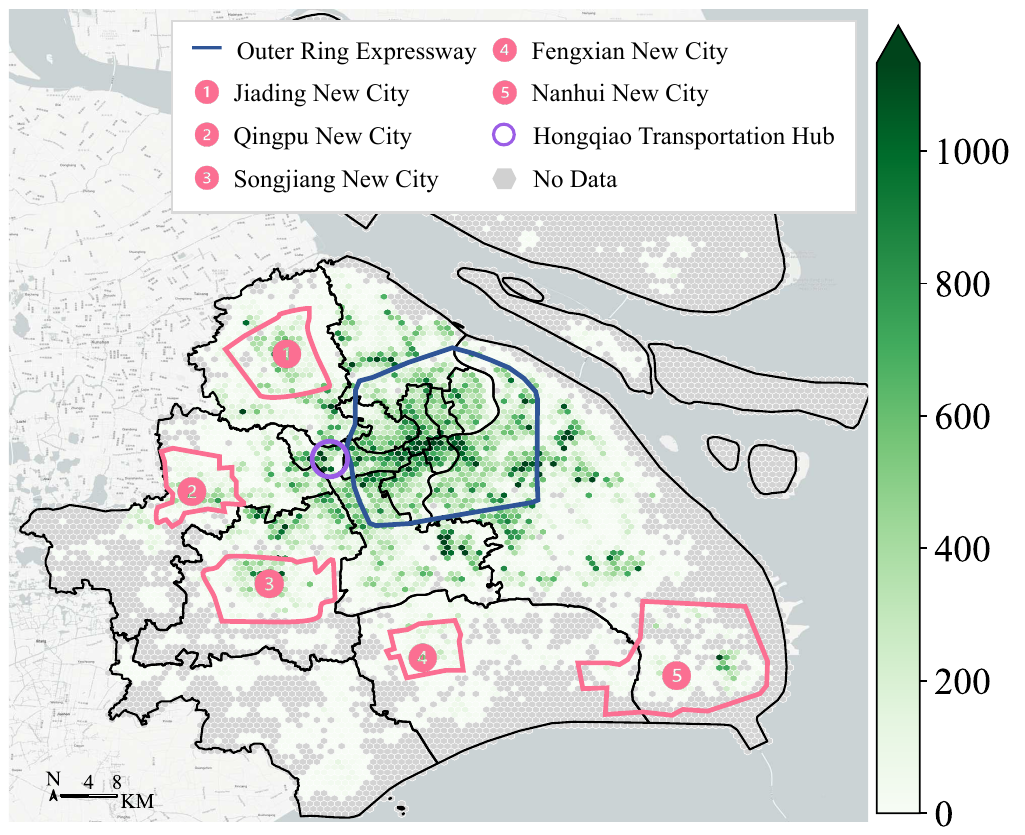}
        \caption{First-mile complementary trips}
        \label{fig:First-mile_Complement_Spatial_trips}
    \end{subfigure}
    \hspace{2em}
    \begin{subfigure}[b]{0.4\textwidth}
        \centering
        \includegraphics[height=0.875\textwidth, keepaspectratio]{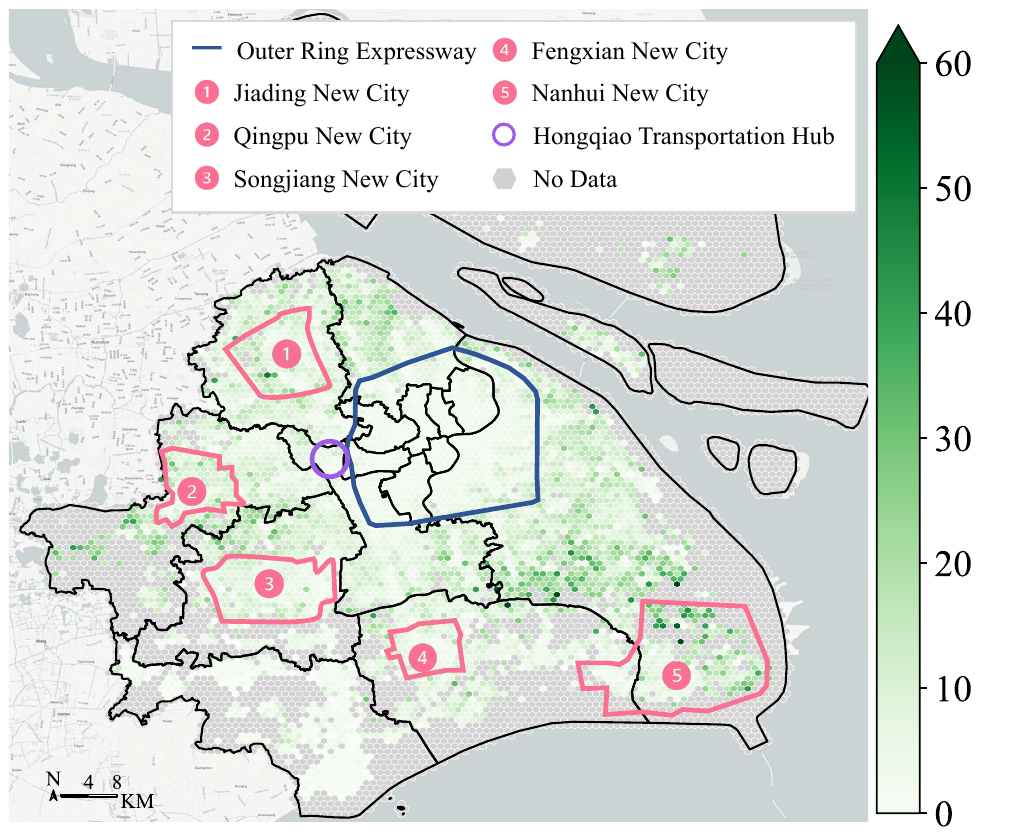}
        \caption{First-mile complementary ratio (\%)}
        \label{fig:First-mile_Complement_Spatial_rates}
    \end{subfigure}
    \vspace{1em} 
    \begin{subfigure}[b]{0.4\textwidth}
        \centering
        \includegraphics[height=0.875\textwidth, keepaspectratio]{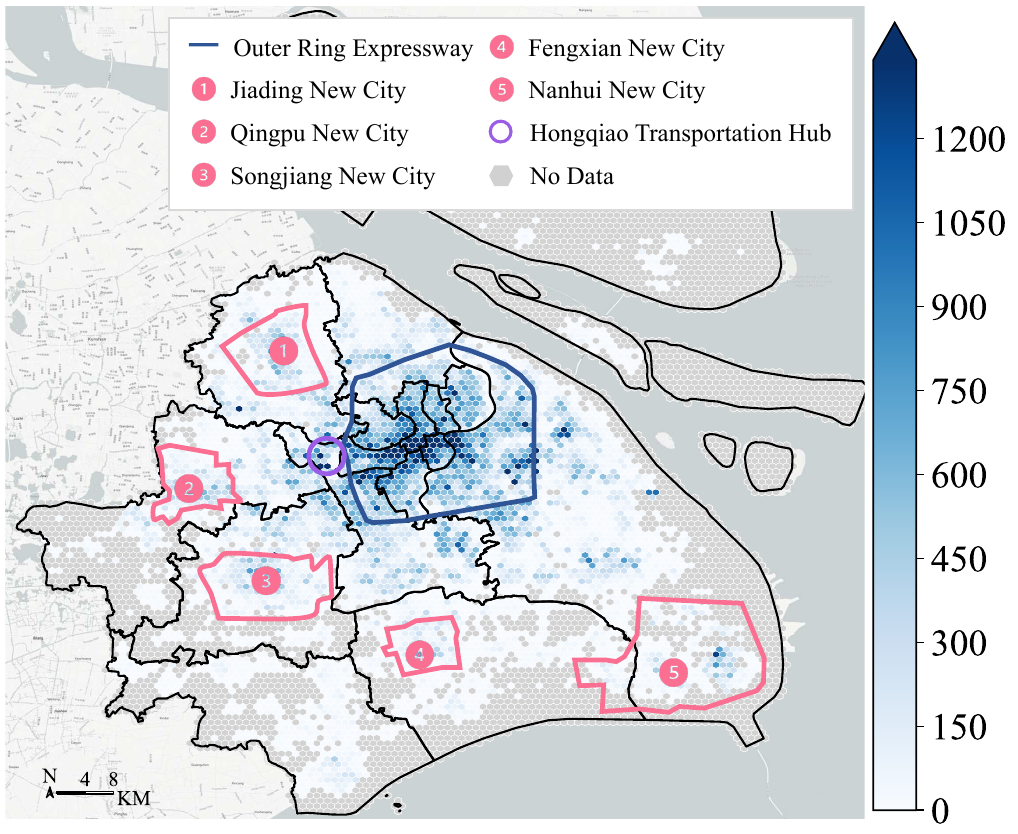}
        \caption{Last-mile complementary trips}
        \label{fig:Last-mile_Complement_Spatial_trips}
    \end{subfigure}
    \hspace{2em}
    \begin{subfigure}[b]{0.4\textwidth}
        \centering
        \includegraphics[height=0.875\textwidth, keepaspectratio]{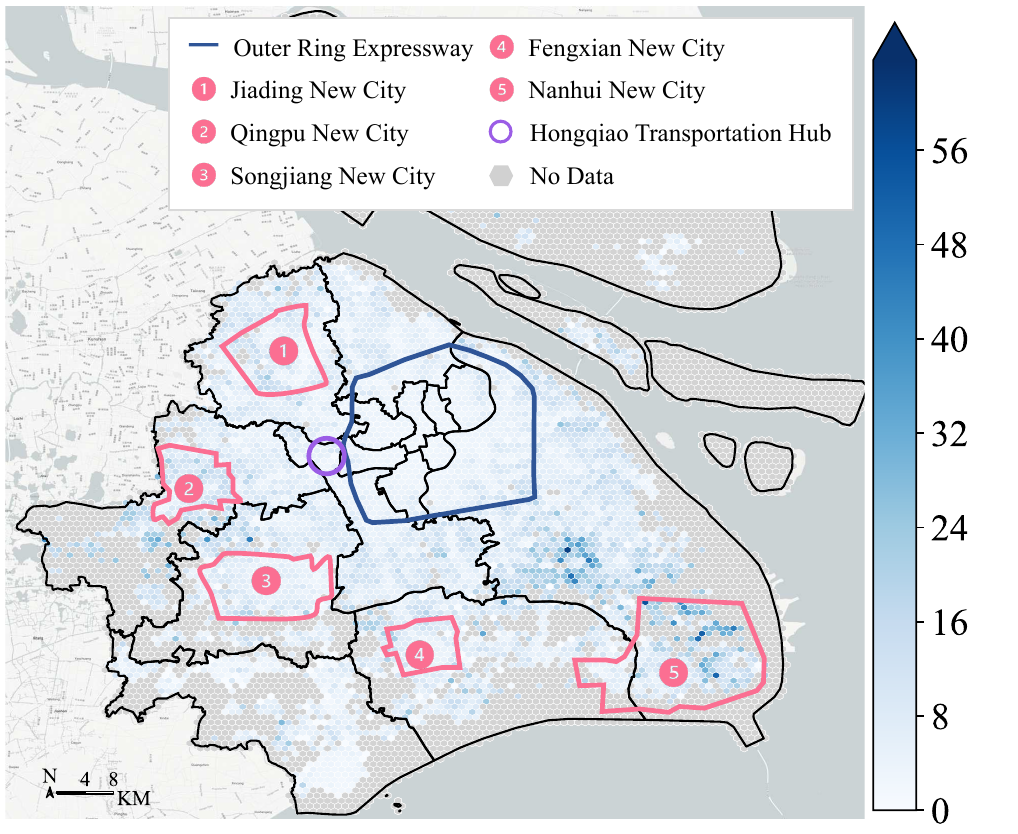}
        \caption{Last-mile complementary ratio (\%)}
        \label{fig:Last-mile_Complement_Spatial_rates}
    \end{subfigure}
    \caption{\centering{Spatial distribution of complementary trips and ratios.}}
    \label{fig:Spatial_complement}
\end{figure}

As shown in \figref{fig:Spatial_complement}, the first- and last-mile complementary trips and their corresponding ratios are spatially distributed in generally similar ways. 
Both first- and last-mile complementary trips are concentrated within the downtown area. 
Nevertheless, the high total volume of TNC trips in the central urban area tends to dilute the complementary ratios. 
This observation of low ratios in the central urban area underscores the necessity of ratio-based analysis. 
Results based solely on trip numbers, such as those in \citep{jinWhatPromotesIntegration2025}, may introduce biases due to varying TNC volumes across urban areas, potentially obscuring true spatial variations in substitutive and complementary relationships between TNC and PT.
On the other hand, in subcenters such as Nanhui and Jiading New Cities, higher first- and last-mile complementary ratios are observed.
A potential reason is that the densely populated residential zones in these subcenters may lead residents to rely on TNCs to access nearby public transit nodes. 

\begin{figure}
    \centering
    \begin{subfigure}[b]{0.4\textwidth}
        \centering
        \includegraphics[height=0.875\textwidth, keepaspectratio]{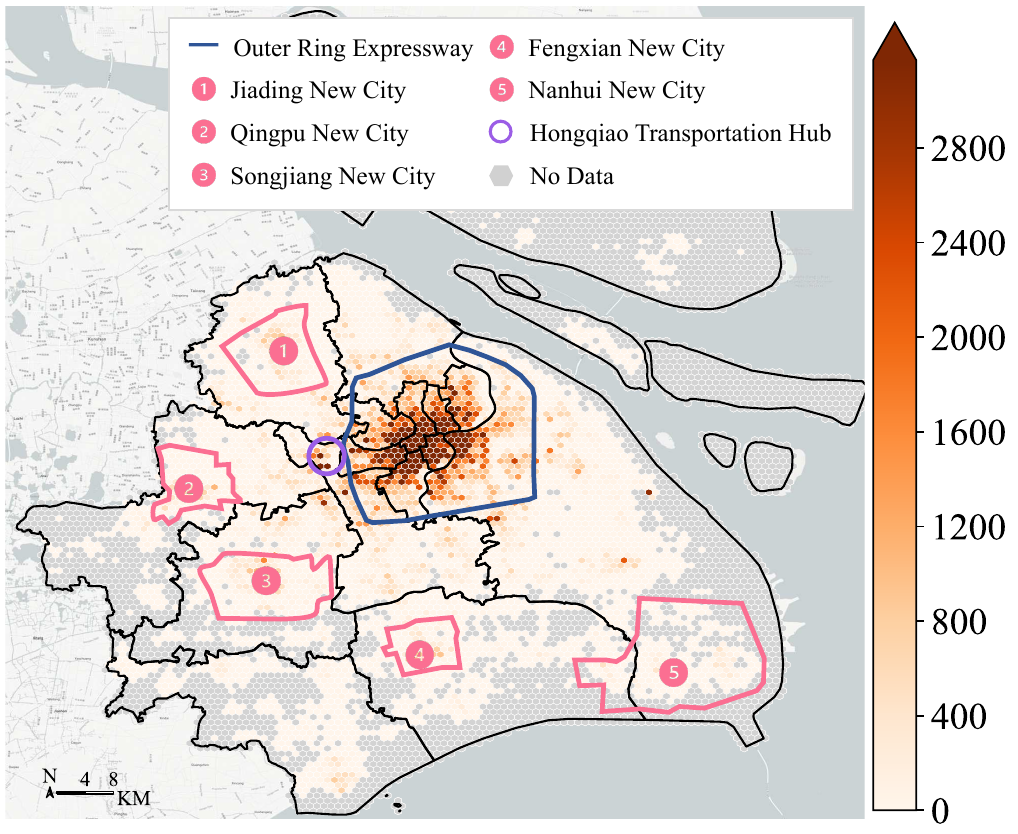}
        \caption{Departure substitutive trips}
        \label{fig:Departure_Substitution_Spatial_trips}
    \end{subfigure}
    \hspace{2em}
    \begin{subfigure}[b]{0.4\textwidth}
        \centering
        \includegraphics[height=0.875\textwidth, keepaspectratio]{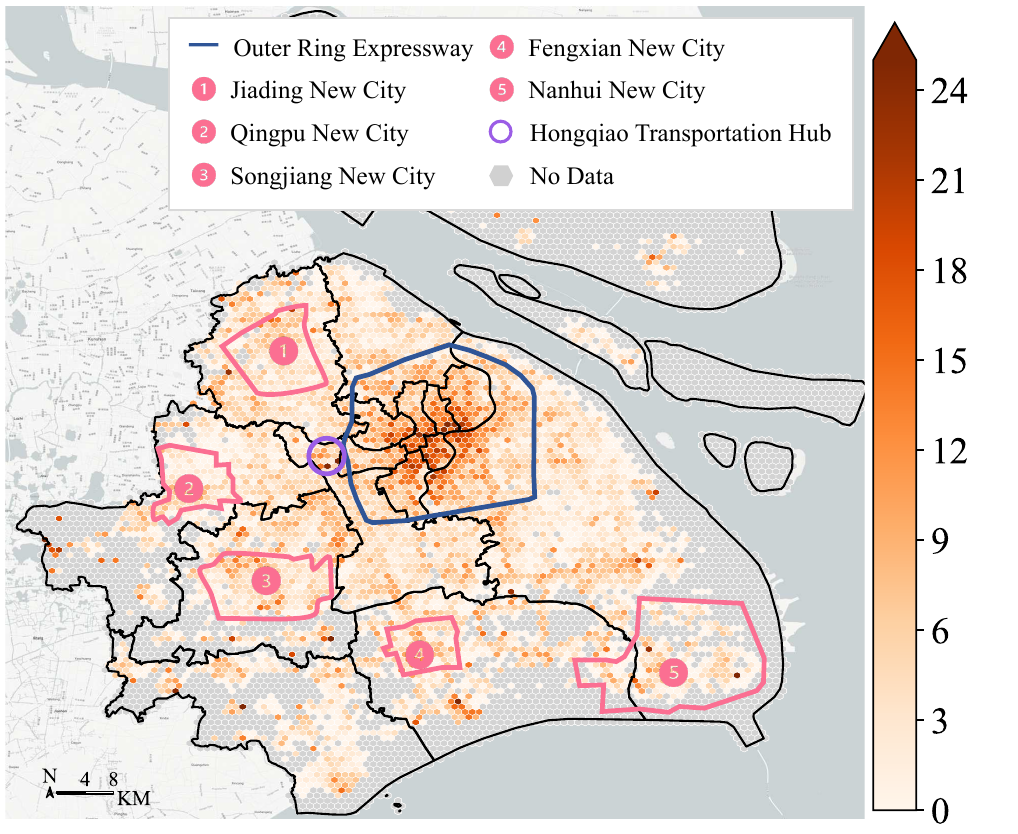}
        \caption{Departure substitutive ratio (\%)}
        \label{fig:Departure_Substitution_Spatial_rates}
    \end{subfigure}
    \vspace{1em} 
    \begin{subfigure}[b]{0.4\textwidth}
        \centering
        \includegraphics[height=0.875\textwidth, keepaspectratio]{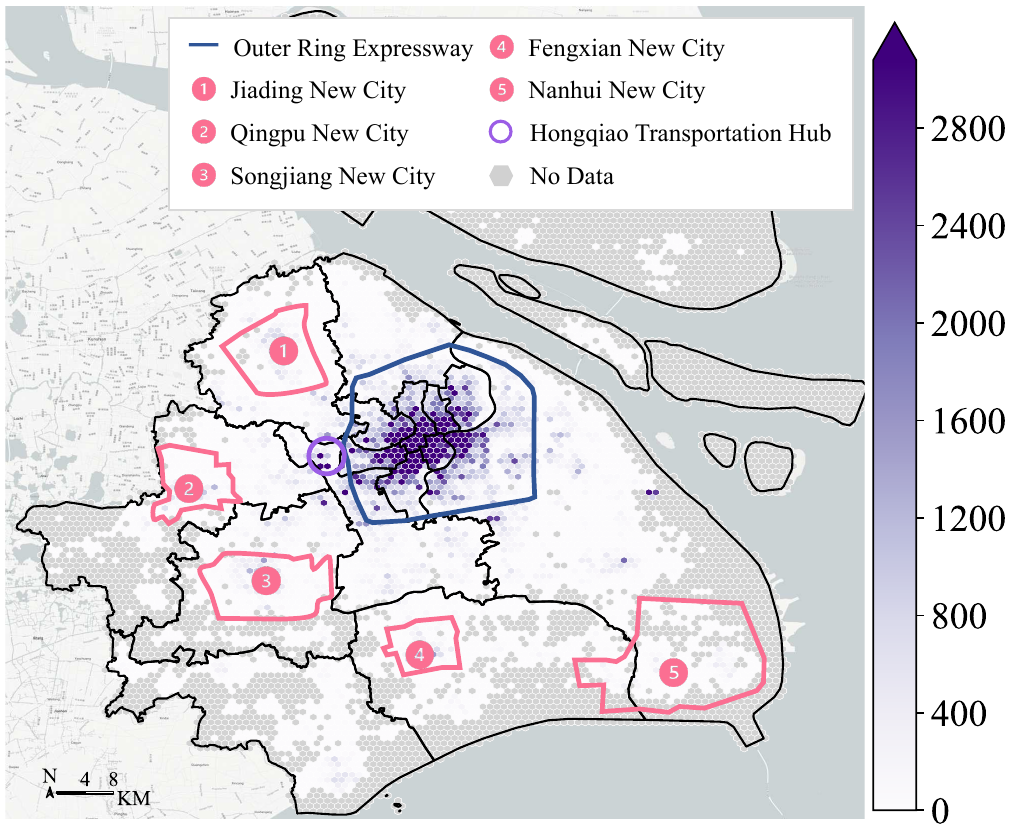}
        \caption{Arrival substitutive trips}
        \label{fig:Arrival_Substitution_Spatial_trips}
    \end{subfigure}
    \hspace{2em}
    \begin{subfigure}[b]{0.4\textwidth}
        \centering
        \includegraphics[height=0.875\textwidth, keepaspectratio]{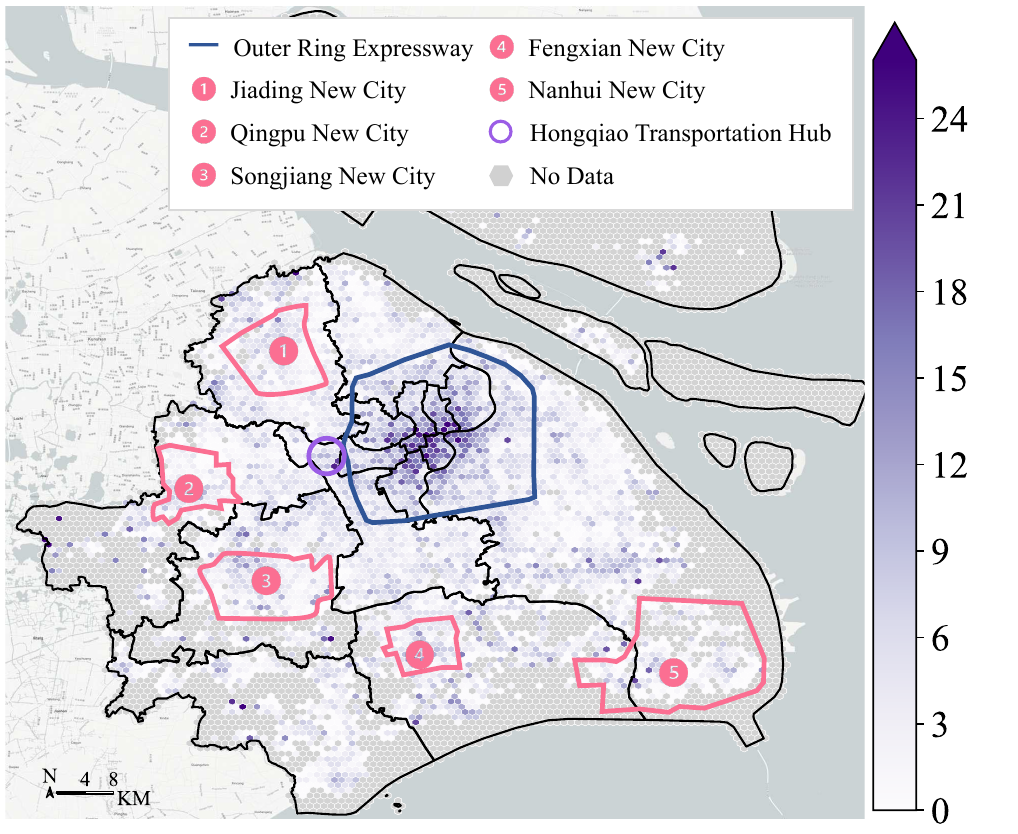}
        \caption{Arrival substitutive ratio (\%)}
        \label{fig:Arrival_Substitution_Spatial_rates}
    \end{subfigure}
    \caption{\centering{Spatial distribution of substitutive trips and ratios.}}
    \label{fig:Spatial_substitution}
\end{figure}

\figref{fig:Spatial_substitution} reveals obvious spatial clustering patterns of substitutive trips in urban areas, particularly around major regional transportation centers (e.g., Shanghai Hongqiao Transport Hub) and commercial zones (e.g., within the Outer Ring Expressway).
These regions typically have dense and well-integrated PT networks, providing more alternatives to ride-hailing. 
In addition, higher substitutive ratios are also observed in the five subcenters in Shanghai, such as the Fengxian New City and the Nanhui New City. 
Although these subcenters are typically equipped with bus depots or subway lines, due to the longer travel distances to the city center, some people prefer ride-hailing services that save both travel times and inconvenient transfers. 
In more remote areas of Shanghai, both departure and arrival substitutive trips and ratios are low, due to limited transit coverage and frequency (i.e., the TNC trips connecting to those areas are more likely to be identified as independent trips).

The above analysis clearly reveals that the complementary and substitutive relationships between TNCs and PT exhibit significant spatial clustering. This indicates that spatial environmental factors, such as land-use characteristics and transit infrastructure, influence the interaction between public transit and ride-hailing services. 
To more rigorously quantify the effects of these factors, the next section adopts a learning-based approach to explore significant environmental determinants of complementary and substitutive ratios.

\section{A learning-based approach for TNC-PT relationship estimation and interpretation}\label{sec: methodology} 
This section introduces our approach to examine the effects of key features on complementary and substitutive ratios. 
In \secref{sec: CatBoost method}, we employ CatBoost, a tree-based gradient boosting model, to build a nonlinear estimation model for the ratios. 
Then, the SHAP method is applied in \secref{sec: SHAP methods} to evaluate and rank the relative importance of each explanatory factor. PDPs are further utilized to visualize the nonlinear relationships.

\subsection{A CatBoost estimation model}\label{sec: CatBoost method}
Various machine learning models were applied to capture the nonlinear relationships between independent variables and outcomes, including GBDT-based algorithms (XGBoost, LightGBM, and CatBoost) and advanced deep learning architectures for tabular data, such as standard MLP, TabNet and FT-Transformer. TabNet and FT-Transformer represent state-of-the-art deep learning approaches that utilize sequential attention and feature tokenization to handle heterogeneous tabular data. We evaluated the performance of all candidate models using mean absolute error (MAE), mean squared error (MSE), and root mean squared error (RMSE). Detailed benchmarking results across these models are summarized in \Tabref{tab:Model Comparison} of \Appref{appendix: Model Comparison}. Comparative analysis shows that CatBoost consistently outperforms both the deep learning architectures and other GBDT models in predicting most dependent variables. Therefore, the subsequent sections exclusively present the model specifications and results derived from CatBoost.

Consider a training dataset $\mathcal{D} = \{(\mathbf{x}_i, y_i), i=1,2,...,n\}$, where $\mathbf{x}_i$ represents the feature vector for the $i$-th sample, $y_i$ the corresponding target value (first- or last-mile complementary ratio, or departure or arrival substitutive ratio), and $n$ is the total number of samples used in the CatBoost model. 
To avoid multicollinearity, explanatory variables with a Variance Inflation Factor (VIF) greater than 10 (e.g., average waiting time and fare) were excluded \citep{gao_spatial_2021, guo_analysis_2022}. 

CatBoost follows the general structure of gradient boosting, which builds an ensemble model iteratively by minimizing the loss function \citep{prokhorenkova2018catboost}. The gradient boosting model is built as
\begin{equation}\label{eq: func_GBDT}
F_m(\mathbf{x}) = F_{m-1}(\mathbf{x}) + \gamma_m h_m(\mathbf{x}),
\end{equation}
where  $F_m(\mathbf{x})$ is the model at the $m$-th iteration,  $h_m(\mathbf{x})$ is the newly added decision tree, and $\gamma_m$ is the learning rate controlling the contribution of $h_m(\mathbf{x})$ . 

To minimize the loss function $L(y, F(\mathbf{x}))$, gradient boosting computes the negative gradient
\begin{equation}\label{eq: the negative gradient}
g_i^{(m)} = - \left. \frac{\partial L(y_i, F(\mathbf{x}_i))}{\partial F(\mathbf{x}_i)} \right|{F(\mathbf{x}_i) = F_{m-1}(\mathbf{x}_i)},
\end{equation}
where the Mean Squared Error (MSE) is used as the loss function
$L(y, F(\mathbf{x})) = \frac{1}{n} \sum_{i=1}^{n} (y_i - F(\mathbf{x}_i))^2$. One of the major challenges in gradient boosting is that the model inadvertently uses future information while training, leading to overfitting and poor generalizability. 
Traditional gradient boosting algorithms calculate residuals using all previous trees, which might introduce biases. 
CatBoost solves this problem by Ordered Boosting, which ensures that, when training the $m$-th tree, the model only uses residuals from samples that precede the current sample. Mathematically, the gradient for each sample $\mathbf{x}_i$ is computed as:
\begin{equation}
g_i^{(m)} = - \left. \frac{\partial L(y_i, F(\mathbf{x}_i))}{\partial F(\mathbf{x}_i)} \right|{F(\mathbf{x}_i) = F_{m-1}^{(i)}(\mathbf{x}_i)}
\end{equation}
where  $F_{m-1}^{(i)}(\mathbf{x}_i)$  is a model trained only on data points preceding  $\mathbf{x}_i$. This prevents data leakage and ensures robust generalization.

Each new tree  $h_m(\mathbf{x})$ is trained to approximate the residual:
\begin{equation}
h_m(\mathbf{x}) = \arg\min_h \sum_{i=1}^{n} (g_i^{(m)} - h(\mathbf{x}_i))^2.
\end{equation}

Then, the optimal coefficient $\gamma_m$  is determined by:
\begin{equation}
\gamma_m = \arg\min_{\gamma} \sum_{i=1}^{n} L(y_i, F_{m-1}(\mathbf{x}_i) + \gamma h_m(\mathbf{x}_i)).
\end{equation}

This iterative process continues until convergence or a predefined stopping criterion is met. After training $M$ trees, the final prediction for a sample $\mathbf{x}$ is given by
\begin{equation}\label{eq: func_final}
F(x) = F_0 + \sum_{m=1}^{M}\gamma_m h_m(x),
\end{equation}
where $F_0$ is the initial prediction (e.g., the mean of the target values). 

To test the robustness of the Catboost results, this study employs a spatial 5-fold cross-validation process to randomly select 80\% of samples for training, and the remaining 20\% for testing. The hyperparameter optimization for CatBoost was performed using Optuna \citep{akiba2019optuna}, which automatically explores parameter configurations through Bayesian search. Key parameters and their ranges are: the number of iterations (100–1000), learning rate (0.001–0.1), tree depth (1–10), bagging temperature (0.0–1.0), random strength (0.0–1.0), column sampling ratio (0.05–1.0), and minimum data in leaf (10–100); see \cite{akiba2019optuna, agheliHowDoesDistraction2025, fanFeatureImportanceBasedMultiLayer2024} for a detailed explanation of these parameters. The final configuration achieves enhanced predictive performance while mitigating overfitting risks. 

\subsection{Model interpretation methods}\label{sec: SHAP methods}
The results of the CatBoost model are put into SHAP for interpretation, which exhibits strong interpretability on machine learning models \citep{lundbergUnifiedApproachInterpreting2017a}.
SHAP leverages concepts from cooperative game theory to provide consistent and fair variable importance values, enhancing the interpretability of complex models \citep{lundbergUnifiedApproachInterpreting2017a}. Specifically, SHAP considers the features (explanatory variables) listed in \Tabref{Table: variables and definitions} as ``contributors'' to the dependent variable (complementary or substitutive ratio). The SHAP value for a feature $j$, denoted as $\phi_j$, is defined by
\begin{equation}\label{eq: shap_calc}
    \phi_j = \sum_{S \subseteq N \setminus \{j\}} \frac{|S|! \, (|N| - |S| - 1)!}{|N|!} \left[ f(S \cup \{j\}) - f(S) \right],
\end{equation}
where $N$ is the set of all features and $\left | N \right |$ is its size. $S$ is any subset of $N$ excluding feature $j$ and $\left | S \right |$ is its size. 
$f(S)$ and $f(S \cup \{j\})$ represent the model's predictions considering the features in subset $S$ without and with feature $j$ added, respectively.

After the SHAP values for all features are obtained, we assess the overall importance of each feature by calculating its mean absolute SHAP value. This value represents the average magnitude of an explanatory variable's impact on the dependent variable across all samples and is calculated as:
\begin{equation}
    A_{j}=\frac{1}{n} \sum_{i=1}^{n}\left|\phi_{j}^{(i)}\right|,
\end{equation}
where $A_{j}$ is the mean absolute SHAP value for variable $j$, $\phi_{j}^{(i)}$ is the SHAP value for variable $j$ in individual prediction $i$, and $n$ is the total number of samples used in the CatBoost model.

Sequentially, to compare the importance across all variables, we calculate the relative importance of feature $j$ by normalizing its mean absolute SHAP value against the sum of mean absolute SHAP values for all features:
\begin{equation}\label{eq: impor_calc}
     P_{j}=\frac{A_{j}}{\sum_{k=1}^{{|N|}}A_{k}} \times 100\%,
\end{equation}

Variable importance refers to the relative contribution of a feature to the predictive power of the model, analogous to the absolute value of the $\beta$ coefficient in linear regression models.
However, variable importance does not indicate the sign (positive or negative) of the feature’s relative contribution. 
Therefore, the Partial Dependence Plot (PDP) is introduced as a tool to reveal the impact of feature changes on predictions and to provide more information about its directionality \citep{friedman2001greedy}.
Specifically, the partial dependence value for feature $j$ at the value $x_j$  can be expressed using the following formula: 
\begin{equation}\label{eq:PDP}
    \text{PDP}_{j}\left(x_j\right)=\frac{1}{n} \sum_{i=1}^{n} f\left(x_{j}, \mathbf{x}_{-j}^{(i)}\right),
\end{equation}
where $n$ is the total number of samples, $f\left(x_{j}, \mathbf{x}_{-j}^{(i)}\right)$ is the model's prediction when feature $j$ is set to $x_j$ and other features take the values from the $i$-th sample, $\mathbf{x}_{-j}^{(i)}$ denotes the vector of all feature values in the $i$-th sample except for feature $j$.

\section{Model results and analysis}\label{sec: results}

\secref{sec:CatBoost model} identifies and reports the key predictive variables that significantly contribute to explaining the four ratios defined in \Secref{sec: Dependent and independent variables}. 
\secref{sec: Partial dependence plot} presents the PDP curves that depict the effects of key factors on these ratios.

\subsection{Variable importance}\label{sec:CatBoost model}
\begin{figure}[t]
    \centering
    \begin{subfigure}[b]{0.95\textwidth}
        \centering
        \includegraphics[width=.475\textwidth]{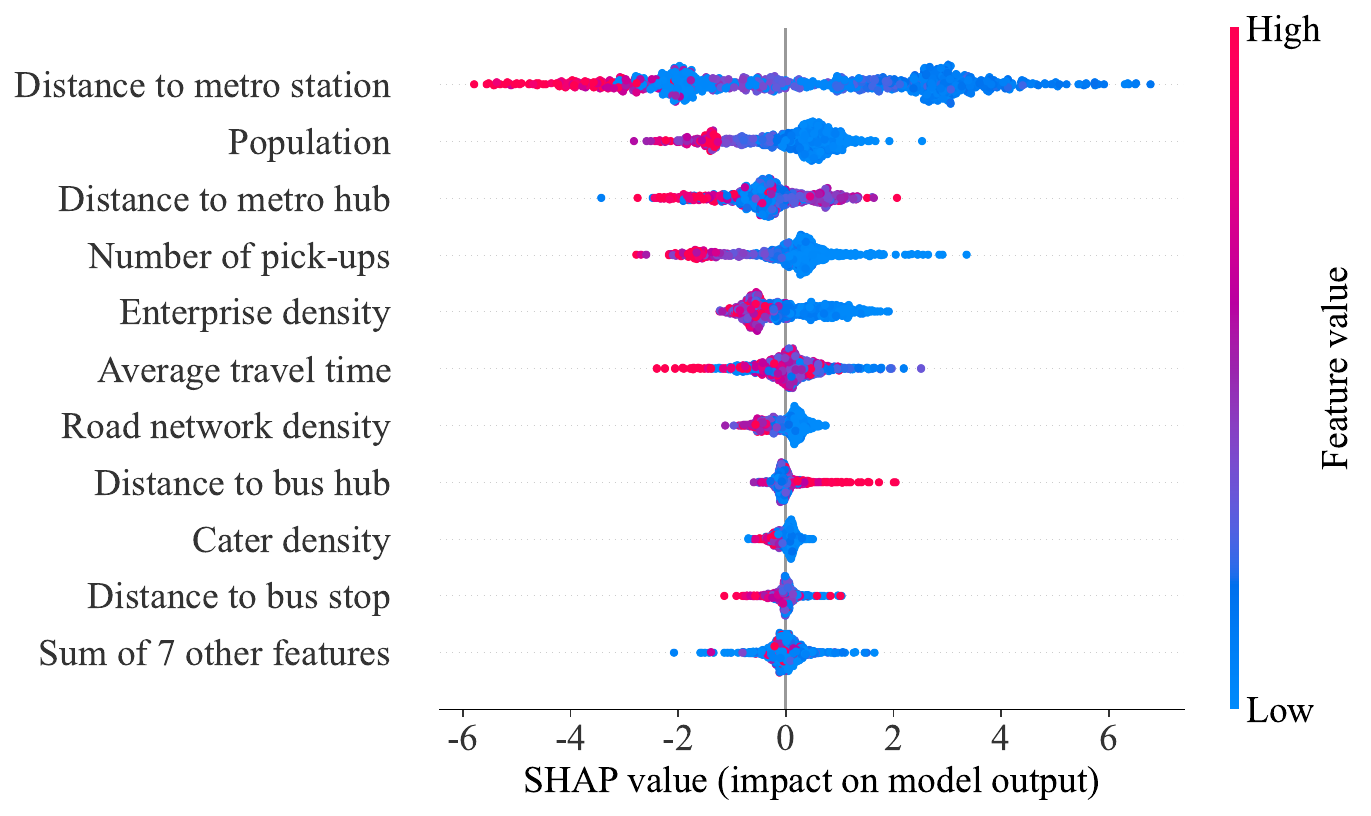}
        \includegraphics[width=.475\textwidth]{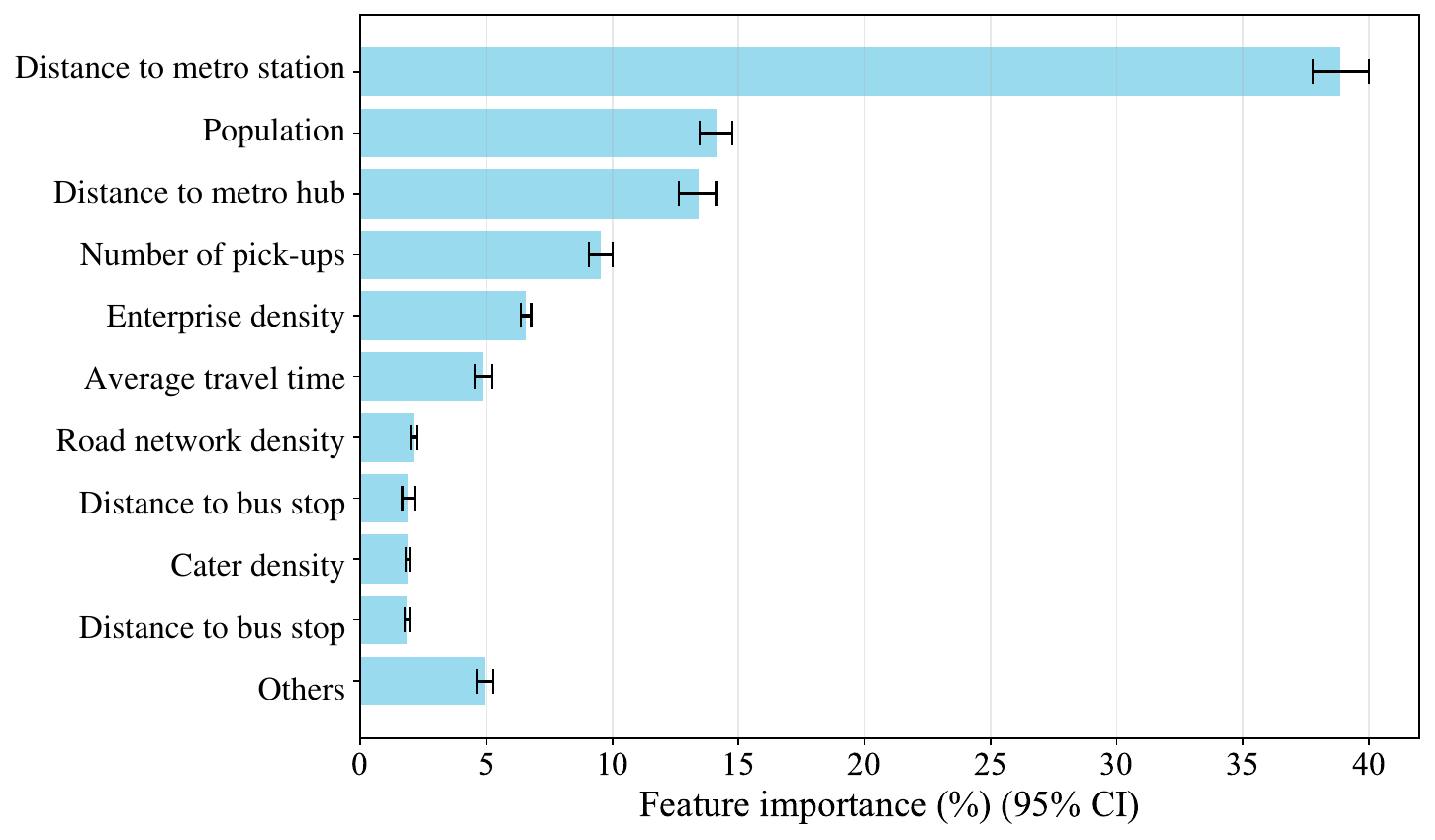}
        \caption{First-mile complementary ratio}
        \label{fig:First-mile complementary ratios SHAP}
    \end{subfigure}
    \begin{subfigure}[b]{0.95\textwidth}
        \centering
        \includegraphics[width=.475\textwidth]{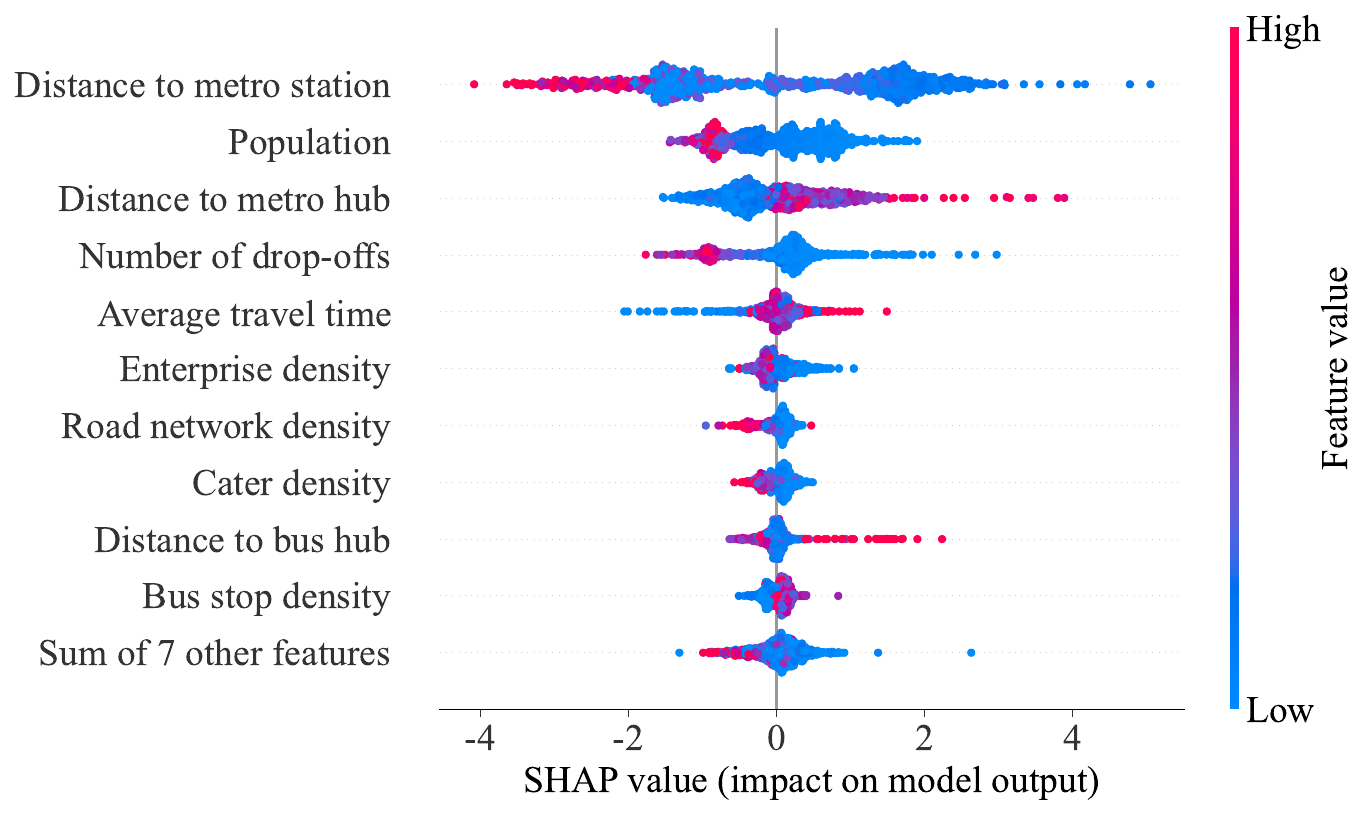}
        \includegraphics[width=.475\textwidth]{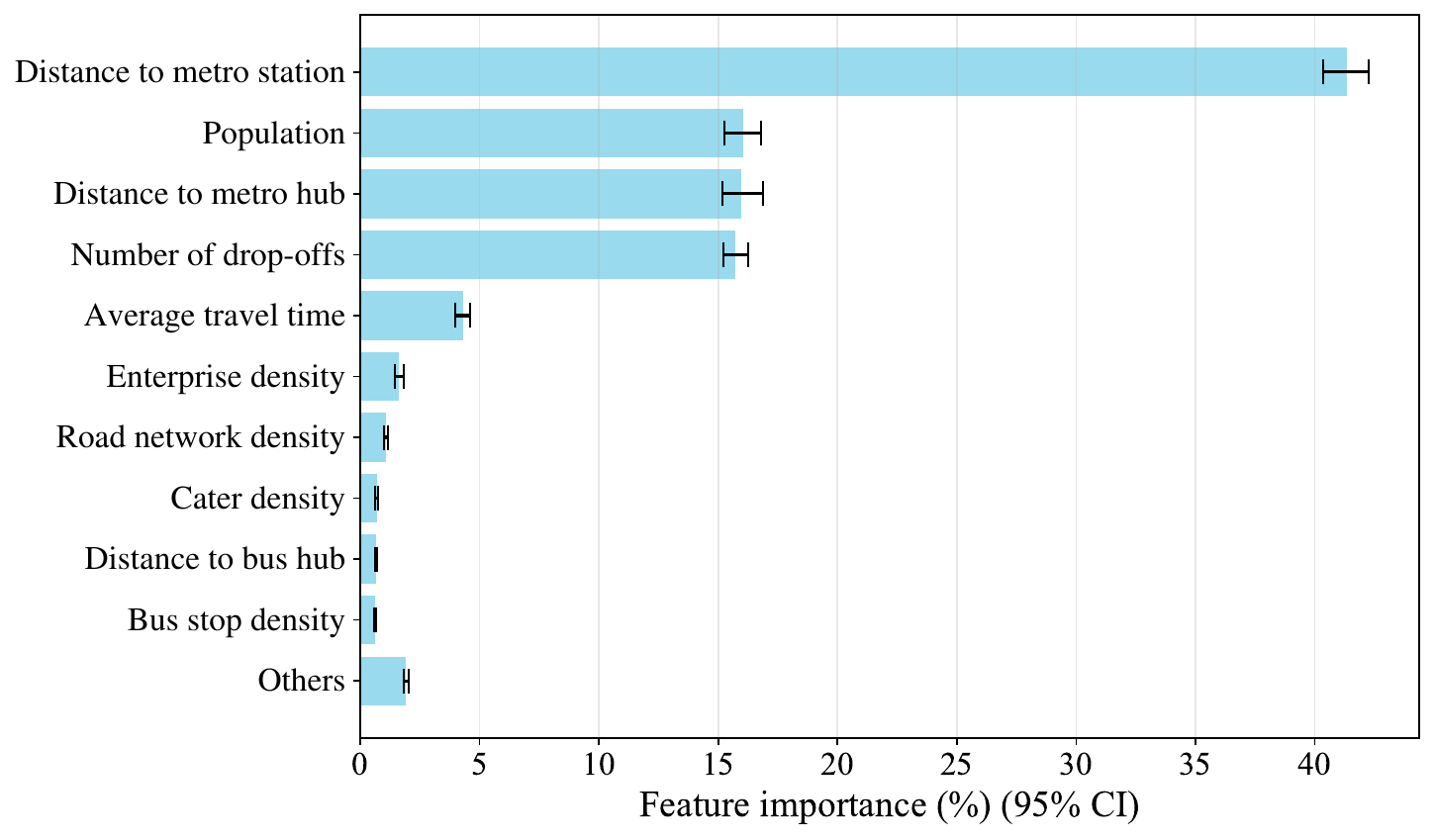}
        \caption{Last-mile complementary ratio}
        \label{fig:Last-mile complementary ratios SHAP}
    \end{subfigure}
    \begin{subfigure}[b]{0.95\textwidth}
        \centering
        \includegraphics[width=.475\textwidth]{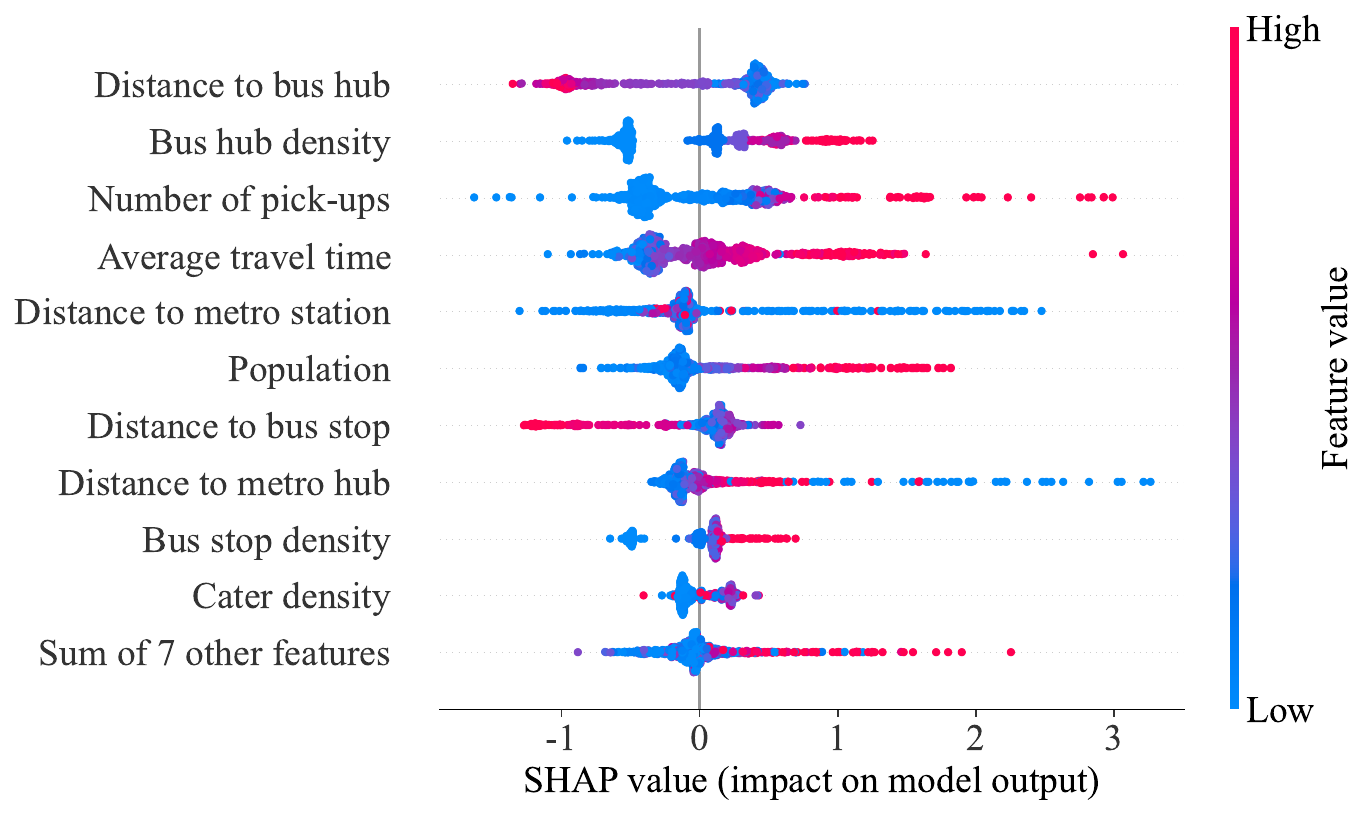}
        \includegraphics[width=.475\textwidth]{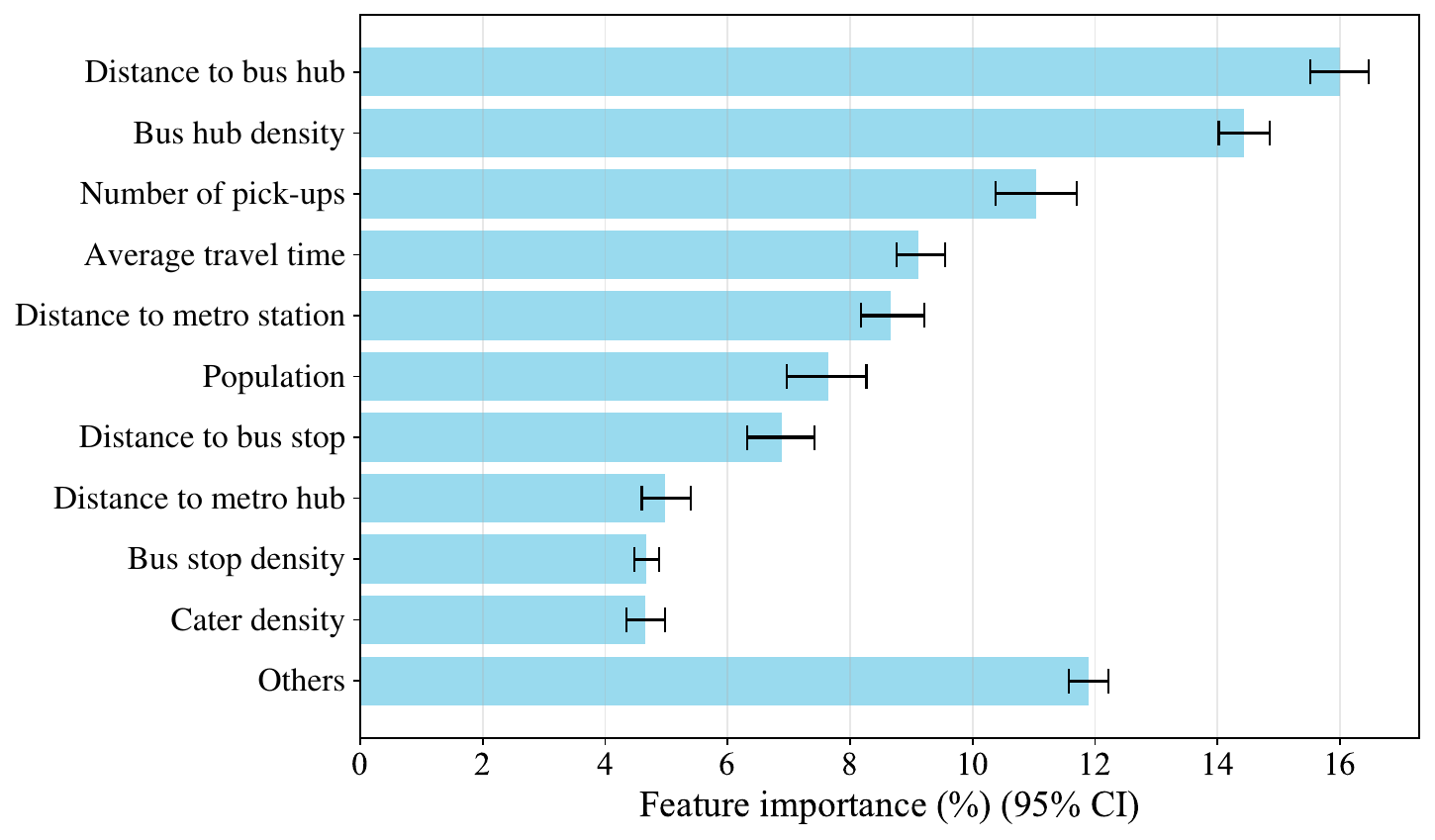}
        \caption{Departure substitutive ratio}
        \label{fig:Departure substitutive ratios SHAP}
    \end{subfigure}
    \begin{subfigure}[b]{0.95\textwidth}
        \centering
        \includegraphics[width=.475\textwidth]{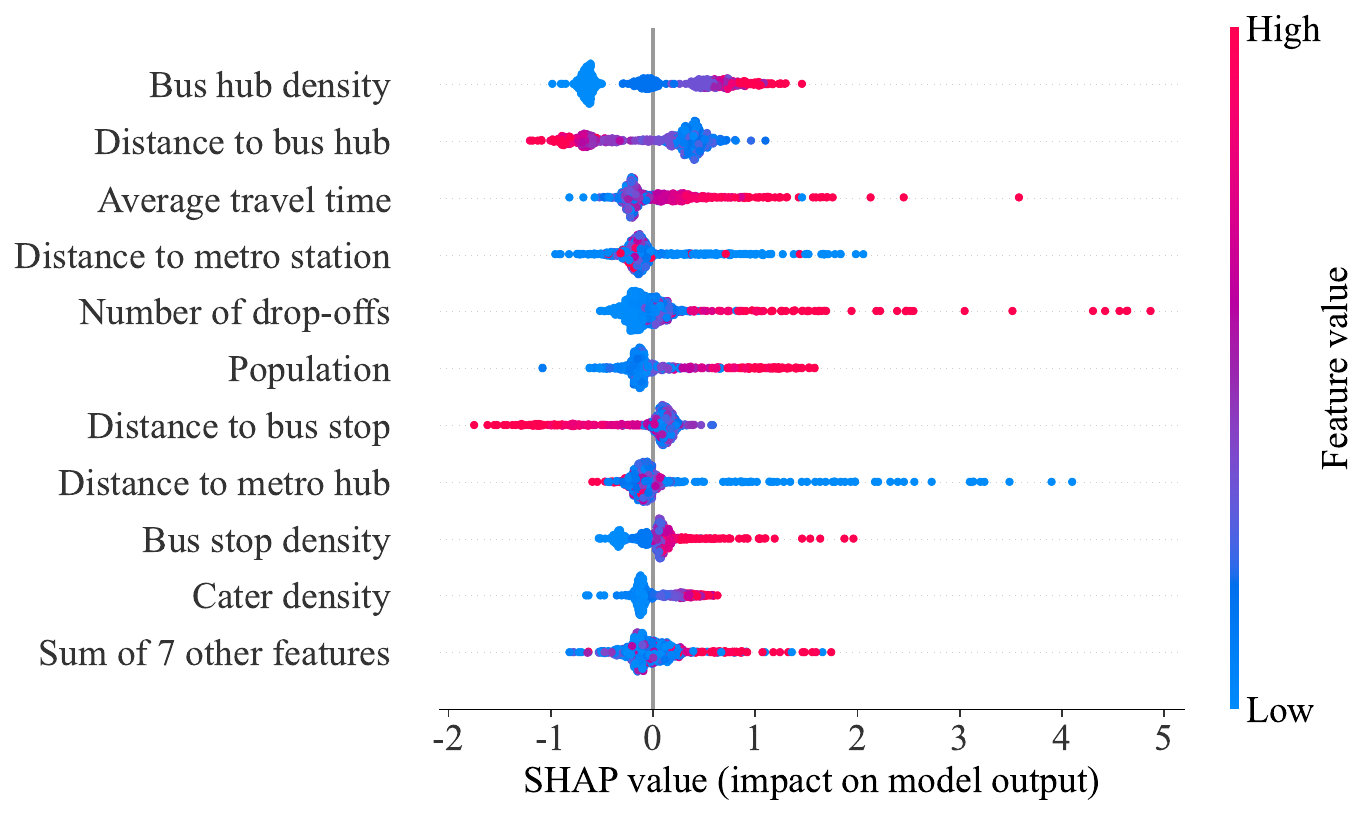}
        \includegraphics[width=.475\textwidth]{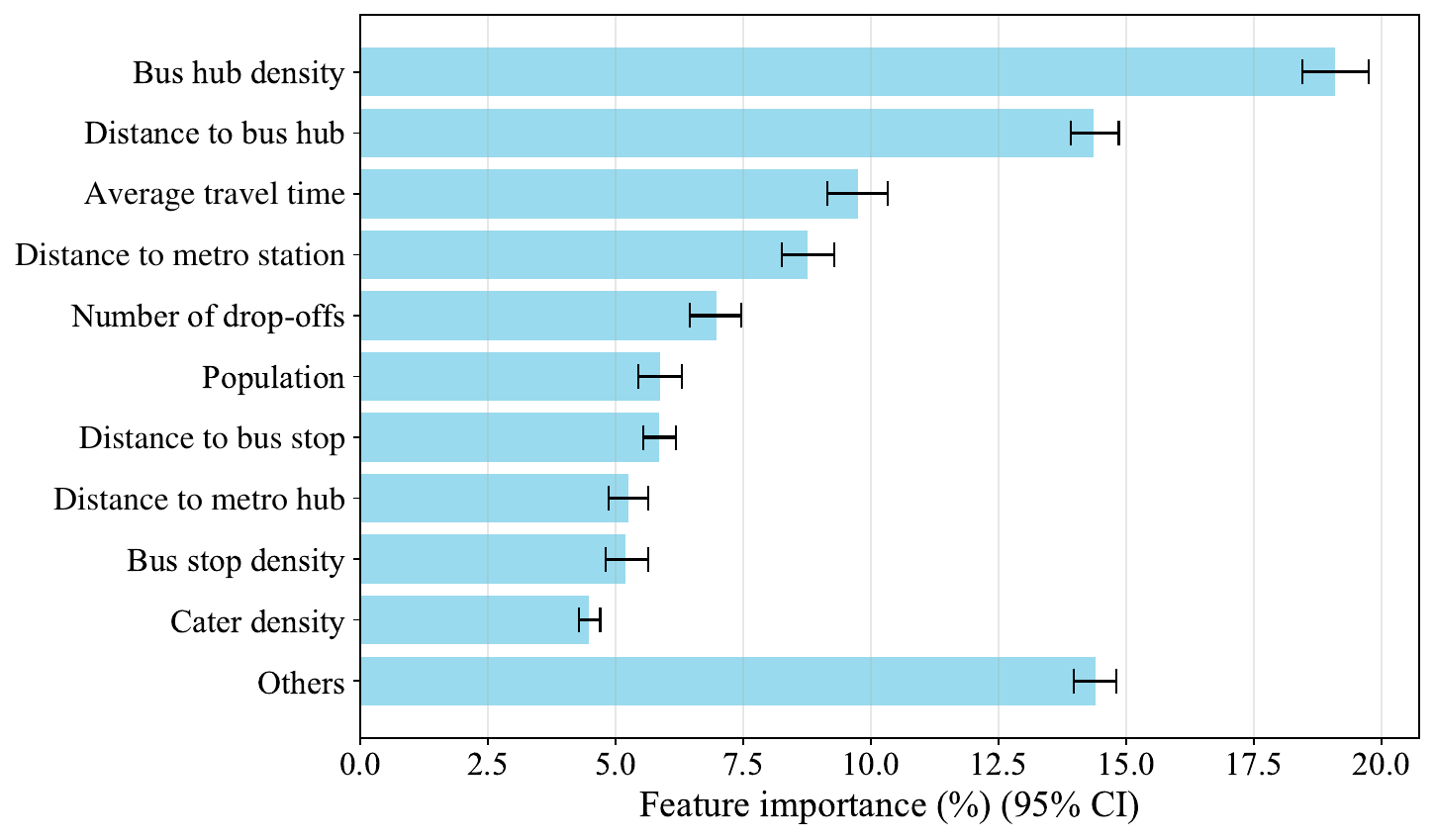}
        \caption{Arrival substitutive ratio}
        \label{fig:Arrival substitutive ratios SHAP}
    \end{subfigure}
    \caption{{SHAP beeswarm diagrams (left). The color represents the value of the feature from low to high. The abscissa denotes each feature's contribution to the output. Charts of the relative importance (right) represent the relative contributions of the features to the model output.}}
    \label{fig:SHAP value scatter plot}
\end{figure}

Independent variables’ SHAP values (calculated by \eqref{eq: shap_calc}) corresponding to the four dependent variables (first- and last-mile complementary ratios, departure and arrival substitutive ratios) and their relative importance (calculated by \eqref{eq: impor_calc}) are shown in \Figref{fig:SHAP value scatter plot}. 
In the beeswarm diagrams on the left, the color of each dot indicates the value of the corresponding feature, ranging from blue (low values) to red (high values). 
Each dot’s horizontal position represents the SHAP value of the corresponding feature, indicating its contribution to the prediction of a specific data point. 
Thus, one can see how large or small feature values (reflected by the dot color) have a positive or negative impact on the prediction (reflected by the SHAP value on the x-axis). 
The beeswarm format piles up data points to reflect how often a given SHAP value occurs.
The bar charts on the right illustrate the relative variable importance for each feature, which is calculated by \eqref{eq: impor_calc}. To ensure the robustness and stability of these rankings, the importance is determined by tracking how frequently each variable appears in the top positions across 1,000 bootstrap resamples. The error bars represent the 95\% confidence intervals derived from the bootstrap process.
Both the beeswarm diagrams and bar charts only display the top 10 most important variables to highlight the key factors influencing the predictions.

Regarding the first-mile complementary ratio, the distance to the nearest single-line metro station has the greatest impact, contributing above 30\%. The beeswarm diagram further illustrates that high values of the variable (represented by red dots) are generally associated with lower SHAP values, indicating that the variable contributes negatively to the first-mile complementary ratio. 
The distance to the nearest multi-line metro hub appears to be another key factor. 
Interestingly, high values of this variable exhibit both negative and positive SHAP values. 
Moreover, areas with dense populations or a higher number of ride-hailing pick-ups tend to have lower first-mile complementary ratios, a pattern consistent with \Figref{fig:First-mile_Complement_Spatial_rates}. 
The negative SHAP values observed in areas with high densities of enterprises and catering establishments may indicate that ride-hailing in these areas faces intensive competition against alternative first-mile options, such as bike-sharing services or pedestrian facilities \citep{wu2020optimal, luo2021joint}.

As for the last-mile complementary ratio, the top 10 contributing variables and their rankings of contribution are highly similar to those of the first-mile complementary ratio, except that the ``number of pick-ups'' is replaced by ``number of drop-offs'', which is consistent with the distinct nature of first- and last-mile complementary trips. 
The distance to the nearest single-line metro station is still the most influential factor. 
And its SHAP value distribution is also similar to that observed for the first-mile complementary ratio.
Additionally, for the distance to the nearest multi-line metro hub, the red dots are located on the right side of the beeswarm diagram, indicating that high values of this variable contribute positively to the last-mile complementary ratio. 
Note that this contrasts with the same variable's mixed contribution to the first-mile complementary ratio \footnote{
Here is a possible reason for this disparity. First-mile complementary trips peak in the mornings but not evenings (see \figref{fig:Temporal First-mile_Complement}), more likely serving the connections between passengers' homes and transit stations in the morning rush. In contrast, last-mile complementary trips peak in both mornings and evenings (see \figref{fig:Temporal Last-mile_Complement}), implying that those trips in morning rush hours may connect transfer stations or farther stations, instead of the nearest stations, to the passengers' workplaces (see our detailed explanation on this issue in \secref{sec: temporal TNC-PT}). This could explain why the contribution of 'distance to metro hub' to the last-mile complementary ratio is higher than to the first-mile complementary ratio, and why the former contribution is generally positive while the latter is mixed.}.

On the contrary, bus-related variables have the most significant impact on substitutive ratios. For the departure substitutive ratio, the distance to the nearest bus hub emerges as the most influential factor, contributing nearly 16\% of the effect. The SHAP diagram indicates that shorter distances to bus hubs correspond to positive SHAP values, suggesting that TNC trips originating from areas near bus hubs are more likely to replace PT.
Bus hub density has the second largest contribution (above 14\%), with higher densities generally associated with increased substitutive ratios.
This is probably because TNC passengers prefer using ride-hailing over buses due to the latter's low speed and unreliable service. 
The number of TNC pick-ups and population density both play a positive role.
Average travel time also makes a positive contribution, indicating that ride-hailing services maintain a competitive edge in long-distance trips. Given the effects of bus hubs and average travel time, it is advisable to introduce long-distance express bus services to connect regional transportation centers with the central urban area, thereby offering a viable alternative to TNC-based trips \citep{gu2016exploring, fan2018optimal, mei2021planning}.
In terms of the arrival substitutive ratio, most variables exhibit similar effects and SHAP distributions. 
Notably, bus hub density now becomes the most influential variable (above 18\%), followed by the distance to the nearest bus hub.

\FloatBarrier

\subsection{Partial dependence plots}\label{sec: Partial dependence plot}
\begin{figure}
    \centering
    \begin{subfigure}[b]{0.96\textwidth} 
        \centering
        \includegraphics[width=.32\textwidth]{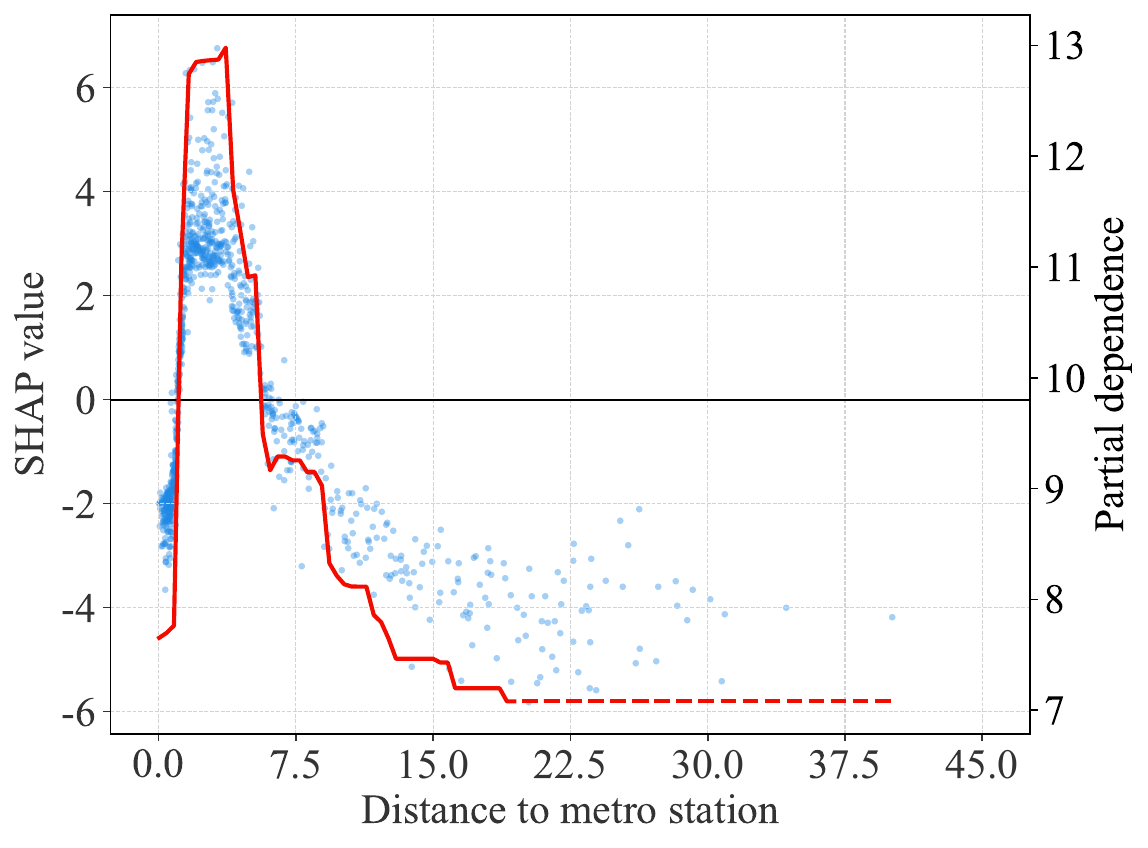}
        \includegraphics[width=.32\textwidth]{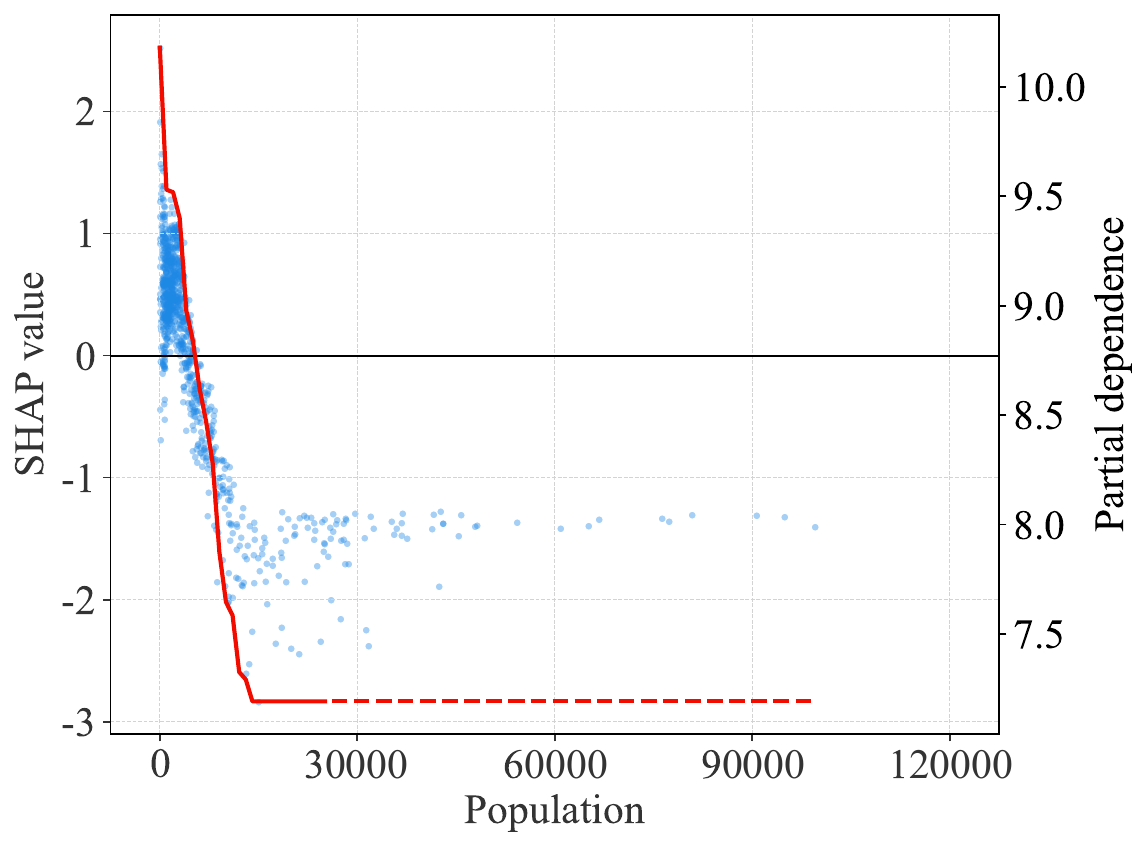} 
        \includegraphics[width=.32\textwidth]{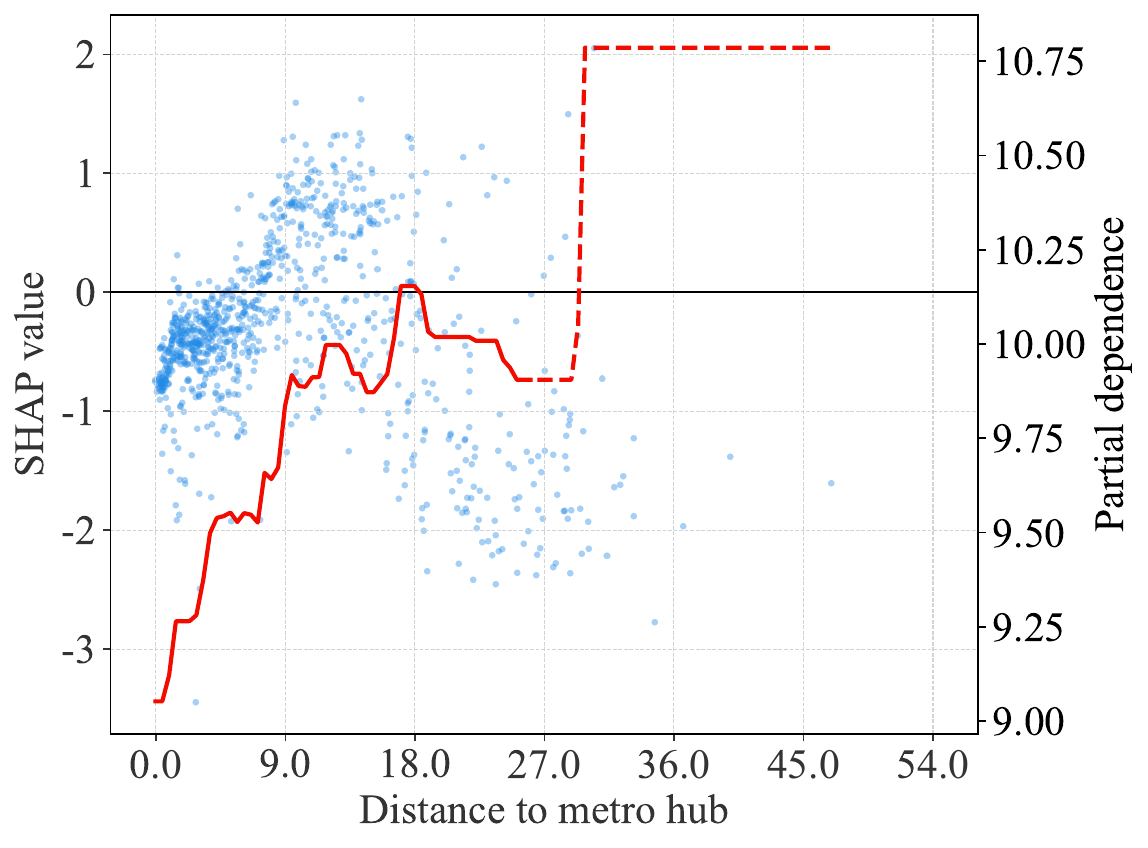} 
        \caption{First-mile complementary ratio}
        \label{fig:First-mile complement partial dependence plot}
    \end{subfigure}
    \begin{subfigure}[b]{0.96\textwidth} 
        \centering
        \includegraphics[width=.32\textwidth]{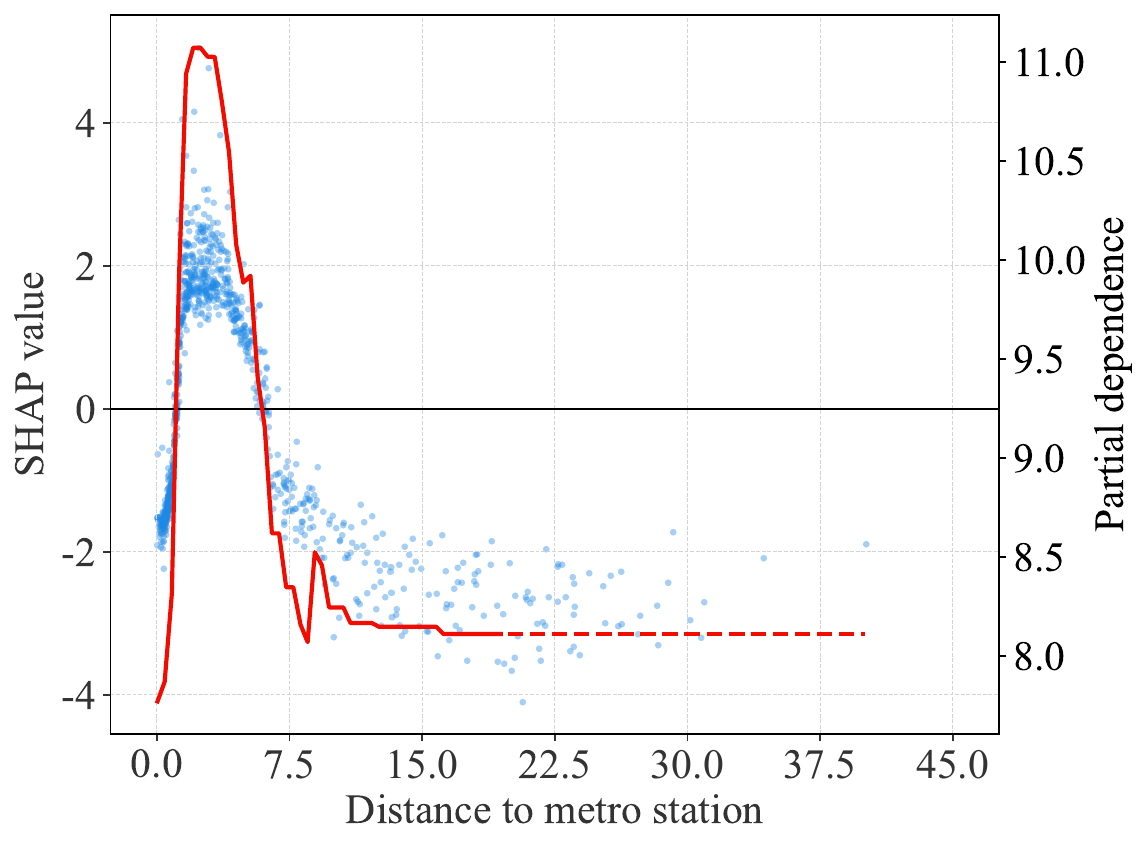}
        \includegraphics[width=.32\textwidth]{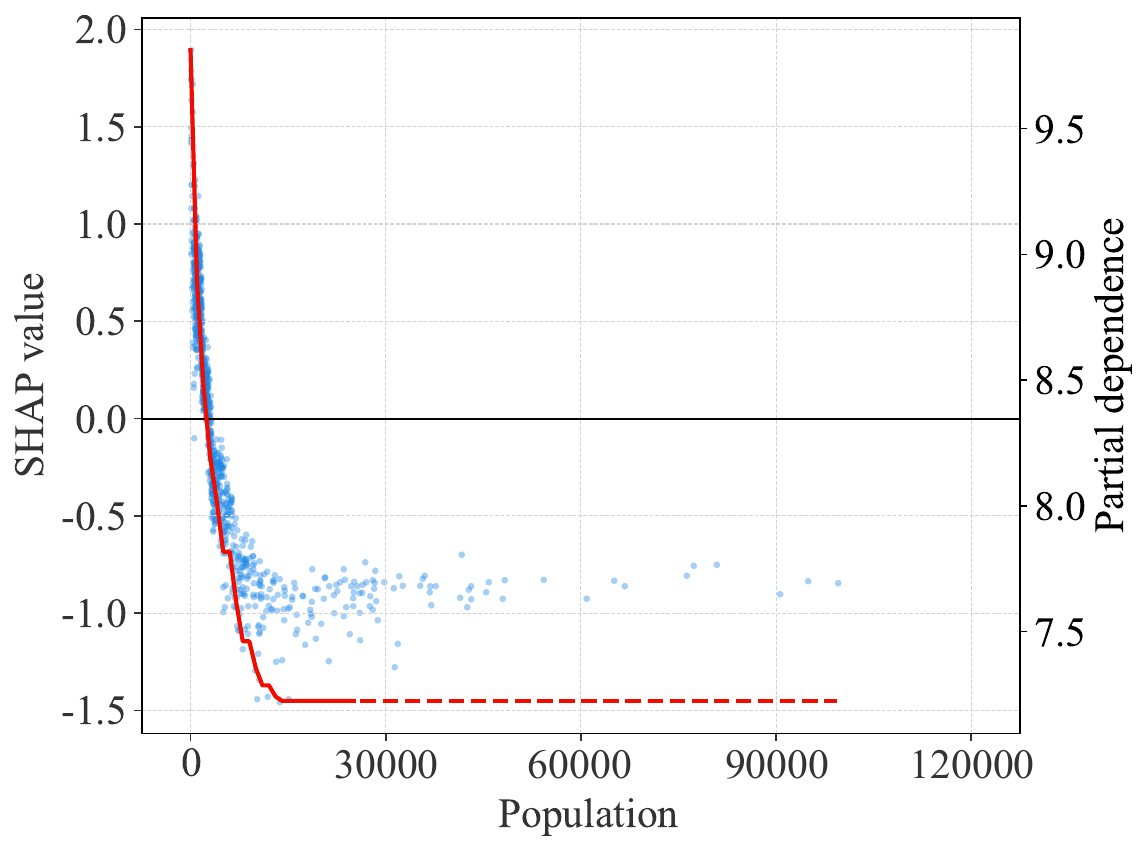} 
        \includegraphics[width=.32\textwidth]{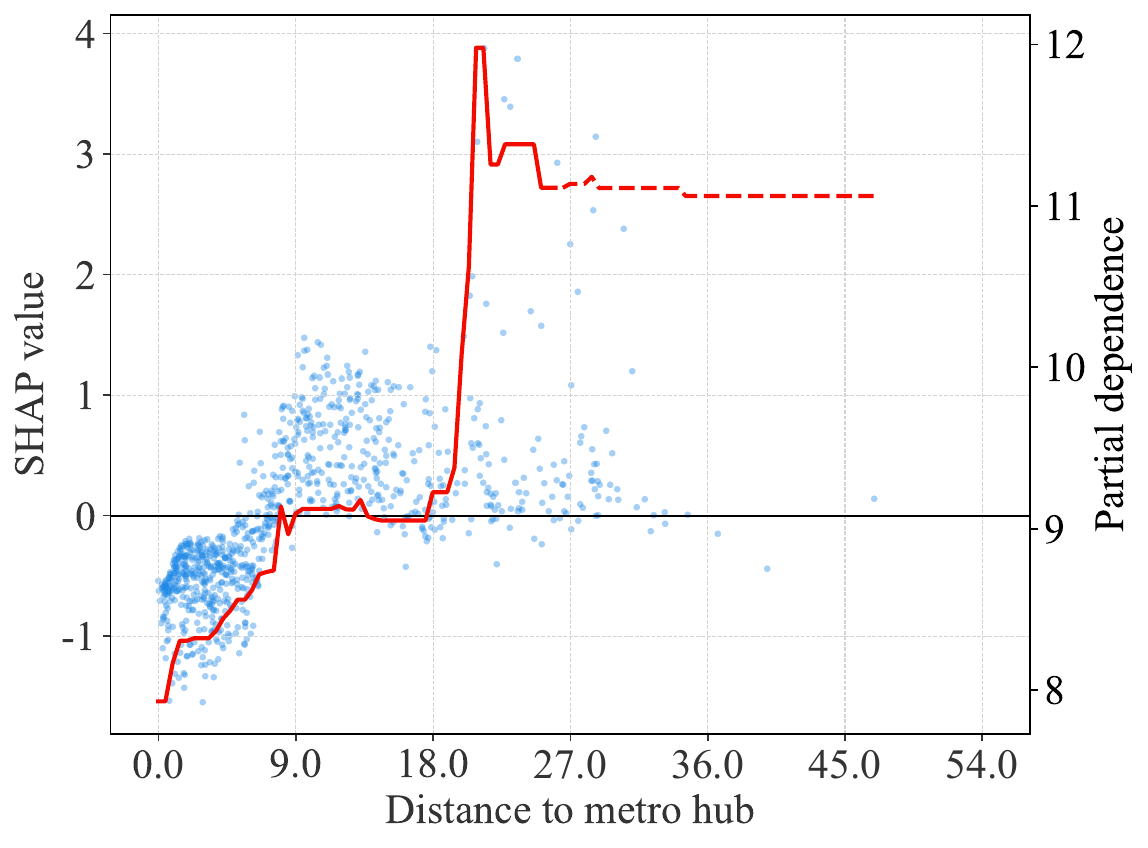} 
        \caption{Last-mile complementary ratio}
        \label{fig:Last-mile complement partial dependence plot}
    \end{subfigure}
    \begin{subfigure}[b]{0.96\textwidth} 
        \centering
        \includegraphics[width=.32\textwidth]{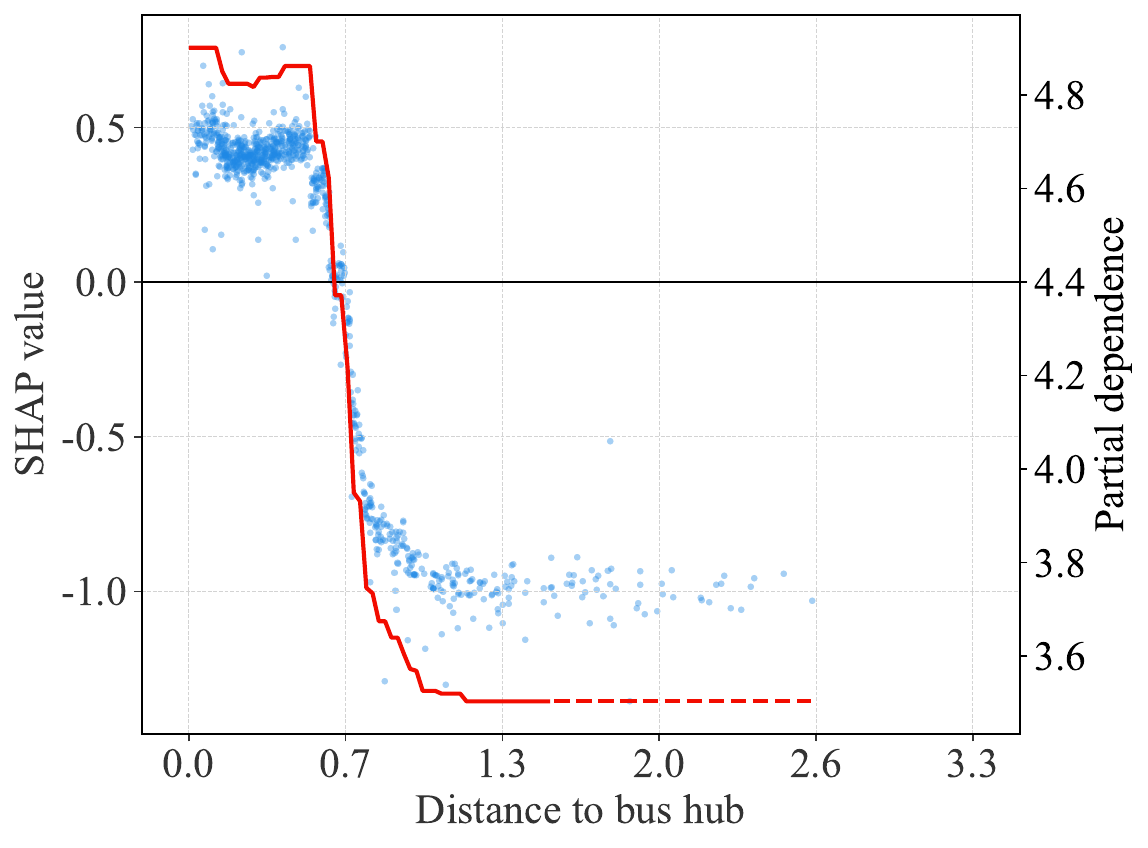}
        \includegraphics[width=.32\textwidth]{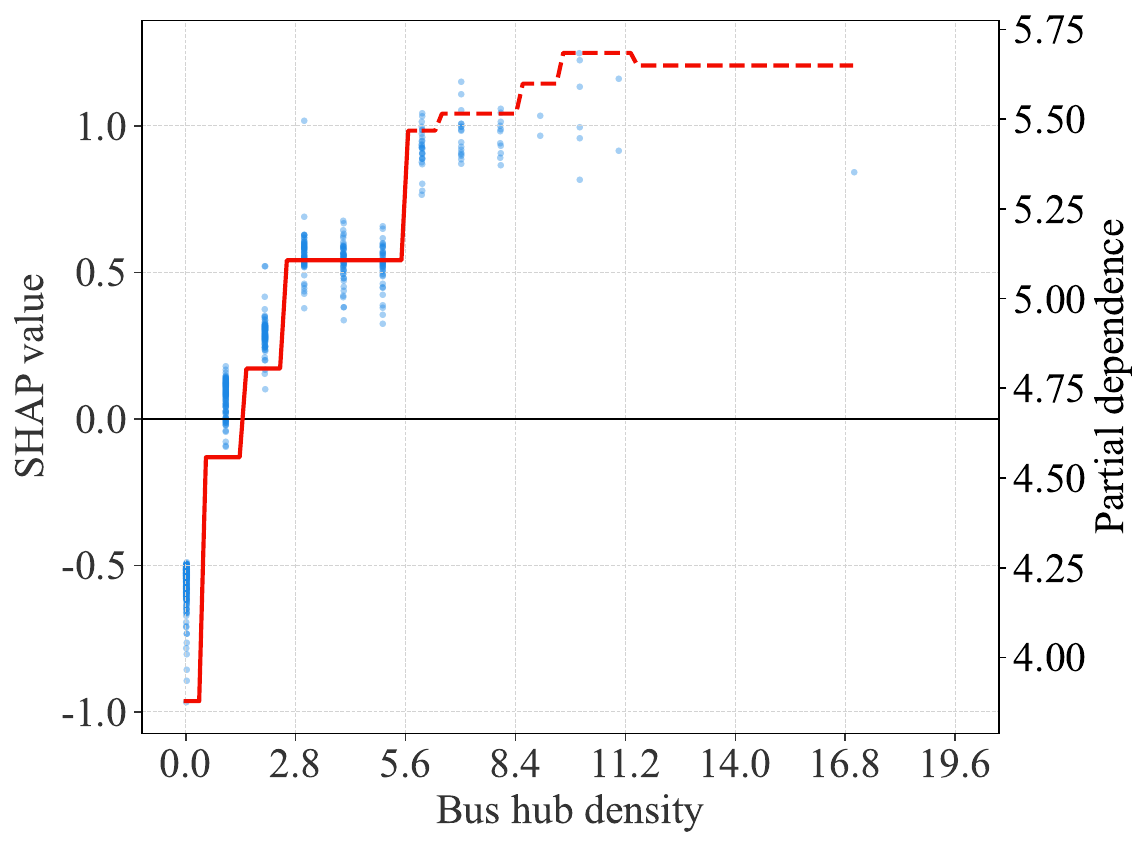} 
        \includegraphics[width=.32\textwidth]{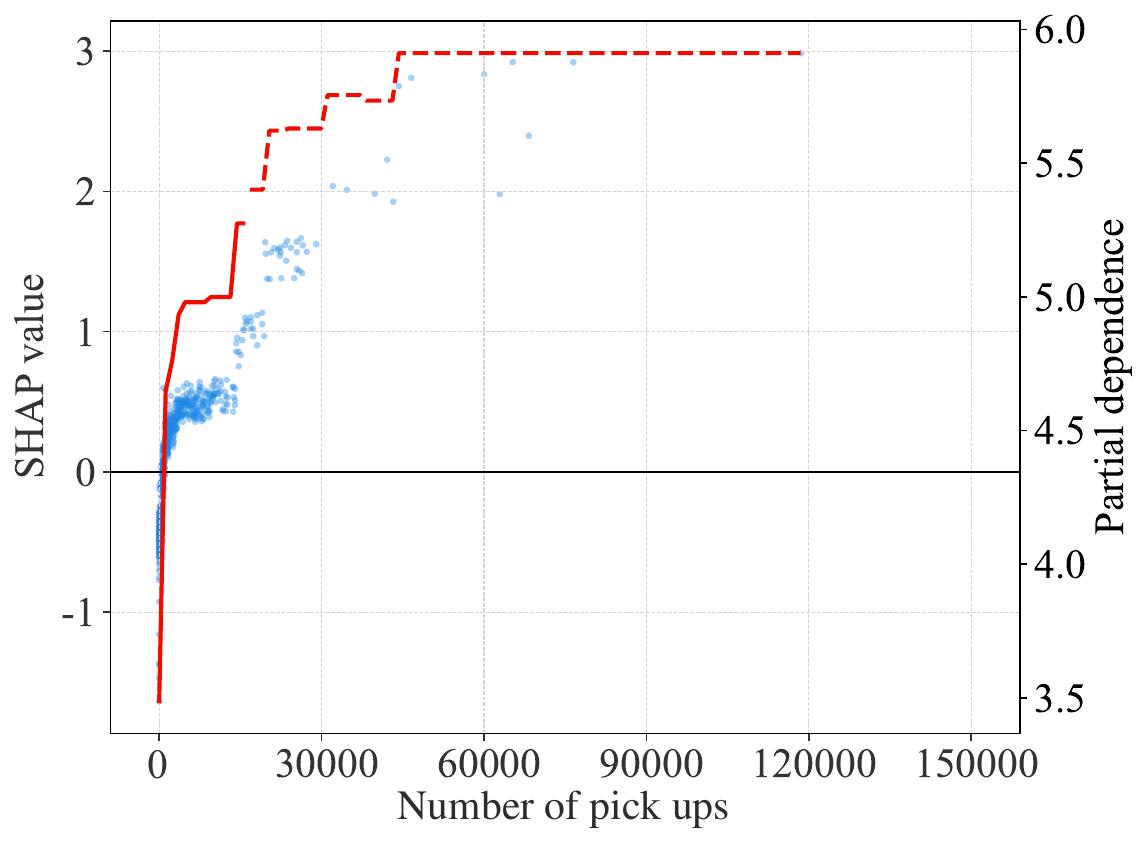} 
        \caption{Departure substitutive ratio}
        \label{fig:Departure substitution partial dependence plot}
    \end{subfigure}
    \begin{subfigure}[b]{0.96\textwidth} 
        \centering
        \includegraphics[width=.32\textwidth]{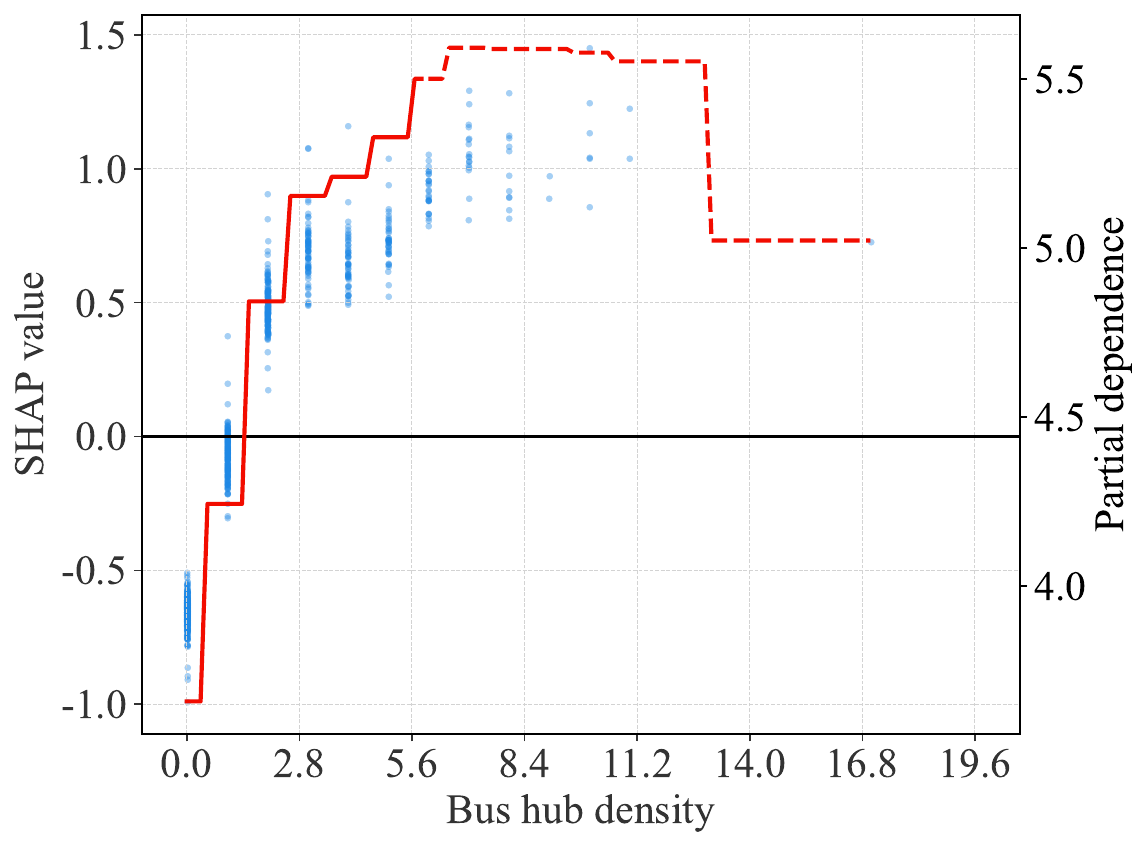}
        \includegraphics[width=.32\textwidth]{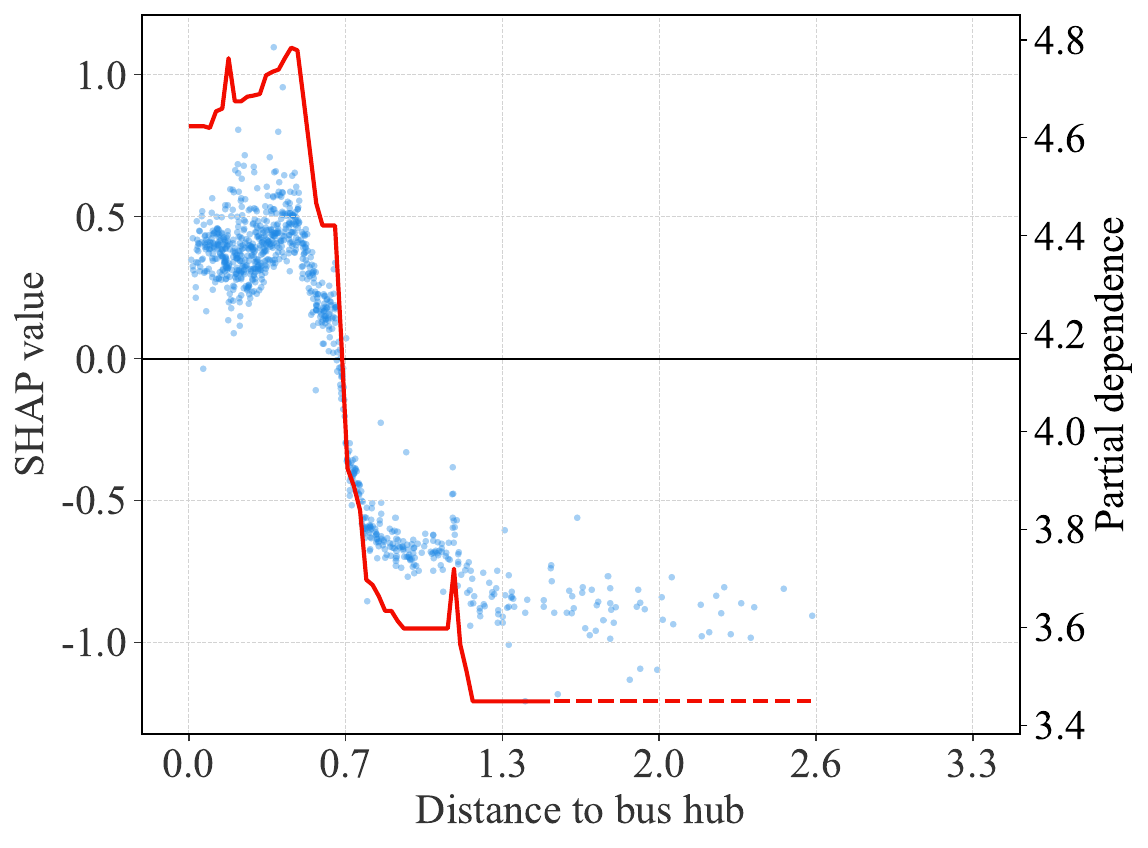} 
        \includegraphics[width=.32\textwidth]{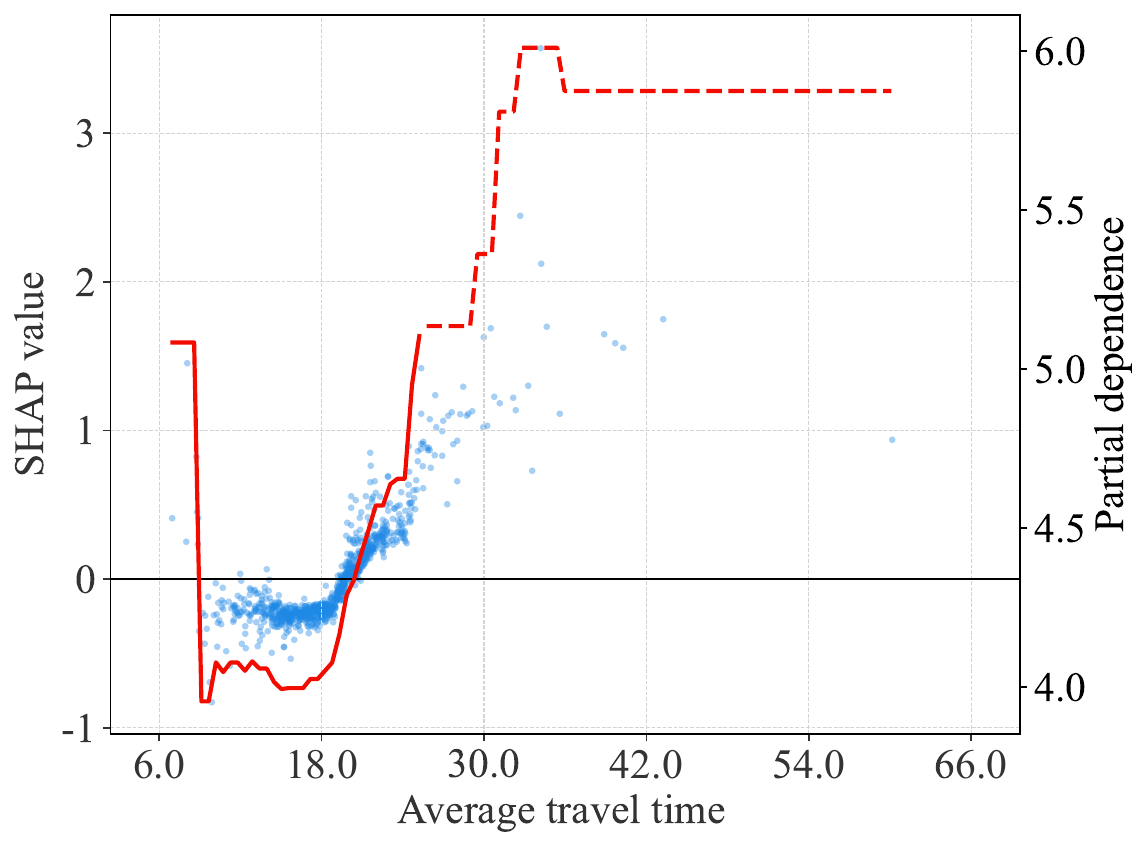} 
        \caption{Arrival substitutive ratio}
        \label{fig:Arrival substitution partial dependence plot}
    \end{subfigure}
    \caption{\centering{SHAP value scatter and partial dependence plots.}}
    \label{fig:SHAP one-dimensional plot}
\end{figure}

Unlike linear spatial regression models, one of the main advantages of machine learning models is excellent nonlinear fitting.
This study selects the top three contributing variables for each dependent variable to analyze their nonlinear effects. 
Results are presented in \manyfigref{fig:First-mile complement partial dependence plot}{fig:Arrival substitution partial dependence plot} for the four dependent variables, respectively. In each subfigure, three diagrams are ranked from left to right by the contribution of the corresponding explanatory variable. In each diagram, the blue dots represent the SHAP values (corresponding to the left y-axis) regarding how different variable values contribute to the model prediction in each observation. 
The red line represents the partial dependence curve (corresponding to the right y-axis), which illustrates how the model’s predicted outcome varies on average when only this particular variable changes (calculated by \eqref{eq:PDP}). 
The red lines are shown in solid and dashed patterns to indicate regions with dense and sparse data points, respectively. 
Considering that the data points are sparse at higher values of each variable, the range for each variable is divided into 100 intervals, with observations exceeding the 95th percentile represented by the dashed segments, and those within this threshold shown by the solid ones. 
Note that estimations may be less reliable in the sparse regions. Therefore, our discussion will primarily focus on the relationships depicted by the solid lines.

As shown in \twofigref{fig:First-mile complement partial dependence plot} {fig:Last-mile complement partial dependence plot}, the SHAP value scatter plots and PDP curves of key variables on both the first- and last-mile complementary ratios are similar. We thus group them in the following discussion.
For the distance to the nearest single-line metro station, its PDPs display an inverted U-shape pattern, revealing that the variable has a strong nonlinear relationship with both ratios. 
Specifically, PDP curves show a sharp increase as the distance grows from 0 to 1.5 km. They then stabilize and peak between 1.5 km and 3 km, indicating the optimal range for TNC trip integration with PT. Beyond 3 km, the curves decline and eventually level off.
A potential explanation for this intriguing trend is that using TNCs as a feeder for metro travel is cost-effective only when the station is within a certain range from passengers' trip origins or destinations.
The combined cost of a short TNC trip and the metro fare is considerably lower than that of taking a TNC service for the entire journey. 
However, as the distance to the metro station increases, the cost advantage of a TNC-metro trip diminishes. For example, if one has to take a taxi for 10 km to reach the metro and then transfer for another 5 km, the total cost is similar to using a taxi for the entire 15 km. 
In other words, the financial incentive to use TNC as a feeder diminishes.
Nonlinear relationships are also present between the population density and both complementary ratios. As population density increases, the corresponding SHAP values and PDP curves decrease sharply and eventually stabilize at a low level beyond approximately 30,000 people per $\text{km}^2$.

Interestingly, the impacts on the two complementary ratios from the distance to the nearest multi-line metro hub differ markedly from those from the distance to the nearest single-line metro station.  
The PDP curves for the former variable demonstrate a generally increasing trend rather than the inverted U-shape pattern seen for the latter variable. 
Specifically, the PDP curves initially increase when the distance is less than 9 km, and then remain roughly flat between 9 km and 18 km. 
Few observations exist when the distance exceeds 18 km, rendering the analysis less reliable. 
These distinct PDP patterns for the two distance variables suggest that different types of metro stations influence TNC-PT integration in unique ways. 
Multi-line metro hubs attract both local and transfer passengers. The latter, especially those undertaking long-distance trips (e.g., cross-district commutes), are typically less sensitive to the distance threshold for using TNCs as feeders. 
For them, taking a TNC directly to or from the transfer station to combine with metro travel offers cost savings while reducing the transfer inconvenience simultaneously. 
In contrast, single-line metro stations primarily cater to passengers with nearby ODs. 
For these users, the decision to use TNC as a feeder is primarily determined by the station’s proximity. In other words, if the distance is very short, they are likely to walk; whereas if it is too long, they tend to use TNCs traveling directly to their destinations, resulting in an inverted U-shaped relationship. 

Regarding both departure and arrival substitutive ratios, the PDP curves of the distance to the nearest bus hub (the left one in \Figref{fig:Departure substitution partial dependence plot} and the middle one in \Figref{fig:Arrival substitution partial dependence plot}) display a three-phase pattern. 
For distances between 0 and 0.5 km, the curves remain relatively flat with high SHAP values, indicating TNC's strong substitution effect.
Beyond this range, both the SHAP values and the PDP curves decline sharply and eventually level off. Once the distance exceeds the range, bus services can no longer serve as an effective substitute for TNCs under the framework discussed in \figref{fig: Framework recognition} and \secref{sec: TNC-PT relationship recognition} due to the excessive walking distance.
Bus hub density also exhibits nonlinear relationships, where the PDP curves increase initially and then stabilize when it comes to nearly 6 stops per $\text{km}^2$.
For other variables, the number of TNC pick-ups shows a direct positive relationship with the departure substitutive ratio. 
Additionally, the average travel time exhibits a distinct nonlinear effect on the arrival substitutive ratio. 
Specifically, from 6 to 9 minutes, the data is limited.
From roughly 10 to 18 minutes, the arrival substitutive ratio remains relatively low. 
As the average travel time increases from 18 to 30 minutes, the ratio rises sharply, indicating that longer TNC journeys have greater substitutive effects.

These nonlinear effects of key explanatory variables on different TNC-PT relationships provide a new point of view to understand the complicated relationship between public transport infrastructure and TNC travel patterns. Meanwhile, these results evidently demonstrate the superiority of machine learning models for exploring how the built environment affects TNC-PT relationships.

A potential concern is whether the identification of these determinants and their nonlinear patterns are sensitive to the thresholds applied for substitutive ratios. 
To verify this, we conducted additional robustness checks by varying the two high-elasticity parameters, namely walking distance and travel time difference.
In order to avoid an overly long main body, the complete results are provided in the supplementary analysis available at \url{https://arxiv.org/abs/2510.19745}.
The results show that, although the substitutive ratio changes across threshold settings, the SHAP-based importance rankings and PDP patterns remain highly stable. 
This indicates that the mechanistic insights provided by our machine learning framework are robust to parameter selection.

\section{Discussion}\label{sec: policy}
This discussion extends the preceding analysis by providing deeper insights into travel behavior dynamics and their policy implications. \secref{sec: OD patterns} advances the spatial analysis from an origin/destination focus in previous sections to a more comprehensive OD flow perspective to reveal the end-to-end spatial patterns of TNC-PT interactions. Next, motivated by the finding in \secref{sec: Partial dependence plot} that bus and metro variables have distinct impacts, \secref{sec: types of complementary trips} further investigates the nuances of complementarity by separating trips connecting to bus versus metro services. 

\subsection{OD patterns of complementary and substitutive trips}\label{sec: OD patterns}
\begin{figure}
    \centering
    \begin{subfigure}[b]{0.4\textwidth}
        \centering
        \includegraphics[height=0.875\textwidth, keepaspectratio]{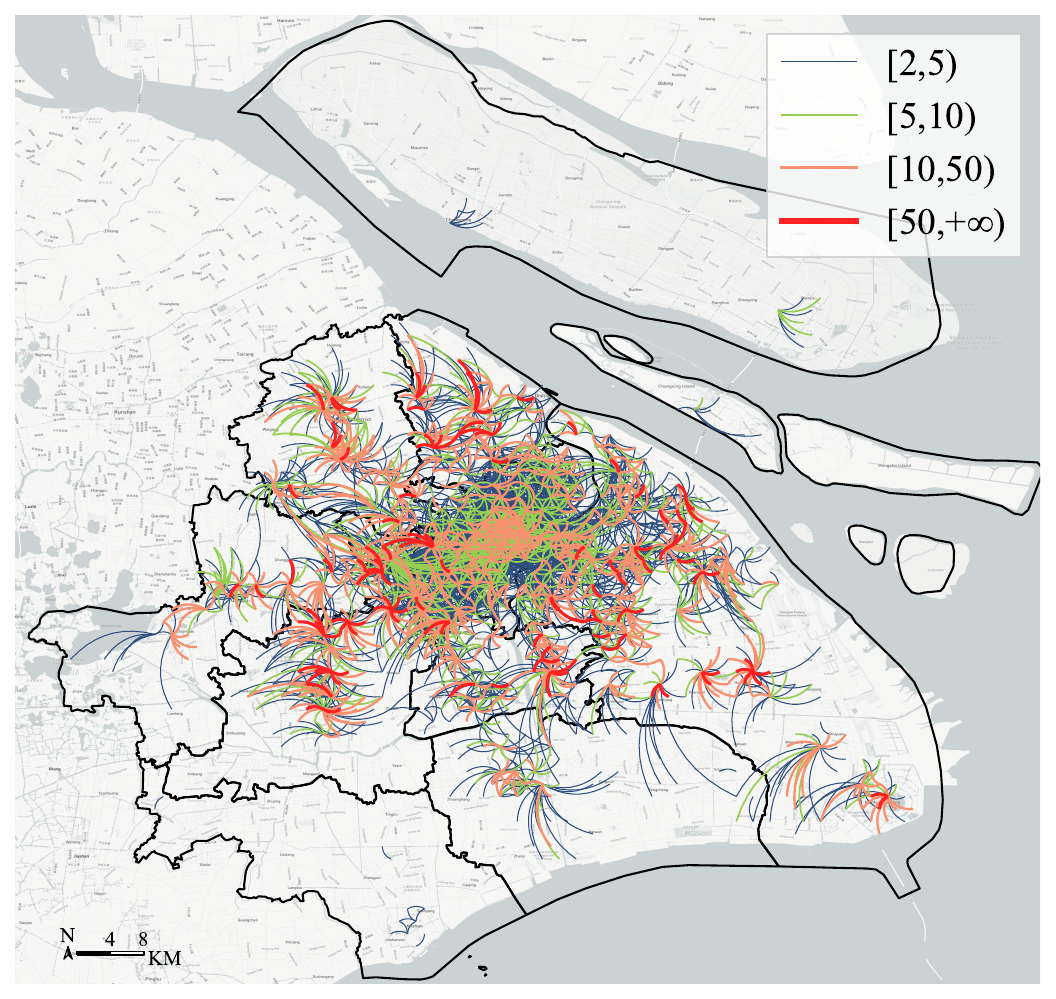}
        \caption{First-mile (grid level)}
        \label{fig:First-mile (grid level)}
    \end{subfigure}
        \hspace{2em}
    \begin{subfigure}[b]{0.4\textwidth}
        \includegraphics[height=0.875\textwidth, keepaspectratio]{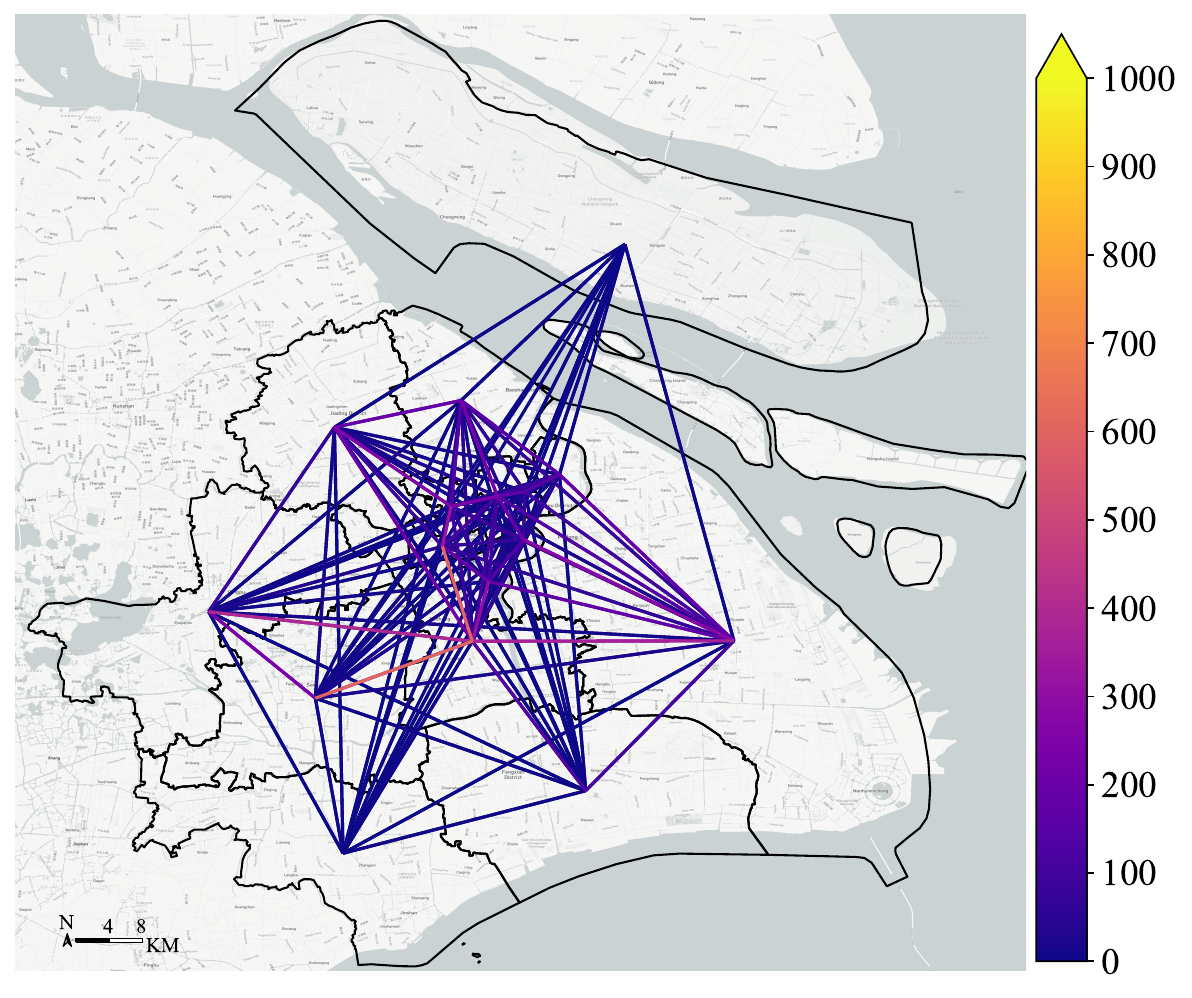}
        \caption{First-mile (district level)}
        \label{fig:First-mile (district level)}
    \end{subfigure}
    \begin{subfigure}[b]{0.4\textwidth}
        \centering
        \includegraphics[height=0.875\textwidth, keepaspectratio]{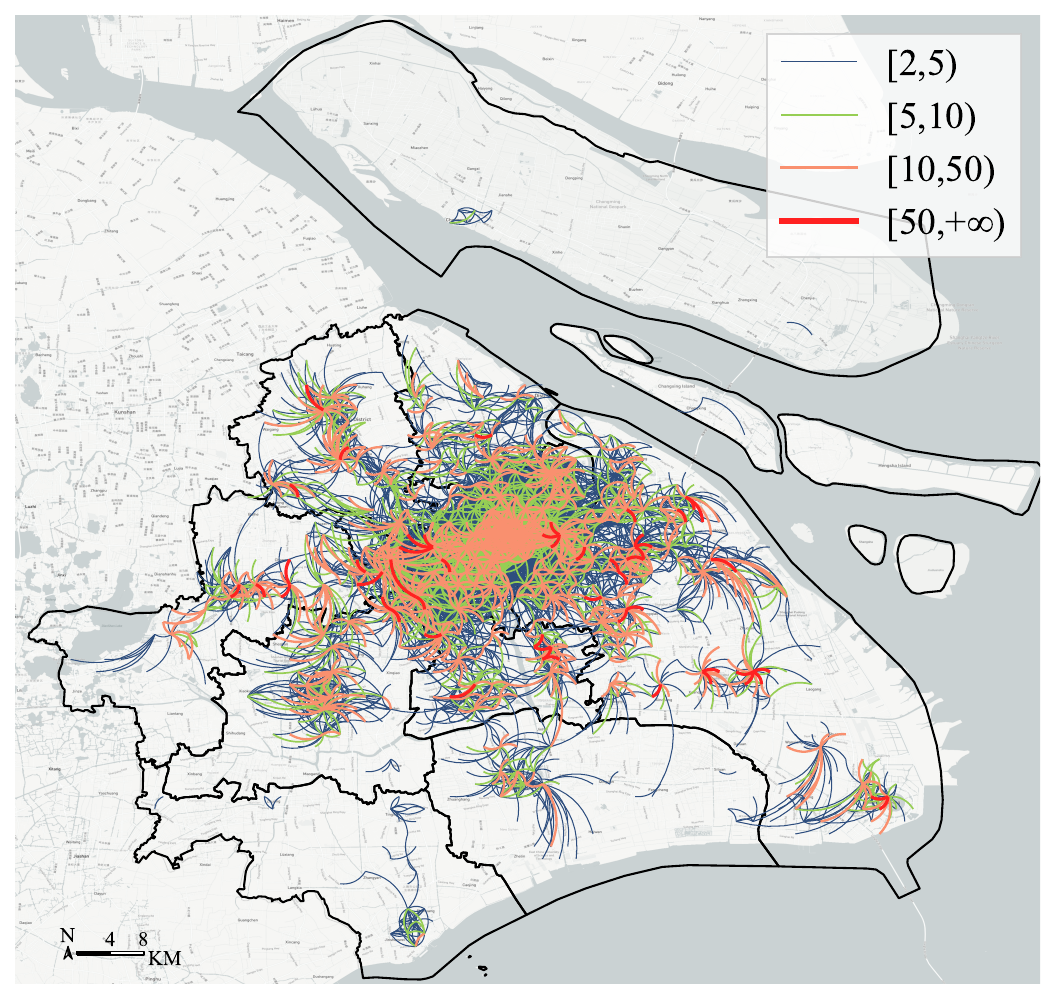}
        \caption{Last-mile (grid level)}
        \label{fig:Last-mile (grid level)}
    \end{subfigure}
        \hspace{2em}
    \begin{subfigure}{0.4\textwidth}
        \includegraphics[height=0.875\textwidth, keepaspectratio]{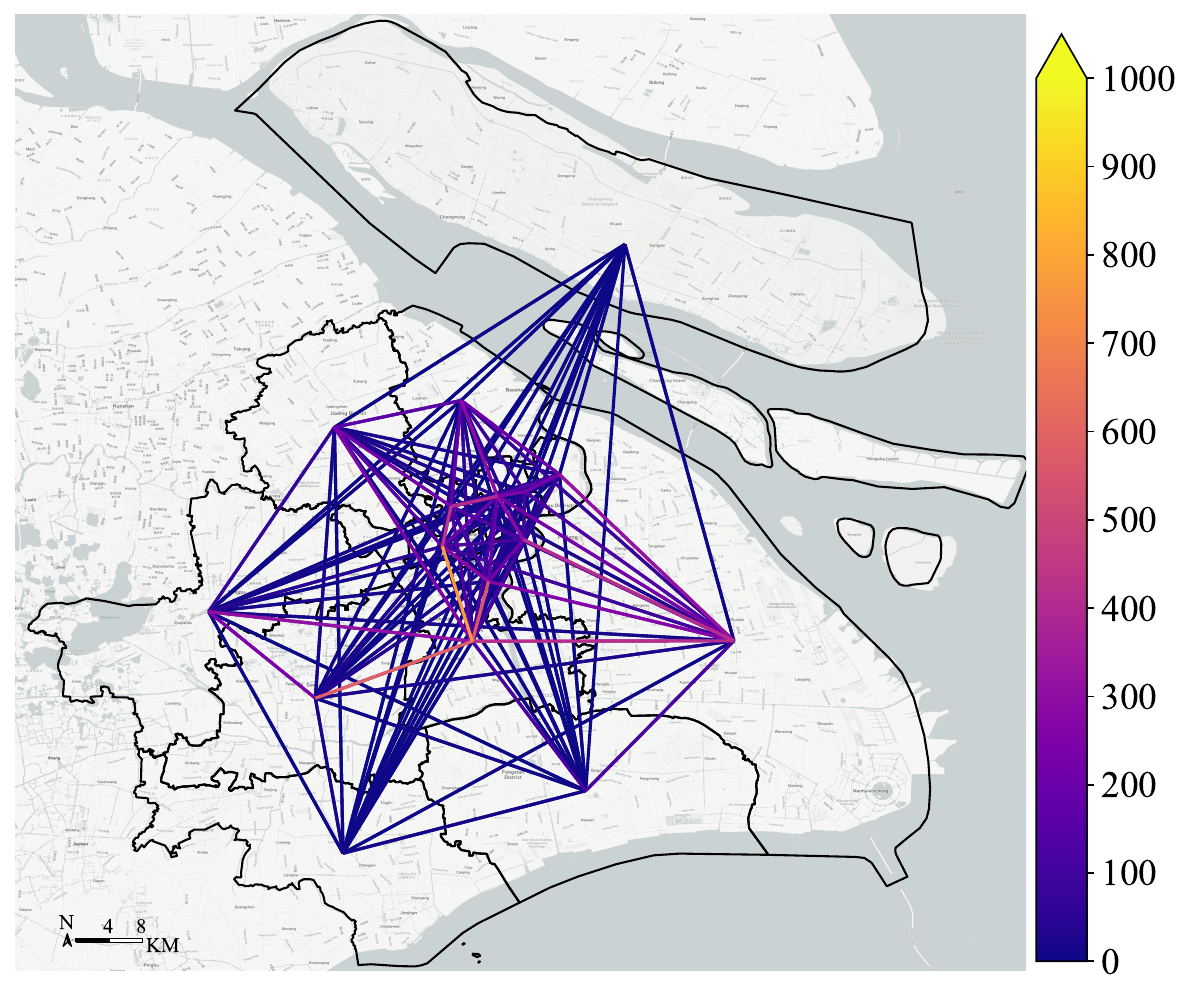}
        \caption{Last-mile (district level)}
        \label{fig:Last-mile (district level)}
    \end{subfigure}
    \caption{\centering{Daily OD flows of complementary trips.}}
    \label{fig:OD patterns of complementary trips}
\end{figure}

\Figref{fig:OD patterns of complementary trips} shows daily OD flows of first- and last-mile complementary trips. 
The left displays the grid-level daily OD flows, while the right aggregates these flows to show travel across different administrative districts in Shanghai. 
As shown in \twofigref{fig:First-mile (grid level)}{fig:Last-mile (grid level)}, most high-volume OD flows (red lines) are short-distance trips.
In addition to the city center, these trips are also concentrated around suburban metro transfer or terminal stations, radiating outward from these stations.
This spatial clustering suggests an opportunity for TNCs to develop specialized, demand-responsive services, such as dedicated shuttles or carpools, to bridge the first- and last-mile gap between these suburban locations and nearby metro transfer stations \citep{fan2024optimal}.
At the cross-district level, \twofigref{fig:First-mile (district level)}{fig:Last-mile (district level)} reveal that the largest complementary trip flows are concentrated in Minhang and Pudong districts. These districts are notable as they host Shanghai’s largest high-speed railway station (Hongqiao Railway Station) and its largest international airport (Pudong International Airport), respectively. The concentration of trips in these areas underscores that a coordinated TNC and public transit system is particularly advantageous for connecting long-distance travelers to railway stations or airports. 

\begin{figure}
    \centering
    \begin{subfigure}[b]{0.4\textwidth}
        \centering
        \includegraphics[height=0.875\textwidth, keepaspectratio]{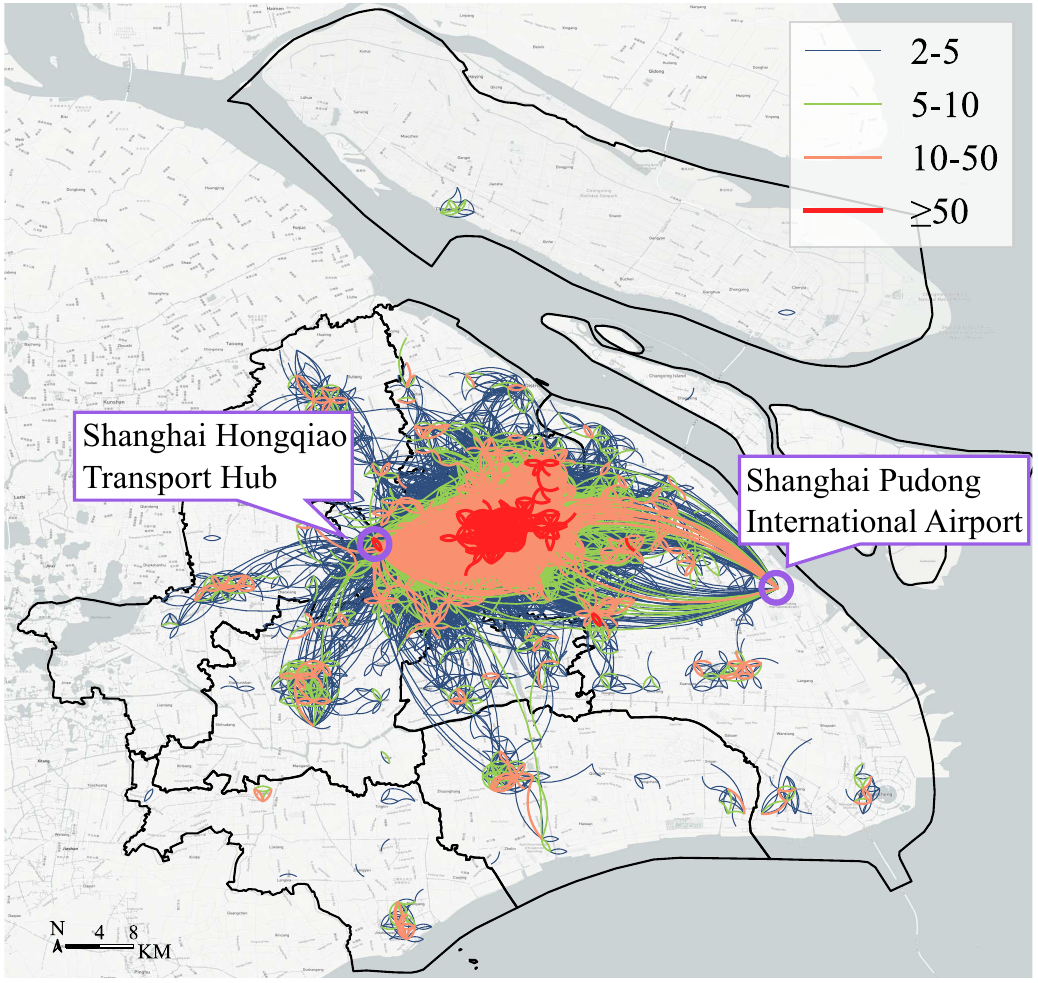}
        \caption{Substitutive trips (grid level)}
        \label{fig: substitutive trips (grid level)}
    \end{subfigure}
    \hspace{2em}
    \begin{subfigure}[b]{0.4\textwidth}
        \centering
        \includegraphics[height=0.875\textwidth, keepaspectratio]{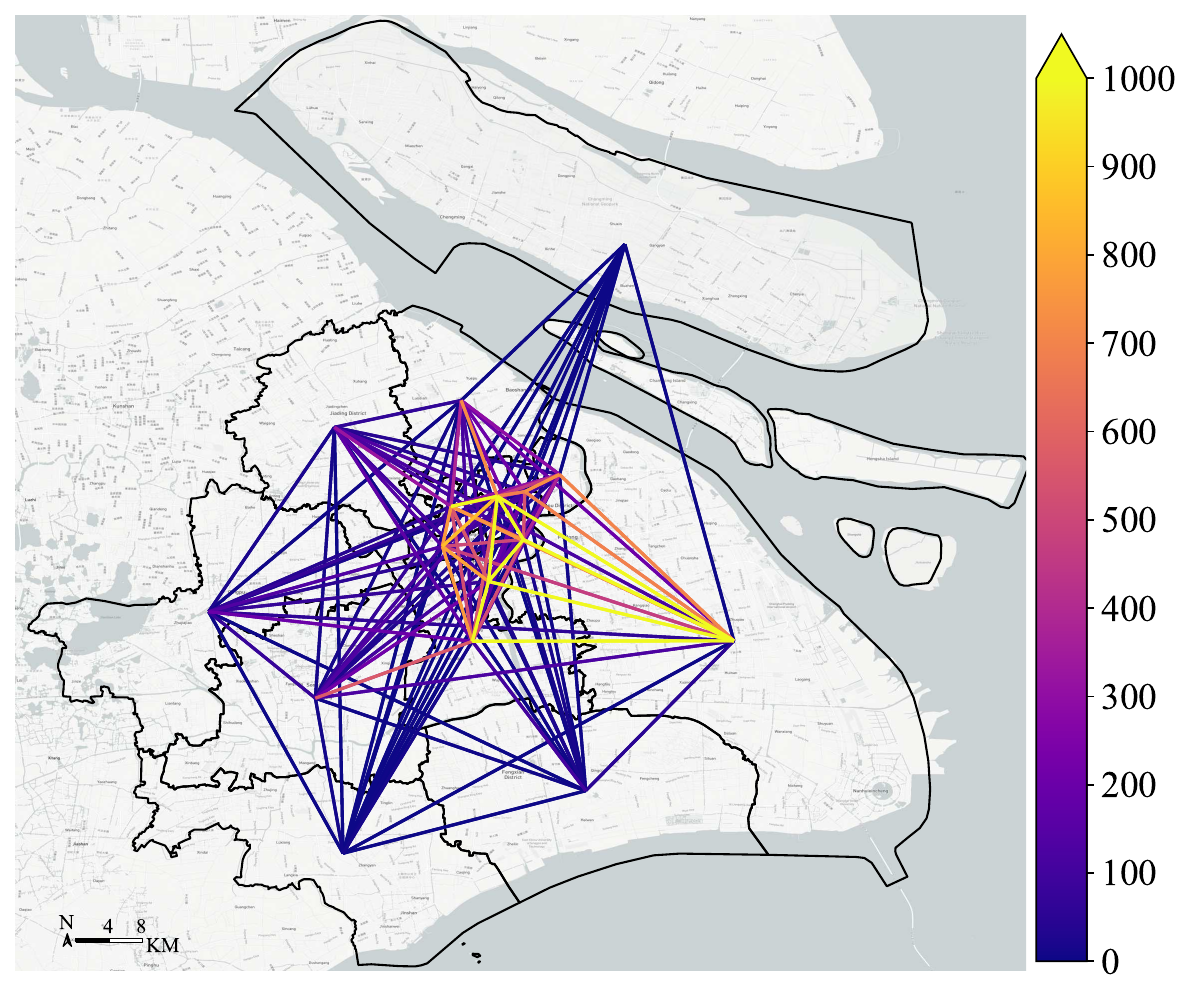}
        \caption{Substitutive trips (district level)}
        \label{fig: substitutive trips (district level)}
    \end{subfigure}
    \caption{\centering{Daily OD flows of substitutive trips.}}
    \label{fig: OD patterns of substitutive trips}
\end{figure}

\Figref{fig: OD patterns of substitutive trips} shows the OD flows of substitutive trips. 
In dense urban areas, more TNC services have PT alternatives, which aligns with the findings in \secref{sec: spatial TNC-PT}. 
Notably, compared to complementary trips, substitutive trips are clearly more clustered around the two major regional transportation centers: the Hongqiao Transport Hub and the Pudong International Airport. 
District-level OD flows further reveal that Pudong District (home to the Pudong International Airport) and Minhang District (home to the Hongqiao Transport Hub) register the highest substitutive trip volumes,  particularly in flows to and from the city center. 
This pattern reveals the fact that passengers traveling to or from these major regional transportation centers are high-speed rail or air travelers who are less sensitive to travel costs and prioritize rapid, reliable, and convenient connections to catch their trains or flights, especially when traveling with family members and luggage.
Based on these findings, transit planners may enhance the existing high-speed metro connections and dedicated bus shuttle services to these regional transportation centers to promote the use of PT services for connection. 

\subsection{Complementary trips further classified by connected PT stations}\label{sec: types of complementary trips}
\begin{figure}
    \centering
    \begin{subfigure}[b]{0.7\textwidth}
        \centering
        \includegraphics[width=\textwidth]{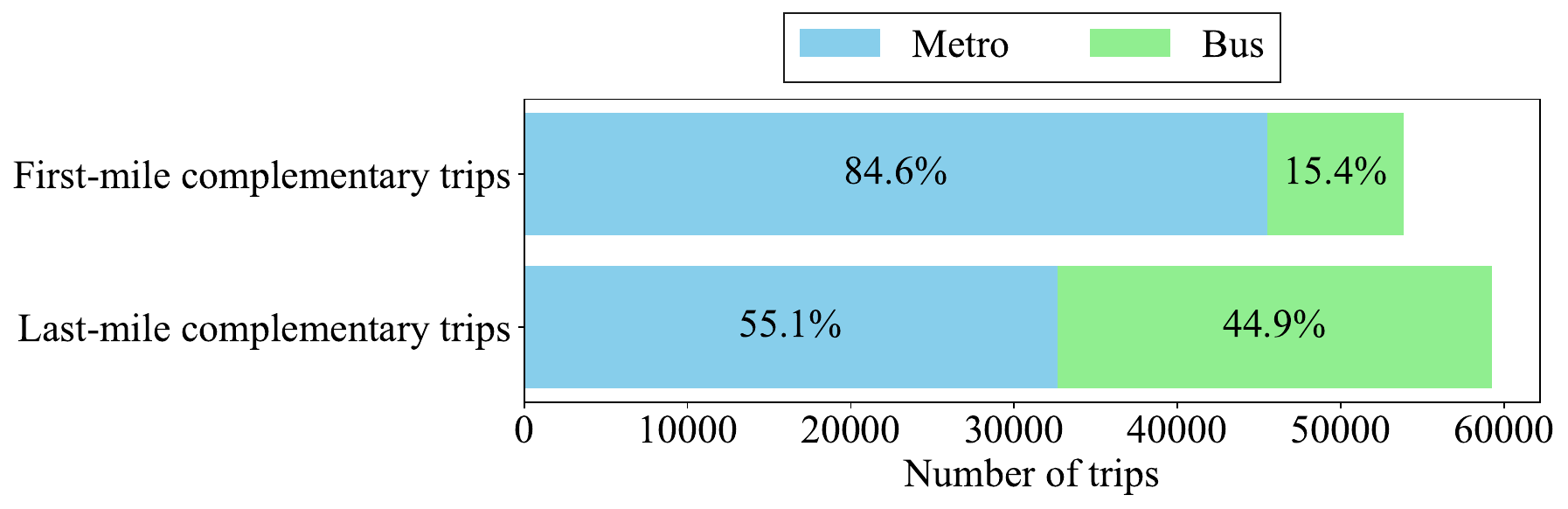}
        \caption{Percentage of complementary trips for bus and metro}
        \label{fig:Percentage_complement_bus_and_metro}
    \end{subfigure}
    \begin{subfigure}[b]{0.7\textwidth}
        \centering
        \includegraphics[width=\textwidth]{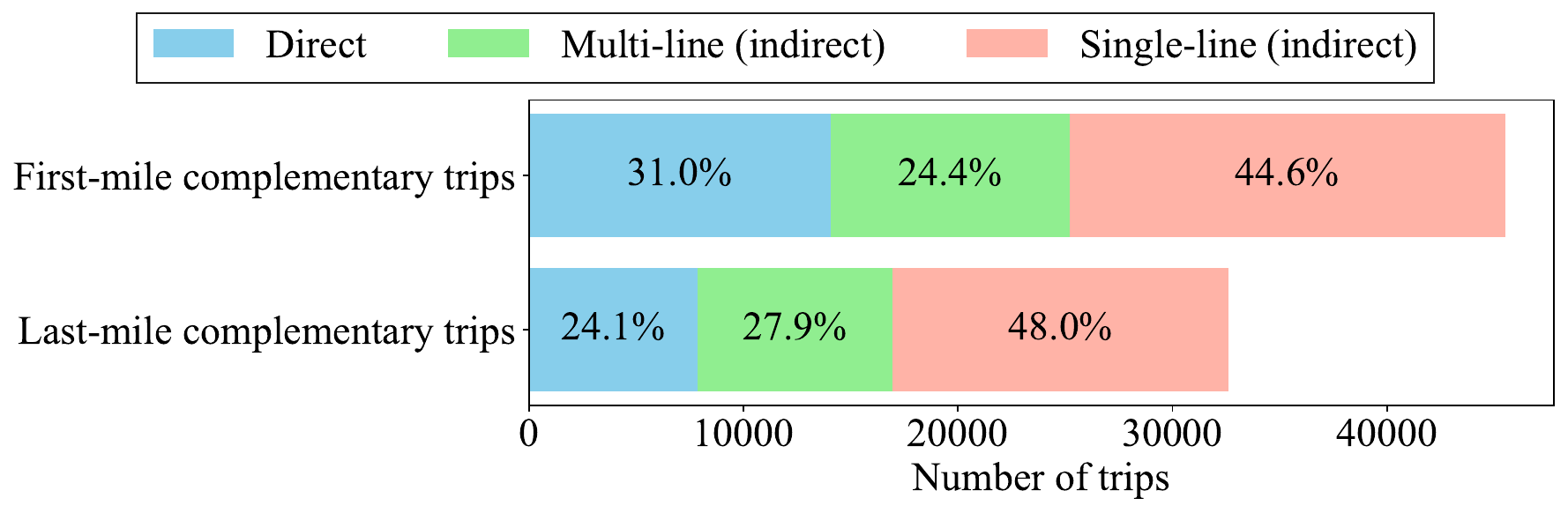}
        \caption{Types of complementary trips to metro}
        \label{fig:Types_complement_metro}
    \end{subfigure}
    \caption{\centering{Types of complementary trips.}}
    \label{fig:Types of complementary trips}
\end{figure}

\Figref{fig:Percentage_complement_bus_and_metro} further divides each of the first- and last-mile complementary trips into two trip types based on whether they connect to bus or metro services. For first-mile complementary trips, 84.6\% connect to metro versus 15.4\% to buses, whereas for last-mile trips the split is 55.1\% to metro and 44.9\% to buses. 
The significant disparity between the compositions of first- and last-mile trips might be due to the fact that last-mile trips have an evening peak while first-mile trips don't (see our findings in \secref{sec: temporal TNC-PT}). In the evening peak, last-mile TNC trips are more likely connected to bus trips because commuters are typically less time-sensitive during evening peak periods.

\Figref{fig:Types_complement_metro} further categorizes metro-connected complementary trips by whether the connection involves the station closest to the origin or destination.
For first-mile complementary trips, a “direct” connection means that the TNC trip destination is the nearest metro station to the trip’s origin. 
For last-mile complementary trips, it means that the TNC trip origin is the station closest to the destination. 
Other trips are classified as “indirect” and further divided into those involving multi-line transfer hubs or further single-line stations.
Specifically, first-mile trips consist of 31.0\% direct, 24.4\% indirect with multi-line stations, and 44.6\% indirect with single-line stations, while last-mile trips are 24.1\% direct, 27.9\% indirect with multi-line stations and 48.0\% indirect with single-line stations. 
The high proportion of indirect trips—nearly 70\% for first-mile and over 75\% for last-mile—is a significant finding.
It reveals that a majority of travelers bypass their nearest metro station, instead choosing to connect to a more distant station (likely on a different line) or a multi-line hub to reduce transfers in the PT network and improve overall travel convenience, which echoes the PDP of the distance to the nearest multi-line metro hub in \secref{sec: Partial dependence plot}.

These findings are also consistent with existing theories and findings.
 Prior studies suggest that major transit hubs, due to their higher accessibility and network connectivity, exert stronger attraction than sign-line stops \citep{Bertolini01051999}. 
 At the same time, the transfer-penalty literature shows that travelers may be willing to accept a longer access trip if doing so reduces transfers \citep{guoAssessingCostTransfer2011, garcia-martinezTransferPenaltiesMultimodal2018}. 
 In this light, the high share of indirect complementary trips observed in our study suggests that many travelers do not simply connect to the nearest station, but instead choose a more attractive hub or a farther station that offers a more convenient overall path. 

\section{Conclusions}\label{sec: conclusion}
Leveraging a dataset comprising over 16 million TNC trips collected in Shanghai in September 2022, the research employed a data-driven framework to identify the relationship between TNCs and PT. In addition to the unique dataset in terms of its size, coverage, and timing, our work distinguishes itself from previous research by incorporating a few modeling innovations, such as the use of location labels instead of spatial coordinate matching to identify complementary trips, which improves classification accuracy. Novel machine learning models are employed to capture complex nonlinear effects of various spatial factors on TNC-PT dynamics. By applying these models to the ratios of complementary and substitutive trips, rather than their absolute volumes, our approach effectively isolates the primary drivers that shape TNCs' functional role relative to public transit. Leveraging the unique strengths of both our dataset and analytical framework, this study yields critical insights that revise the current understanding of TNC-PT interactions while uncovering previously undocumented findings.

First and foremost, this study identifies a significant increase in the complementary role (9.22\%) of TNCs and a relative decrease in their substitutive role (9.06\%) compared to early-stage markets \citep{kongHowDoesRidesourcing2020, meredith-karamRelationshipRidehailingPublic2021, liuInvestigatingRelationshipsRidesourcing2025}. 
While the specific substitutive ratio is subject to classification thresholds, our sensitivity analysis indicates that even under very loose threshold settings, the substitutive ratio rises to at most 22.12\%, ensuring that this main conclusion remains robust.
Temporally, travelers often choose TNC for the first-mile connection during morning rush hours only, whereas the last-mile TNC trips exhibit two peak periods, consistent with the daily commute pattern. On the other hand, substitutive ratios were high throughout the day, reaching their peak in the evening. Spatially, complementary ratios are lower in the urban core and higher in suburban areas, while substitutive ratios show an opposite trend. Furthermore, powered by CatBoost modeling and analyses using SHAP values and PDPs, we identify the nonlinear effects of several key explanatory variables on TNC-PT relationships, such as the distance to the nearest metro stations and bus stop density. For example, the contributions to the first- or last-mile complementary ratios by the distance to the nearest \textit{single-line metro station} show an inverted U-shape pattern (while the contributions by the distance to the nearest \textit{multi-line metro hub} are clearly different); see \secref{sec: Partial dependence plot} for the details.

Our findings offer profound managerial insights for urban transportation authorities and policymakers. 
For example, the unveiled shift in competitive Chinese TNC markets necessitates a reevaluation of ride-hailing market regulations. In short, regulatory bodies should transition from the current approach of limiting TNCs' competition with PT towards fostering and enhancing collaboration and synergy between ride-hailing services and public transport.
Specifically, the following suggestions are proposed to policymakers.
First, in suburban areas and regions with inadequate transit coverage, TNCs should be viewed as an effective extension of the public transit system, not a threat. 
Public authorities could incentivize TNC platforms to provide cost-effective, on-demand feeder services by offering targeted subsidies, designating dedicated pick-up/drop-off zones at metro stations, or promoting integrated payment systems \citep{fan2024optimal}. 
This approach may prove more economically viable than extending traditional fixed-route bus services in low-density areas \citep{zhen_feeder_2024,zhenConnector2026}. 
Second, our findings reveal passengers' high valuation of travel convenience, with many bypassing their nearest station for a more distant transfer station to minimize transfers within the PT network. 
This indicates opportunities to develop novel integrated services such as premium shuttles that connect major residential areas to multi-line metro hubs \citep{jin2024multi, fan_novel_2025}. 
This finding also underscores the importance of integrating these route preferences into design models for unified transit and ride-hailing service networks \citep{luo_integrated_2023}
Finally, for trips to and from major regional transportation centers like airports and high-speed railway stations, the substitutive effect of TNCs remains strong, indicating that PT is less appealing. To alleviate traffic congestion at these critical nodes, authorities should enhance the competitiveness of transit services, for instance, by improving the operational convenience, speed, and reliability of express metro lines or airport shuttles.

Admittedly, our study has its limitations. Notably, the classification framework for substitutive trips relies on empirical rules that incorporate subjective parameter values. 
While such a rule-based approach is common in the literature, it inevitably introduces a degree of potential bias into the results \citep{kongHowDoesRidesourcing2020, meredith-karamRelationshipRidehailingPublic2021, catsDichotomyHowRidehailing2022, pereiraRidehailingTransitAccessibility2024}. 
Our sensitivity analyses also demonstrated the existence of parameter-related biases; see \Appref{appendix: Parameter elasticity analysis}. 
Moreover, despite our rigorous label filtering procedure, a small fraction of trips classified as complementary may still represent terminal trips rather than actual transfers, due to the high density of urban facilities integrated with major transit stations in Shanghai. However, compared with the coordinate-only approaches commonly used in previous studies, our label-based strategy is substantially more conservative and better able to reduce such misclassification.
When applied to other study areas, one should be careful in choosing the classification parameter values and interpreting the results.
Finally, while the comparison with prior Shanghai-based studies supports a shift in TNC-PT interactions, these results are inferred from regional data. Care should be taken when extrapolating this trend to other cities where longitudinal data are currently unavailable.
Future research could extend this work in several promising directions. For example, future work could integrate stated preference surveys to uncover the underlying behavioral mechanisms and trade-offs (e.g., time vs. cost, comfort vs. convenience) that drive travelers' choices between substitution and complement, which are not revealed in the present dataset.
Incorporating emerging modes like shared e-bikes and autonomous vehicles into the analysis would provide a more holistic understanding of the present and future multi-modal transportation ecosystem. Ultimately, key findings from these extensions could inform the development of optimization models for designing more synergistic and socially equitable transit systems.

\section*{Acknowledgment}
This work was partially supported by a grant from the Research Grants Council of the Hong Kong Special Administrative Region, China (PolyU P0051967/RGC 15237624). Dr. Xiaotong Sun would like to thank the support provided by the National Natural Science Foundation of China (72201073). The authors are grateful for the data support provided by the Shanghai Electric Vehicle Public Data Collecting, Monitoring, and Research Center. In addition, the authors express their sincere gratitude to Mr. Ruiguo Zhong in HKUST(GZ) for his invaluable suggestions and expertise in coding.

\appendix
\section*{Appendix}
\section{Identification of complementary trips using passenger-provided location labels}\label{appendix: Identification of complementary trips}
\setcounter{figure}{0}
\renewcommand{\thefigure}{A\arabic{figure}}
In the Shanghai ride-hailing market, pick-up and drop-off points are typically selected from standardized, predefined location labels generated by the platform’s electronic map system, rather than being entered as fully free text.
As illustrated in \Figref{fig:Example of standardized location labels}, when a passenger inputs a keyword such as ``Nanjing West Road'', the app returns a structured list of nearby candidate labels, including metro stations and their entrances/exits, street names, commercial buildings, shopping malls, and hotels. The passenger then selects one final label as the actual pick-up or drop-off point. This standardized label-selection mechanism provides richer semantic information than coordinates alone and makes it possible to infer whether a TNC trip is intended to connect with public transit. If a passenger's actual destination is a nearby hotel, shopping mall, or other non-transit facility, selecting the station label would usually increase walking distance or create confusion for the driver, making it a less convenient choice than simply selecting the specific hotel or mall label provided in the app’s list. 
This issue is particularly relevant for metro stations, because different entrances of the same station can be hundreds of meters apart, rendering selecting a generic station label rather than an accurate POI a suboptimal choice.

\begin{figure}
    \centering
    \begin{subfigure}[h]{0.4\textwidth}
        \centering
        \includegraphics[width=\linewidth,keepaspectratio]{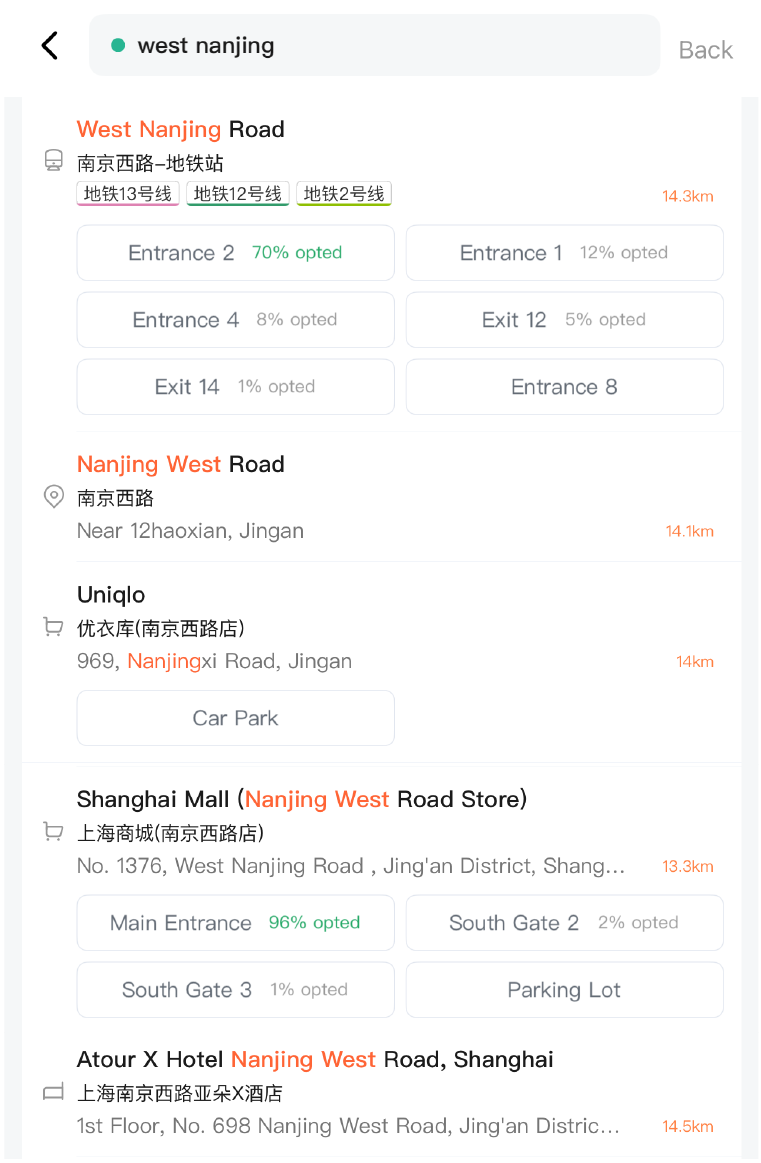}
        \caption{Example of standardized location labels}
        \label{fig:Example of standardized location labels}
    \end{subfigure}
    \begin{subfigure}[h]{0.4\textwidth}
        \centering
        \includegraphics[width=\linewidth,keepaspectratio]{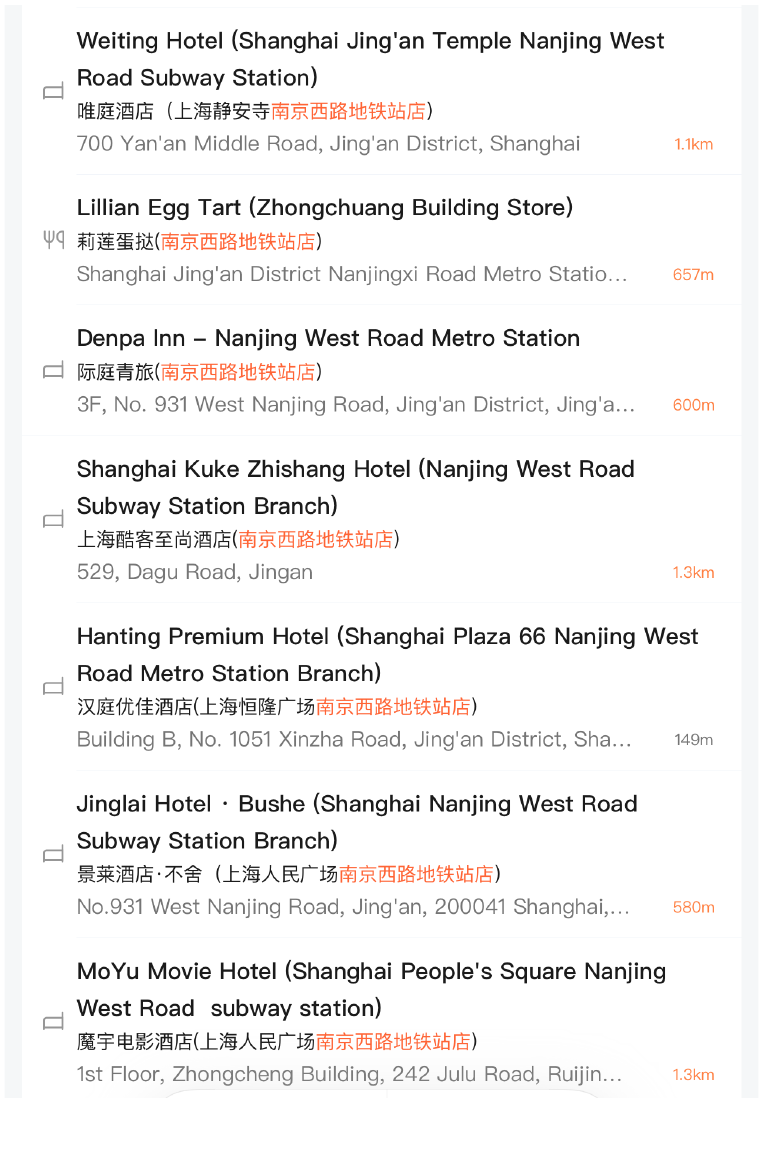}
        \caption{Example of non-transit location labels whose names contain station keywords}
        \label{fig:Example of non-transit location labels whose names contain station keywords}
    \end{subfigure}
    \caption{\centering{Examples of passenger-provided location labels in TNC apps.}}
    \label{fig:Examples of labels}
\end{figure}

Based on this data feature, we developed a label-based screening procedure to identify first- and last-mile complementary trips. 
\begin{itemize}
    \item Step 1. We retrieved a comprehensive database of all metro station and bus stop names across Shanghai using the Amap API \citep{amap_amap_2024}.
    \item Step 2. For each passenger-selected pick-up and drop-off label, we performed string matching against the database obtained in Step 1 to detect whether the label contained the names of the transit stations. Trips whose origin labels matched transit stations were marked as candidate last-mile complementary trips, while trips whose destination labels matched transit stations were marked as candidate first-mile complementary trips.
    \item Step 3. To avoid falsely treating nearby non-transit facilities as transit connections, we excluded labels that contained commercial, entertainment, or residential suffixes, such as ``Hotel'', ``Mall'', or ``Restaurant'', even when they also contained a station name. \Figref{fig:Example of non-transit location labels whose names contain station keywords} shows an example of non-transit location labels whose names contain station keywords (e.g., ``Weiting Hotel (Shanghai Jing’an Temple Nanjing West Road Subway Station)'').
\end{itemize}
This refined approach significantly mitigates the overestimation of complementary trips. The coordinate-only method commonly used in previous literature assume any trip ending within a specific radius of a station is complementary \citep{kongHowDoesRidesourcing2020, meredith-karamRelationshipRidehailingPublic2021}. 
As shown in \Figref{fig:Example of standardized location labels}, the coordinate-only method would classify non-transit labels, such as ``Atour X Hotel Nanjing West Road, Shanghai'', as complementary trips once they fall within a station catchment area. 
Our analysis shows this leads to substantial bias. 
When we applied a standard coordinate-only method to our dataset, the first-mile complementary ratio rose to 7.93\% and the last-mile complementary ratio rose to 10.49\%, which are 3.5 and 5.7 percentage points higher than our label-based estimates. To the best of our knowledge, this study is the first to use passenger-provided location labels in large-scale ride-hailing data to identify TNC–PT complementarity in this way.

\section{Parameter elasticity in substitution trip classification}\label{appendix: Parameter elasticity analysis}
\setcounter{figure}{0}
\renewcommand{\thefigure}{B\arabic{figure}}
To assess the robustness of the parameter values used in our classification framework for substitution trips, particularly those used in Conditions 3 through 6 in \secref{sec: TNC-PT relationship recognition}, an elasticity analysis was conducted. 
This analysis evaluates the percentage change in the substitutive ratio divided by the percentage change in a certain parameter's value, providing a measure of sensitivity. 
The parameters examined here are thresholds for the walking access distance (Condition 3), travel time saving (Condition 4), number of transfers (Condition 5), and PT-to-TNC cost ratio (Condition 6), with default values of 400 meters, 15 minutes, 2, and 0.5, respectively.

\begin{figure}
    \centering
    \includegraphics[width=.75\textwidth]{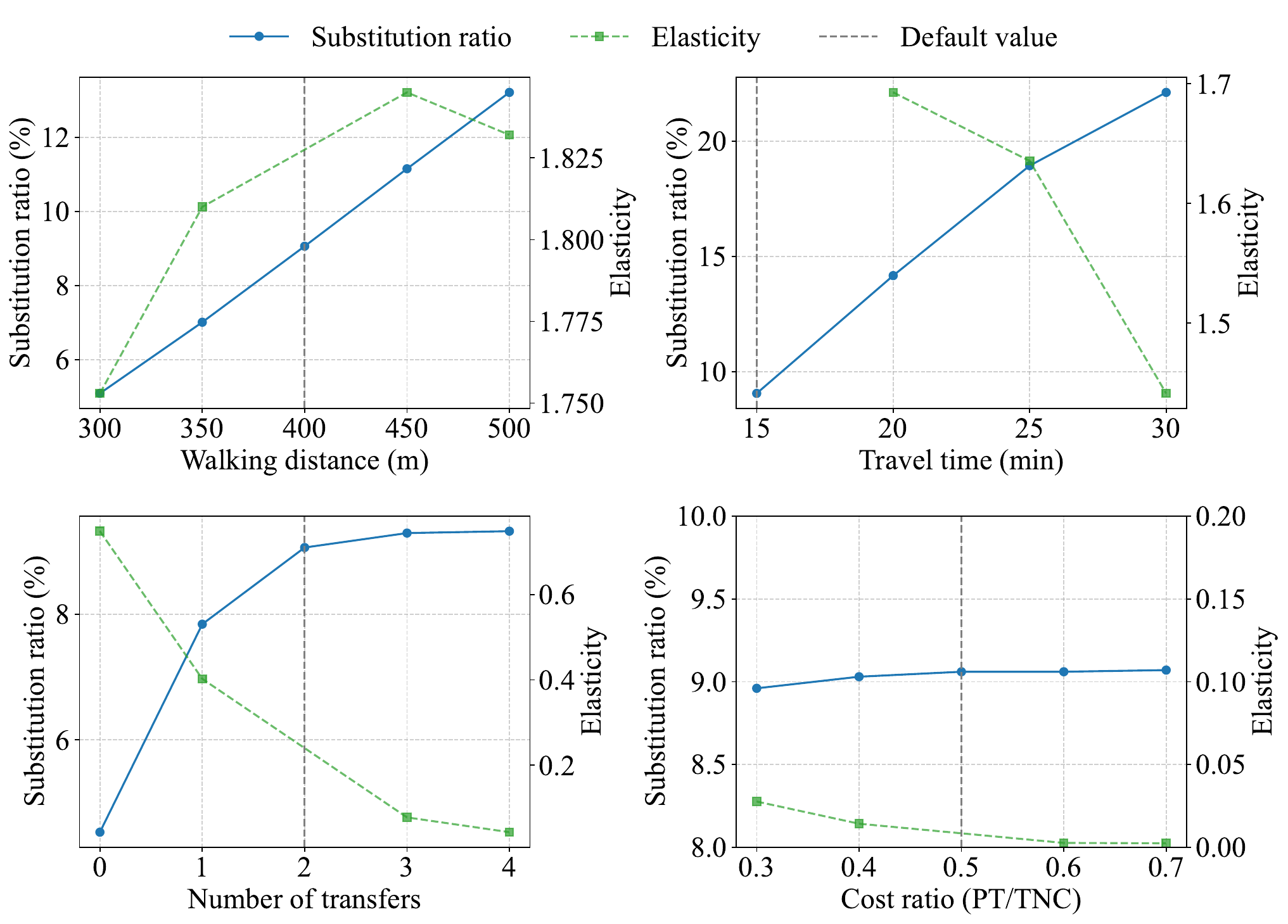}
    \caption{\centering{Elasticity of different parameters in defining substitutive trips.}}
    \label{fig: elasticity}
\end{figure}

The results are presented in \figref{fig: elasticity}, which indicates the substitution ratio is more elastic with respect to the walking distance and travel time thresholds (elasticity $>1$). 
Specifically, the substitutive ratio continuously increases as the walking distance threshold grows from 300 to 500 meters and as the travel time difference threshold increases from 15 to 30 minutes. 
In addition, the number of transfers and the cost ratio show low elasticity, indicating that the selection of these two parameters does not significantly affect the calculation of the substitutive ratio.
Hence, in this study, we adopt the values commonly used in the literature for these two thresholds \citep{kongHowDoesRidesourcing2020, meredith-karamRelationshipRidehailingPublic2021, liuInvestigatingRelationshipsRidesourcing2025}, with the precaution that the actual substitutive ratio may be sensitive to the parameter selection.  

\section{Comparison of ML models}\label{appendix: Model Comparison}
\setcounter{table}{0}
\renewcommand{\thetable}{C\arabic{table}}

\begin{table}[width=.98\linewidth,cols=5,pos=h]\footnotesize\rmfamily
\centering
\caption{Benchmarking results across models for four dependent variables (with 95\% CI).}
\label{tab:Model Comparison}
\begin{tabular*}{\tblwidth}{@{} llccc@{} } 
\toprule
Dependent variable & Model & MAE & MSE & RMSE \\
\midrule
\multirow{5}{*}{\makecell[l]{First-mile\\ complement ratio}} 
& CatBoost & 3.842 (3.566--4.115) & 32.574 (26.282--40.080) &5.707 (5.127--6.331) \\
& XGBoost & 3.881 (3.620--4.147) & 32.827 (26.683--40.281) & 5.730 (5.166--6.347) \\
& LightGBM & 3.916 (3.640--4.189) & 34.238 (27.967--41.944) & 5.851 (5.288--6.476) \\
& FT-Transformer & 4.025 (3.751--4.283) & 34.650 (28.444--42.342) & 5.886 (5.333--6.507) \\
& TabNet & 4.198 (3.920--4.488) & 37.340 (30.725--45.016) & 6.111 (5.543--6.709) \\
& MLP & 4.277 (4.010--4.552) & 37.762 (31.061--45.844) & 6.145 (5.573--6.771) \\
\midrule
\multirow{5}{*}{\makecell[l]{Last-mile\\ complement ratio}} 
& CatBoost & 3.329 (3.093--3.572) & 25.132 (19.875--32.244) & 5.013 (4.458--5.678) \\
& XGBoost & 3.294 (3.059--3.551) & 25.115 (19.955--31.800) & 5.012 (4.467--5.639) \\
& LightGBM & 3.314 (3.076--3.567) & 25.446 (20.099--32.344) & 5.044 (4.483--5.687) \\
& FT-Transformer & 3.372 (3.136--3.623) & 25.648 (20.251--32.860) & 5.064 (4.500--5.732) \\
& TabNet & 3.428 (3.198--3.678) & 25.967 (20.185--33.757) & 5.096 (4.493--5.810) \\
& MLP & 3.495 (3.263--3.741) & 25.736 (20.541--32.870) & 5.073 (4.532--5.733) \\
\midrule
\multirow{5}{*}{\makecell[l]{Departure\\ substitution ratio}} 
& CatBoost & 2.254 (2.108--2.418) & 10.768 (8.558--13.562) & 3.281 (2.925--3.683) \\
& XGBoost & 2.259 (2.112--2.423) & 10.812 (8.545--13.541) & 3.288 (2.923--3.680) \\
& LightGBM & 2.260 (2.111--2.422) & 10.941 (8.651--13.784) & 3.308 (2.941--3.713) \\
& FT-Transformer & 2.270 (2.113--2.439) & 11.355 (8.886--14.248) & 3.370 (2.981--3.775) \\
& TabNet & 2.313 (2.154--2.486) & 11.425 (9.042--14.326) & 3.380 (3.007--3.785) \\
& MLP & 2.417 (2.248--2.579) & 12.052 (9.590--14.822) & 3.472 (3.097--3.850) \\
\midrule
\multirow{5}{*}{\makecell[l]{Arrival\\ substitution ratio}} 
& CatBoost & 2.110 (1.981--2.254) & 9.196 (7.549--11.090) & 3.033 (2.747--3.330) \\
& XGBoost & 2.144 (2.009--2.290) & 9.479 (7.723--11.475) & 3.079 (2.779--3.387) \\
& LightGBM & 2.161 (2.021--2.313) & 9.756 (8.018--11.793) & 3.123 (2.832--3.434) \\
& FT-Transformer & 2.203 (2.060--2.354) & 10.089 (8.238--12.210) & 3.176 (2.870--3.494) \\
& TabNet & 2.175 (2.017--2.350) & 10.957 (8.008--14.928) & 3.310 (2.830--3.864) \\
& MLP & 2.318 (2.139--2.533) & 14.418 (8.895--23.800) & 3.797 (2.982--4.879) \\
\bottomrule
\end{tabular*}
\end{table}

\printcredits

\clearpage
\section*{Supplementary analysis}
\subsection*{Robustness of the ML interpretations under alternative threshold settings}
We conducted supplemental experiments to examine whether the machine-learning findings related to the substitutive ratio are sensitive to the high-elasticity parameters. Since our sensitivity analysis showed that the number of transfers threshold and cost-ratio threshold have relatively low sensitivity, these additional tests focused only on walking distance and travel time difference. Specifically, using the manuscript’s baseline setting of 400 m walking distance and 15 min travel time difference as the reference, we conducted additional experiments with walking-distance thresholds of 300 m, 350 m, 450 m, and 500 m, and travel-time-difference thresholds of 20 min, 25 min, and 30 min.

The results of these additional analyses are presented in \twofigref{fig:Stability of ranking (walk) departure substitutive ratio}{fig:stability_travel PDPs arrival substitutive ratio}. Among them, Figs. S1, S3, S5, and S7 report the ranked relative contribution percentages of all variables to the departure/arrival substitutive ratios under different parameter settings, analogous to Figs. 8c and 8d in the revised manuscript. Figs. S2, S4, S6, and S8 report the PDPs of the top two variables under different parameter settings, analogous to Figs. 9c and 9d in the revised manuscript. These results show that, under the vast majority of threshold settings, ``Distance to bus hub'' and ``Bus hub density'' remain the top two most important determinants for both departure and arrival substitutive ratios, which is fully consistent with the baseline findings reported in the manuscript. In addition, the PDPs of these two variables exhibit nearly identical nonlinear patterns across all settings. For example, ``Distance to bus hub'' consistently shows a sharp decline in substitution ratios after approximately 0.7 km, regardless of changes in the threshold values. This indicates that, although the absolute value of the substitutive ratio changes with the thresholds, the underlying mechanisms and key drivers identified by the ML framework remain highly stable.

\begin{landscape}
\setcounter{figure}{0}
\renewcommand{\thefigure}{S\arabic{figure}}
\begin{figure}[!t]
    \centering
    \begin{subfigure}[h]{0.65\textwidth}
        \centering
        \includegraphics[width=\linewidth,keepaspectratio]{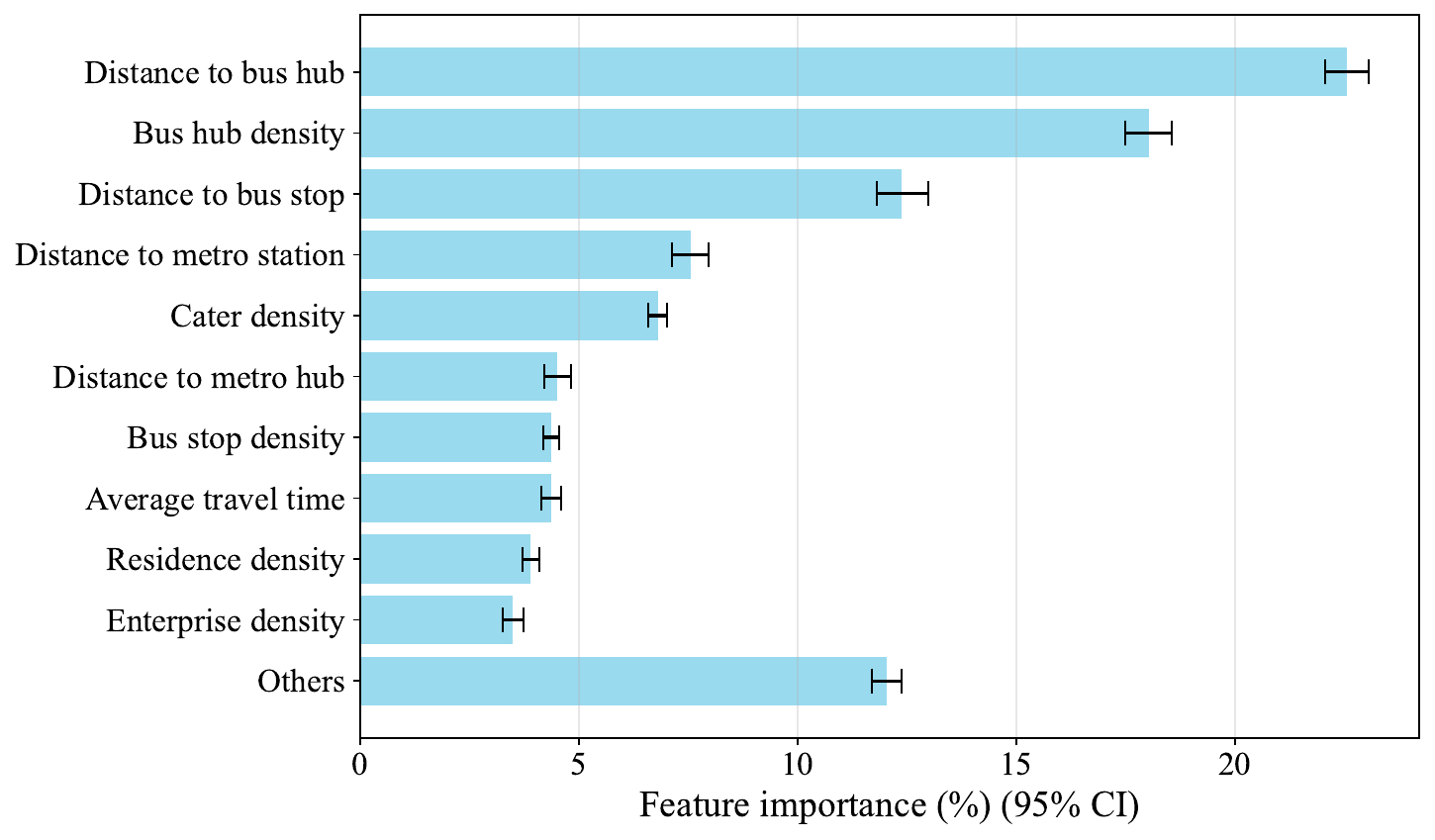}
        \caption{(300 m, 15-min, 2, 0.5)}
        \label{fig:300m departure substitutive ratio}
    \end{subfigure}
    \begin{subfigure}[h]{0.65\textwidth}
        \centering
        \includegraphics[width=\linewidth,keepaspectratio]{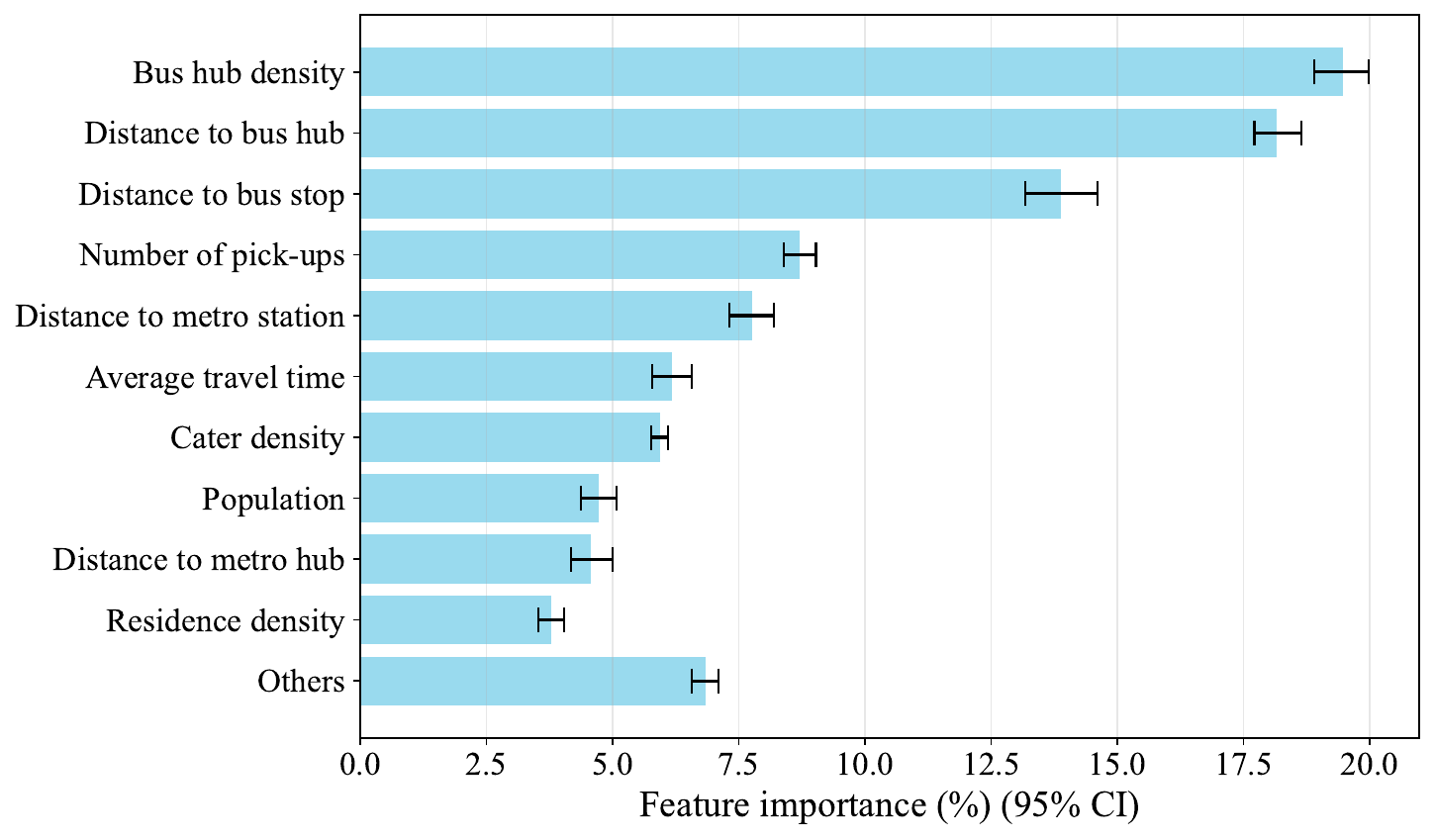}
        \caption{(350 m, 15-min, 2, 0.5)}
        \label{fig:350m departure substitutive ratio}
    \end{subfigure}
    \begin{subfigure}[h]{0.65\textwidth}
        \centering
        \includegraphics[width=\linewidth,keepaspectratio]{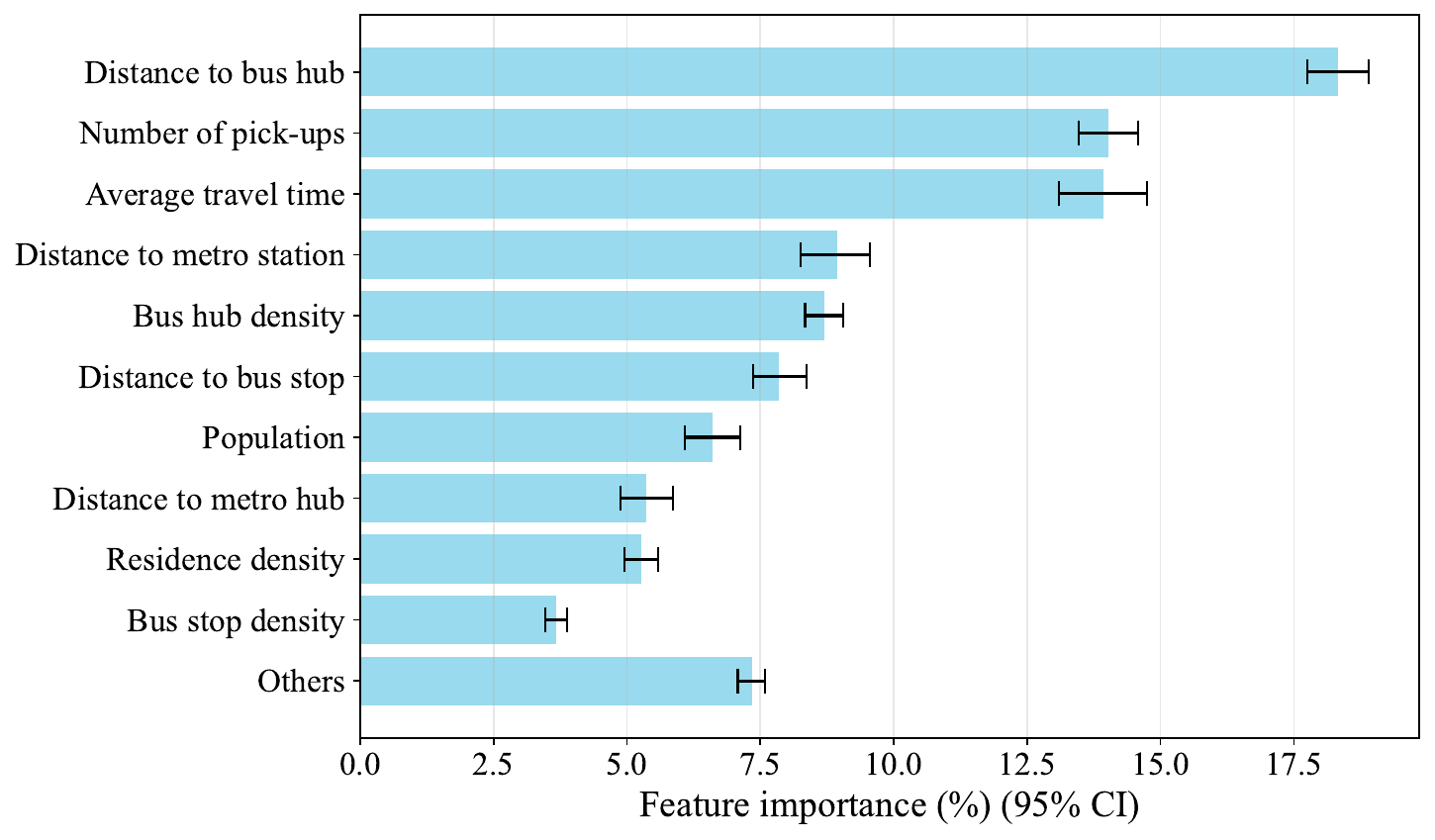}
        \caption{(450 m, 15-min, 2, 0.5)}
        \label{fig:450m departure substitutive ratio}
    \end{subfigure}
    \begin{subfigure}[h]{0.65\textwidth}
        \centering
        \includegraphics[width=\linewidth,keepaspectratio]{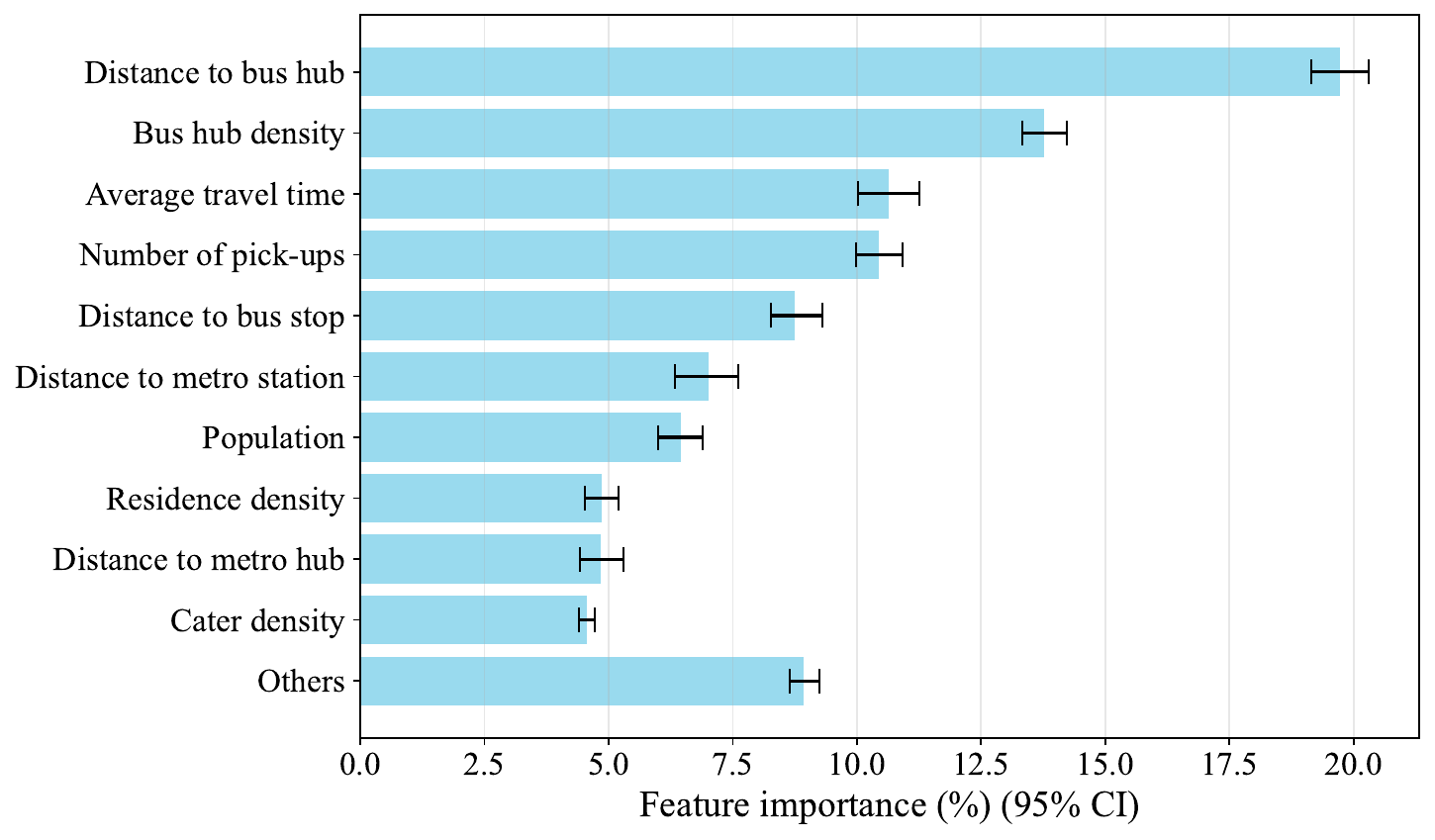}
        \caption{(500 m, 15-min, 2, 0.5)}
        \label{fig:500m departure substitutive ratio}
    \end{subfigure}
    \caption{\centering{Variable importance ranking (different walking distance for departure substitutive ratio).}}
    \label{fig:Stability of ranking (walk) departure substitutive ratio}
\end{figure}

\begin{figure}[!t]
    \centering
    \begin{subfigure}[t]{0.65\textwidth}
        \centering
        \begin{subfigure}[t]{0.5\linewidth}
            \centering
            \includegraphics[width=\linewidth]{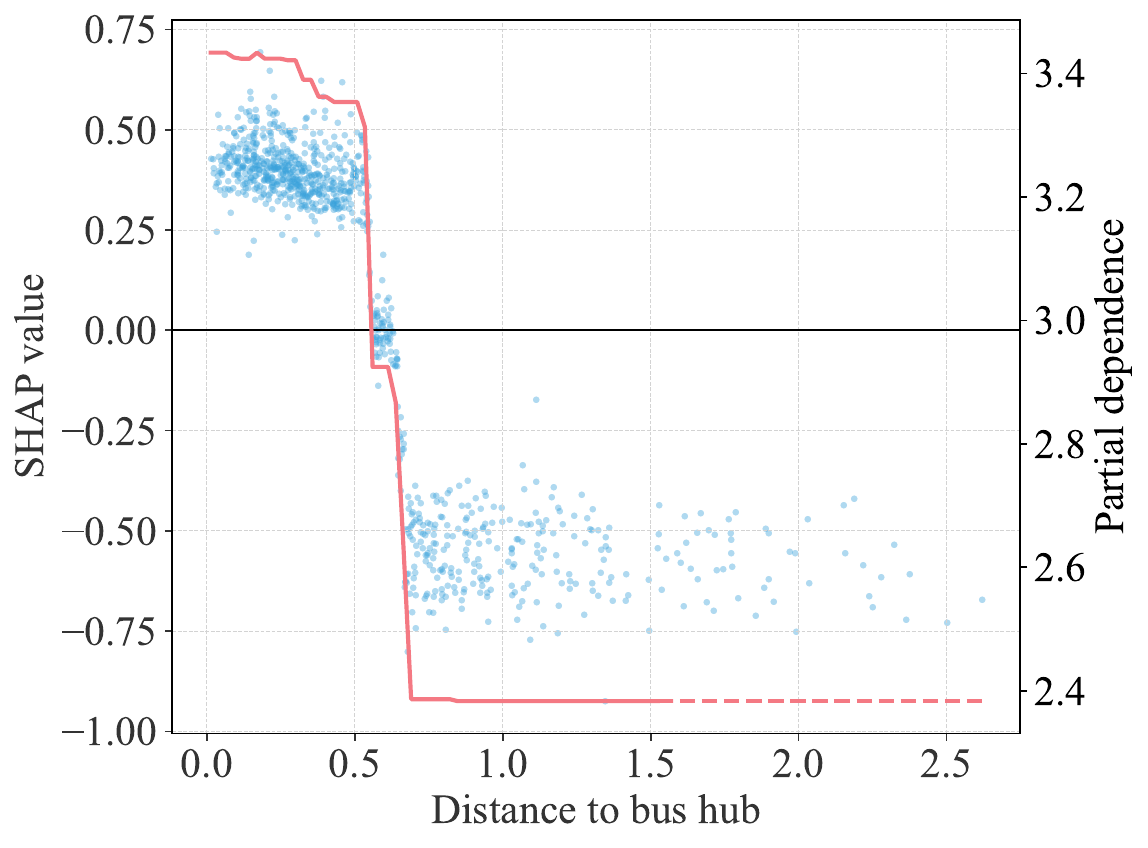}
            \caption*{Distance to bus hub}
        \end{subfigure}\hfill
        \begin{subfigure}[t]{0.5\linewidth}
            \centering
            \includegraphics[width=\linewidth]{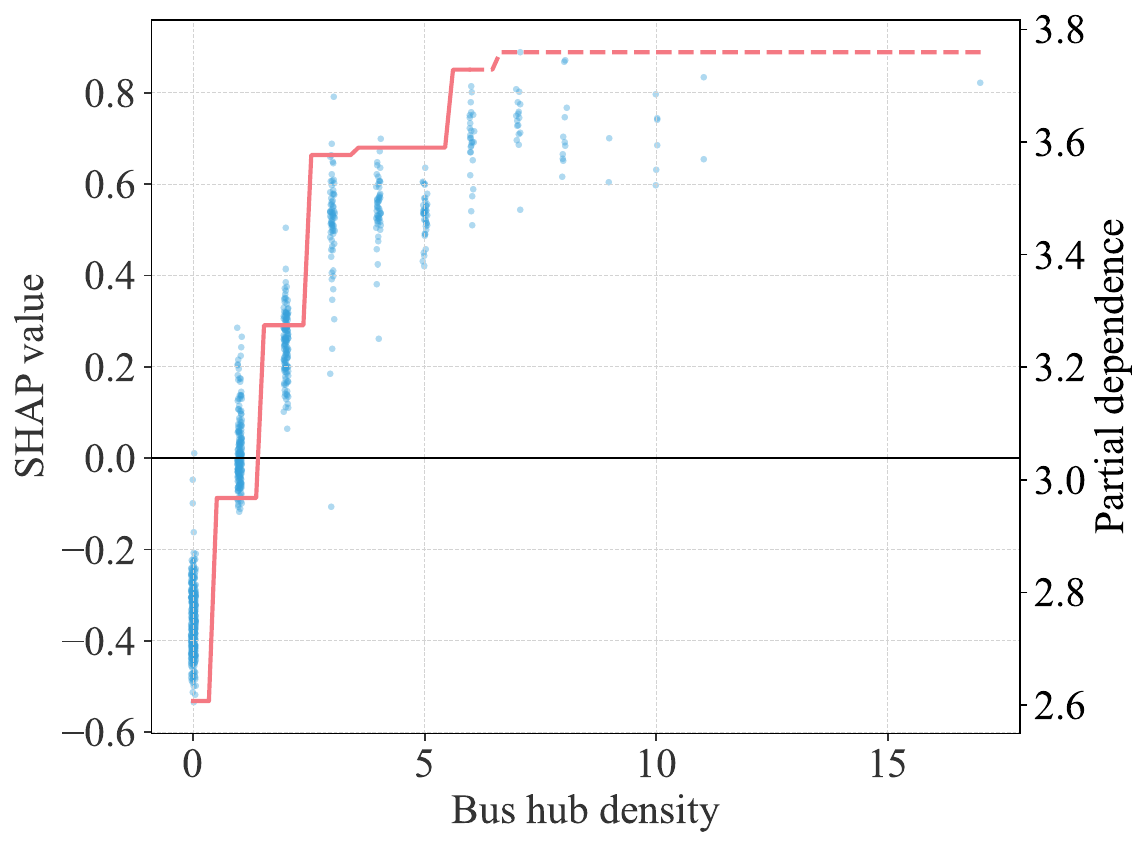}
            \caption*{Bus hub density}
        \end{subfigure}
        \caption{(300 m, 15-min, 2, 0.5)}
        \label{fig:300m_dep_group}
    \end{subfigure}
    \begin{subfigure}[t]{0.65\textwidth}
        \centering
        \begin{subfigure}[t]{0.5\linewidth}
            \centering
            \includegraphics[width=\linewidth]{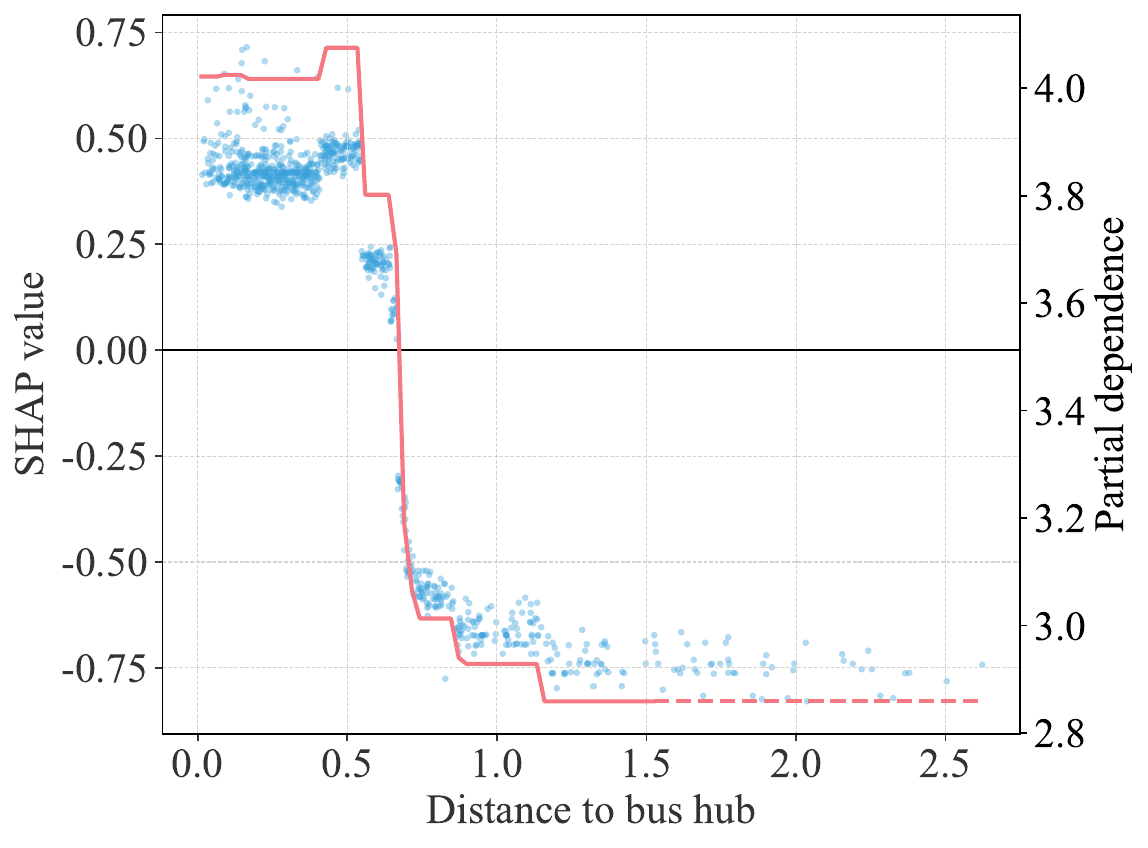}
            \caption*{Distance to bus hub}
        \end{subfigure}\hfill
        \begin{subfigure}[t]{0.5\linewidth}
            \centering
            \includegraphics[width=\linewidth]{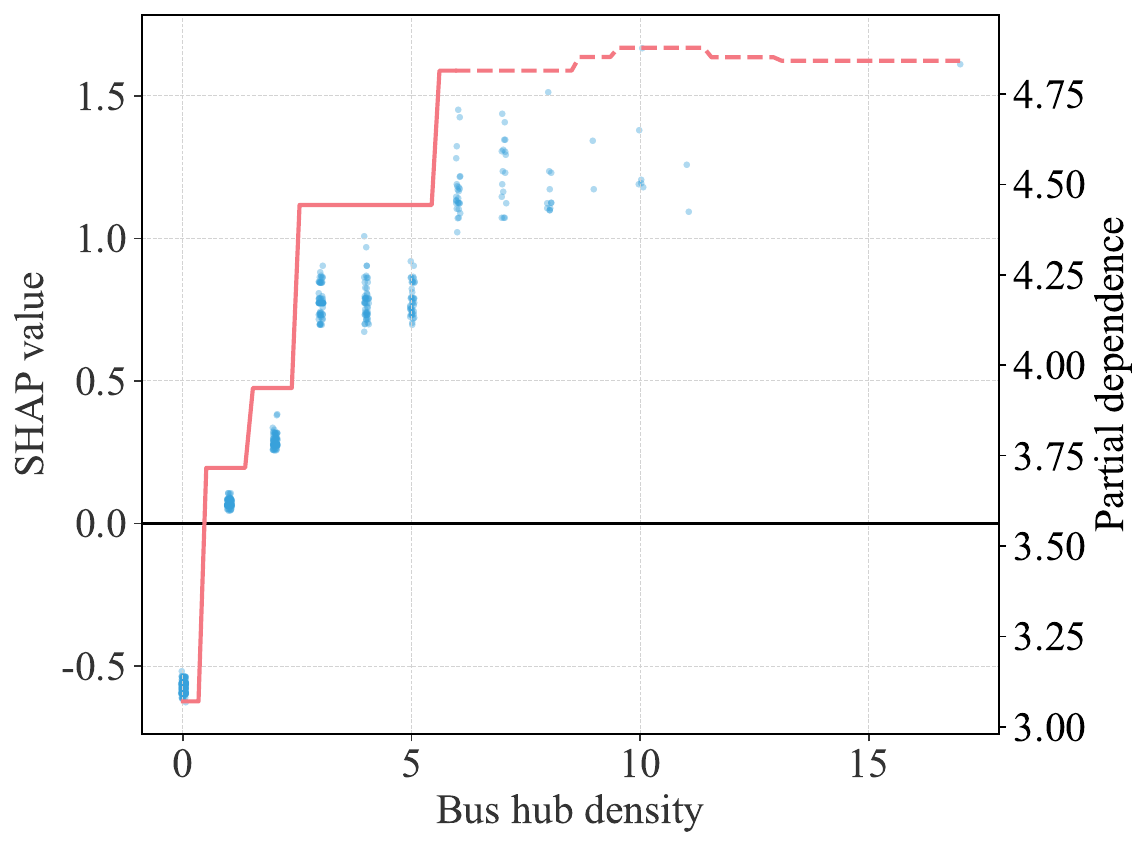}
            \caption*{Bus hub density}
        \end{subfigure}
        \caption{(350 m, 15-min, 2, 0.5)}
        \label{fig:350m_dep_group}
    \end{subfigure}
    \begin{subfigure}[t]{0.65\textwidth}
        \centering
        \begin{subfigure}[t]{0.5\linewidth}
            \centering
            \includegraphics[width=\linewidth]{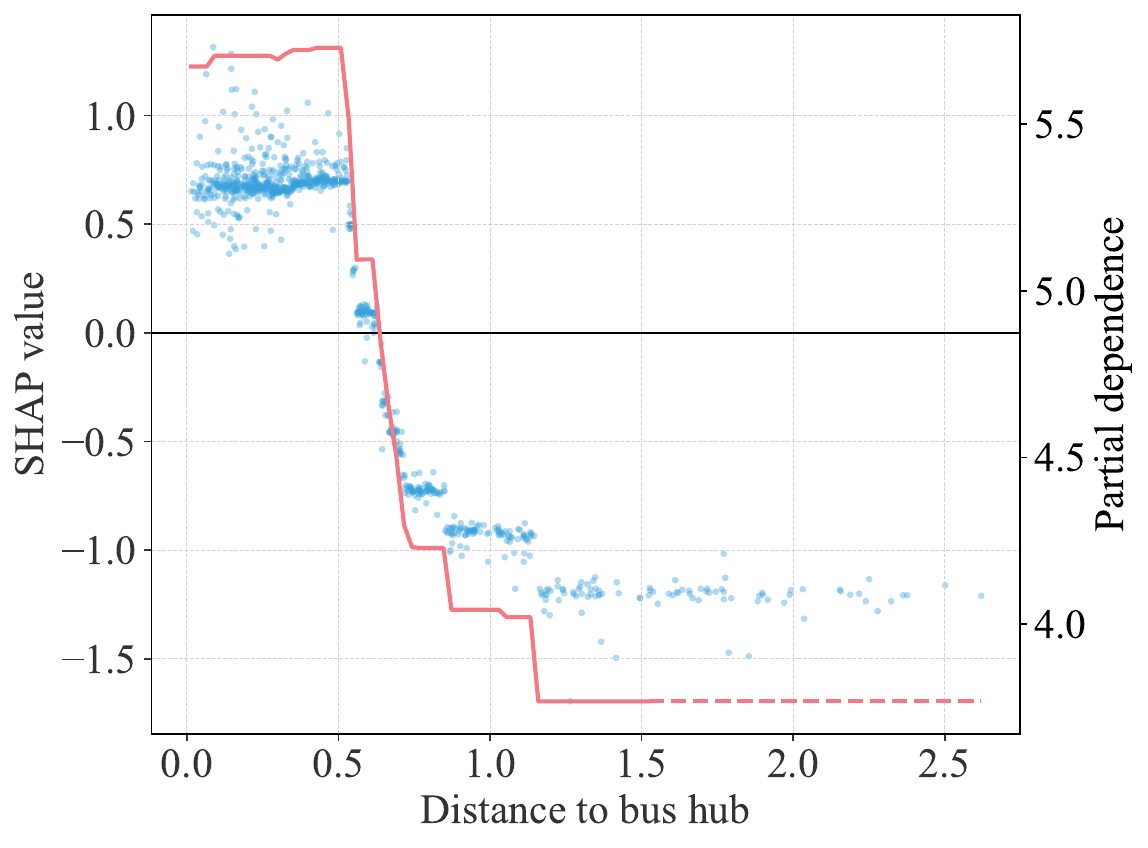}
            \caption*{Distance to bus hub}
        \end{subfigure}\hfill
        \begin{subfigure}[t]{0.5\linewidth}
            \centering
            \includegraphics[width=\linewidth]{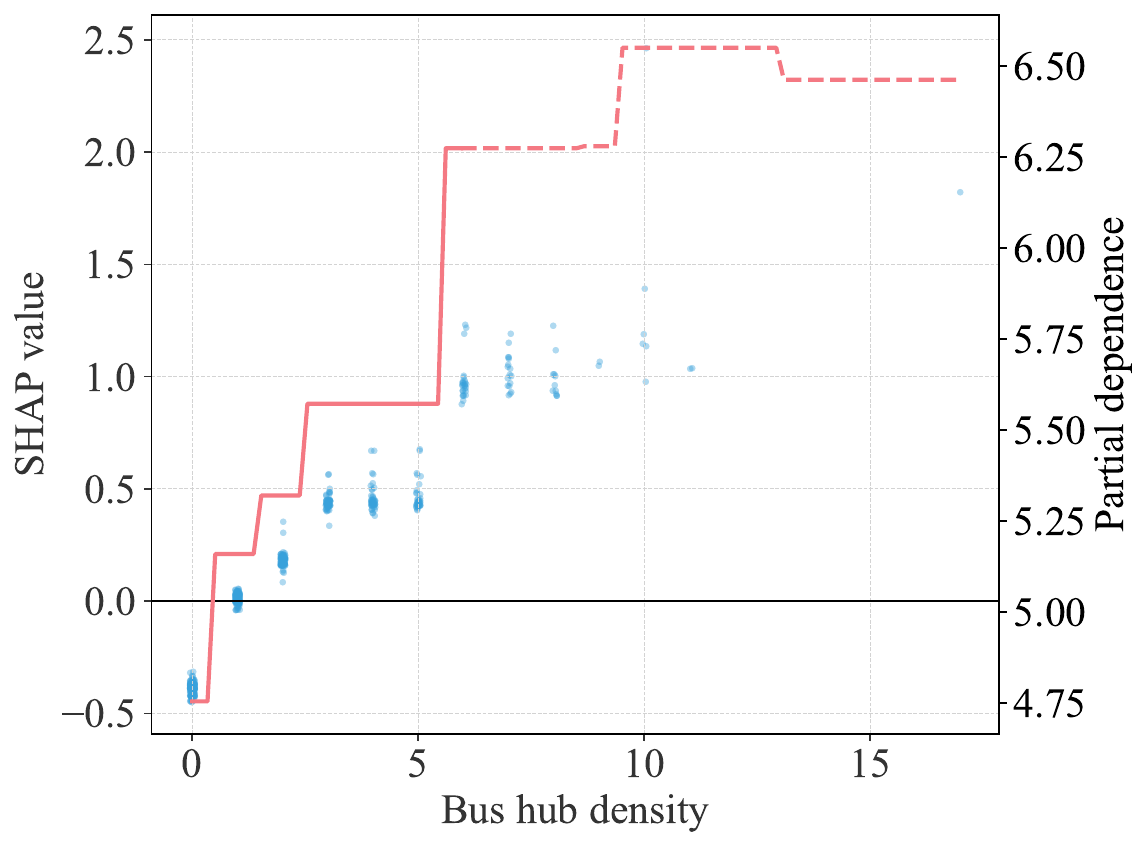}
            \caption*{Bus hub density}
        \end{subfigure}
        \caption{(450 m, 15-min, 2, 0.5)}
        \label{fig:450m_dep_group}
    \end{subfigure}
    \begin{subfigure}[t]{0.65\textwidth}
        \centering
        \begin{subfigure}[t]{0.5\linewidth}
            \centering
            \includegraphics[width=\linewidth]{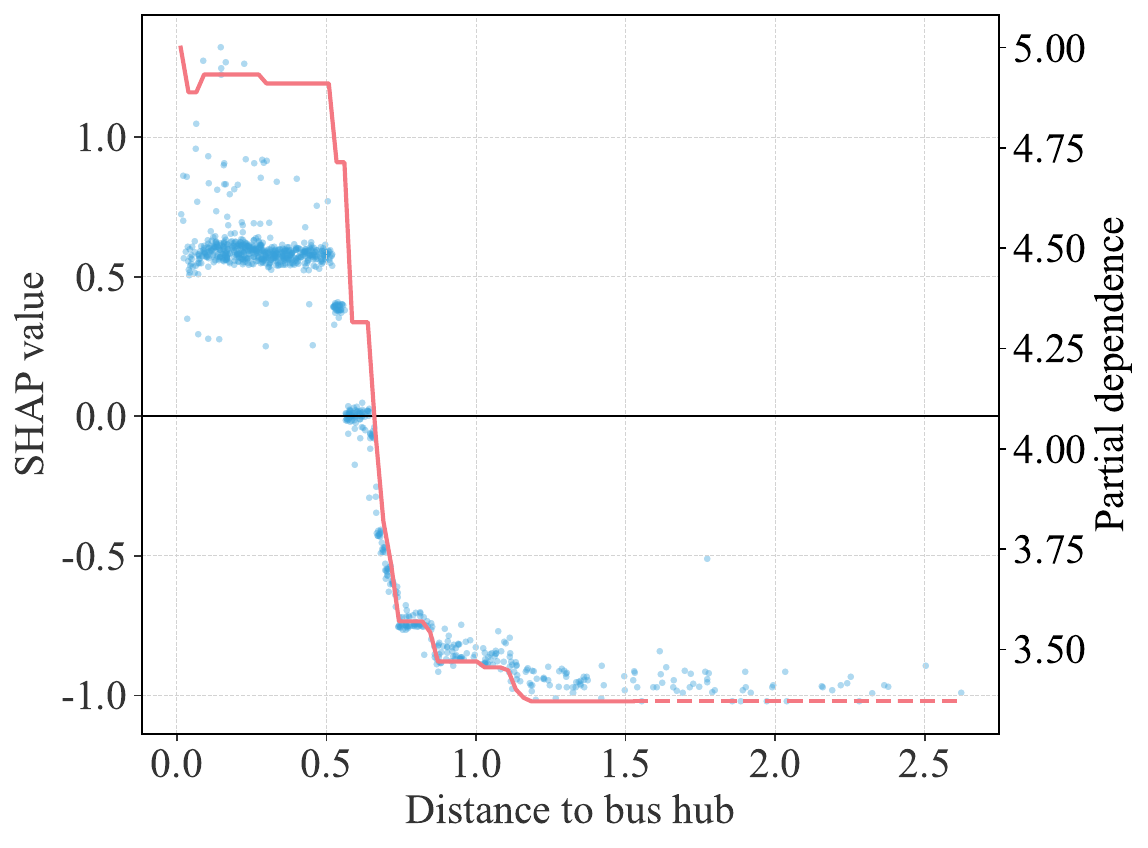}
            \caption*{Distance to bus hub}
        \end{subfigure}\hfill
        \begin{subfigure}[t]{0.5\linewidth}
            \centering
            \includegraphics[width=\linewidth]{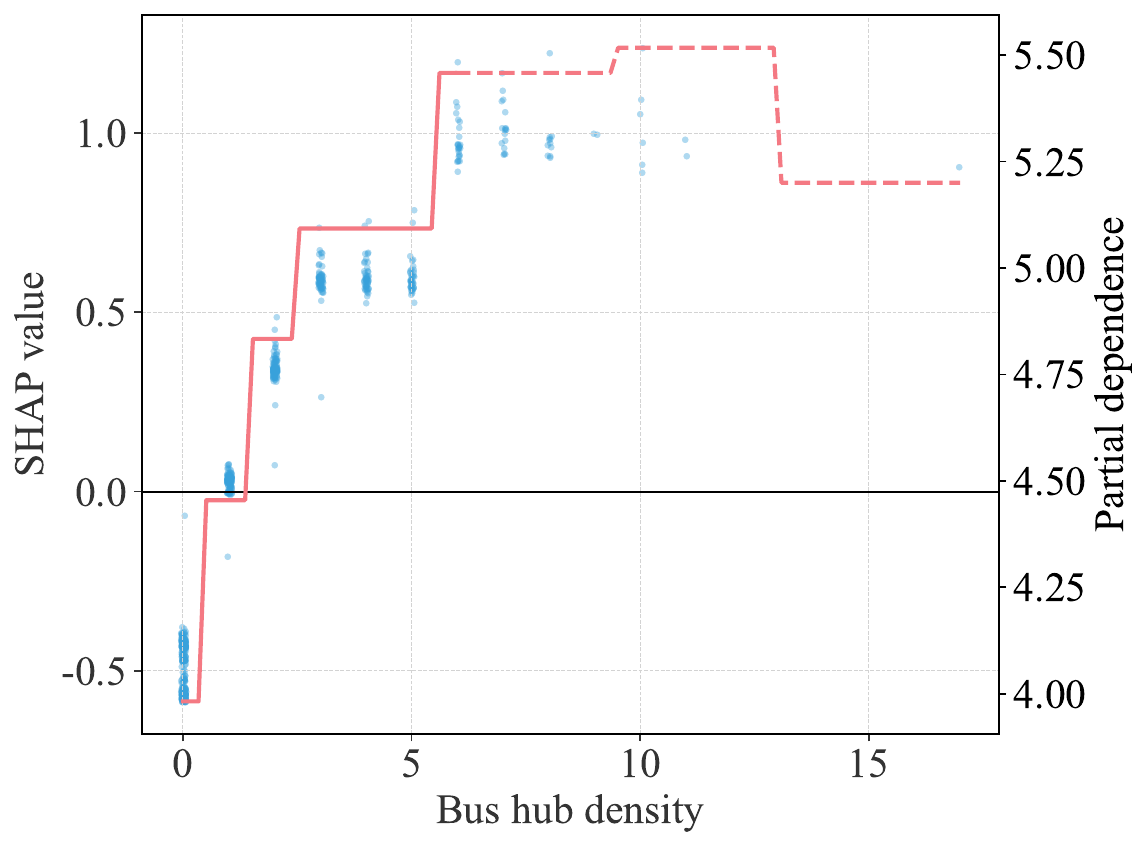}
            \caption*{Bus hub density}
        \end{subfigure}
        \caption{(500 m, 15-min, 2, 0.5)}
        \label{fig:500m_dep_group}
    \end{subfigure}
\caption{\centering{PDPs for top-2 variables (different walking distance for departure substitutive ratio).}}
\label{fig:stability_walk PDPs departure substitutive ratio}
\end{figure}

\begin{figure}[!t]
    \centering
    \begin{subfigure}[h]{0.65\textwidth}
        \centering
        \includegraphics[width=\linewidth,keepaspectratio]{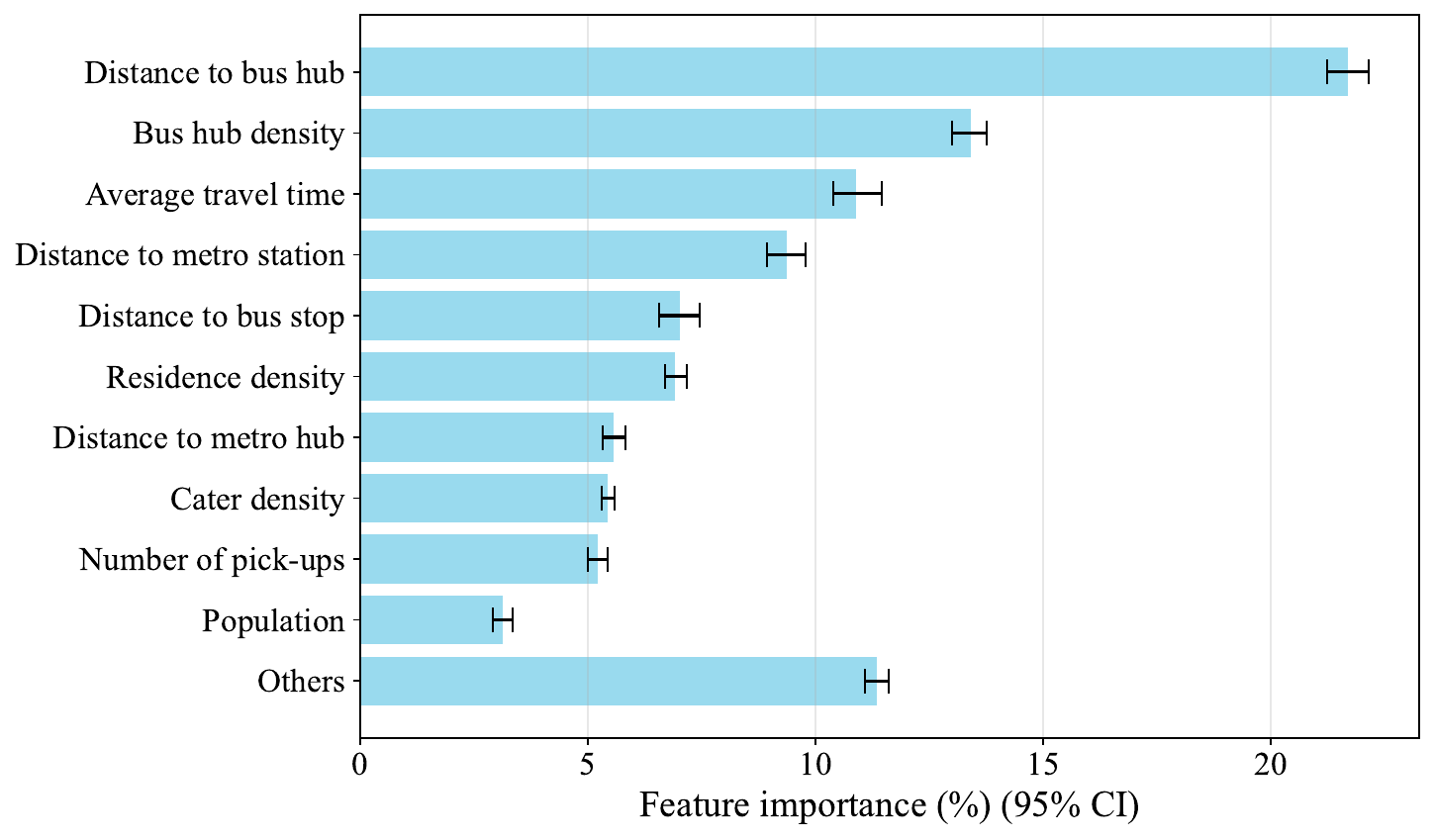}
        \caption{(400 m, 20-min, 2, 0.5)}
        \label{fig:20-min departure substitutive ratio}
    \end{subfigure}
    \begin{subfigure}[h]{0.65\textwidth}
        \centering
        \includegraphics[width=\linewidth,keepaspectratio]{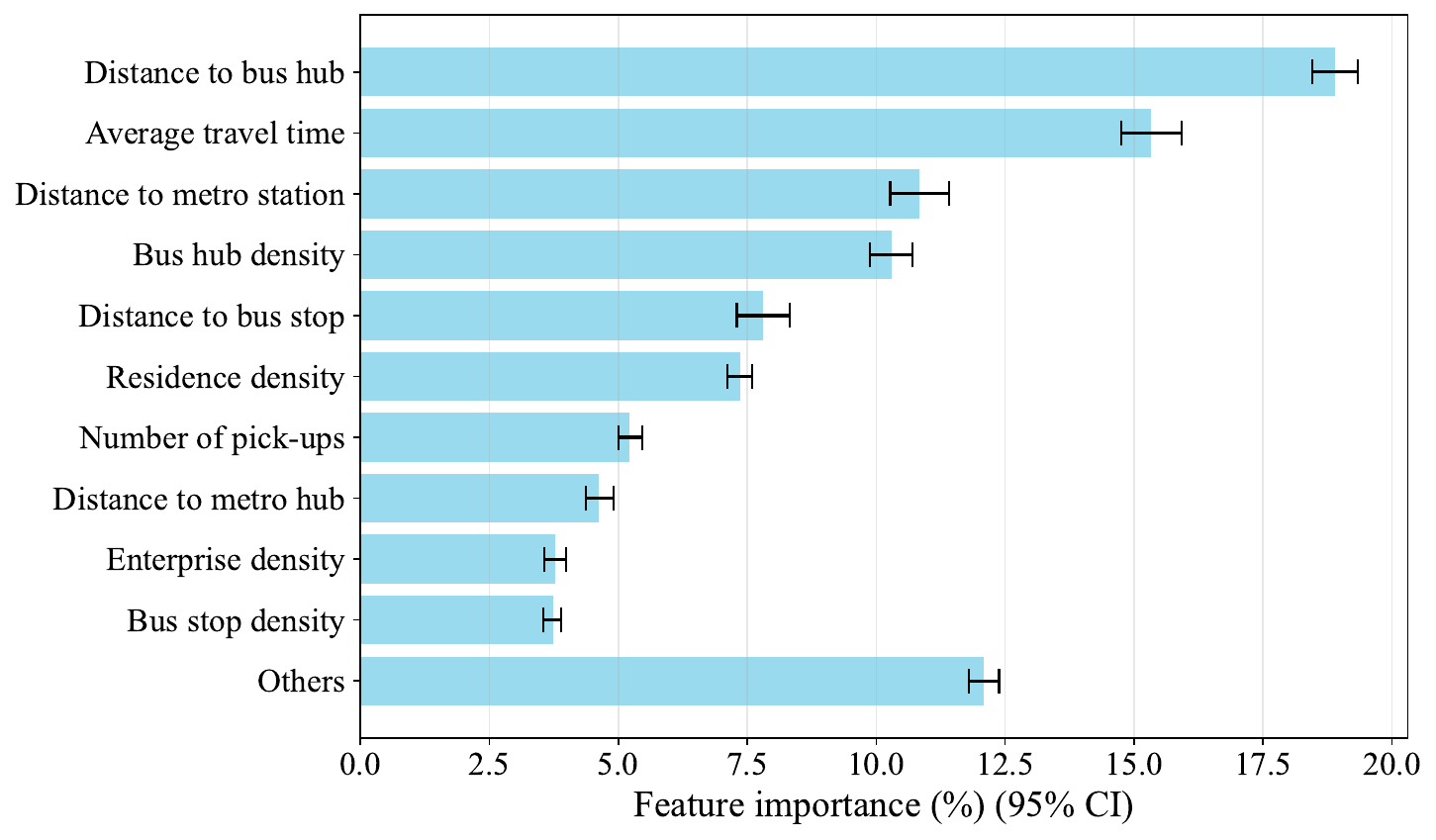}
        \caption{(400 m, 25-min, 2, 0.5)}
        \label{fig:25-min departure substitutive ratio}
    \end{subfigure}
    \begin{subfigure}[h]{0.65\textwidth}
        \centering
        \includegraphics[width=\linewidth,keepaspectratio]{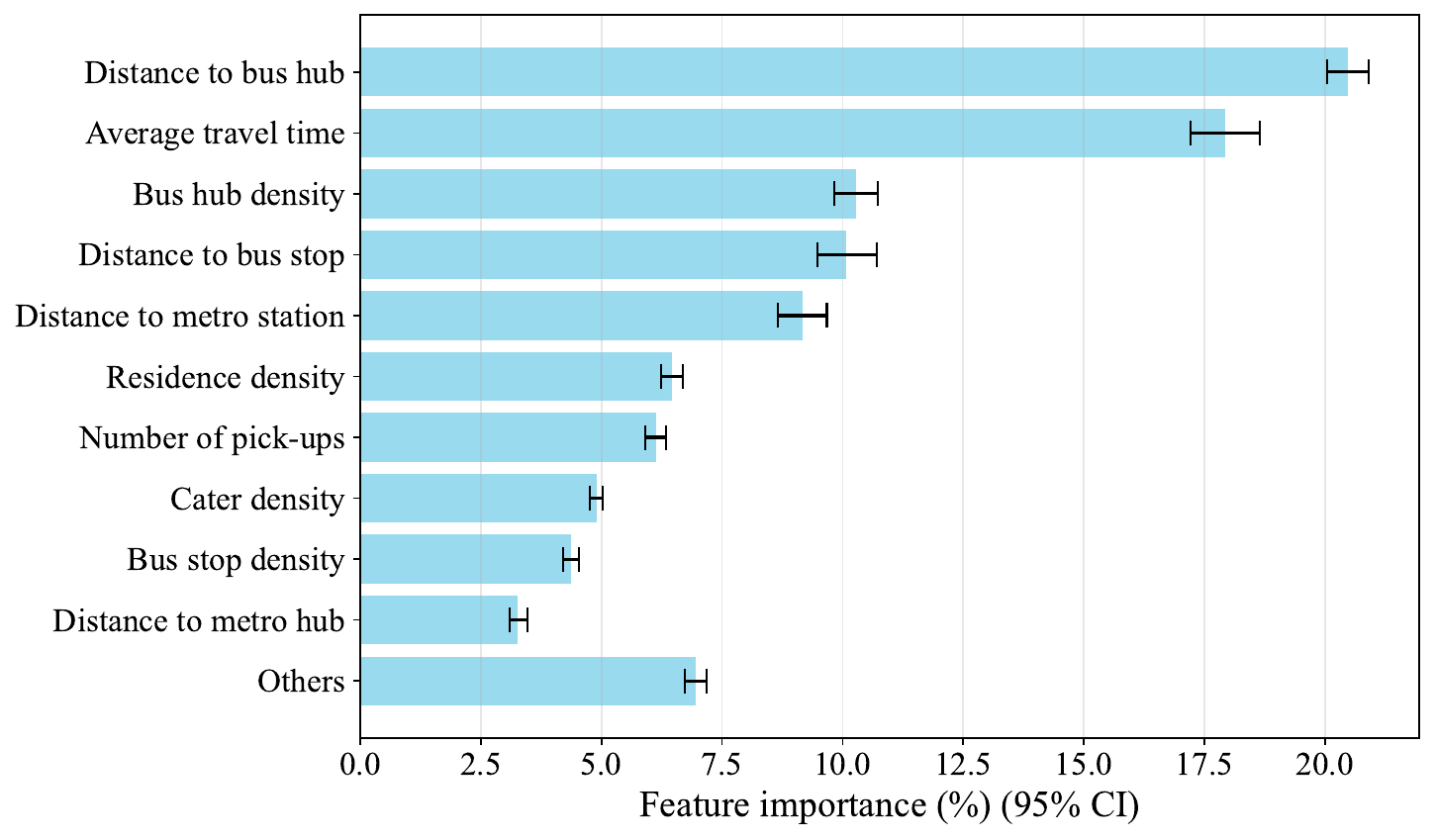}
        \caption{(400 m, 30-min, 2, 0.5)}
        \label{fig:30-min departure substitutive ratio}
    \end{subfigure}
    \caption{\centering{Variable importance ranking (different travel time difference for departure substitutive ratio).}}
    \label{fig:Stability of ranking (travel) departure substitutive ratio}
\end{figure}

\begin{figure}[!t]
    \centering
    \begin{subfigure}[t]{0.65\textwidth}
        \centering
        \begin{subfigure}[t]{0.5\linewidth}
            \centering
            \includegraphics[width=\linewidth]{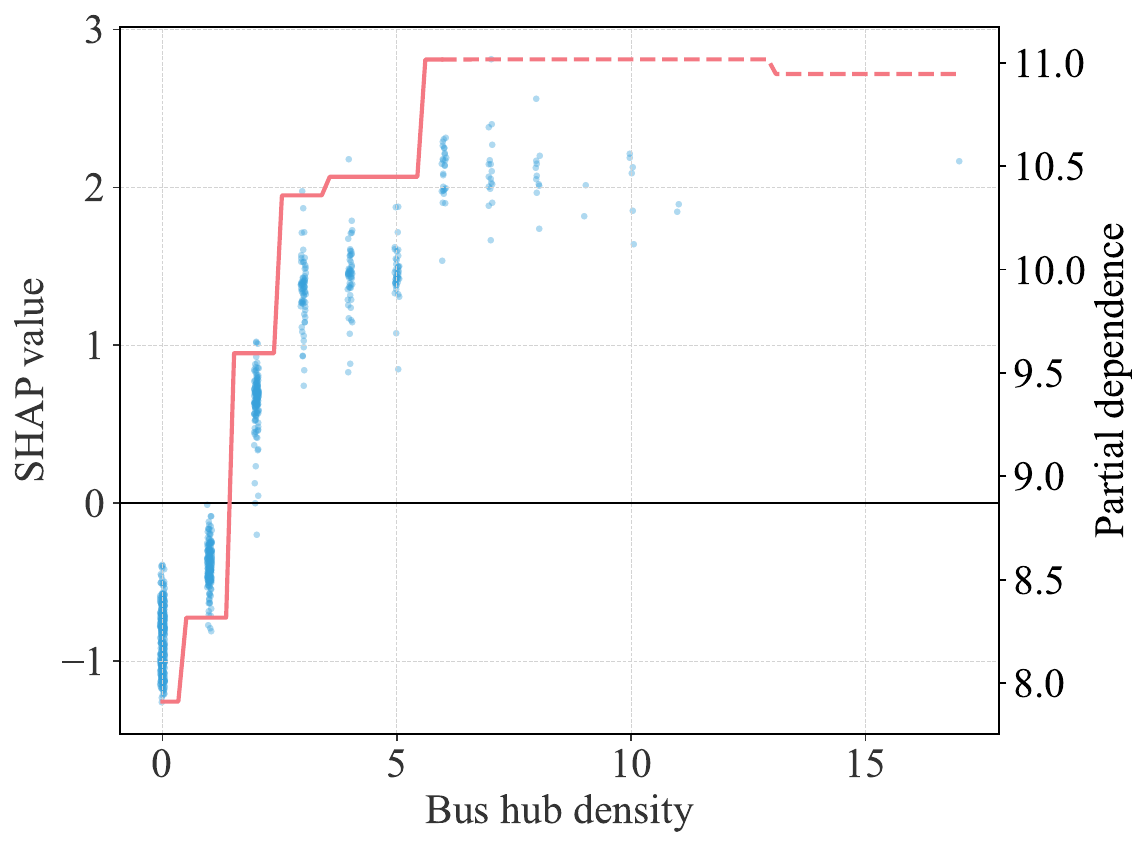}
            \caption*{Distance to bus hub}
        \end{subfigure}\hfill
        \begin{subfigure}[t]{0.5\linewidth}
            \centering
            \includegraphics[width=\linewidth]{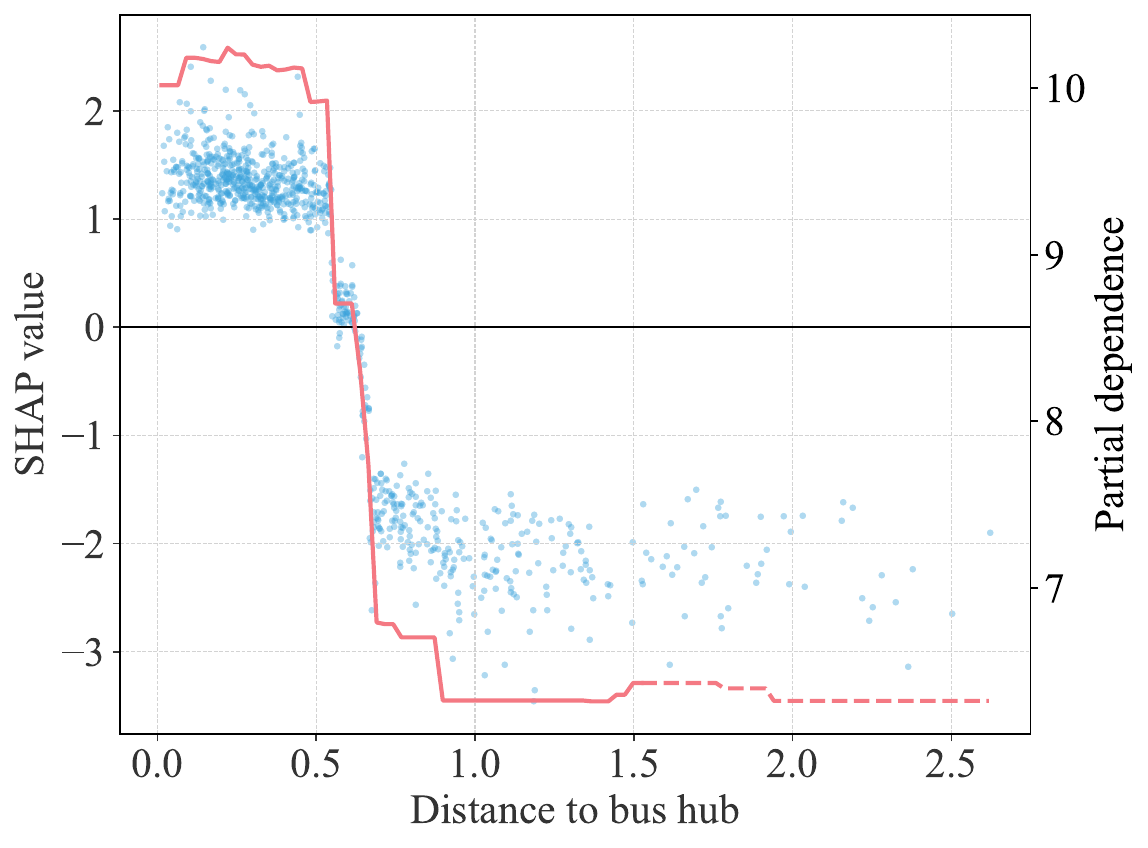}
            \caption*{Bus hub density}
        \end{subfigure}
        \caption{(400 m, 20-min, 2, 0.5)}
        \label{fig:20min_dep_group}
    \end{subfigure}
    \begin{subfigure}[t]{0.65\textwidth}
        \centering
        \begin{subfigure}[t]{0.5\linewidth}
            \centering
            \includegraphics[width=\linewidth]{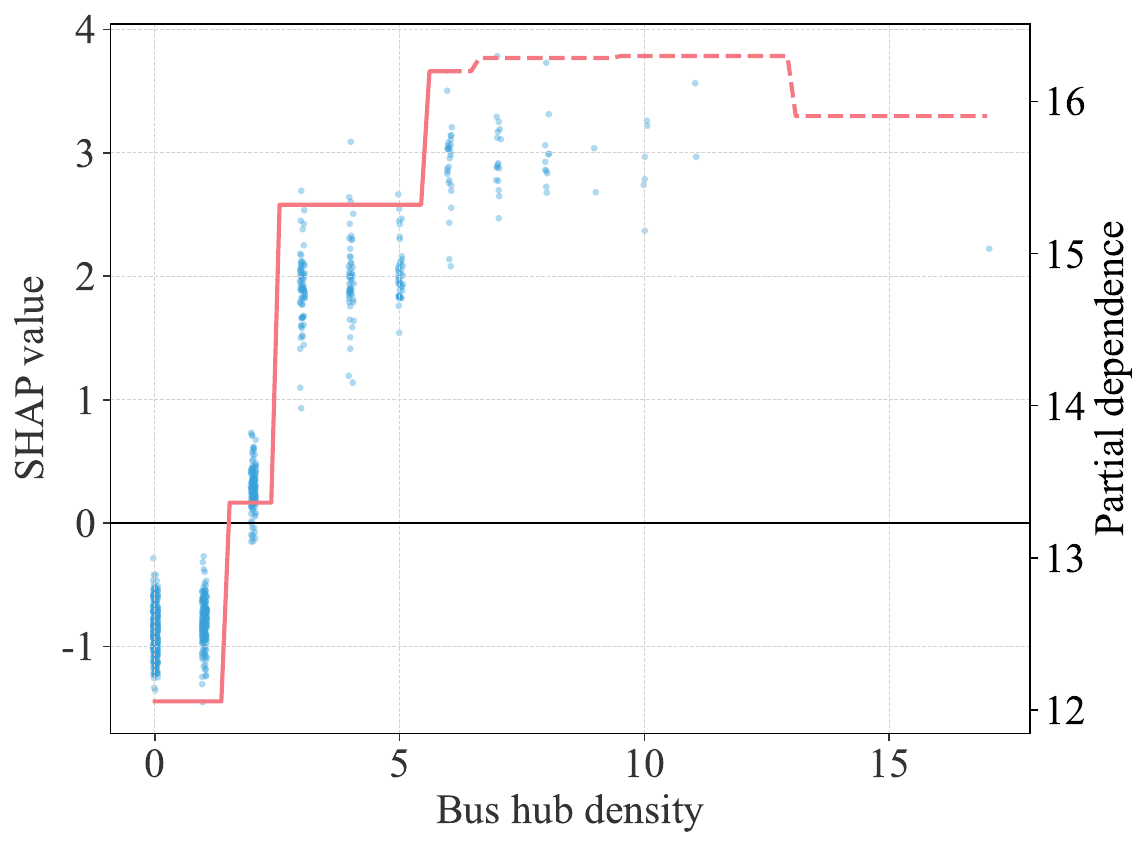}
            \caption*{Distance to bus hub}
        \end{subfigure}\hfill
        \begin{subfigure}[t]{0.5\linewidth}
            \centering
            \includegraphics[width=\linewidth]{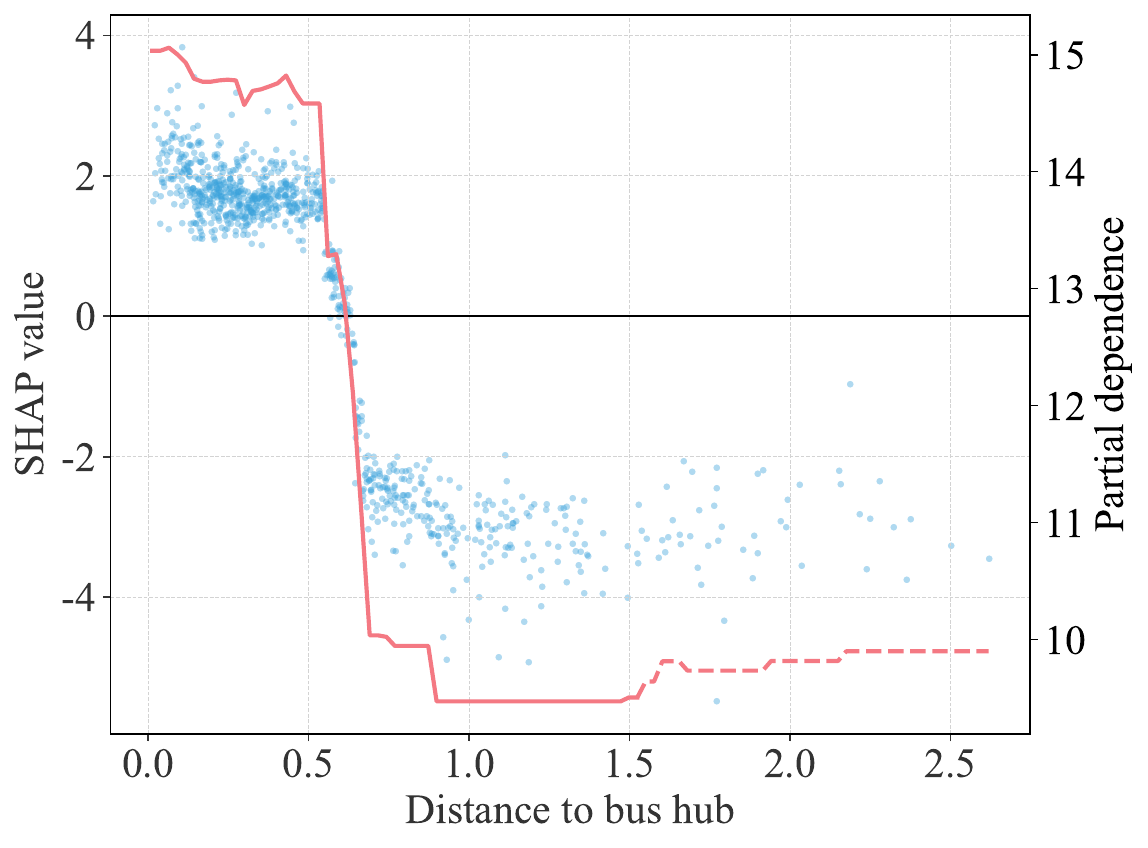}
            \caption*{Bus hub density}
        \end{subfigure}
        \caption{(400 m, 25-min, 2, 0.5)}
        \label{fig:25min_dep_group}
    \end{subfigure}
    \begin{subfigure}[t]{0.65\textwidth}
        \centering
        \begin{subfigure}[t]{0.5\linewidth}
            \centering
            \includegraphics[width=\linewidth]{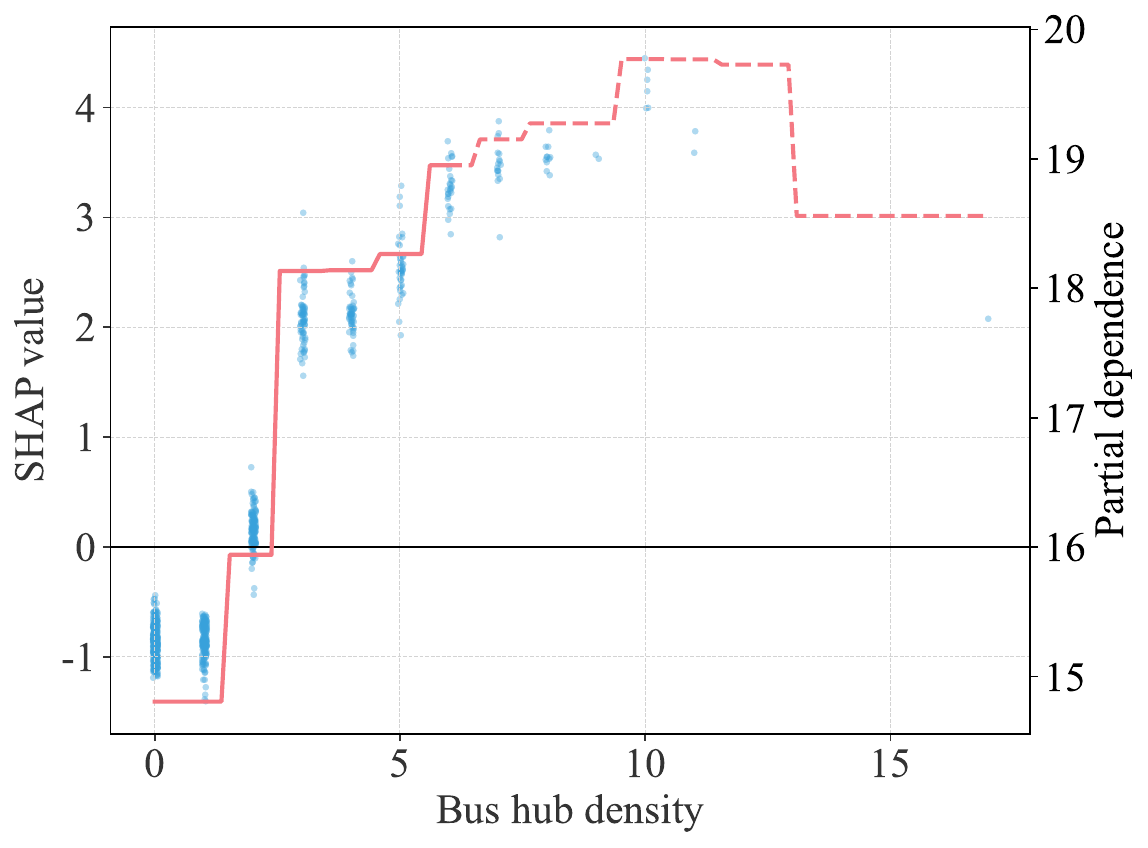}
            \caption*{Distance to bus hub}
        \end{subfigure}\hfill
        \begin{subfigure}[t]{0.5\linewidth}
            \centering
            \includegraphics[width=\linewidth]{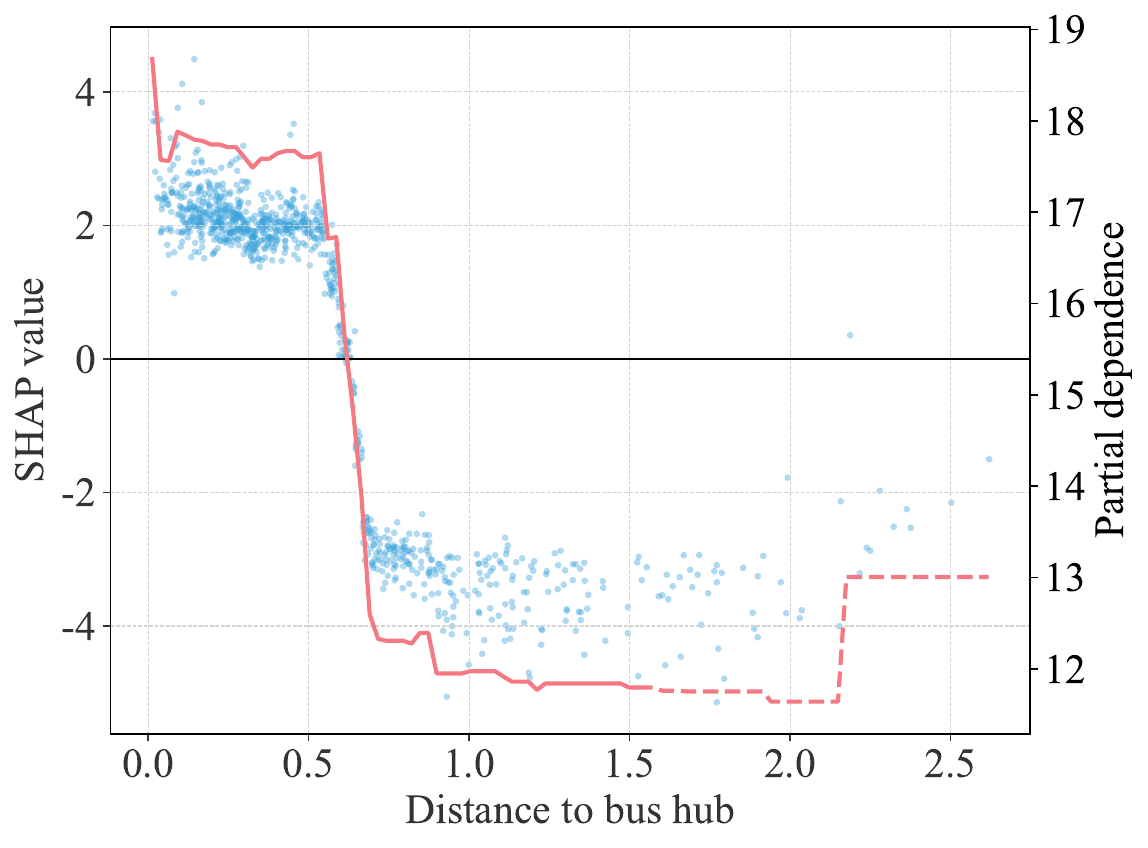}
            \caption*{Bus hub density}
        \end{subfigure}
        \caption{(400 m, 30-min, 2, 0.5)}
        \label{fig:30min_dep_group}
    \end{subfigure}
\caption{\centering{PDPs for top-2 variables (different travel time difference for departure substitutive ratio).}}
\label{fig:stability_travel PDPs departure substitutive ratio}
\end{figure}

\begin{figure}[!t]
    \centering
    \begin{subfigure}[h]{0.65\textwidth}
        \centering
        \includegraphics[width=\linewidth,keepaspectratio]{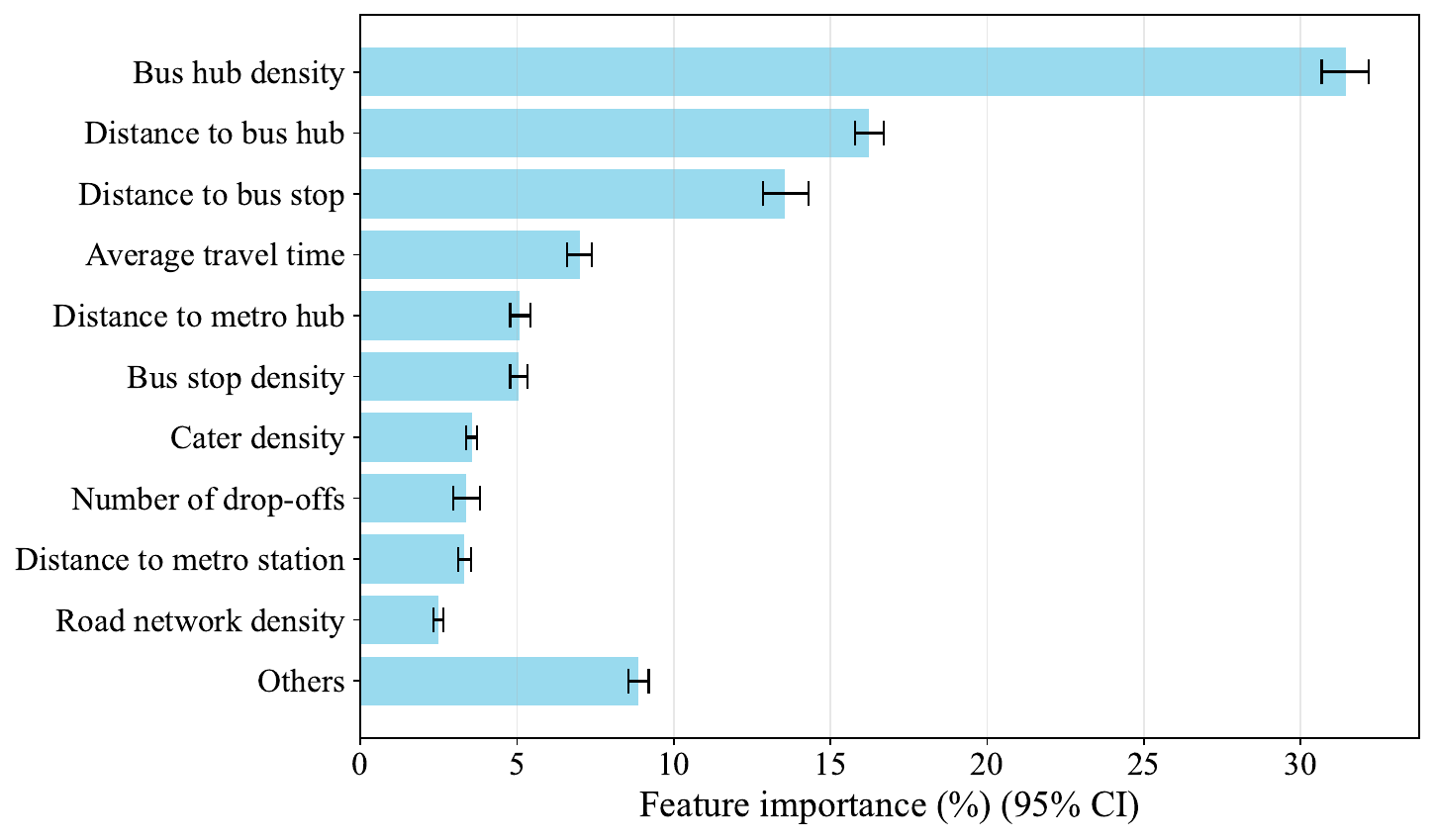}
        \caption{(300 m, 15-min, 2, 0.5)}
        \label{fig:300m arrival substitutive ratio}
    \end{subfigure}
    \begin{subfigure}[h]{0.65\textwidth}
        \centering
        \includegraphics[width=\linewidth,keepaspectratio]{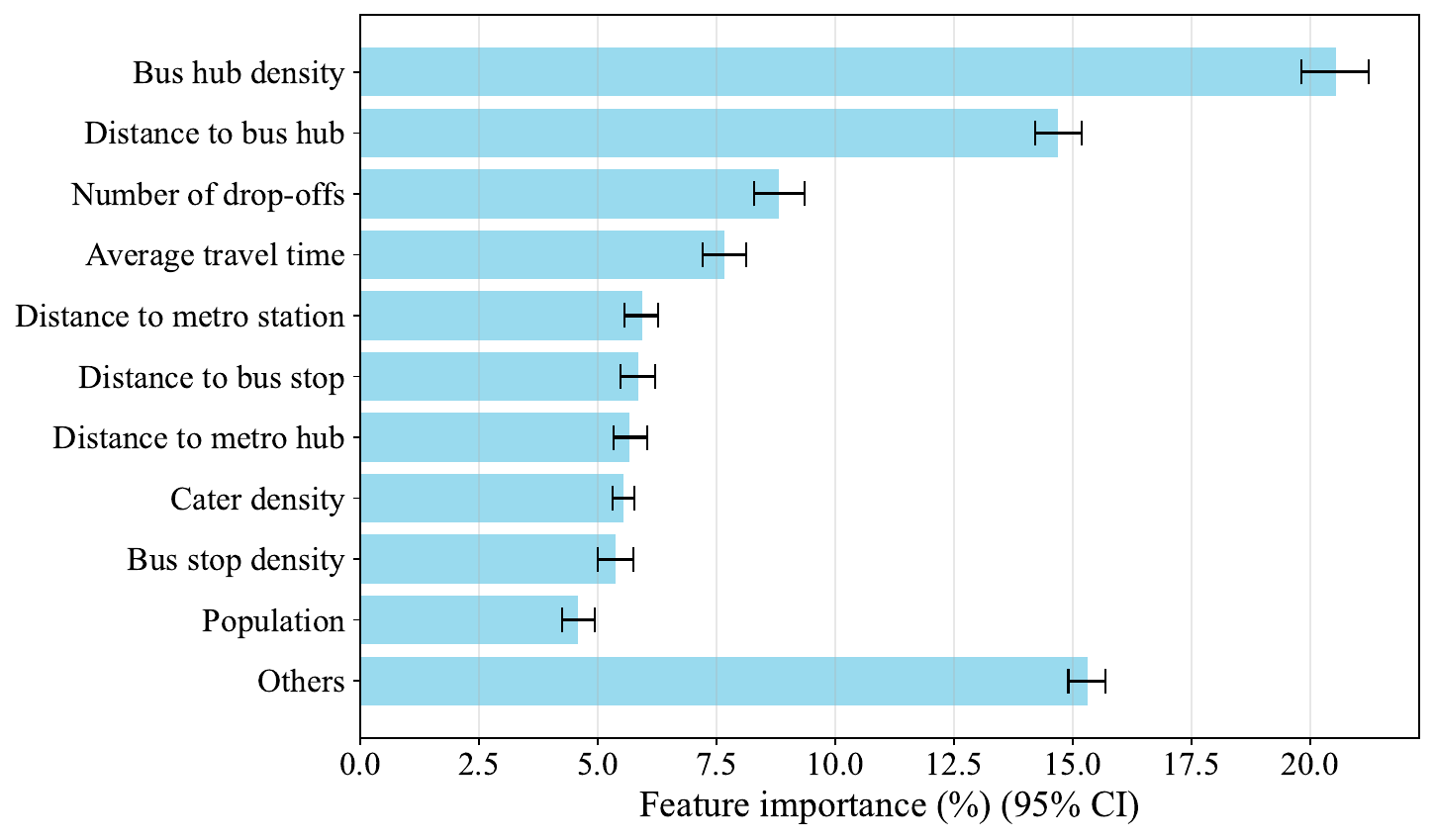}
        \caption{(350 m, 15-min, 2, 0.5)}
        \label{fig:350m arrival substitutive ratio}
    \end{subfigure}
    \begin{subfigure}[h]{0.65\textwidth}
        \centering
        \includegraphics[width=\linewidth,keepaspectratio]{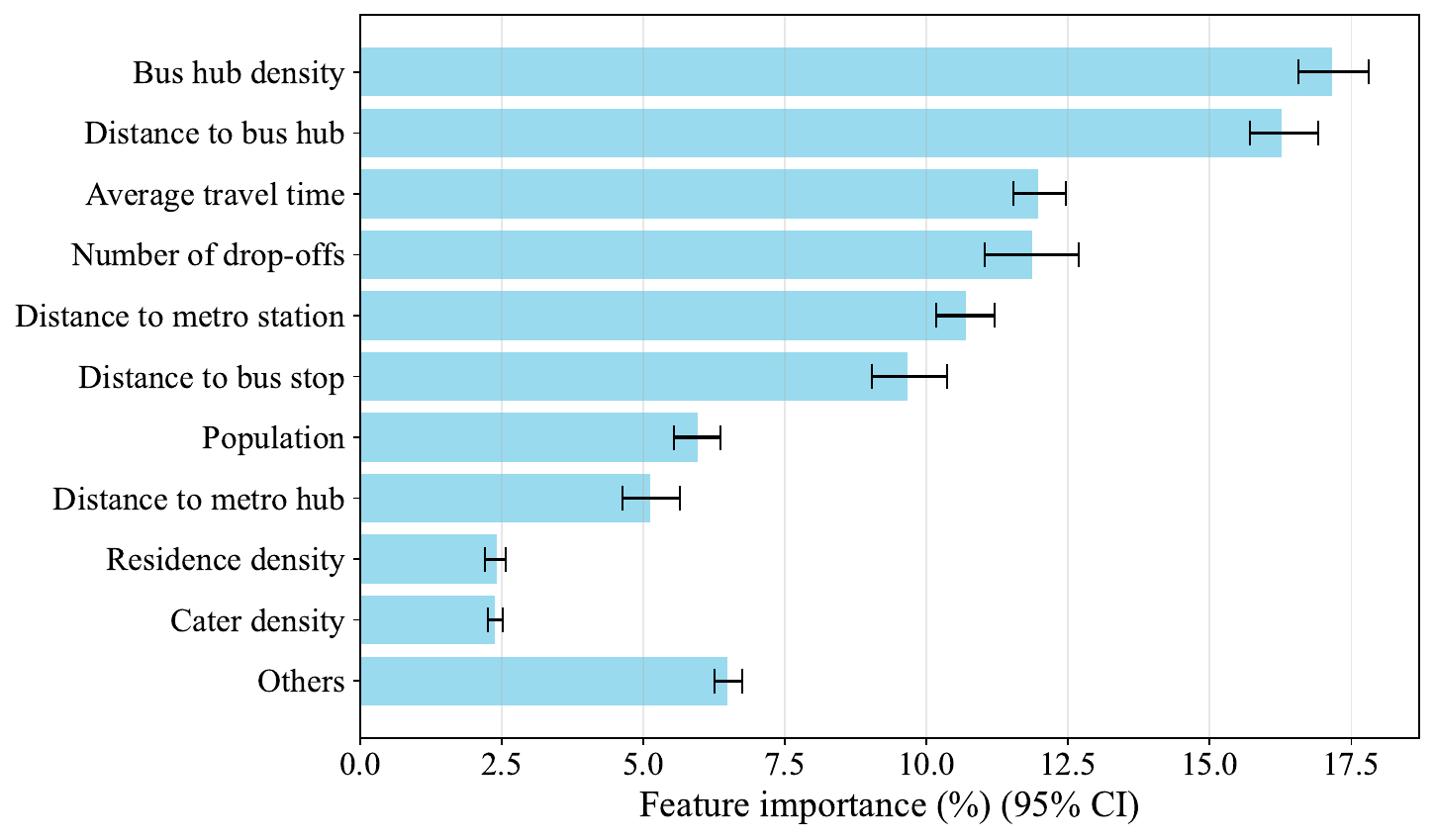}
        \caption{(450 m, 15-min, 2, 0.5)}
        \label{fig:450m arrival substitutive ratio}
    \end{subfigure}
    \begin{subfigure}[h]{0.65\textwidth}
        \centering
        \includegraphics[width=\linewidth,keepaspectratio]{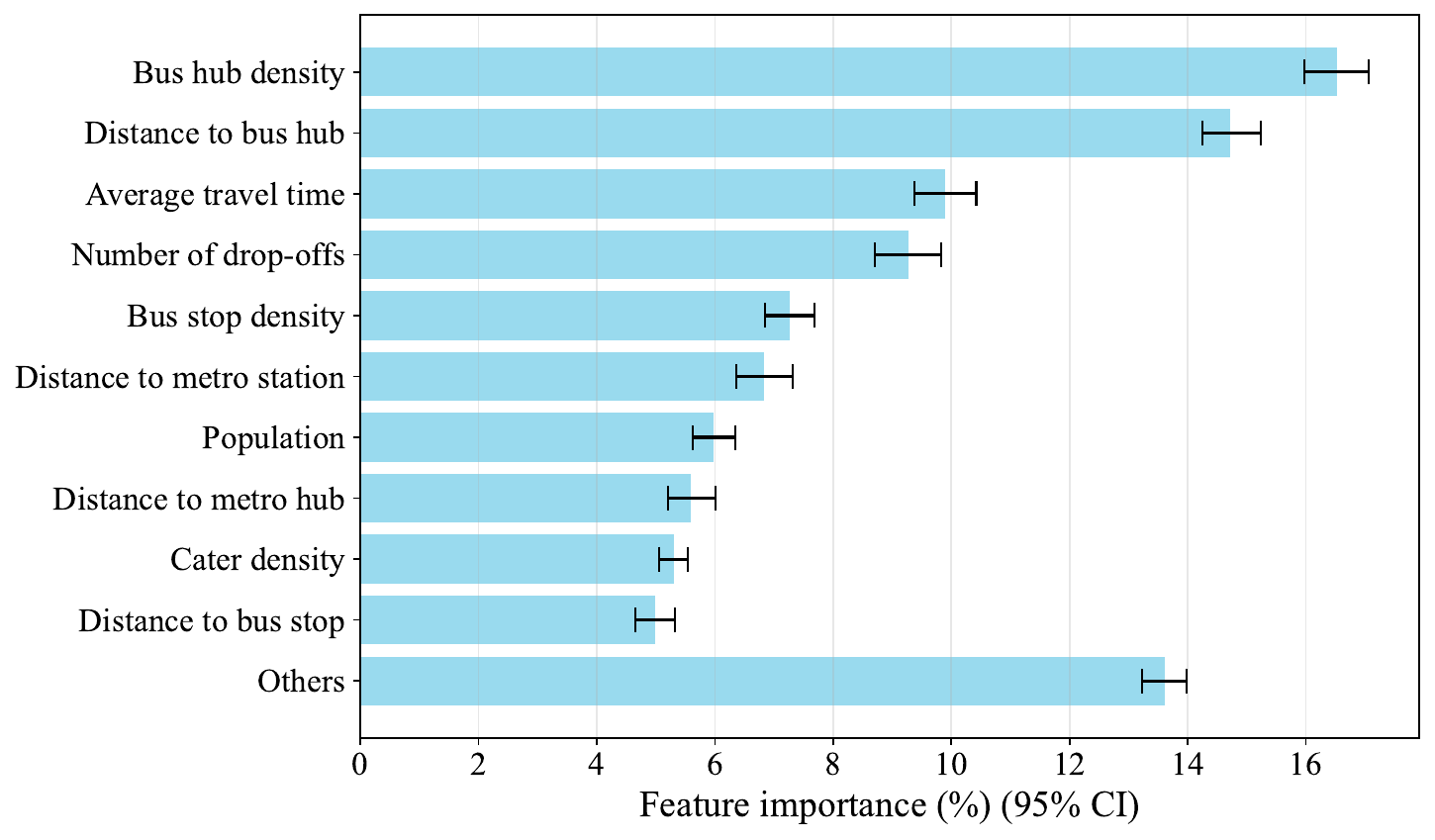}
        \caption{(500 m, 15-min, 2, 0.5)}
        \label{fig:500m arrival substitutive ratio}
    \end{subfigure}
    \caption*{\centering{Variable importance ranking (different walking distance for arrival substitutive ratio).}}
    \label{fig:Stability of ranking (walk) arrival substitutive ratio}
\end{figure}

\begin{figure}[!t]
    \centering
    \begin{subfigure}[t]{0.65\textwidth}
        \centering
        \begin{subfigure}[t]{0.5\linewidth}
            \centering
            \includegraphics[width=\linewidth]{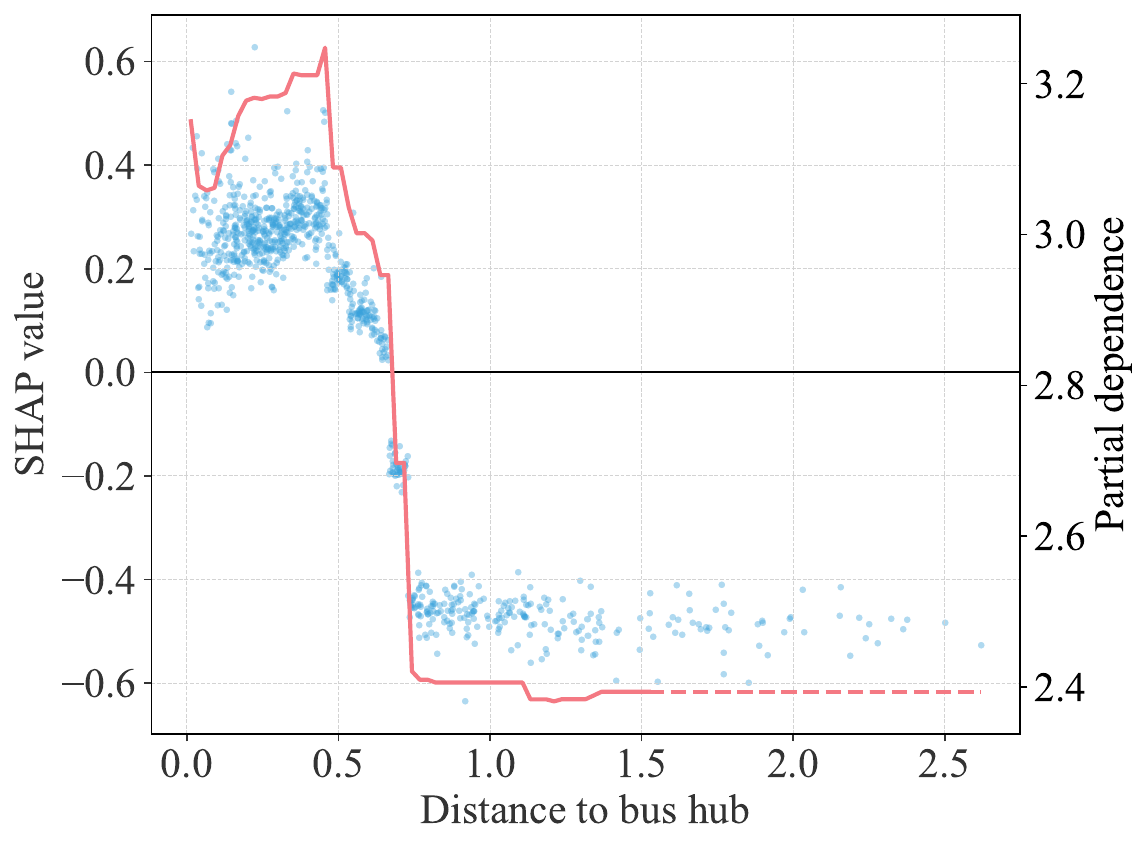}
            \caption*{Distance to bus hub}
            \label{fig:300m_arr_dist}
        \end{subfigure}\hfill
        \begin{subfigure}[t]{0.5\linewidth}
            \centering
            \includegraphics[width=\linewidth]{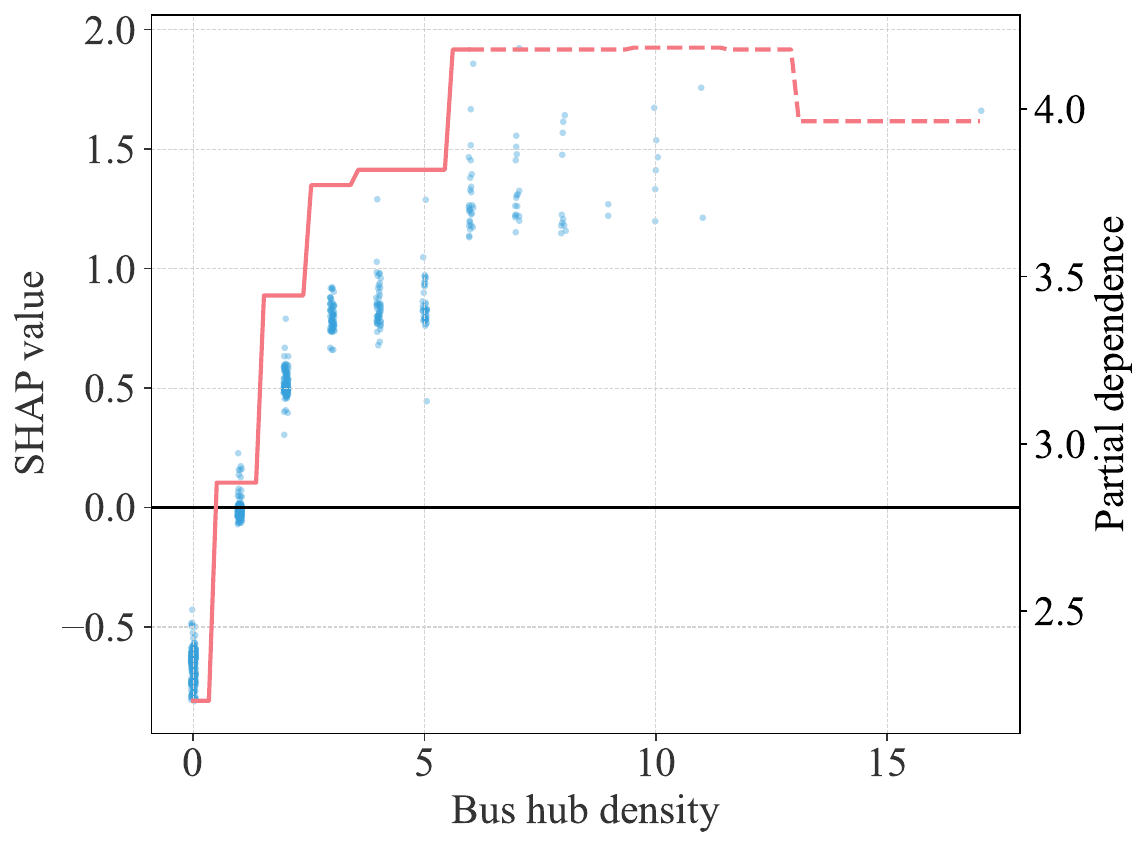}
            \caption*{Bus hub density}
            \label{fig:300m_arr_density}
        \end{subfigure}

        \caption{(300 m, 15-min, 2, 0.5)}
        \label{fig:300m_arr_group}
    \end{subfigure}
    \begin{subfigure}[t]{0.65\textwidth}
        \centering

        \begin{subfigure}[t]{0.5\linewidth}
            \centering
            \includegraphics[width=\linewidth]{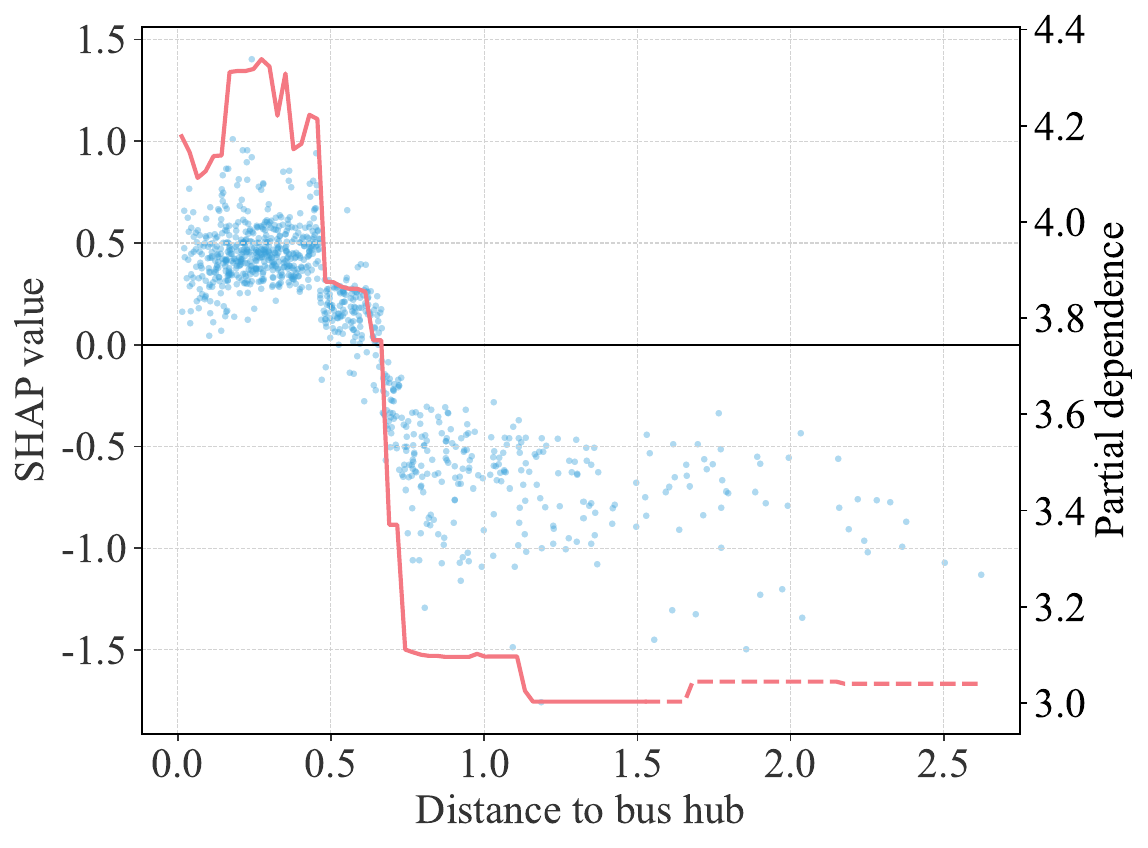}
            \caption*{Distance to bus hub}
            \label{fig:350m_arr_dist}
        \end{subfigure}\hfill
        \begin{subfigure}[t]{0.5\linewidth}
            \centering
            \includegraphics[width=\linewidth]{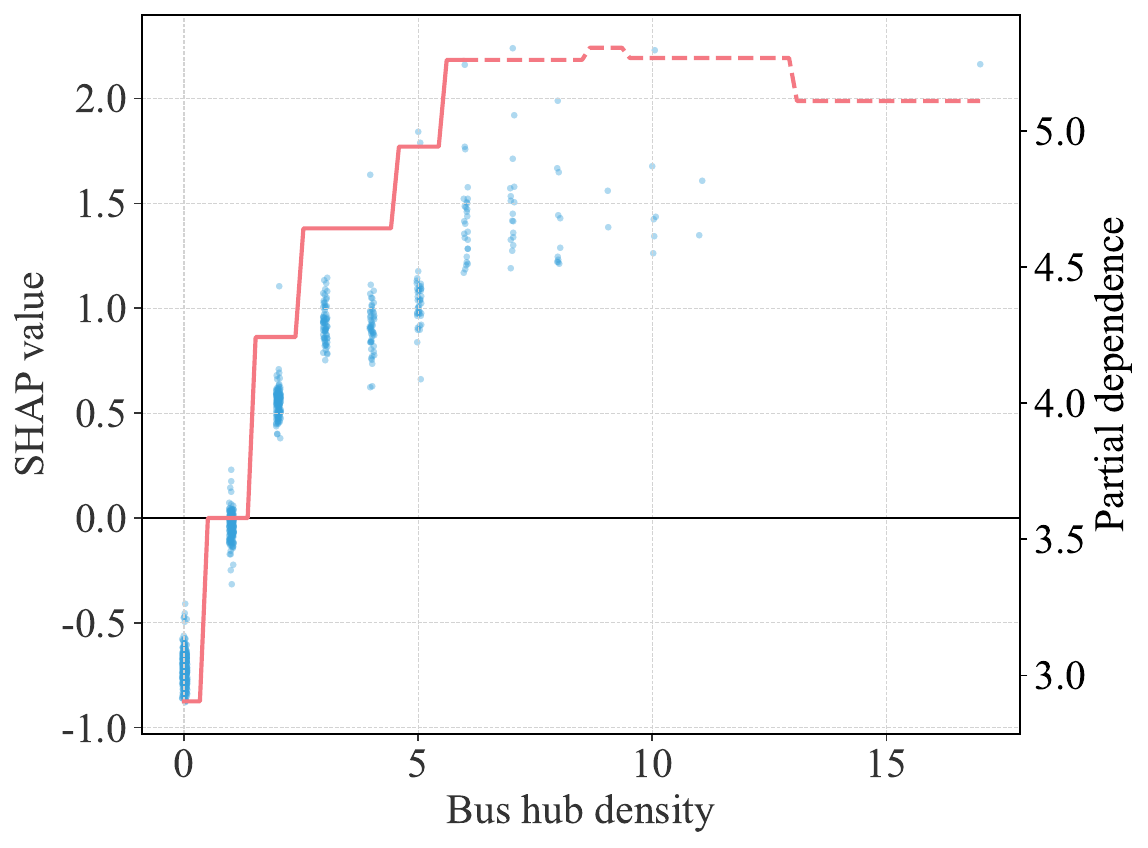}
            \caption*{Bus hub density}
            \label{fig:350m_arr_density}
        \end{subfigure}
        \caption{(350 m, 15-min, 2, 0.5)}
        \label{fig:350m_arr_group}
    \end{subfigure}
    \begin{subfigure}[t]{0.65\textwidth}
        \centering
        \begin{subfigure}[t]{0.5\linewidth}
            \centering
            \includegraphics[width=\linewidth]{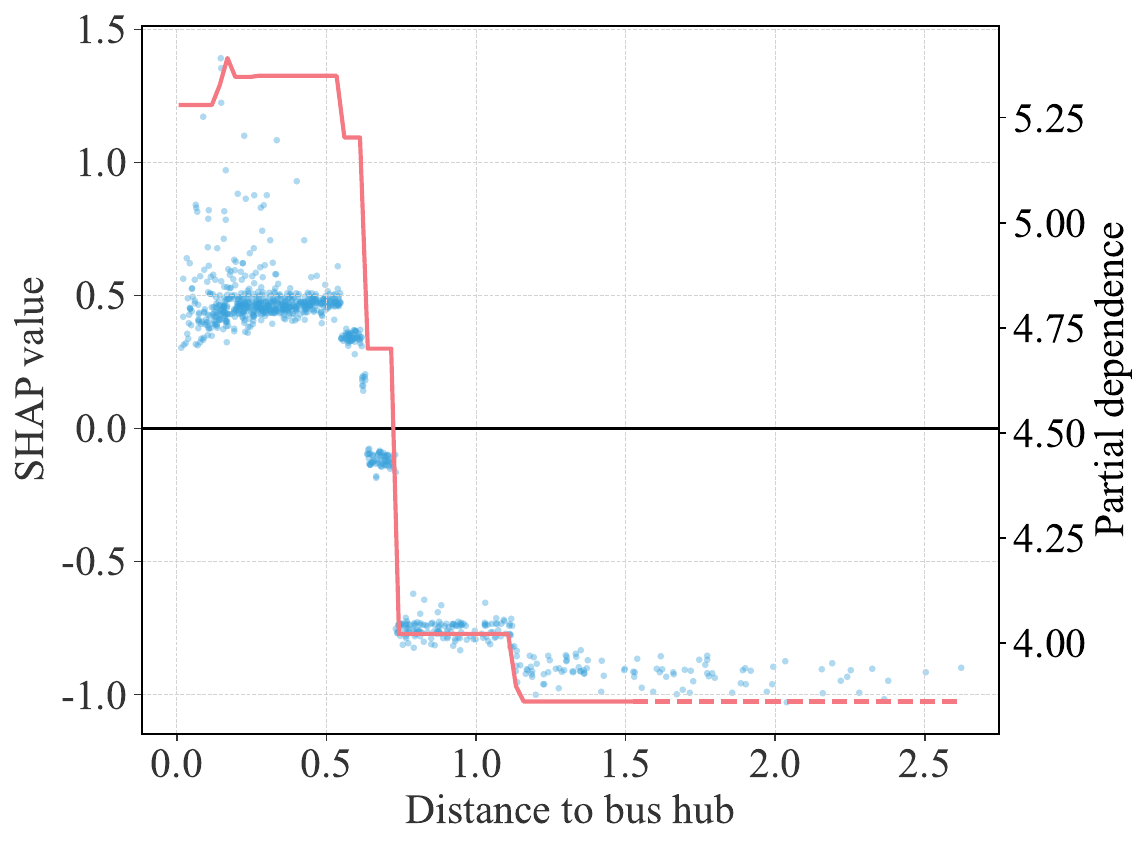}
            \caption*{Distance to bus hub}
            \label{fig:450m_arr_dist}
        \end{subfigure}\hfill
        \begin{subfigure}[t]{0.5\linewidth}
            \centering
            \includegraphics[width=\linewidth]{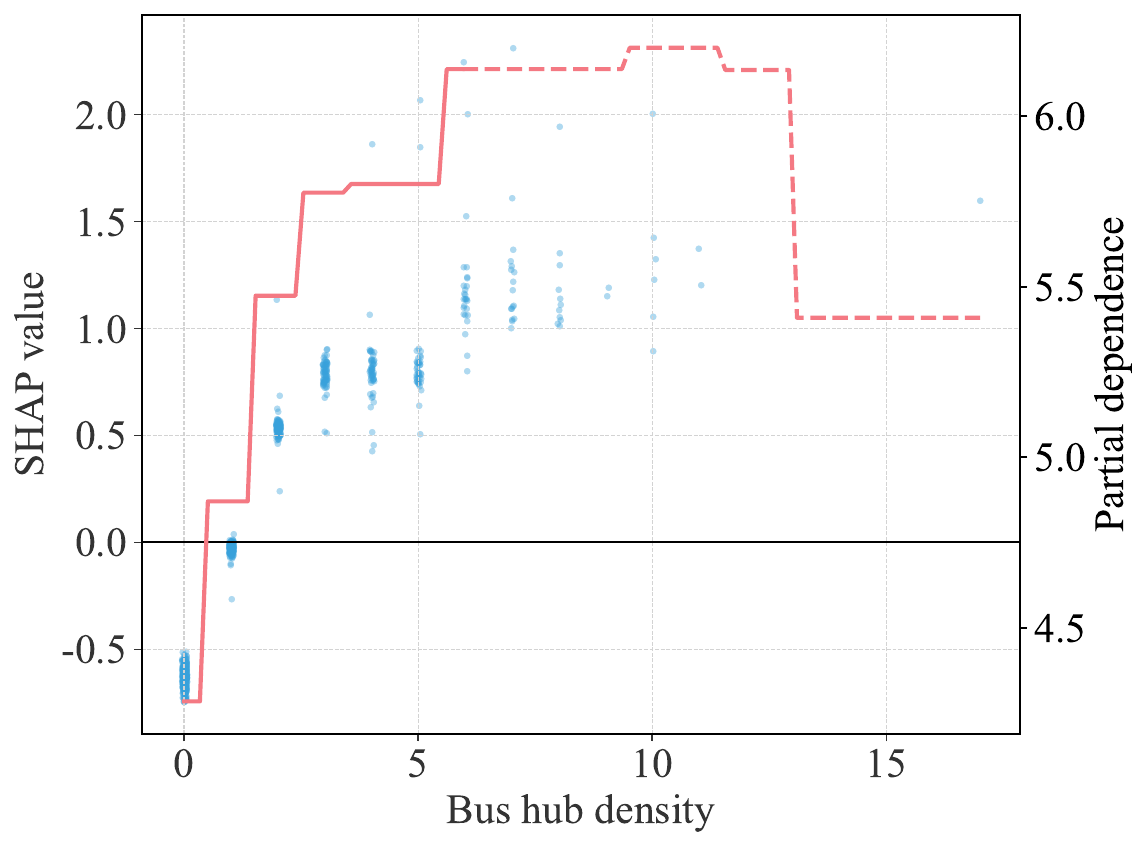}
            \caption*{Bus hub density}
            \label{fig:450m_arr_density}
        \end{subfigure}
        \caption{(450 m, 15-min, 2, 0.5)}
        \label{fig:450m_arr_group}
    \end{subfigure}
    \begin{subfigure}[t]{0.65\textwidth}
        \centering
        \begin{subfigure}[t]{0.5\linewidth}
            \centering
            \includegraphics[width=\linewidth]{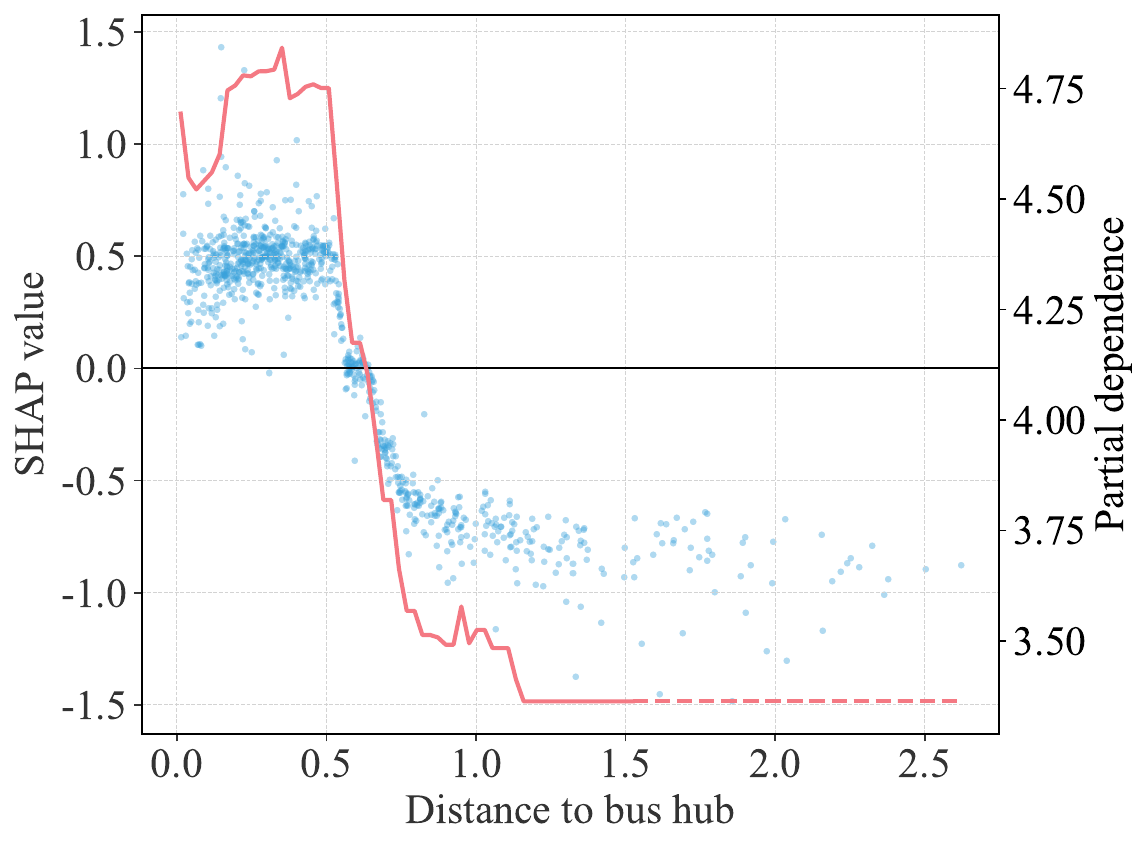}
            \caption*{Distance to bus hub}
            \label{fig:500m_arr_dist}
        \end{subfigure}\hfill
        \begin{subfigure}[t]{0.5\linewidth}
            \centering
            \includegraphics[width=\linewidth]{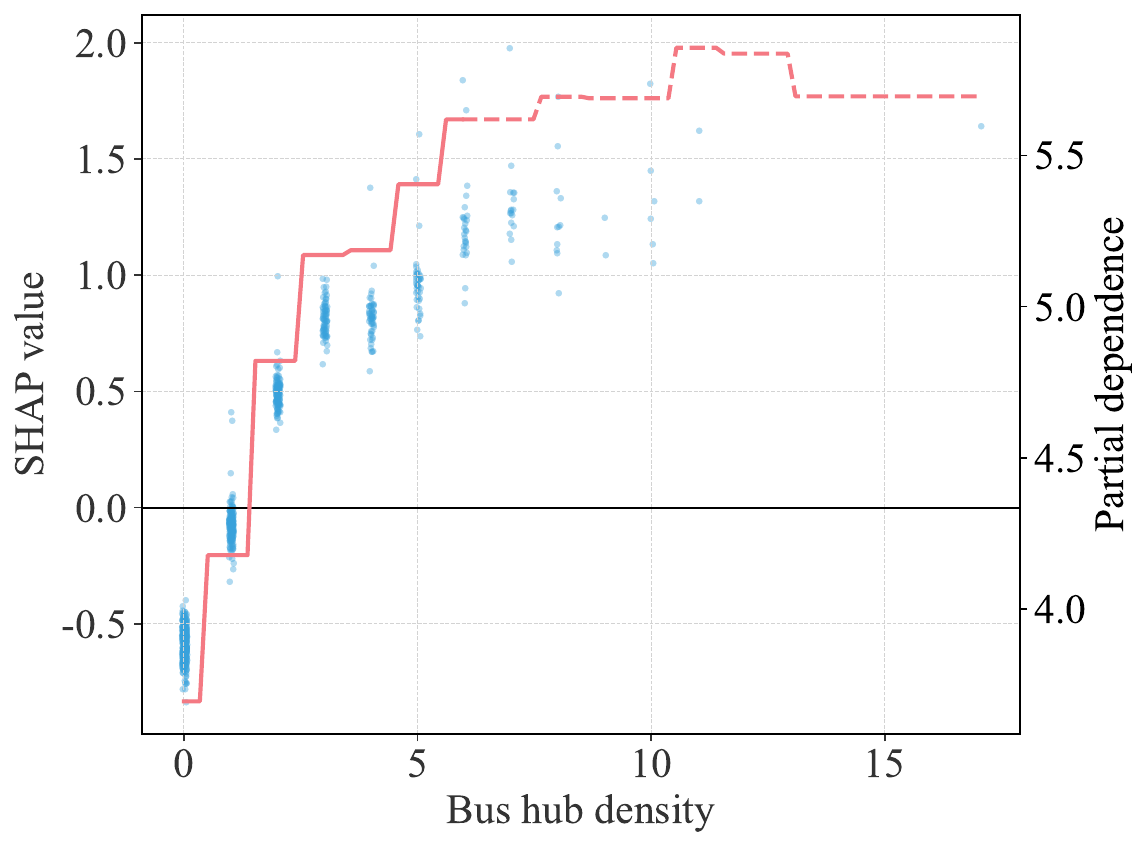}
            \caption*{Bus hub density}
            \label{fig:500m_arr_density}
        \end{subfigure}
        \caption{(500 m, 15-min, 2, 0.5)}
        \label{fig:500m_arr_group}
    \end{subfigure}
\caption{\centering{PDPs for top-2 variables (different walking distance for arrival substitutive ratio).}}
\label{fig:stability_walk PDPs arrival substitutive ratio}
\end{figure}

\begin{figure}[!t]
    \centering
    \begin{subfigure}[h]{0.65\textwidth}
        \centering
        \includegraphics[width=\linewidth,keepaspectratio]{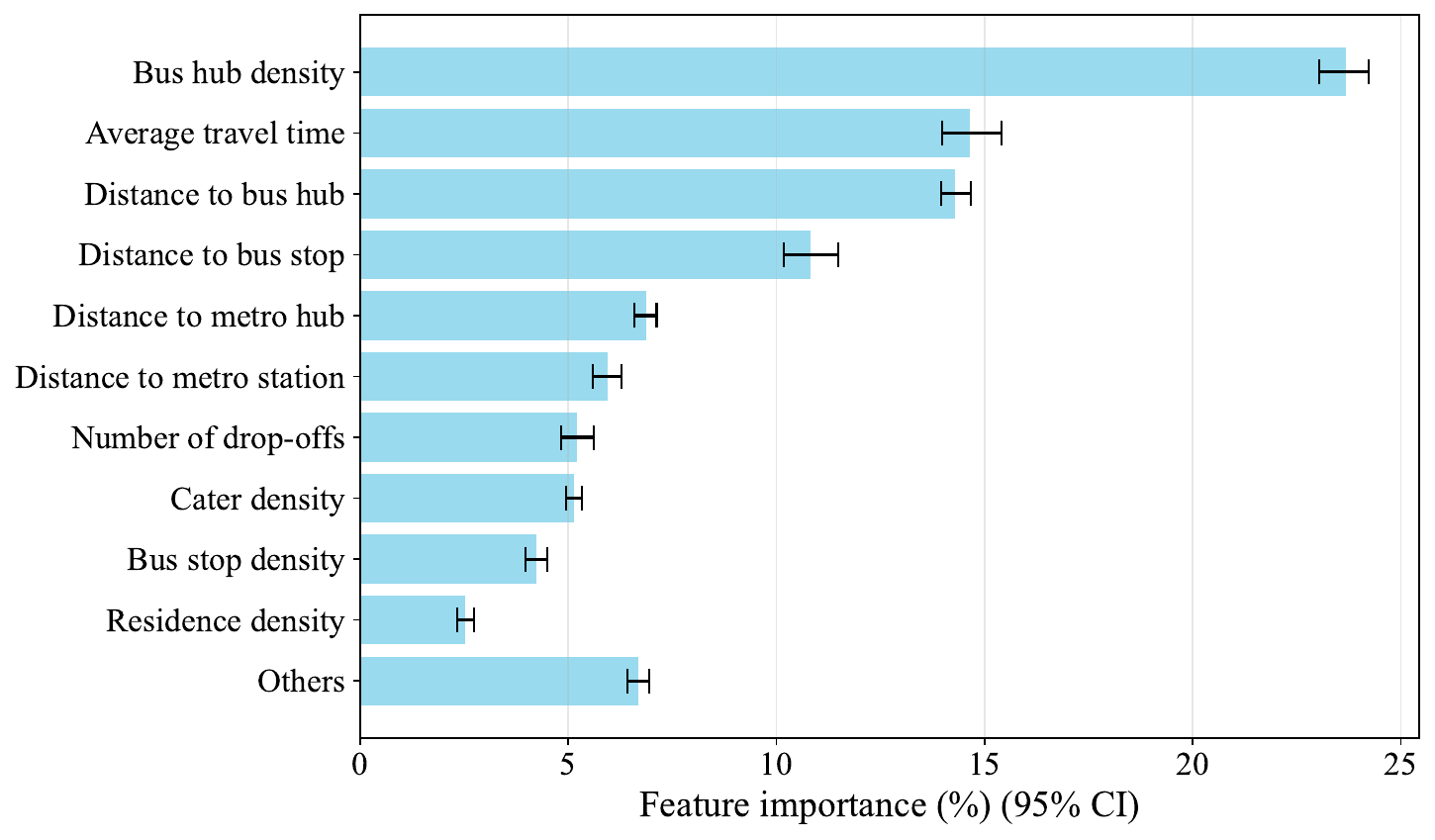}
        \caption{(400 m, 20-min, 2, 0.5)}
        \label{fig:20-min arrival substitutive ratio}
    \end{subfigure}
    \begin{subfigure}[h]{0.65\textwidth}
        \centering
        \includegraphics[width=\linewidth,keepaspectratio]{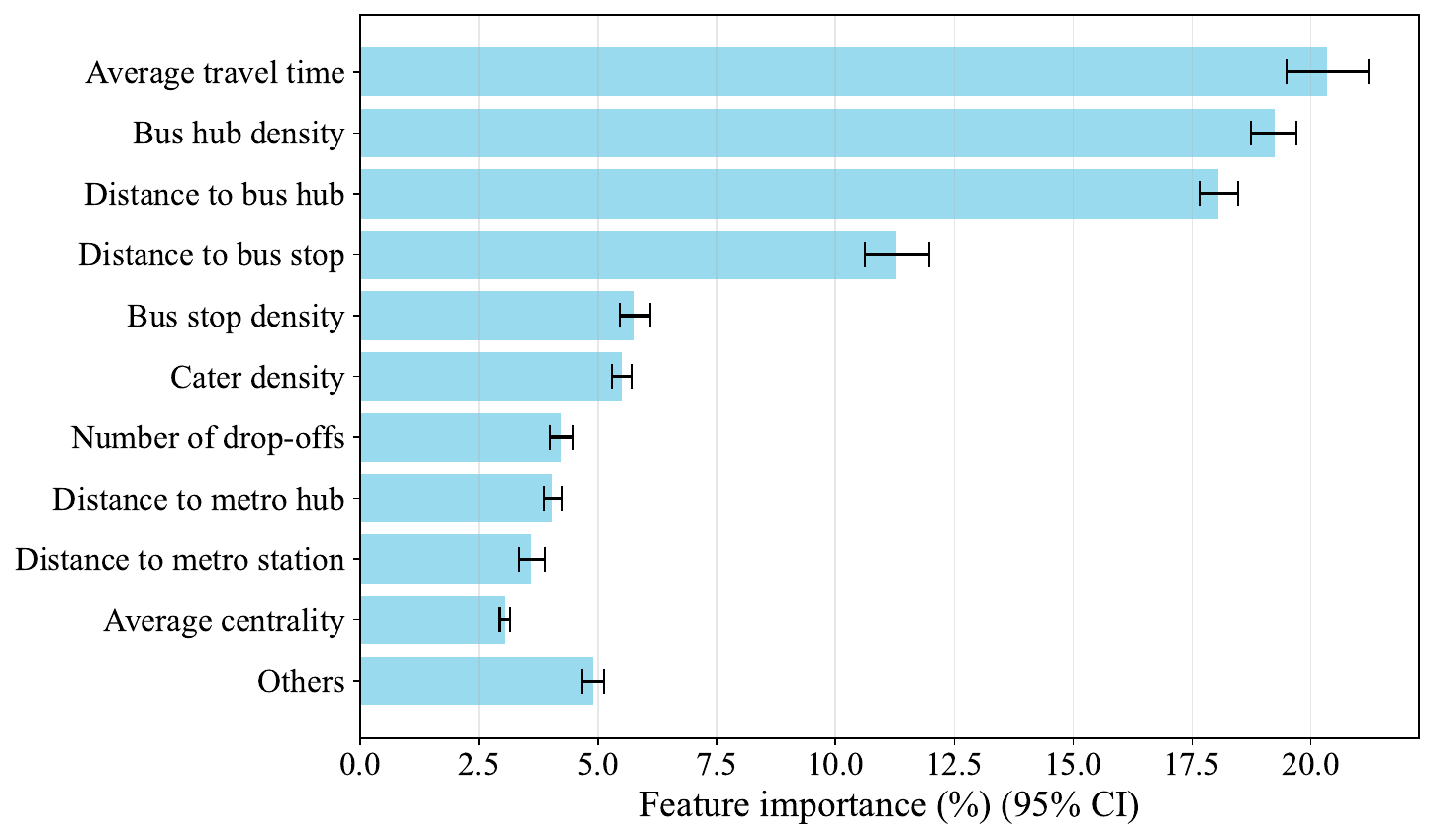}
        \caption{(400 m, 25-min, 2, 0.5)}
        \label{fig:25-min arrival substitutive ratio}
    \end{subfigure}
    \begin{subfigure}[h]{0.65\textwidth}
        \centering
        \includegraphics[width=\linewidth,keepaspectratio]{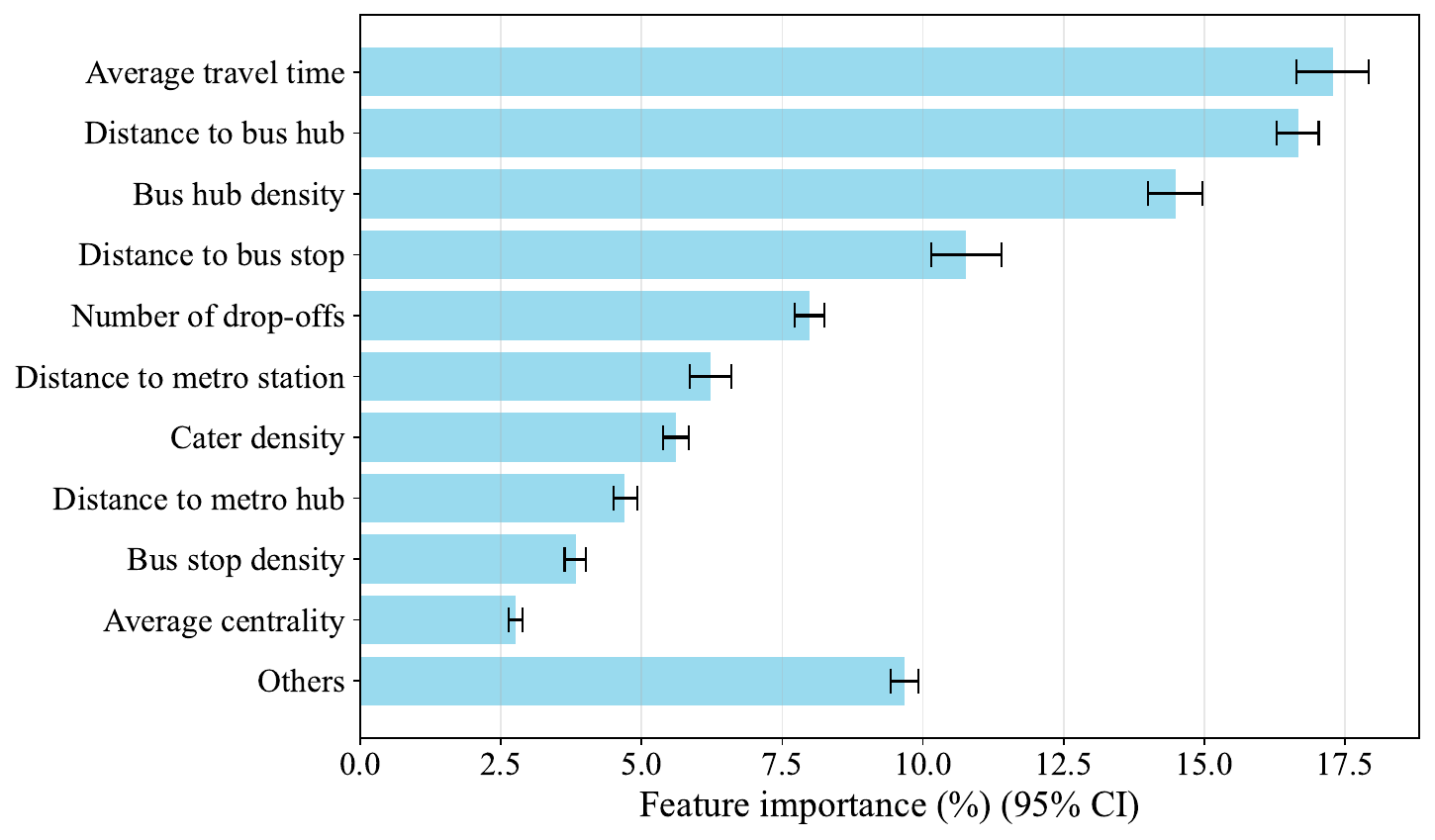}
        \caption{(400 m, 30-min, 2, 0.5)}
        \label{fig:30-min arrival substitutive ratio}
    \end{subfigure}
    \caption{\centering{Variable importance ranking (different travel time difference for arrival substitutive ratio).}}
    \label{fig:Stability of ranking (travel) arrival substitutive ratio}
\end{figure}

\begin{figure}[!t]
    \centering
    \begin{subfigure}[t]{0.65\textwidth}
        \centering
        \begin{subfigure}[t]{0.5\linewidth}
            \centering
            \includegraphics[width=\linewidth]{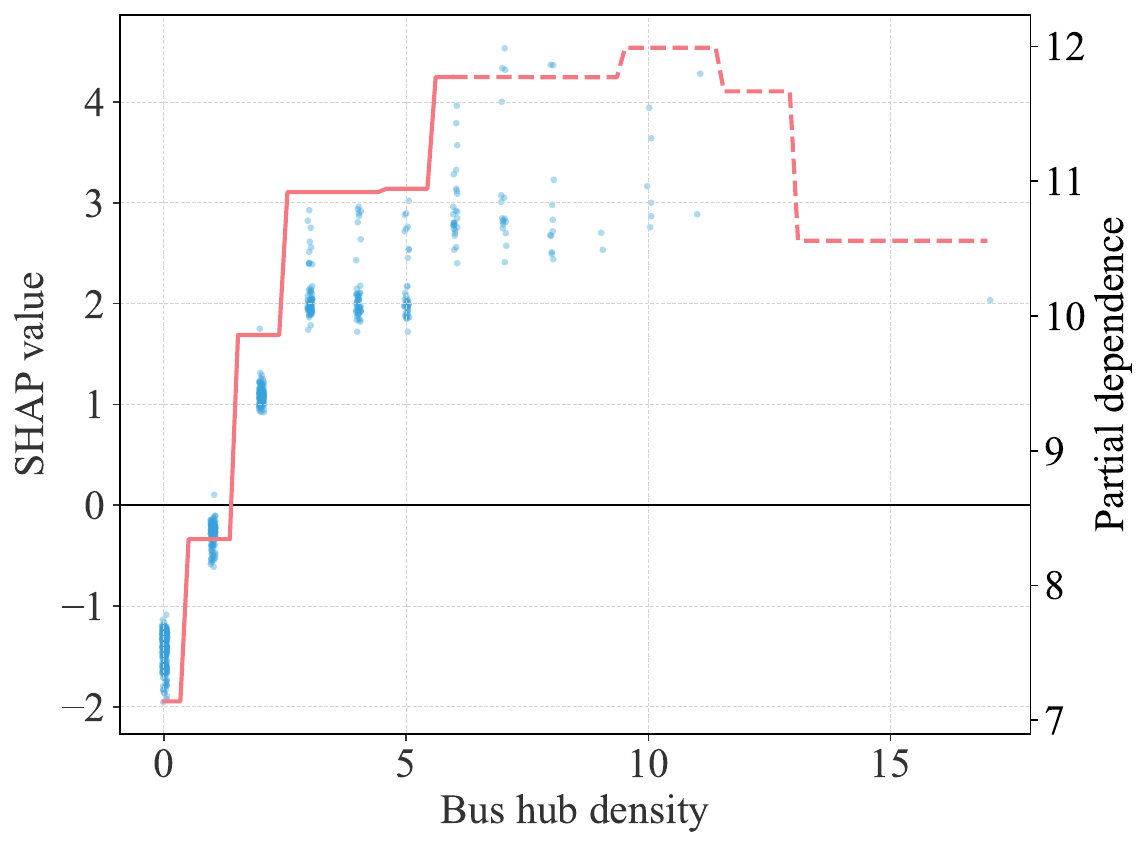}
            \caption*{Distance to bus hub}
        \end{subfigure}\hfill
        \begin{subfigure}[t]{0.5\linewidth}
            \centering
            \includegraphics[width=\linewidth]{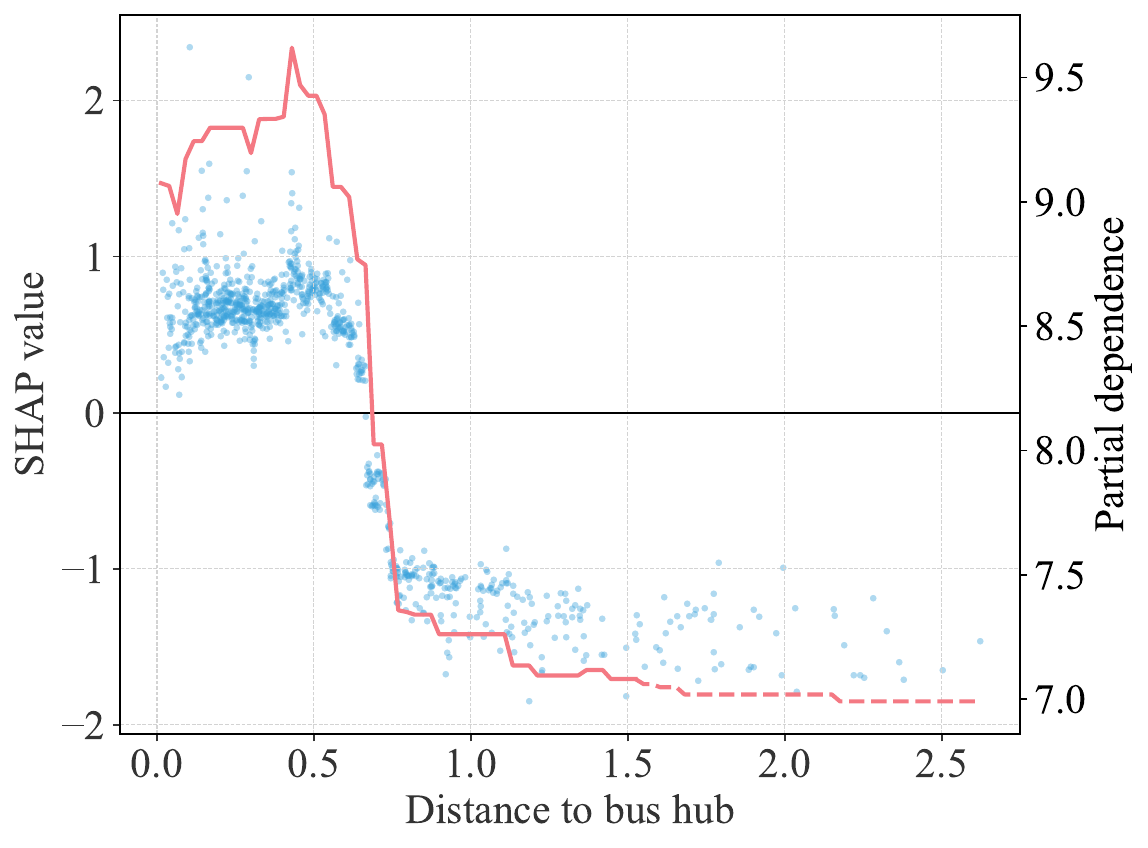}
            \caption*{Bus hub density}
        \end{subfigure}
        \caption{(400 m, 20-min, 2, 0.5)}
        \label{fig:20min_arr_group}
    \end{subfigure}
    \begin{subfigure}[t]{0.65\textwidth}
        \centering
        \begin{subfigure}[t]{0.5\linewidth}
            \centering
            \includegraphics[width=\linewidth]{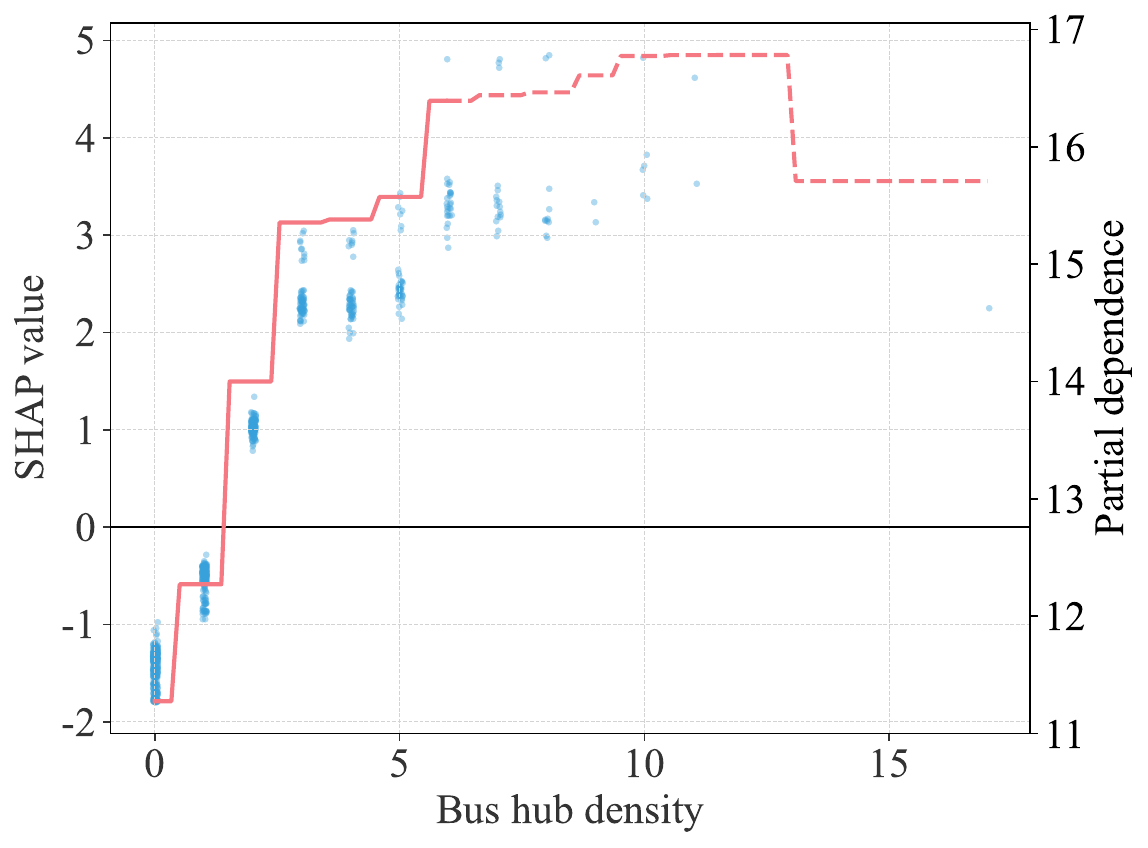}
            \caption*{Distance to bus hub}
        \end{subfigure}\hfill
        \begin{subfigure}[t]{0.5\linewidth}
            \centering
            \includegraphics[width=\linewidth]{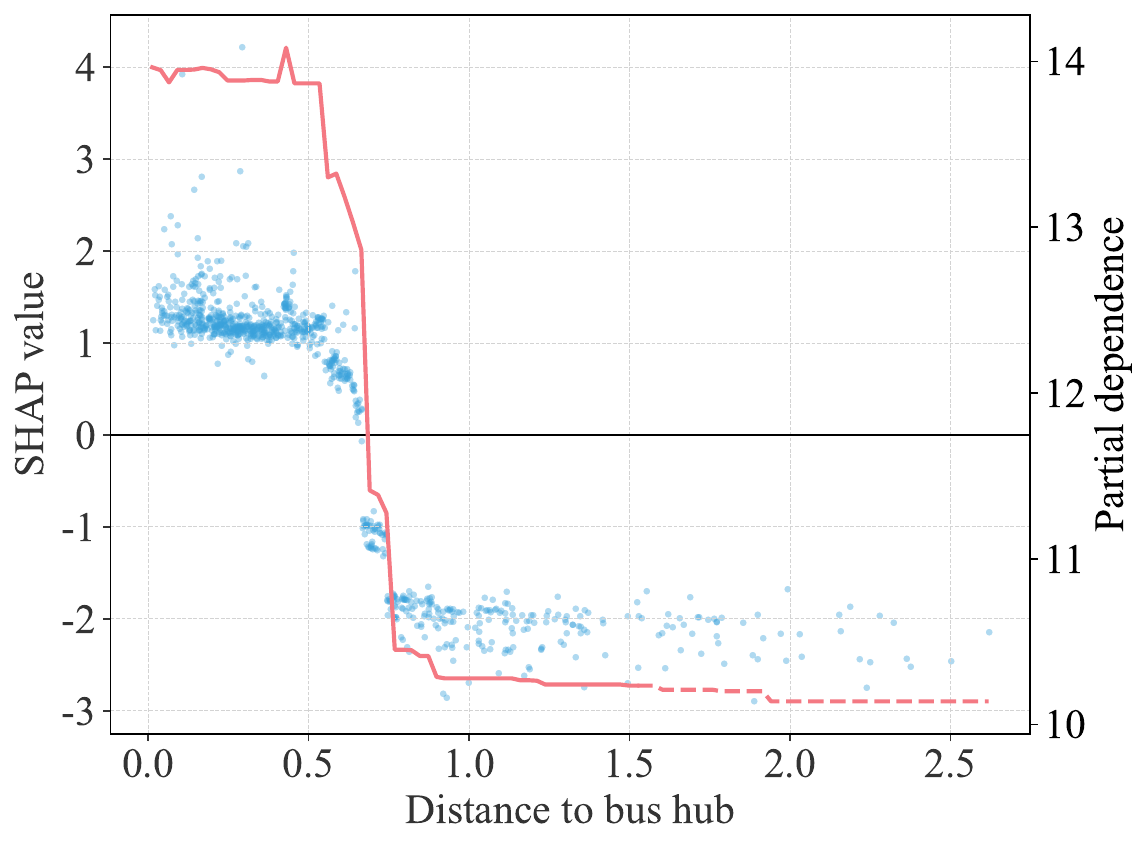}
            \caption*{Bus hub density}
        \end{subfigure}
        \caption{(400 m, 25-min, 2, 0.5)}
        \label{fig:25min_arr_group}
    \end{subfigure}
    \begin{subfigure}[t]{0.65\textwidth}
        \centering
        \begin{subfigure}[t]{0.5\linewidth}
            \centering
            \includegraphics[width=\linewidth]{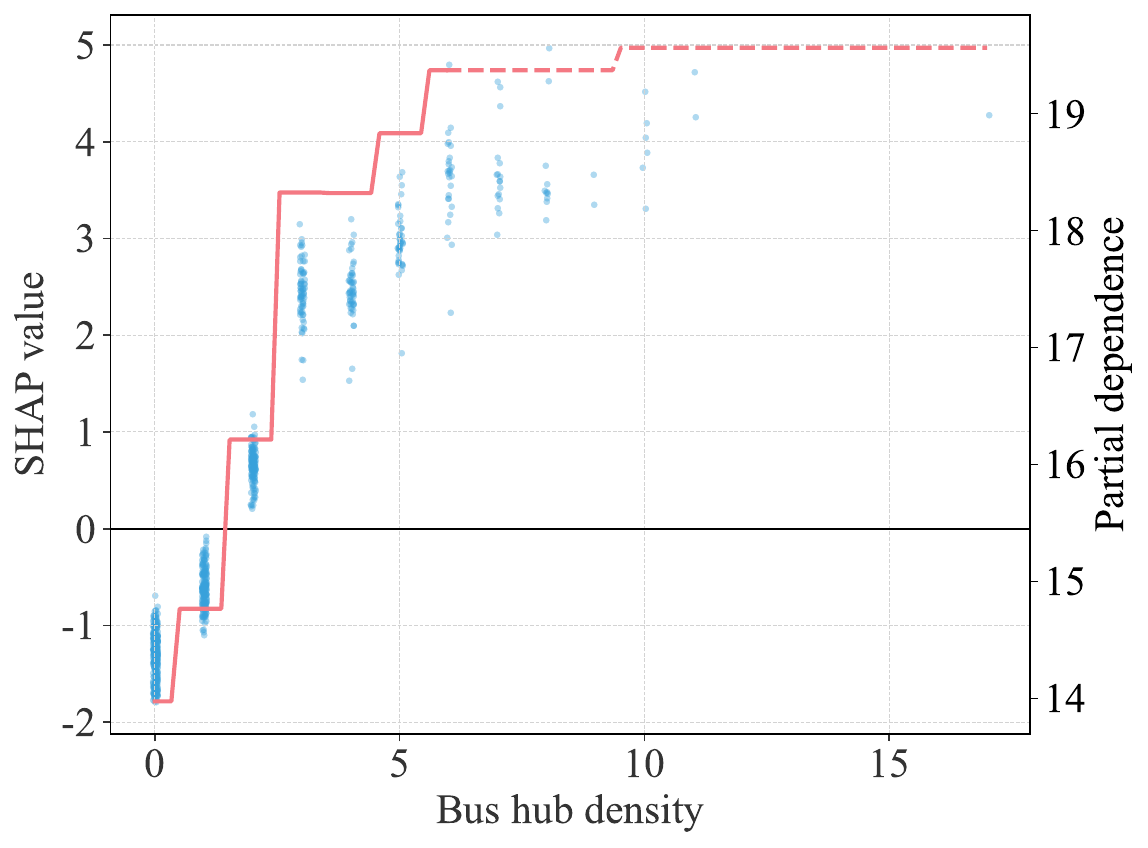}
            \caption*{Distance to bus hub}
        \end{subfigure}\hfill
        \begin{subfigure}[t]{0.5\linewidth}
            \centering
            \includegraphics[width=\linewidth]{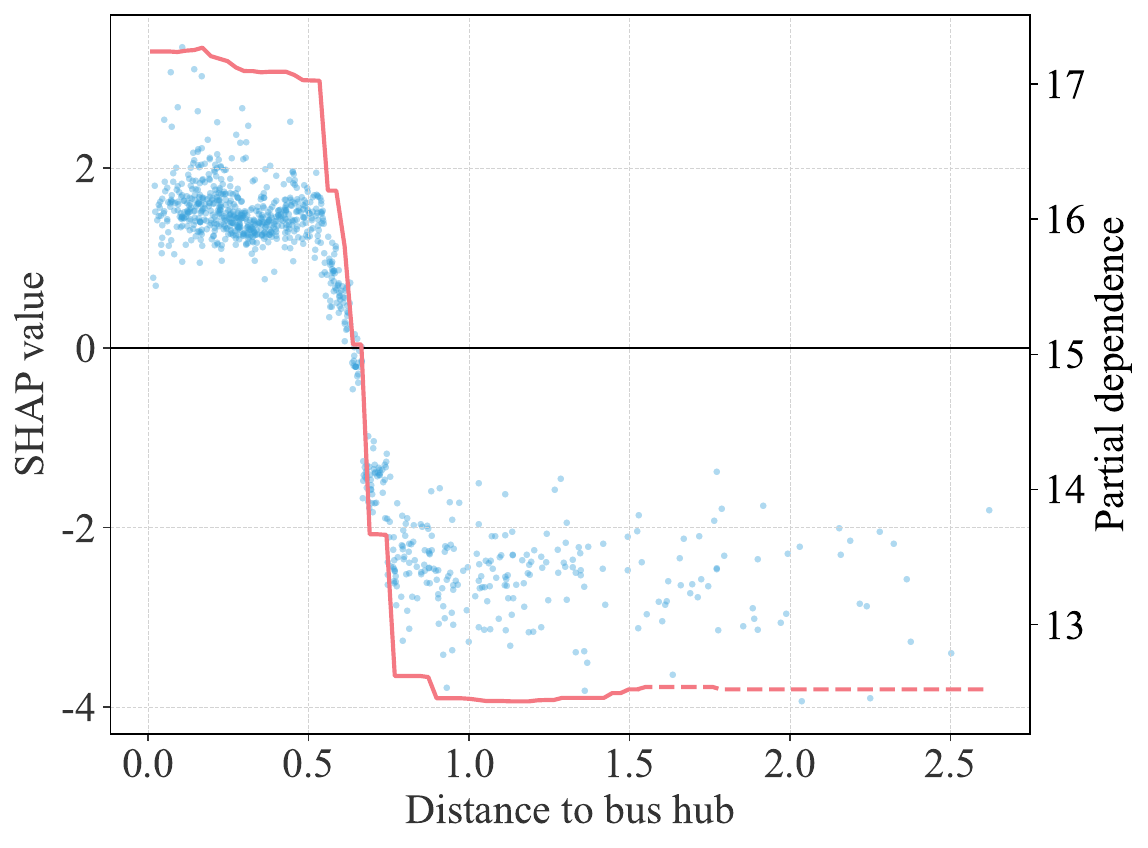}
            \caption*{Bus hub density}
        \end{subfigure}
        \caption{(400 m, 30-min, 2, 0.5)}
        \label{fig:30min_arr_group}
    \end{subfigure}
\caption{\centering{PDPs for top-2 variables (different travel time difference for arrival substitutive ratio).}}
\label{fig:stability_travel PDPs arrival substitutive ratio}
\end{figure}

\end{landscape}

\subsection*{Robustness of the ML interpretations under TabNet model}
\begin{figure}[h]
    \centering
    \begin{subfigure}[h]{0.48\textwidth}
        \centering
        \includegraphics[width=\linewidth,keepaspectratio]{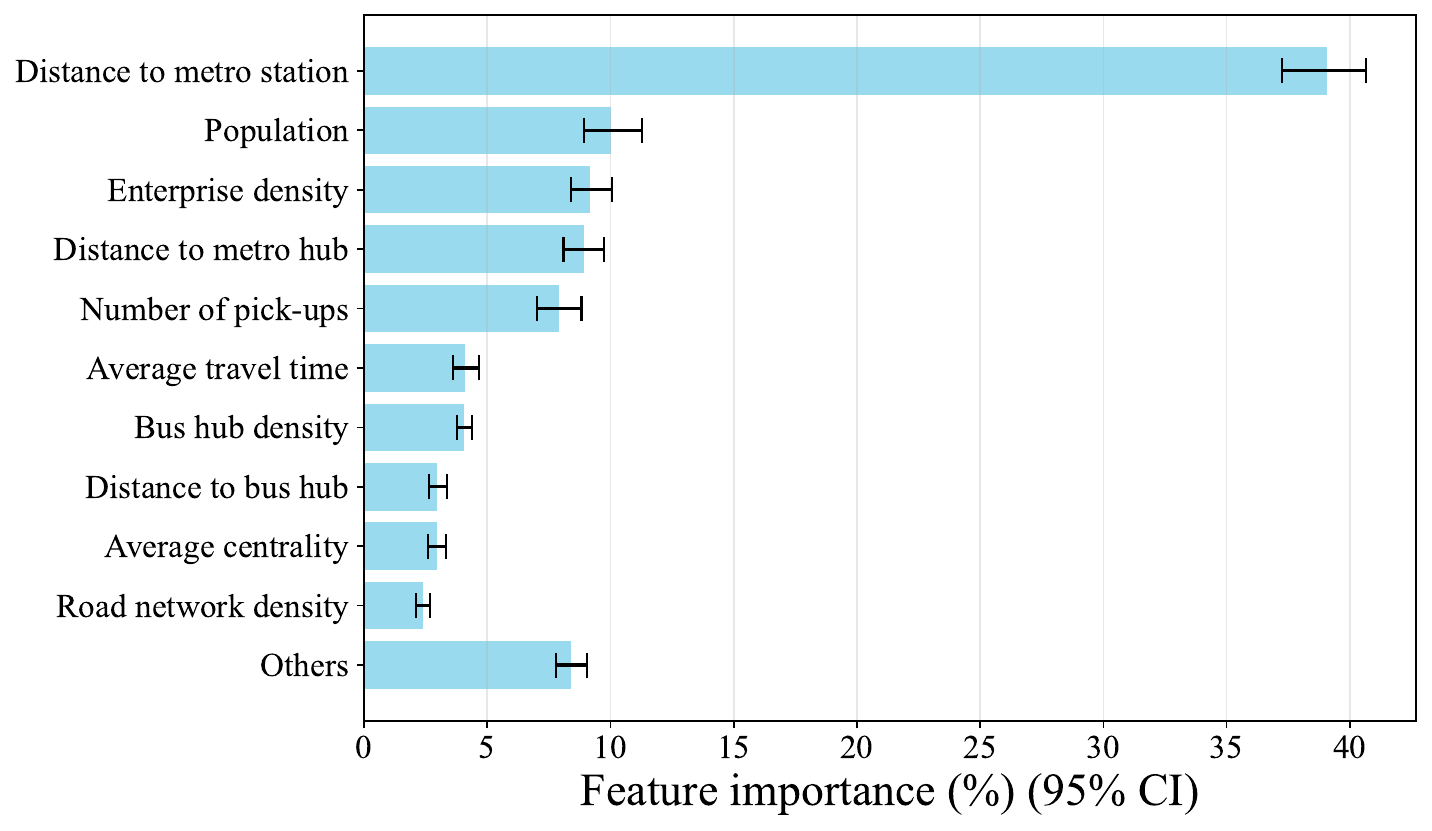}
        \caption{First-mile complementary ratio}
        \label{fig:TabNet_First-mile complement ratio}
    \end{subfigure}
    \begin{subfigure}[h]{0.48\textwidth}
        \centering
        \includegraphics[width=\linewidth,keepaspectratio]{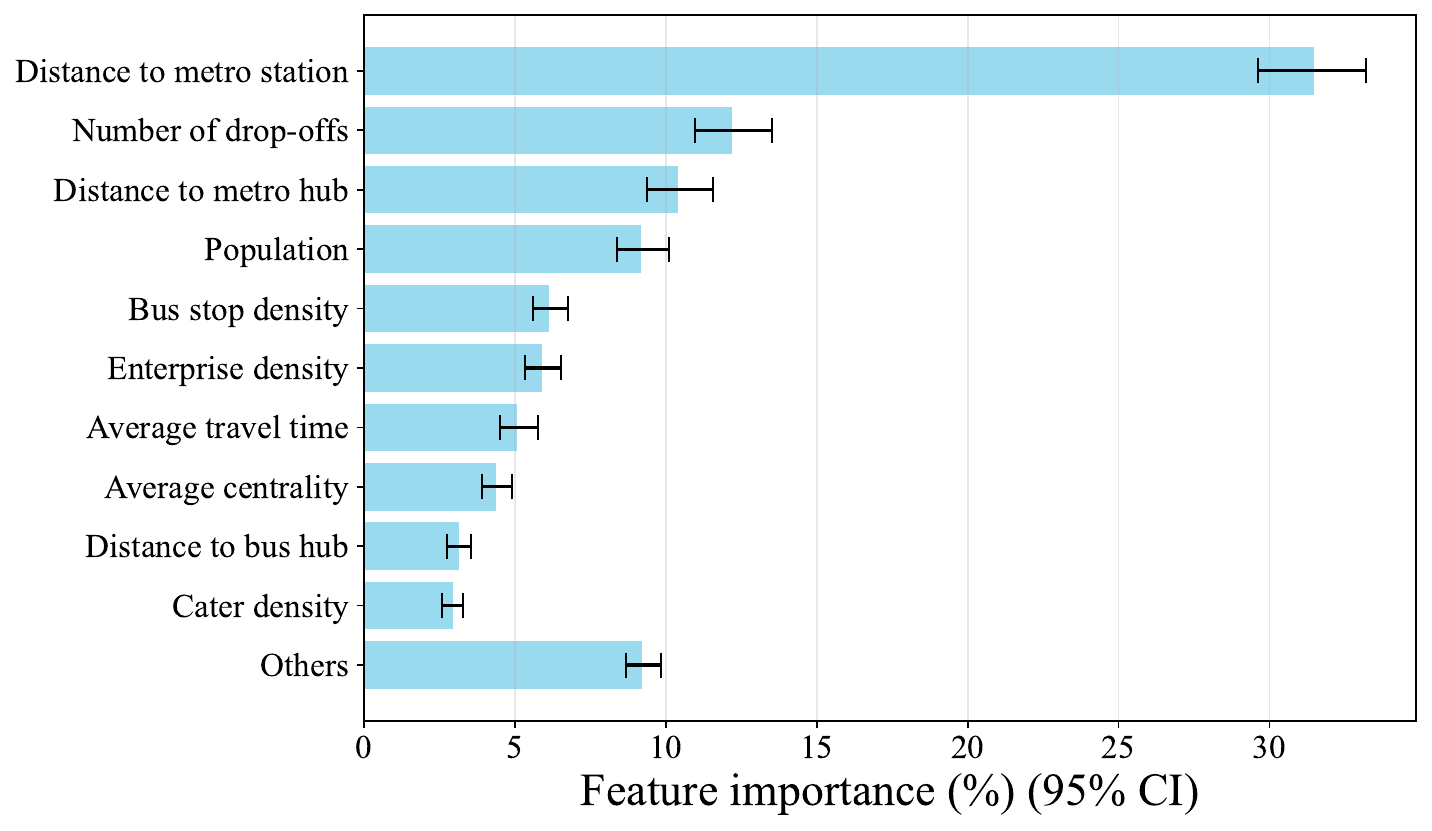}
        \caption{Last-mile complementary ratio}
        \label{fig:TabNet_Last-mile complement ratio}
    \end{subfigure}
    \begin{subfigure}[h]{0.48\textwidth}
        \centering
        \includegraphics[width=\linewidth,keepaspectratio]{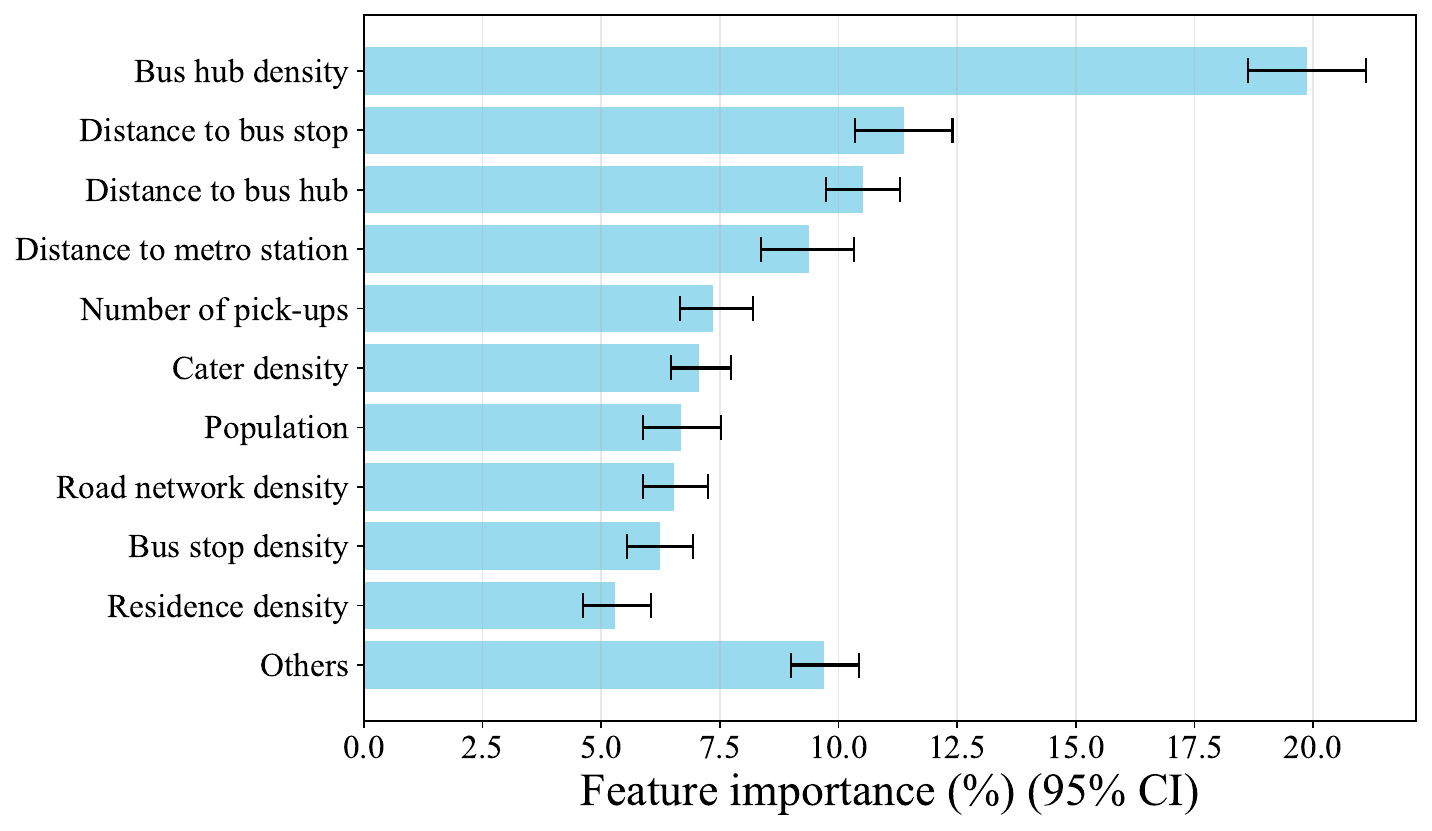}
        \caption{Departure substitutive ratio}
        \label{fig:TabNet_Departure substitution ratio}
    \end{subfigure}
    \begin{subfigure}[h]{0.48\textwidth}
        \centering
        \includegraphics[width=\linewidth,keepaspectratio]{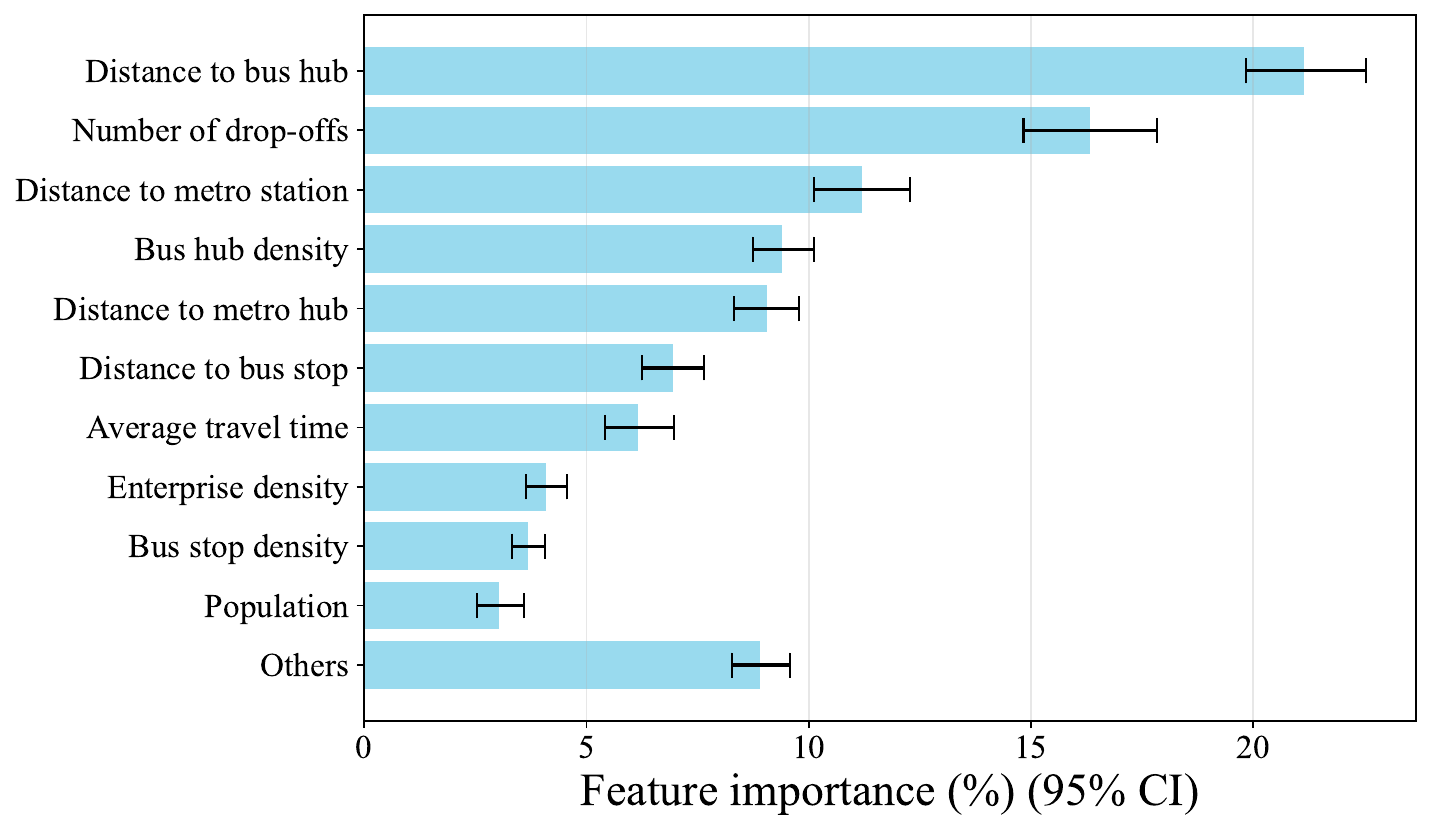}
        \caption{Arrival substitutive ratio}
        \label{fig:TabNet_Arrival substitution ratio}
    \end{subfigure}
    \caption{\centering{Variable importance ranking under TabNet model.}}
    \label{fig:Variable importance ranking under TabNet model}
\end{figure}

\begin{figure}[h]
    \centering
    \begin{subfigure}[h]{0.48\textwidth}
        \centering
        \includegraphics[width=\linewidth,keepaspectratio]{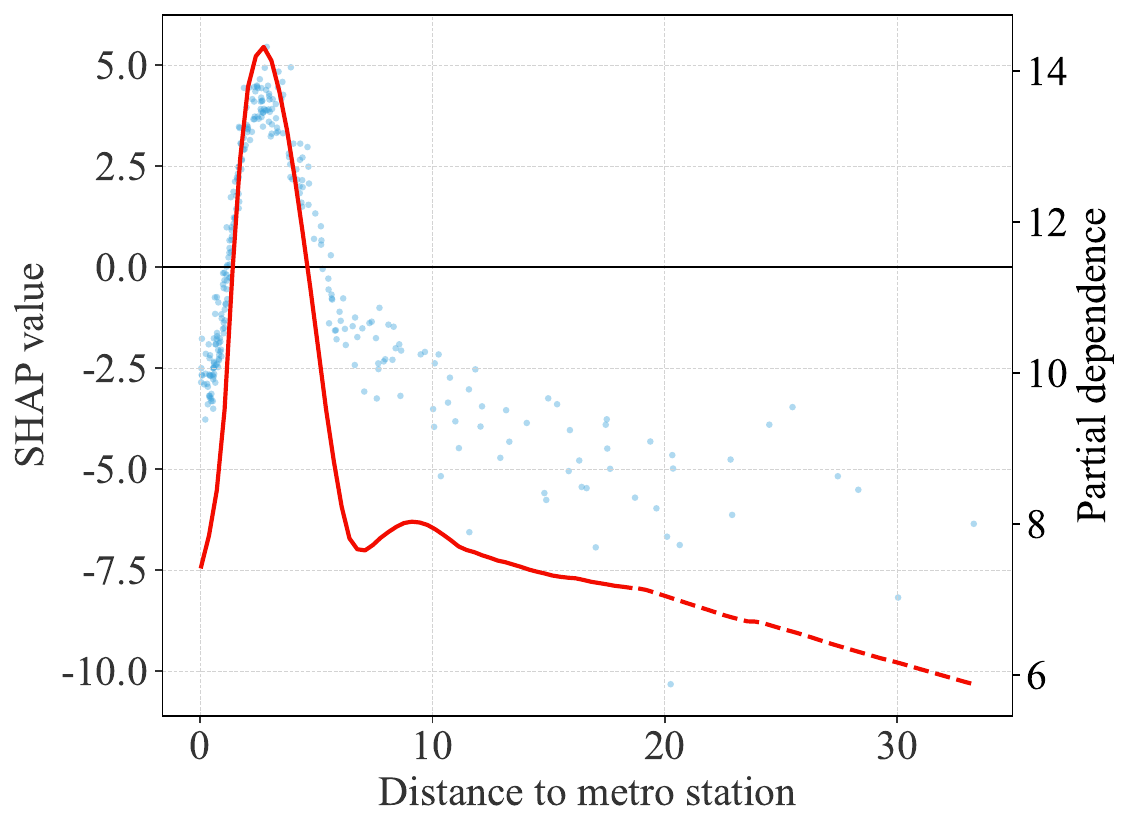}
        \caption{First-mile complementary ratio}
        \label{fig:TabNet_First-mile complement ratio PDP}
    \end{subfigure}
    \begin{subfigure}[h]{0.48\textwidth}
        \centering
        \includegraphics[width=\linewidth,keepaspectratio]{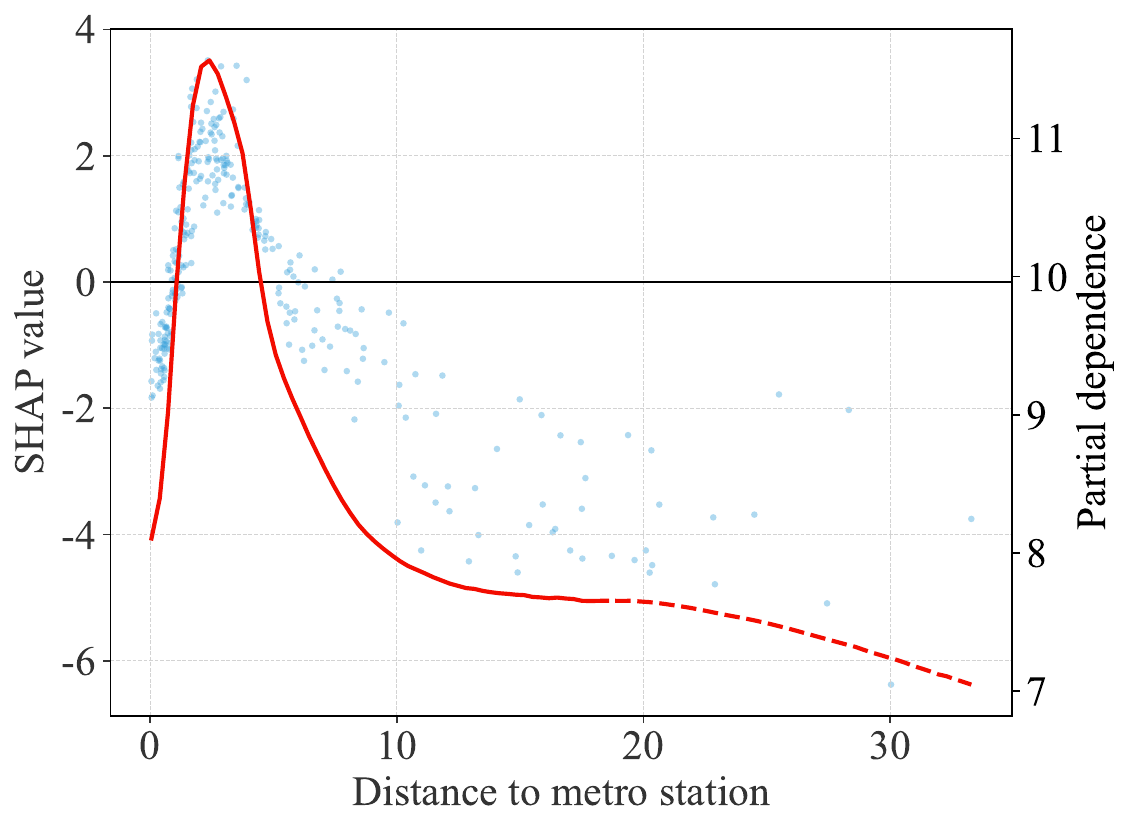}
        \caption{Last-mile complementary ratio}
        \label{fig:TabNet_Last-mile complement ratio PDP}
    \end{subfigure}
    \begin{subfigure}[h]{0.48\textwidth}
        \centering
        \includegraphics[width=\linewidth,keepaspectratio]{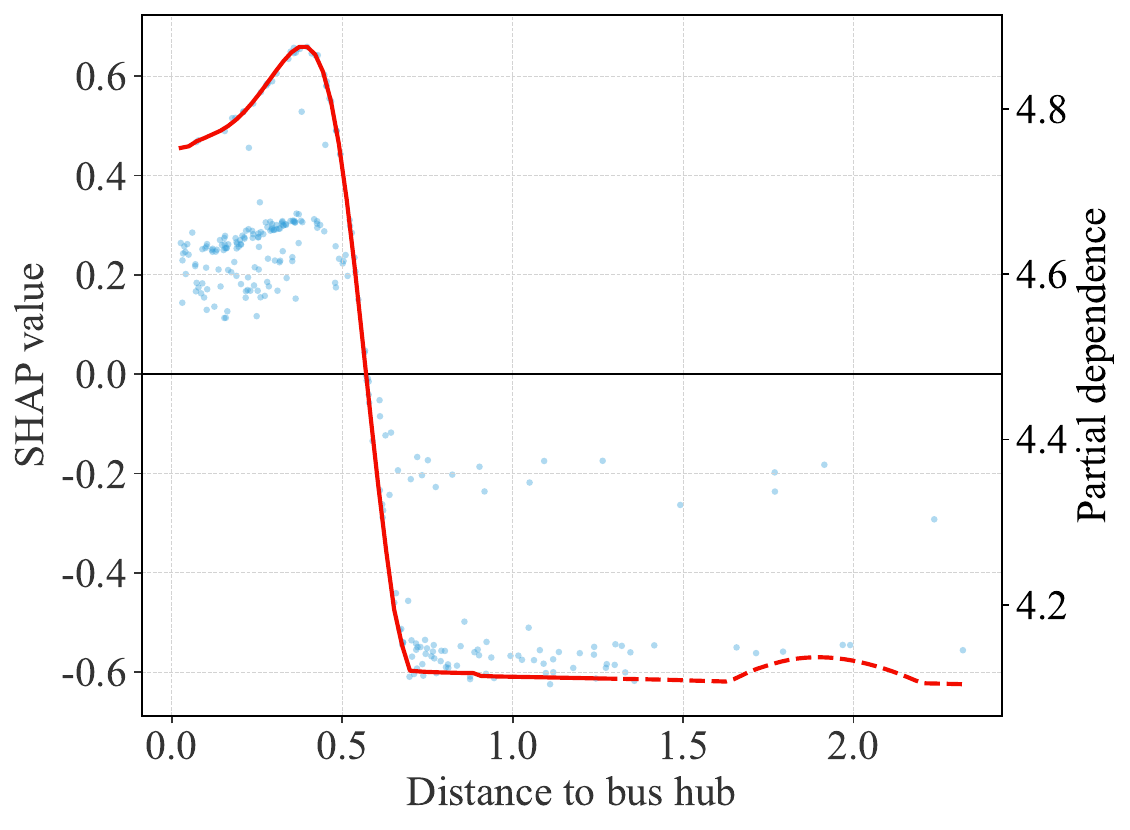}
        \caption{Departure substitutive ratio}
        \label{fig:TabNet_Departure substitution ratio PDP}
    \end{subfigure}
    \begin{subfigure}[h]{0.48\textwidth}
        \centering
        \includegraphics[width=\linewidth,keepaspectratio]{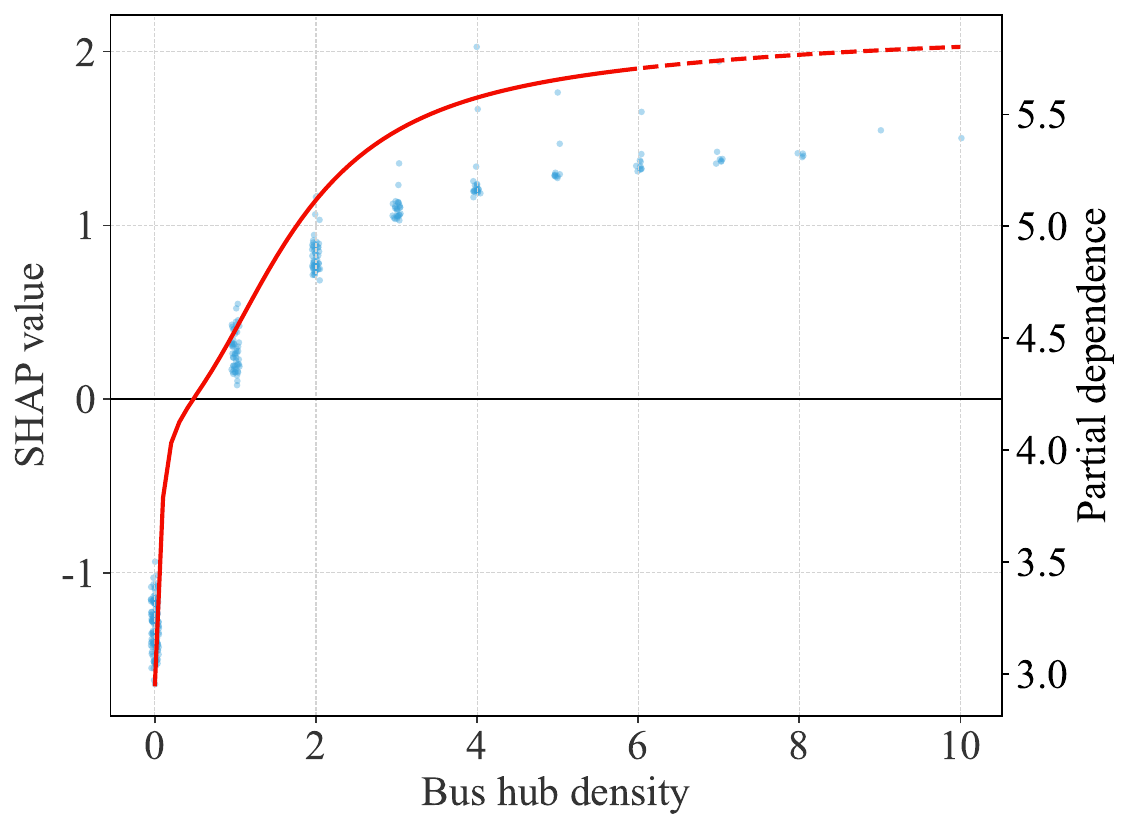}
        \caption{Arrival substitutive ratio}
        \label{fig:TabNet_Arrival substitution ratio PDP}
    \end{subfigure}
\caption{PDPs for top-1 variables under TabNet model.}
\label{fig:PDPs for top-1 variables under TabNet model}
\end{figure}

To further validate the robustness of our results, we implemented several representative deep learning models for tabular data, including TabNet \citep{arik2021tabnet}, FT-Transformer \citep{gorishniy2021revisiting}, and a standard MLP, to re-examine the nonlinear relationships. In terms of predictive accuracy, the benchmarking results in \Tabref{tab:Model Comparison} show that CatBoost still maintains a slight edge in MAE, MSE, and RMSE over the tested deep learning models.
 More importantly, the interpretability results, specifically variable rankings and nonlinear trends, do not materially change.
 Taking TabNet as an example, the SHAP variable importance rankings (shown in \Figref{fig:Variable importance ranking under TabNet model}) reveal that the top five most influential variables remain almost identical to those identified by CatBoost (Fig. 8 of the manuscript), with only minor shifts in relative order.
 Likewise, the PDP patterns of key variables (\Figref{fig:PDPs for top-1 variables under TabNet model}) such as ``Distance to metro station'', ``Distance to bus hub'', and ``Bus hub density'' are nearly identical to those from CatBoost (Fig. 9 of the manuscript).

 These results suggest that the nonlinear relationships identified by CatBoost and SHAP are not model-specific, but robust across multiple deep learning methods for tabular data.

\bibliographystyle{cas-model2-names}

\bibliography{customized_ref.bib, compete_complement.bib}


\bio{}

\endbio

\end{document}